\documentclass[hyper,a4paper]{JHEP3}
\usepackage{latexsym}
\usepackage{graphicx}
\usepackage{amssymb,amsmath}

\def \nqs#1#2{(\ref{#1})--(\ref{#2})}
\def \fig#1{Fig.~\ref{#1}}

\def\slash#1{\setbox0=\hbox{$#1$}#1\hskip-\wd0\dimen0=5pt\advance
       \dimen0 by-\ht0\advance\dimen0 by\dp0\lower0.5\dimen0\hbox
         to\wd0{\hss\sl/\/\hss}}
\newcommand\unitmatrix{ {\mathchoice 
{\rm 1\mskip-4mu l}{\rm 1\mskip-4mu l}{\rm 1\mskip-4.5mu l} {\rm 1\mskip-5mu l}}} 

\newcommand\bsm{\bar{B}_s\to\mu^+\mu^-}

\newcommand\bsg{\bar{B}\to X_s\gamma}
\newcommand\bbb{\bar{B}_s-B_s}
\newcommand\bbbd{\bar{B}_d-B_d}
\newcommand\bsll{\bar{B}\to X_s l^+ l^-}

\newcommand\dll{\delta_{LL}^{d}} \newcommand\drr{\delta_{RR}^{d}}
\newcommand\dlr{\delta_{LR}^{d}} \newcommand\drl{\delta_{RL}^{d}}
\newcommand\ull{\delta_{LL}^{u}}

\newcommand\dxy{\delta_{XY}^{d}}

\newcommand\mdbare{m_{d}^{(0)}} % \newcommand\mdbare{m_{d_0}}
\newcommand\mubare{m_{u}^{(0)}} % \newcommand\mubare{m_{u_0}}
\newcommand\muphys{m_{u}}
\newcommand\mdphys{m_{d}} % \newcommand\mdphys{\overline{m}_{d}}

\newcommand\mbphys{m_{b}} % \newcommand\mbphys{\overline{m}_{b}}
\newcommand\msphys{m_{s}} % \newcommand\msphys{\overline{m}_{s}}
\newcommand\ydint{Y_d^{(0)o}}
\newcommand\yuint{Y_u^{(0)o}}
\newcommand\ylint{Y_l^{(0)o}}
\newcommand\ydbsckm{\hat{Y}_d^{(0)}}

\newcommand\yupsckm{Y_u^{(0)}}
\newcommand\ydpsckm{Y_d^{(0)}}

\newcommand\ybpsckm{Y_b^{(0)}}
\newcommand\yspsckm{Y_s^{(0)}}

\newcommand\GdL{\Gamma_{d\,L}}
\newcommand\GdR{\Gamma_{d\,R}}
\newcommand\GuL{\Gamma_{u\,L}}
\newcommand\GuR{\Gamma_{u\,R}}
\newcommand\Keff{K^{\rm eff}}

\newcommand\mgl{m_{\widetilde{g}}}
\newcommand\msq{m_{\widetilde{q}}}
\newcommand\tanb{\tan\beta} \newcommand\sinb{\sin\beta}
\newcommand\cosb{\cos\beta} \newcommand\cotb{\cot\beta}
\newcommand\delmbs{\Delta M_{B_s}}
\newcommand\delmbd{\Delta M_{B_d}}

\newcommand\EWSym{{\rm SU}(2)_L\times{\rm U}(1)_Y}

\newcommand\xdl{x_{\widetilde{d}_L}}     \newcommand\xdr{x_{\widetilde{d}_R}}
\newcommand\xdrl{x_{\widetilde{d}_{RL}}} \newcommand\xd{x_{\widetilde{d}}}
\newcommand\ydl{y_{\widetilde{d}_L}}     \newcommand\ydr{y_{\widetilde{d}_R}}
\newcommand\ydrl{y_{\widetilde{d}_{RL}}} \newcommand\yd{y_{\widetilde{d}}}

     \newcommand\xur{x_{\widetilde{u}_R}}
 
\newcommand\yul{y_{\widetilde{u}_L}}     \newcommand\yur{y_{\widetilde{u}_R}}

\newcommand\eps{\epsilon}
\newcommand\epst{\epsilon_3}
\newcommand\epsg{\epsilon_s}
\newcommand\epsgp{\epsilon_s^{\prime}}
\newcommand\epsy{\epsilon_Y}
\newcommand\epsj{\epsilon_j}
\newcommand\epsi{\epsilon_i}
\newcommand\epsyII{\epsilon_Y Y_t^2}
\newcommand\BLOfact{1+\epst\tanb}
\newcommand\BLOfacg{1+\epsg\tanb}
\newcommand\BLOfacj{1+\epsj\tanb}
\newcommand\BLOfaci{1+\epsi\tanb}

\newcommand\ps{\,\mbox{ps}}

\newcommand{\newc}{\newcommand}
\newc\eg{{\it {e.g.}}}  \newc\vs{{\it {vs.}}}   \newc\etal{{\it {et al.}}}
\newc\etc{{\it {etc.}}} \newc\ie{{\it {i.e.}}}
\newcommand\tev{\,\mbox{TeV}}
\newcommand\gev{\,\mbox{GeV}}
\newcommand\mev{\,\mbox{MeV}}

\title{Probing the Flavour Structure of Supersymmetry Breaking With Rare B--Processes -- A Beyond Leading Order Analysis}

\author{John Foster\\
Dipartimento di Fisica, Universit\`a di Padova, Via F. Marzolo 8, I--35131, Padua, Italy\\
Department of Physics and Astronomy, University of Sheffield, 
Sheffield S3 7RH, UK\\
E-mail: \email{john.foster@pd.infn.it}}
\author{Ken-ichi Okumura\\
Department of Physics, Kyushu University, Fukuoka 812--8581, Japan\\
Department of Physics, KAIST, Daejeon, 305--701, Korea\\
E--mail: \email{okumura@higgs.phys.kyushu-u.ac.jp}}
\author{Leszek Roszkowski\\
Department of Physics and Astronomy, University of Sheffield, 
Sheffield S3 7RH, UK\\
E--mail: \email{l.roszkowski@sheffield.ac.uk}}

\abstract{
In the framework of minimal supersymmetry with general flavour mixing
in the squark sector we consider dominant beyond leading order (BLO)
effects in the rare processes $\bsg$, $\bsm$ and $\bbb$ mixing. We
present analytic expressions for corrected vertices, which are
applicable in general, and provide a recipe for the inclusion
of the dominant and subdominant BLO effects in existing LO calculations.
We also derive similar expressions in the mass insertion approximation.
We
investigate in more detail the focusing effect pointed out in our
earlier work, which, at large $\tanb$ and $\mu>0$, leads to a reduced
supersymmetric contribution to the above processes.
We also find that, in some cases, flavour dependence, that
accidentally cancels at leading order, can reappear at BLO. 
We further
include electroweak corrections, 
which, while generally subdominant, in some cases may have a
substantial effect.
For example, their contribution to the charged Higgs vertex in $\bsg$
can be of the order of 20\% at BLO. They can also reduce the
contribution of LL insertions to
$\bsm$ and $\bbb$ mixing by up to 20\%, even at the LO. 
We also analyse radiative generation of CKM elements and find the possibility
that the CKM matrix elements $K_{ts}$ and $K_{cb}$ can be generated
entirely by LR insertions. This work constitutes the first complete
analysis of dominant BLO effects in the GFM scenario.
}

\preprint{{\sf KAIST--TH 2005/08}\\ {\sf KYUSHU--HET--83}}

\begin{document}

%%%%%%%%%%%%%%%%%%%%%%%%%%%%%%%%%%%%%%%%%%%%%%%%%%%%%%%%%%%%%%%%%
\section{Introduction}
\label{int}
%%%%%%%%%%%%%%%%%%%%%%%%%%%%%%%%%%%%%%%%%%%%%%%%%%%%%%%%%%%%%%%%%

Flavour physics, in both the leptonic and hadronic sectors, currently
provides one of the best hopes of discovering, or at least
constraining, new physics beyond the Standard Model (SM).  In the hadronic
sector in particular, decays mediated by flavour changing neutral 
currents (FCNC) play an important role as the
Glashow--Iliopoulos--Maiani (GIM) mechanism~\cite{GIM} ensures that both SM and
contributions due to beyond the SM (BSM) physics enter at the
one--loop level. It is therefore possible that such contributions can
be comparable to the SM ones, or even completely dominate the
behaviour of the underlying process.  Once one takes into account the
increasingly accurate experimental data that is being gathered at both
dedicated flavour physics experiments, as well as the B--physics
programmes operating at collider experiments, useful constraints can
often be placed on the parameters and mass scale of a given model of
new physics.
Conversely, for some processes, like $\bsm$, a measurement of a branching
ratio at the Tevatron would immediately indicate a detection of BSM physics.

One of the most compelling extensions of the Standard Model
is the Minimal Supersymmetric Standard Model (MSSM)~\cite{Soft:Rev}. The
non--renormalization theorem of the underlying supersymmetric
theory can explain the stability of scalar potentials in
theories involving two different hierarchies. Additionally, the 
MSSM provides a viable cold dark matter candidate 
(namely the lightest supersymmetric particle),
a natural scheme for gauge coupling unification, and is usually compatible
with the precision electroweak data currently available. 

Softly broken low--energy supersymmetry (SUSY), however, like most new
physics schemes, allows for the possibility that contributions to FCNC
and CP violating processes can exceed SM expectations by orders of
magnitude (the flavour and CP problems). The source of the flavour
problem in the MSSM is primarily due to the arbitrary nature soft
supersymmetry breaking terms~\cite{Soft:Rev}.

The most common
approach to these problems is to assume that the underlying
theory obeys the conditions imposed by minimal flavour violation
(MFV)~\cite{AGIS:bdec}. The definition of MFV, presented in~\cite{AGIS:bdec},
is that flavour violation is determined completely by the
structure of the usual Yukawa couplings. In other words, the
mixings among the down and up squarks are governed by the CKM matrix.
In the MSSM
this restricts the form the soft terms can take (see~\cite{AGIS:bdec}
for the exact expressions). One popular scheme, that respects
MFV, is that the soft terms are universal at some high scale 
associated with supersymmetry (SUSY) breaking, like the grand
unified scale or the Planck scale (a parameterisation
used, for example, in the Constrained MSSM). However, this hypothesis
is not renormalization group invariant. Flavour violating
terms are induced, via running from the
high scale $\Lambda$ to the characteristic mass scale
of the squarks $\mu_{SUSY}$,
that are proportional to $\log(\Lambda^2/\mu_{SUSY}^2)/(4\pi^2)$~\cite{HKS:bdec}.
It should be noted however, that, provided
the theory satisfies MFV up to the scale $\Lambda$ (a rather
strong assumption), all FCNC transitions remain proportional
to the appropriate CKM matrix elements and the resulting low
energy theory still satisfies the MFV hypothesis. However, once
seeds of non--universality are introduced at the high scale
it is possible that they can become amplified by running.

This provides motivation to generalise to 
a broader framework, namely general flavour mixing (GFM) in the
sfermion sector. In general, the flavour structure of the
soft terms is not protected by any
symmetry and can be rather arbitrary.  
One simple example is that a degree of non--universality
can be allowed 
in the squark soft terms (beyond that allowed by
MFV). In this case additional effects are possible that are proportional,
essentially, to the degree of splitting between the entries for
each generation. 

Deviations from MFV can easily appear in a variety of SUSY models.
In theories with SUSY breaking mediated by supergravity, for example,
it is possible to induce a wide range of flavour violating effects~\cite{CLP:sugra}
once one proceeds beyond the simplest minimal SUGRA scheme~\cite{Soft:Rev}.
Grand unified theories involving right handed neutrinos, like
the minimal ${\rm SO}(10)$ models with a specific family structure,
often lead to additional sources of flavour violation
due to the interactions that exist, at the unification
scale, between right--handed down squarks and
neutrinos~\cite{BGOO1:drr,Moroi1:drr,Moroi2:drr,BGOO2:drr,CMM:drr}.
Experimental limits and results are therefore especially helpful
when restricting the possible mixings between the various
generations and constraining these models. 

In this paper we shall concern ourselves chiefly with
flavour violation between the second and third generations.
FCNC processes involving such transitions have been studied
in detail in the context of the SM and the short--distance
contributions to a wide variety of processes have typically
been calculated to NLO in the SM (the evaluation of long--distance
effects is another matter). In the case of $\bsg$ these efforts
have resulted in a very good agreement between SM calculations
and experimental results with relatively little room left
for new physics. When placing constraints on a given model
it is useful to have a calculation that is of a similar
accuracy to the SM contribution. In the MSSM complete
NLO calculations, however, are rather complicated as
additional two--loop diagrams involving gluinos need to be evaluated.
It is, however, possible to include the effects that are
large once one proceeds beyond the LO (BLO). Such effects are
typically classified as being proportional to either $\tanb$
or large logs. Such BLO analyses have been performed in
MFV~\cite{DGG:bsg,CGNW:bsg,AGIS:bdec,DP:bdec,BCRS:bdec}
and, more recently, in GFM~\cite{OR:bsg,OR2:bsg,FOR1:bdec}.
In GFM, in particular, a focusing effect 
was found in~\cite{OR:bsg} that gave rise to significant shifts
in the allowed regions of parameter space compared to a
LO analysis. (A similar effect appears also in the MFV scheme but is
much weaker~\cite{OR:bsg,OR2:bsg}.) Basically, in many
cases of phenomenological interest (\eg, large $\tanb$ and $\mu>0$),
SUSY contributions to $\bsg$ are significantly reduced compared to the
LO approximation. A similar effect was also found in the decay $\bsm$
and $\bbb$ mixing~\cite{FOR1:bdec}. 

The aim of this paper is to present the first complete analysis of 
dominant BLO effects in general flavour mixing in the case of three
processes. The first, $\bsg$, has been discussed previously
in~\cite{OR:bsg,OR2:bsg}. However, here we shall include the additional
corrections that arise when one includes charginos
and neutralinos in the resummation procedure discussed in~\cite{OR2:bsg}. In
particular, we include contributions arising from higgsino exchange,
that are proportional to the Yukawa couplings of the third generation,
and the additional contributions to the charged Higgs vertex that were
discussed in the context of MFV in~\cite{BCRS:bdec}. 

The other two processes we shall consider are the decay $\bsm$
and $\bbb$ mixing. These processes have not been observed yet but
have come under a lot
of theoretical scrutiny lately due to the large contributions
possible in the large $\tanb$ regime. In this paper we discuss
the GFM contributions to both processes in detail, highlighting
the effects that appear once one proceeds beyond the LO. 

In all the three cases, we shall present the analytic 
expressions required to implement BLO corrections in the GFM scenario
for possibly large deviations from the MFV
scheme. However, since these general expressions are often rather
complicated, we shall also derive expressions in the the mass
insertion approximation (MIA), allowing the BLO effects to be shown
explicitly. In both cases we will provide an explicit recipe for
including the BLO effects into the existing LO expressions.
Whilst we shall not include such effects
in the forthcoming analysis, the formalism we shall present
should, with relatively little modification, be applicable
to the CP violating case.

The paper is organised as follows. In section~\ref{GA} we
summarise the formalism employed in this paper, giving
complete expressions for all the corrected masses and
vertices used in our calculation. In section~\ref{MIA}
we present analytic expressions for these masses and
vertices in the MIA. In section~\ref{bsg} we discuss
the decay $\bsg$ providing analytic expressions for the
BLO corrections to supersymmetric and electroweak contributions
in the MIA. In sections~\ref{bsm} and~\ref{bbb} we perform
a similar analyses for the decay $\bsm$ and $\bbb$ mixing,
respectively. Finally, in section~\ref{NRes} we present
our numerical analysis.

%%%%%%%%%%%%%%%%%%%%%%%%%%%%%%%%%%%%%%%%%%%%%%%%%%%%%%%%%%%%%%%%%
\section{Beyond Leading Order Effects and General Flavour Mixing}
\label{GA}
%%%%%%%%%%%%%%%%%%%%%%%%%%%%%%%%%%%%%%%%%%%%%%%%%%%%%%%%%%%%%%%%%

The influence of $\tanb$ enhanced effects on the down quark masses, the
charged Higgs and neutral Higgs vertex are known to be large. It is
therefore essential, especially when working in the large $\tanb$ regime,
that such contributions are taken into account (and resummed
if necessary).

In this section, we shall follow the method first developed 
in~\cite{OR:bsg,OR2:bsg} and generalise it to include the
additional effects that appear once the contributions of chargino and
neutralino loops are taken into account. It should be noted
that the analysis below encompasses both MFV and the GFM scenario
and can be easily extended to include, for example, CP violation or
flavour violation in the leptonic sector.

%%%%%%%%%%%%%%%%%%%%%%%%%%%%%%%%%%%%%%%%%%%%%%%%%%%%%%%%%%%%%%%%%
\subsection{The Framework}
\label{GA:FR}
%%%%%%%%%%%%%%%%%%%%%%%%%%%%%%%%%%%%%%%%%%%%%%%%%%%%%%%%%%%%%%%%%

Once the supersymmetric particles have been integrated out, the
effective Lagrangian describing the quark mass terms, at some
scale $\mu<M_{SUSY}$, in the physical super--CKM basis (SCKM) is
given by
\begin{align}
-\mathcal{L}_{q}^{\rm mass}=
\bar{d}_R\left(\mdbare+\delta m_d\right)d_L
+\bar{u}_R\left(\mubare+\delta m_u\right)u_L+h.c.,
\label{GA:masst}
\end{align}
where $d_{L,R}$ and $u_{L,R}$ denote the down and up components
of the left and right quark fields, respectively.\footnote{It should
be noted that, throughout this section, we shall
adopt matrix notation and suppress flavour indices unless
otherwise specified.} In the physical SCKM basis
the quark mass matrices are, by definition, diagonal and it is possible
to make the identifications
\begin{align}
\mdphys=&\mdbare+\delta m_d={\rm diag}\left(m_d,m_s,m_b\right),\\
\muphys=&\mubare+\delta m_u={\rm diag}\left(m_u,m_c,m_t\right),
\label{GA:mdphys}
\end{align}
where $m_{d,s,b}$ and $m_{u,c,t}$ denote the physical masses
of the down and up--type quarks respecitively.
The bare mass matrix $\mdbare$ is related to the $3\times 3$
Yukawa couplings $Y_{d,u}^{\left(0\right)}$ derived from the
superpotential in the usual manner,
\begin{align}
m_{d,u}^{(0)}=v_{d,u} Y_{d,u}^{(0)}.
\label{GA:mdbare1}
\end{align}
where $v_{d,u}=\langle H^0_{d,u}\rangle$. Finally,
$\delta m_d$ and $\delta m_u$ denote the radiative corrections
to the quark masses induced by integrating out the SUSY
particles~\cite{mass:ref,HRS:bdec,COPW:bdec,BRP:CKM}.
The corrections have the form\footnote{As we allow the
inclusion of electroweak effects we will
not assume proportionality to the strong coupling
constant here, unlike in~\cite{OR:bsg,OR2:bsg}.}
\begin{align}
\delta m_d=\Sigma^{d}_{m\,L}+\frac{1}{2}\Sigma_{v\,R}^{d}\mdbare
+\frac{1}{2}\mdbare\Sigma_{v\,L}^{d}.
\label{GA:deltamd}
\end{align}
$\delta m_u$ is given by a similar formula after one performs
the substitution $d\to u$. The $3\times 3$ hermitian matrices
$\Sigma_{v\,L,R}^{d,u}$ and the $3\times 3$ complex matrices
$\Sigma_{m\,L}^{d,u}$ denote the contributions arising from
wavefunction and mass corrections due to two point diagrams
involving gluinos, charginos, neutralinos
and squarks. (Full expressions will be given later in the text.)

Before discussing how the radiative corrections $\delta m_d$ are
calculated, it will be useful to consider the
transformation from the interaction basis to the physical SCKM
basis. In the interaction basis, the MSSM superpotential is
\begin{equation}
W_F=
-\mu \hat{H}_d \hat{H}_u+\ylint \hat{H}_d \hat{L}^o \hat{E}^o
+\ydint \hat{H}_d \hat{Q}^o \hat{D}^o-\yuint \hat{H}_u \hat{Q}^o \hat{U}^o,
\label{GA:sup}
\end{equation}
$\hat{Q}^o$ and $\hat{L}^o$ are the quark and lepton
${\rm SU}\left(2\right)$ doublet superfields, $\hat{D}^o$,
$\hat{U}^o$ and $\hat{E}^o$ denote the singlet superfields
and $\hat{H}_u$ and $\hat{H}_d$ are the two Higgs doublets
that appear in the MSSM (for more details see,
for example,~\cite{Soft:Rev}), while $Y_{l,d,u}^o$
are the appropriate $3\times 3$ Yukawa mass matrices in
that basis.

The soft SUSY breaking terms are also usually introduced
in the interaction basis. The Lagrangian for the bilinear
soft SUSY breaking terms is given by
\begin{align}
-\mathcal{L}_{\widetilde{q}\, {\rm soft}}^{o\,{\rm mass}}=&
+\tilde{d}_L^{o\dag}m_{Q}^2\tilde{d}_L^{o}
+\tilde{d}_R^{o\dag}m_{D}^2\tilde{d}_R^{o}
+\left[\tilde{d}_L^{o\dag}\left(v_d A_d^{\ast}\right)\tilde{d}_R^{o}
+h.c.\right]
\nonumber\\
&+\tilde{u}_L^{o\dag}m_{Q}^2\tilde{u}_L^{o}
+\tilde{u}_R^{o\dag}m_{U}^2\tilde{u}_R^{o}
+\left[\tilde{u}_L^{o\dag}\left(v_u A_u^{\ast}\right)\tilde{u}_R^{o}
+h.c.\right],
\label{GA:stl}
\end{align}
where $m_{Q}^2$, $m_{D}^2$ and $m_{U}^2$ are, in general, arbitrary
$3\times 3$ hermitian matrices. The trilinear terms $A_d$ and $A_u$,
on the other hand, are arbitrary $3\times 3$ complex matrices. We
have not assumed that the trilinear soft terms are
proportional to the appropriate Yukawa coupling. (We discuss an
alternative parameterisation, that can be used in the GFM scenario,
in appendix~\ref{dlrapp}.)

Transforming the quark fields from the interaction basis to
the physical SCKM basis involves performing unitary transformations
on both the left and right handed fields such that
\begin{align}
d_R&= V_{d_R}d_R^o,
&d_L&= V_{d_L}d_L^o,
\label{GA:qtransd}
\\
u_R&= V_{u_R}d_R^o,
&u_L&= V_{u_L}d_L^o.
\label{GA:qtransu}
\end{align}
The bare mass matrix is then related to the Yukawa couplings
defined in the interaction basis via the relation
\begin{align}
\mdbare=&V_{d_R} v_d \ydint V_{d_L}^{\dagger},
&
\mubare=&V_{u_R} v_u \yuint V_{u_L}^{\dagger},
\label{GA:mdbare2}
\end{align}
$\mdbare$ and $\mubare$ appear in all quantities derived from the
superpotential~\eqref{GA:sup} not subject to the corrections
\eqref{GA:deltamd}, such as the couplings of supersymmetric
particles. The CKM matrix $K$ is related to the transformations
\nqs{GA:qtransd}{GA:qtransu} in the usual manner
\begin{align}
K=V_{u_L}V_{d_L}^{\dag}.
\label{GA:CKM}
\end{align}

As the radiative corrections $\delta m_{d,u}$ are calculated in the
SCKM basis, it is necessary to consider how the transformations
\nqs{GA:qtransd}{GA:qtransu}, when performed on the squark fields,
affect the relevant mass matrices. After transforming to the physical
SCKM basis, the soft terms become
\begin{align}
m_{d,LL}^2=&V_{D_L} m_{Q}^2V_{D_L}^{\dag},
&m_{d,RR}^2=&V_{D_R} m_{D}^2 V_{D_R}^{\dag},
&m_{d,LR}^2=&V_{D_L}\left(v_d A_d^{\ast}\right)V_{D_R}^{\dag},
\nonumber\\
m_{u,LL}^2=&V_{U_L} m_{Q}^2 V_{U_L}^{\dag},
&m_{u,RR}^2=&V_{U_R} m_{U}^2 V_{U_R}^{\dag},
&m_{u,LR}^2=&V_{U_L}\left(v_u A_u^{\ast}\right)V_{U_R}^{\dag}.
\label{GA:SCKMst}
\end{align}
The $6\times 6$ down squark mass matrix $\mathcal{M}_{\tilde{d}}^2$
may then be written in the following manner
\begin{align}
\mathcal{M}_{\tilde{d}}^2=
\left(
\begin{array}{ll}
m_{d,LL}^2+F_{d,LL}+D_{d,LL}~~&
m_{d,LR}^2+F_{d,LR}\\
\left(m_{d,LR}^2+F_{d,LR}\right)^\dag &  
m_{d,RR}^2+F_{d,RR}+D_{d,RR}  
\end{array}
\right).
\label{GA:sqmm}
\end{align}
(The up squark mass matrix may be similarly defined by substituting
$d$ with $u$.) The $F$--terms that appear in~\eqref{GA:sqmm} are
given by
\begin{align}
F_{d,LL}&=\mdbare{}^{\dagger}\mdbare,
& F_{d,RR}&=\mdbare\mdbare{}^{\dagger},
& F_{d,LR}&=-\mu\tanb\mdbare{}^{\dagger},
\label{GA:Fterms}
\end{align}
and the flavour diagonal $D$--terms are
\begin{align}
D_{d,LL}&=m_Z\cos 2\beta\left(T_{3d}-Q_d \sin^2\theta_W\right)\unitmatrix_3,
&D_{d,RR}=m_Z\cos 2\beta Q_d \sin^2\theta_W\unitmatrix_3.
\label{GA:Dterms}
\end{align}
It should be noted that, in the physical SCKM basis, the $F$--terms
are, in general, not necessarily flavour diagonal as they are derived from the
superpotential and are therefore functions of the bare mass matrix $\mdbare$.

To obtain the physical squark masses $m_{\widetilde{d}_I}$
it is necessary to perform an additional unitary transformation
on the squark fields such that
\begin{align}
\Gamma_{d}\mathcal{M}_{\tilde{d}}^2\Gamma_{d}^{\dag}=
{\rm diag}\left(m_{\tilde{d}_1},\dots,m_{\tilde{d}_6}\right).
\label{GA:GammaDef}
\end{align}
It is conventional to decompose the original $6\times 6$ unitary
matrix $\Gamma_{d}$ into two $6\times 3$ submatrices $\GdL$ and $\GdR$:
\begin{align}
\left(\Gamma_{d}\right)_{Ii}=&\left(\GdL\right)_{Ii},
&\left(\Gamma_{d}\right)_{Ii+3}=&\left(\GdR\right)_{Ii}.
\label{GA:GammaDecomp}
\end{align}
where $I=1,\dots,6$ and $i=1,2,3$.

Departures from the MFV scenario are often parameterised in
terms of the dimensionless quantities
\begin{align}
\left(\dll\right)_{ij}=
&\frac{\left(m_{d,LL}^2\right)_{ij}}
{\sqrt{\left(m_{d,LL}^{2}\right)_{ii}\left(m^{2}_{d,LL}\right)_{jj}}},
&\left(\dlr\right)_{ij}=
&\frac{\left(m_{d,LR}^2\right)_{ij}}
{\sqrt{\left(m_{d,LL}^{2}\right)_{ii}\left(m^{2}_{d,RR}\right)_{jj}}},
\label{GA:dels1}
\\
\left(\drl\right)_{ij}=&
\frac{\left(m_{d,RL}^2\right)_{ij}}
{\sqrt{\left(m_{d,RR}^{2}\right)_{ii}\left(m^{2}_{d,LL}\right)_{jj}}},
&\left(\drr\right)_{ij}=
&\frac{\left(m_{d,RR}^2\right)_{ij}}
{\sqrt{\left(m_{d,RR}^{2}\right)_{ii}\left(m^{2}_{d,RR}\right)_{jj}}}.
\label{GA:dels2}
\end{align}
The soft terms $m_{d,XY}^2$ ($X,Y=L,R$) are given
in~\eqref{GA:SCKMst} and $i,j=1,2,3$. Similar definitions
apply for the up quarks. It should be noted that, since
$m^2_{u,LL}$ and $m^2_{d,LL}$ are related to one another
by ${\rm SU}(2)$ invariance, we have the relation
\begin{equation}
\dll=K^{\dag}\ull K
\label{GA:dllull}
\end{equation}

Let us briefly comment on the basis dependence of these
definitions of $\delta^d_{XY}$. Physical quantities such as
cross--sections and branching ratios are naturally independent
of the basis in which one defines the soft terms.
The basis in which one defines the insertions $\delta^d_{XY}$,
however, is essentially a matter of convenience.
As discussed above, we work in the physical
SCKM basis throughout this analysis and as such the
definition~\nqs{GA:dels1}{GA:dels2} is essentially
the easiest to implement numerically.
Other definitions of $\dxy$ have been used in the literature
before, for example, one might define $\dxy$ in the uncorrected
(bare) SCKM basis where the Yukawa matrices derived from the
superpotential are diagonal (we shall define this basis more formally in
subsection~\ref{BLOeff:OM}).
Transforming between different
bases involves performing additional unitary transformations on the
soft terms~\eqref{GA:SCKMst} and, unless large non--universalities
exist, the differences between the transformed and the
original $\dxy$ are typically rather small.
Below we will
derive many expressions in the MIA where one usually assumes that
the diagonal entries of the soft terms are universal.
In light of the above we expect them
to be applicable to alternative definitions of $\dxy$.
During
our numerical analysis we employ a similar definition for the 
SUSY soft terms to ensure that our formalism remains applicable
to as wide a variety of models as possible.

After defining our framework, let us now move on to the effects
that these corrections have on the electroweak and supersymmetric
vertices.

%%%%%%%%%%%%%%%%%%%%%%%%%%%%%%%%%%%%%%%%%%%%%%%%%%%%%%%%%%%%%%%%%
\subsection{Corrections to Electroweak Vertices}
\label{GA:EW}
%%%%%%%%%%%%%%%%%%%%%%%%%%%%%%%%%%%%%%%%%%%%%%%%%%%%%%%%%%%%%%%%%

Integrating out the supersymmetric particles, coupled with the effect of
transforming between the interaction and the physical SCKM bases, can 
affect the form of the electroweak (\ie~the Higgs and gauge boson)
vertices present in the resulting effective theory.

After transforming to the physical SCKM basis, the W boson vertex has the
following form
\begin{equation}
\mathcal{L}_W=
\bar{u}_L \gamma^{\mu} C_L^W d_L W^+_{\mu}
+\bar{u}_R \gamma^{\mu} C_R^W d_R W^+_{\mu}+h.c.
\label{GA:EW:WVerL}
\end{equation}
The $3\times 3$ coupling matrices $C_L^W$ and $C_R^W$, are given by
\begin{align}
C_L^W&=-\frac{g_2}{\sqrt{2}}\left(K+\frac{1}{2}\Sigma_{v\,L}^{u}K
+\frac{1}{2}K\Sigma_{v\,L}^{d}\right)
+\Delta C_L^W=-\frac{g_2}{\sqrt{2}}\Keff,
\\
C_R^W&=\Delta C_R^W.
\label{GA:EW:WVer}
\end{align}
We employ the notation $\Delta C_{L,R}^X$ to denote the vertex
corrections that arise from three point diagrams when one integrates
out the SUSY particles. Identifying the left handed coupling of
the $W$ boson with the physical CKM matrix $\Keff$, that is measured
from experiment, we have the relation
\begin{align}
\Keff=K+\frac{1}{2}\Sigma_{v\,L}^{u}K
+\frac{1}{2}K\Sigma_{v\,L}^{d}
-\frac{\sqrt{2}}{g_2}\Delta C_L^W.
\label{GA:EW:KM}
\end{align}
The uncorrected CKM matrix $K$ is defined in~\eqref{GA:CKM} and
appears in all vertices not subject to the
corrections~\eqref{GA:EW:KM}.

Now consider the coupling of the $Z$ boson with down quarks
\begin{equation}
\mathcal{L}_{Z^0}=
\bar{d}_L\gamma^{\mu}C_L^Z d_L Z^0_{\mu}
+\bar{d}_R\gamma^{\mu}C_R^Zd_R Z^0_{\mu}.
\label{GA:EW:ZVerL}
\end{equation}
The $3\times 3$ coupling matrices $C_L^Z$ and $C_R^Z$ are given by
\begin{align}
C_L^Z&=\frac{g_2}{2\cos\theta_W}\left(1-\frac{2}{3}\sin^2\theta_W\right)
\left(1+\Sigma_{v\,L}^{d}\right)+\Delta C_L^Z,
\\
C_R^Z&=-\frac{g_2}{2\cos\theta_W}\frac{2}{3}\sin^2\theta_W
\left(1+\Sigma_{v\,R}^{d}\right)+\Delta C_R^Z.
\label{GA:EW:ZVer}
\end{align}
The radiative corrections to $C_L^Z$ and $C_R^Z$ can induce
off--diagonal elements to the coupling that lead to
additional sources of FCNC.

Turning to the Higgs sector, the inclusion of radiative
corrections is especially important. As the coupling between
the Higgs sector and squarks features a dependence on the
soft SUSY breaking terms (rather than only gauge interactions),
the corrected vertices that arise when one integrates out the
coloured SUSY particles can display a non--decoupling property.
Large corrections to the vertices are therefore
feasible for even \tev~scale sparticle masses.

Once one has integrated out the SUSY particles, the charged Higgs
interaction becomes
\begin{equation}
\mathcal{L}_{S^+}=
\bar{u}_R C^{S^+}_L d_L S^++\bar{u}_L C^{S^+}_R d_R S^+
+h.c.,
\label{GA:EW:CVerL}
\end{equation}
where $S^+=H^+,G^+$ and the $3\times 3$ matrices
coupling $C^{S^+}_{L,R}$ are given by
\begin{align}
C^{S^+}_{L}&=\frac{g_2}{\sqrt{2}m_W\sinb}y_{(1)}^{S^+}
\left(\muphys K-\Sigma^{u}_{m\,L} K
-\frac{1}{2}\mubare\Sigma^{u}_{v\,L} K
+\frac{1}{2}\mubare K\Sigma^{d}_{v\,L}\right)
+\Delta C_L^{S^+},\\
C^{S^+}_{R}&=\frac{g_2}{\sqrt{2}m_W\cosb}y_{(2)}^{S^+}
\left(K\mdphys-K\Sigma^{d\dag}_{m\,L}
-\frac{1}{2}K\Sigma^{d}_{v\,L}\mdbare{}^{\dag}
+\frac{1}{2}\Sigma^{u}_{v\,L}K\mdbare{}^{\dag}\right)
+\Delta C_R^{S^+},
\label{GA:EW:CVer}
\end{align}
where $y_{(1)}^{S^+}=\cosb,\sinb$ and
$y_{(2)}^{S^+}=\sinb,-\cosb$.

The neutral Higgs and Goldstone boson interact with the down quarks
in the following way
\begin{align}
\mathcal{L}_{S^0}=
\bar{d}_R C_L^{S^0} d_L S^0+\bar{d}_L C_R^{S^0} d_R S^0,
\label{GA:EW:HVerL}
\end{align}
where $S^0=H^0,h^0,A^0,G^0$ and the effective
vertices $C_R^{S^0}$ and $C_R^{S^0}$ may be written in terms
of the $3\times 3$ matrices
\begin{align}
C_L^{S^0}&=-\frac{g_2}{2m_W\cosb}x_{(1)}^{S^0}\left(\mdphys
-\Sigma_{m\,L}^{d}\right)
+\Delta C_L^{S^0},
\nonumber\\
C_R^{S^0}&=-\frac{g_2}{2m_W\cosb}x_{(1)}^{S^0\ast}\left(\mdphys
-\Sigma_{m\,L}^{d\dag}\right)
+\Delta C_R^{S^0}
\label{GA:EW:HVer}
\end{align}
and $x_{(1)}^{S^0}=\cos\alpha,-\sin\alpha,i\sinb,-i\cosb$.

In the limit where the physical SCKM basis is identical to the bare
SCKM basis (\ie~where $\mdphys=\mdbare$),~\eqref{GA:EW:HVer} is
identical to the diagrammatic result derived in the on--shell formalism used
in~\cite{BEKU1:bsm,BEKU2:bsm}.

%%%%%%%%%%%%%%%%%%%%%%%%%%%%%%%%%%%%%%%%%%%%%%%%%%%%%%%%%%%%%%%%%
\subsection{Corrections to Supersymmetric Vertices}
\label{GA:SU}
%%%%%%%%%%%%%%%%%%%%%%%%%%%%%%%%%%%%%%%%%%%%%%%%%%%%%%%%%%%%%%%%%

As the corrections to the electroweak vertices are calculated
in the physical SCKM basis, it is necessary to discuss
how the supersymmetric interactions are altered once transformed
into this basis. Ignoring the effects of wavefunction
renormalizations,\footnote{We do, however, include these
contributions in our numerical analysis.} that are not enhanced by
$\tanb$, the changes introduced by
transforming to the physical SCKM basis typically amount to the
introduction of $\mdbare$ and $K$ into the various vertices.
For instance, after these replacements have been performed, the
chargino vertex becomes\footnote{Our notation for the supersymmetric
vertices differs slightly from that used in~\cite{OR2:bsg}, broadly speaking,
one may convert between the two by making the substitution $L\leftrightarrow R$. }
\begin{align}
\mathcal{L}_{\chi^{\pm}}=
&\;\sum_{a,i,J}\tilde{u}_J^{\dagger}\left(\bar{\chi}^{-}\right)_{a}
\left[\left(C_{d\,L}\right)_{aJi} P_L
+\left(C_{d\,R}\right)_{aJi} P_R\right]\left(d\right)_i,
\label{GA:SU:Char}
\end{align}
where $a=1,2$, $i=1,2,3$, and $J=1,\dots,6$ and the couplings $C_{d\,L}$ and
$C_{d\,R}$ may be written in terms of the matrices
\begin{align}
\left(C_{d\,L}\right)_{aJi}=&-g_2 V_{a1}^{\ast}\left(\GuL K\right)_{Ji}
+\frac{g_2}{\sqrt{2}m_W\sinb}V_{a2}^{\ast}\left(\GuR\mubare K\right)_{Ji},
\label{GA:SU:CdL}
\\
\left(C_{d\,R}\right)_{aJi}=&
\frac{g_2}{\sqrt{2}m_W\cosb}U_{a2}^{\ast}\left(\GuL K\mdbare{}^{\dagger}\right)_{Ji},
\label{GA:SU:CdR}
\end{align}
where $K$ is defined in terms of the physical CKM matrix by $\Keff$
in \eqref{GA:EW:KM} and the bare masses are given by~\eqref{GA:mdphys}
and its
analogue for the up quarks. The matrices $U$ and $V$ diagonalise the chargino
mass matrix $\mathcal{M}_{\chi^{\pm}}$ such that
\begin{align}
U \mathcal{M}_{\chi^{\pm}} V^{\dag}=
{\rm diag} \left(m_{\chi_1^{\pm}},m_{\chi_2^{\pm}}\right).
\nonumber
\end{align}

The appearance of the bare quark mass matrix $\mdbare$ in these vertices
can lead to large effects in both MFV and GFM models. A full list of
vertices relevant to our calculation appear in Appendix~\ref{SUSYver}.

%%%%%%%%%%%%%%%%%%%%%%%%%%%%%%%%%%%%%%%%%%%%%%%%%%%%%%%%%%%%%%%%%
\subsection{Numerical Aspects}
\label{num}
%%%%%%%%%%%%%%%%%%%%%%%%%%%%%%%%%%%%%%%%%%%%%%%%%%%%%%%%%%%%%%%%%

Let us now discuss how the method discussed above should be
implemented numerically. As we will be investigating values
of up to $\mathcal{O}(1)$ for the flavour violating
parameters $\dxy$ ($X,Y=L,R$)~\nqs{GA:dels1}{GA:dels2}, it is
important to devise a method such that
the effects discussed in section~\ref{GA} are taken into account,
whilst also retaining the numerical accuracy associated with
working in the squark mass basis. Such an
iterative method was proposed in~\cite{OR2:bsg} and it will be useful
for our purposes to briefly summarise it here.

Once the unitary transformations~\eqref{GA:GammaDef} have been
performed on the squark fields the gluino contribution to
$\delta m_d$ becomes~\cite{OR2:bsg}
\begin{align}
\delta m_d=\frac{\alpha_s}{2\pi}\;C_2\left(3\right)\sum_{I=1}^{6}
\left(\GdR^{\ast}\right)_{Ii}\left(\GdL\right)_{Ij}\mgl\,
B_0\left(\mgl^2,m_{\widetilde{d}_I}^2\right).
\nonumber
\end{align}
The Passarino--Veltman function $B_0$ can be found in appendix~\ref{PVfunc}.
Using this relation it is possible to calculate the bare mass matrix
using~\eqref{GA:mdphys}. It should be noted, however,
that $\delta m_d$ contains a dependence on $\mdbare$
as it appears in the squark mass matrix through the $F$--terms
\eqref{GA:Fterms}. It is therefore necessary to
employ an iterative procedure such that $\mdbare$ and the mixing
matrices $\Gamma_{dL,R}$ are determined to the desired level of accuracy.
The inclusion of the effects induced by chargino and neutralino
contributions introduces a dependence on $\mubare$ and the
uncorrected CKM matrix $K$ in the formula for $\delta m_d$. One
must therefore expand and generalise the iterative
procedure presented in~\cite{OR2:bsg} such that these effects
are included as well.

In the first step of the procedure, $\mdbare$, $\mubare$ and $K$ are
set equal to the input parameters $m_d$, $m_u$ and $\Keff$, respectively,
and $\delta m_{d,u}$ is set equal to zero. In the second step, the squark
mass matrices and the supersymmetric couplings are then calculated
with these input values, allowing the evaluation of the radiative
corrections $\Sigma^{d}$, $\Sigma^{u}$ and $\Delta C_L^W$ using
the formulae presented in appendix~\ref{RadCor}. In the third step,
$\delta m_{d,u}$, the bare mass matrices and $K$ are determined
using~\eqref{GA:mdphys} and~\eqref{GA:EW:KM}. The resulting
values are then used as the new input parameters in step two. 
The second and third steps are then repeated until convergence
occurs. The iterative procedure converges rather rapidly
and after $n$ iterations amounts to including the first $n$ terms
that arise in a Taylor expansion in $\tanb$. For an
example of the procedure applied to MFV, see~\cite{OR2:bsg}.
With the final forms of the supersymmetric couplings, uncorrected
CKM matrix and bare mass matrices determined, the corrections to the
$Z$ boson, charged Higgs and neutral Higgs vertices may be calculated
using the formulae presented in section~\ref{GA:EW} and
appendix~\ref{RadCor}. We should emphasise here that we work in
the squark mass basis throughout and therefore include all the
effects that can occur at higher orders in the MIA as well as
the BLO effects described in the previous subsections. In addition,
we include the effects induced by additional electroweak
contributions, light quark effects and $\EWSym$ breaking.

%%%%%%%%%%%%%%%%%%%%%%%%%%%%%%%%%%%%%%%%%%%%%%%%%%%%%%%%%%%%%%%%%
\section{The Mass Insertion Approximation}
\label{MIA}
%%%%%%%%%%%%%%%%%%%%%%%%%%%%%%%%%%%%%%%%%%%%%%%%%%%%%%%%%%%%%%%%%

Expressions for $\mdbare$ and the corrected vertices are well
known in MFV models~\cite{CGNW:Htver,DGG:bsg,BCRS:bdec} and it
will be useful for our purposes to extend these results to the
GFM scenario. The
flavour dependence of analytic expressions
is often rather obscure when flavour violation is communicated
via the matrices~\eqref{GA:GammaDecomp}. To express the
underlying dependence on the off--diagonal elements of the soft
breaking terms it is therefore useful to work in the mass insertion
approximation (MIA). According to this approximation the
off--diagonal elements of $\mathcal{M}_{\tilde{d}}^2$ are
treated as perturbations and enter expressions through mixed
propagators proportional to the relevant element (or insertion).
These insertions are parameterised in terms of the dimensionless
quantities defined in~\nqs{GA:dels1}{GA:dels2}.
Equivalently, one may expand the matrices~\eqref{GA:GammaDecomp}
about the diagonal. 
When  performing actual numerical calculations it is
more advantageous to diagonalise the squark mass
matrices~\eqref{GA:sqmm} using numerical routines
to ensure that higher order terms in the MIA are
included, this is what we shall do in our numerical analysis
presented in
section~\ref{NRes}. 

Before proceeding with our analytic expressions for the bare mass
matrix and the various effective vertices, let us first outline
the approximations we shall use throughout this section.
As we are chiefly concerned in exhibiting the dominant behaviour
displayed by the corrections to the bare mass matrix
and effective vertices, in this section we shall work in the approximation
of vanishing electroweak couplings and typically ignore
$\EWSym$ breaking effects. We therefore mainly concern
ourselves with the effects induced by gluino exchange,
and those that arise from higgsino
exchange that are proportional to the Yukawa couplings of the
third generation. Let us emphasise, however, that during
our numerical analysis we include all the effects that
arise from non--zero electroweak couplings, $\EWSym$ breaking
effects and the effects of the Yukawa couplings of the
first two generations. Concerning the accuracy that we work
to within the MIA, we typically include terms up to
second order in the MIA (unless specified otherwise). We
therefore do not include the effects of multiple diagonal
LR insertions that are proportional to $m^2_{d,LR}-\mu\mdbare\tanb$.
It is possible to resum the effects induced by these insertions
via the method outlined in~\cite{BEKU2:bsm}. In this
method however, some BLO effects are then encoded into
the factors of $\cos\theta_{\widetilde{b}}$ and
$\sin\theta_{\widetilde{b}}$ that appear in the MFV
squark mixing matrices. Since the ultimate aim of this
section is to present analytic expressions that
represent all of the BLO corrections that appear in the
framework presented in section~\ref{GA}, we shall only consider
the effects of at most one flavour diagonal LR insertion.
Converting our expressions to take into account
such effects however should be relatively easy.
Finally, to allow for easy comparison
between our results and those that already exist in the
literature, we have assumed that the trilinear up--type soft
terms are proportional to the appropriate Yukawa
coupling (\ie~$m^2_{u,LR}=m_u A_u$, where $A_u={\rm diag}(A_u,A_c,A_t)$),
although we still do not
assume a similar relation for the down--squark sector,
(it is easy to convert back by making the substitution
$A_t\to \left(m^2_{u,LR}\right)_{33}/m_t$ in the various expressions
that follow).

%%%%%%%%%%%%%%%%%%%%%%%%%%%%%%%%%%%%%%%%%%%%%%%%%%%%%%%%%%%%%%%%%
\subsection{The Bare Mass Matrix in the MIA}
\label{MIA:mdb}
%%%%%%%%%%%%%%%%%%%%%%%%%%%%%%%%%%%%%%%%%%%%%%%%%%%%%%%%%%%%%%%%%

We shall first consider the corrections induced by CKM and GFM
effects on the bare mass matrix $\mdbare$. Loop corrections to the
bare mass matrix in the MFV scenario were derived
in~\cite{HRS:bdec,CGNW:Htver,DGG:bsg}. They were subsequently generalised
to GFM in~\cite{BRP:CKM,OR:bsg,OR2:bsg}.

The bare mass matrix may be determined by evaluating the self energy
corrections that appear~\eqref{GA:deltamd}.
The dominant contributions to $\Sigma^{d}_{m\,L}$
arise from self--energy diagrams involving gluino and chargino exchange,
which are depicted in \fig{MIA:mdb:fig1}.
\FIGURE[t!]{
\includegraphics[angle=0,width=0.40\textwidth]{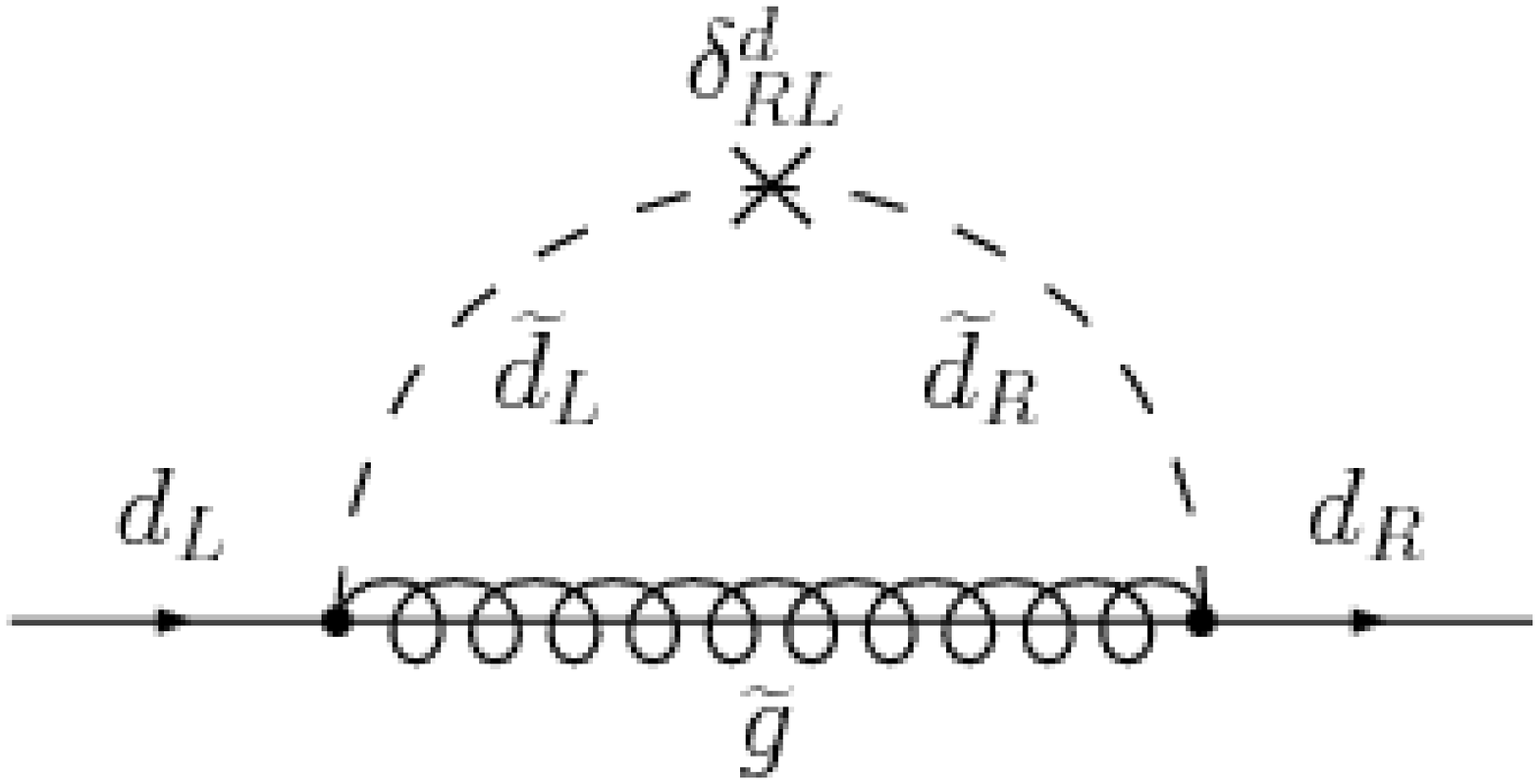}
\includegraphics[angle=0,width=0.40\textwidth]{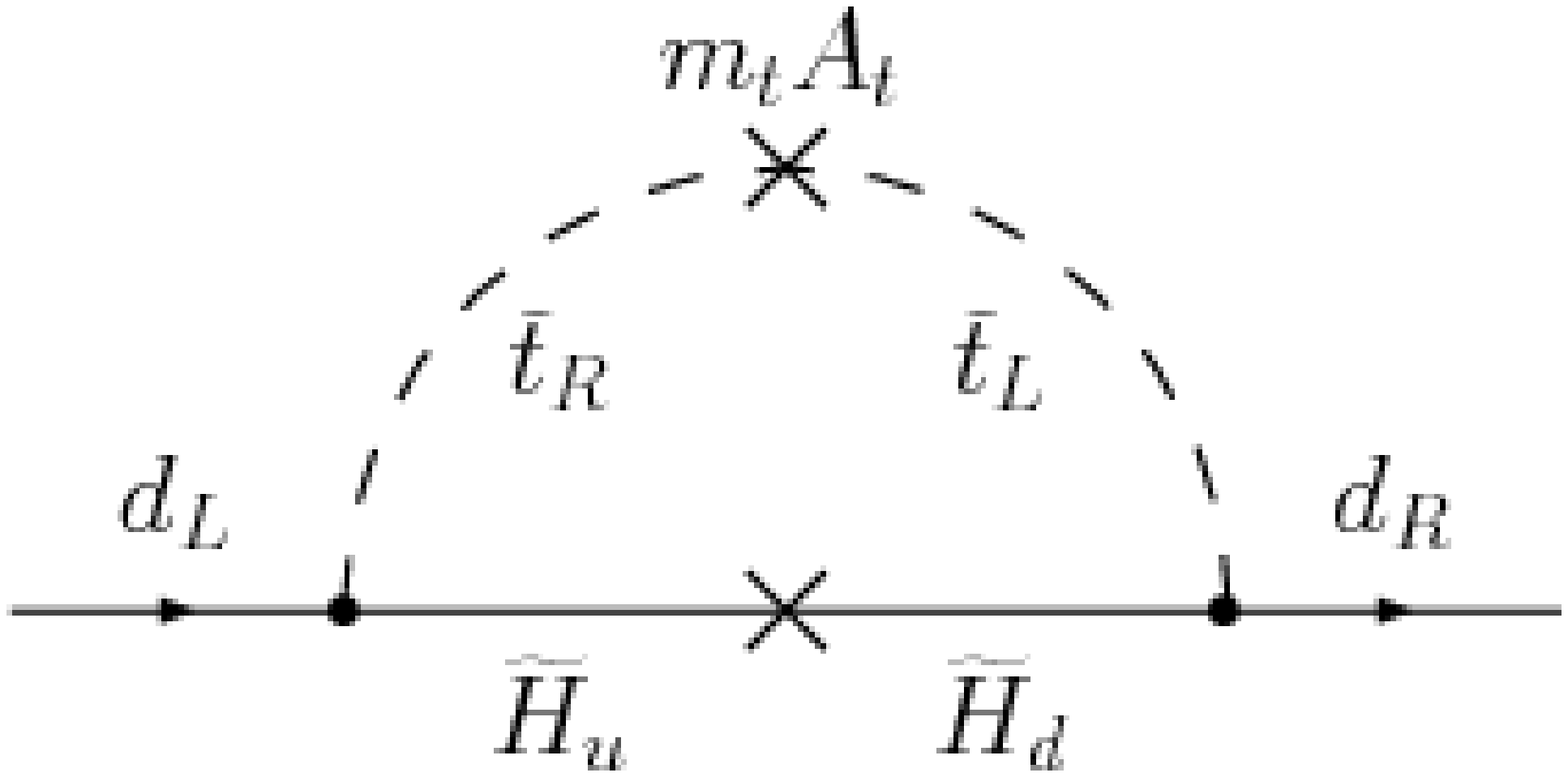}
\caption{The dominant gluino and higgsino diagrams that contribute
to $\Sigma^{d}_{m\,L}$.\label{MIA:mdb:fig1}}
}
On the other hand, the corrections to $\Sigma^{d}_{v\,L,R}$ are rather
small when compared to those that arise from $\Sigma^{d}_{m\,L}$
as they are not enhanced by $\tanb$, nor do they feature a chirality
flip on the gluino line. Coupled with the suppression
factors of $\mdbare$ that accompany them in~\eqref{GA:deltamd},
their omission will not dramatically affect our final results.

To first order in the MIA, the diagonal elements
of $\mdbare$ are given by\footnote{In the following we shall
neglect the flavour diagonal contributions that arise from
the soft terms $m^2_{d,LR}$ unless they $\tanb$ enhanced.
The corrections induced by these terms are included in our
numerical analysis.}
\begin{equation}
\left(\mdbare{}\right)_{ii}=\frac{\left(\mdphys\right)_{ii}}{1+\epsi\tanb},
\label{MIA:mdb:diag}
\end{equation}
where $i=1,2,3$ and $\epsi$ denotes the combined dominant gluino and
chargino contributions
\begin{equation}
\epsi=\epsg+\epsy Y_t^2\delta_{i3},
\label{MIA:mdb:aux1}
\end{equation}
$Y_t$ is the top quark Yukawa coupling and $\delta_{i3}$ is the
usual Kronecker delta function. The coefficients $\epsg$ and
$\epsy$ are given by
\begin{align}
\epsg=&-\frac{\alpha_s}{2\pi}C_2(3)\frac{\mu}{\mgl}
H_2\left(\xdr,\xdl\right),
\\
\epsy=&-\frac{A_t}{16\pi^2\mu}
H_2\left(\yur,\yul\right).
\label{MIA:mdb:aux2}
\end{align}
In the above expressions $\alpha_s$ is the strong coupling
constant and $C_2(3)=4/3$ is the quadratic Casimir operator
for ${\rm SU}(3)$ and $A_t=(A_u)_{33}$.
The loop function $H_2$ is given in appendix~\ref{LF:H}, whilst the
arguments of the function are
\begin{align}
\xdl&=\frac{m_{d,LL}^2}{\mgl^2},
&\yul&=\frac{m_{u,LL}^2}{\mu^2},
\label{MIA:xydef}
\end{align}
the definitions for $\xdr$ and $\yur$ may
be easily obtained by substituting $L$ with $R$ in the above
expressions. It should be noted that the soft terms that appear
in~\eqref{MIA:xydef} are common values of the diagonal entries
of the SUSY soft terms~\eqref{GA:SCKMst}.
\FIGURE[t!]{
\parbox{0.80\textwidth}{
\begin{center}
\includegraphics[angle=0,width=0.40\textwidth]{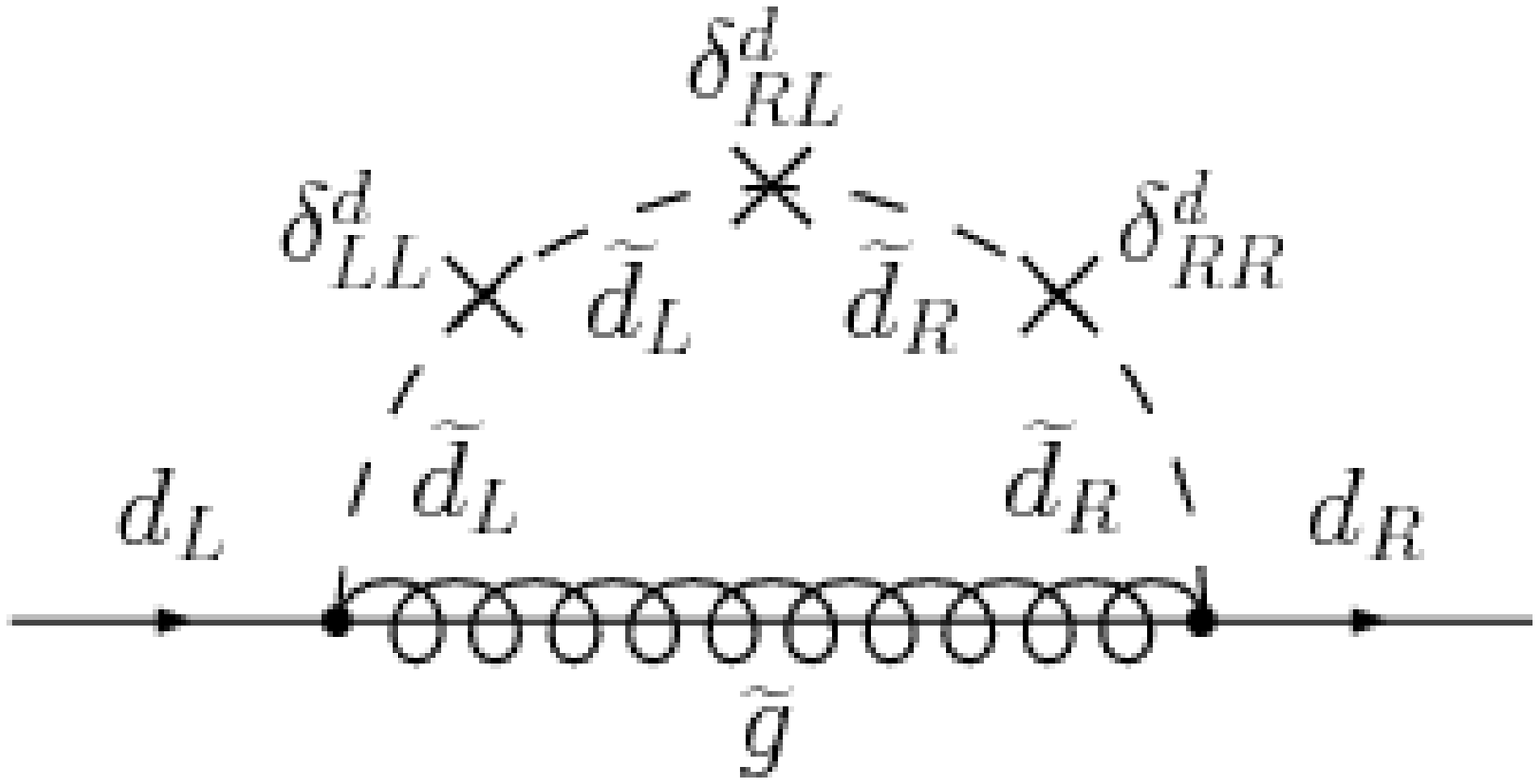}
\end{center}
}
\caption{Additional GFM contributions to the diagonal elements of $\mdbare$.
\label{MIA:mdb:fig2}}
}

It is possible in GFM models to induce large contributions to the
bare down and strange quark masses through diagrams involving three
insertions (\fig{MIA:mdb:fig2})~\cite{Demir:bdec}. For example,
\begin{equation}
\left(\mdbare\right)_{22}=
\frac
{m_s}
{\left(\BLOfacg\right)}
\left[
1-\frac{\mbphys}{m_s}
\frac{\epsilon_4 \tanb \xdr\xdl}
{\left(\BLOfact\right)}\left(\dll\right)_{32}\left(\drr\right)_{23}
\right],
\label{MIA:mdb:diagGFM}
\end{equation}
where $\epsilon_4$ is given by
\begin{equation}
\epsilon_4=-\frac{\alpha_s}{2\pi}C_2(3)\frac{\mu}{\mgl}
H_4\left(\xdr,\xdl,\xdr,\xdl\right).
\label{MIA:mdb:aux3}
\end{equation}
The loop function $H_4$ is given in appendix~\ref{LF:H} whilst
its arguments are given in~\eqref{MIA:xydef}.

Now let us turn to the off--diagonal elements of $\mdbare$. The
diagrams in \fig{MIA:mdb:fig1} and \fig{MIA:mdb:fig3}
illustrate the flavour violating corrections that arise from MFV and GFM
contributions. Evaluating all four diagrams, we find the
contribution
\FIGURE[t!]{
\includegraphics[angle=0,width=0.40\textwidth]{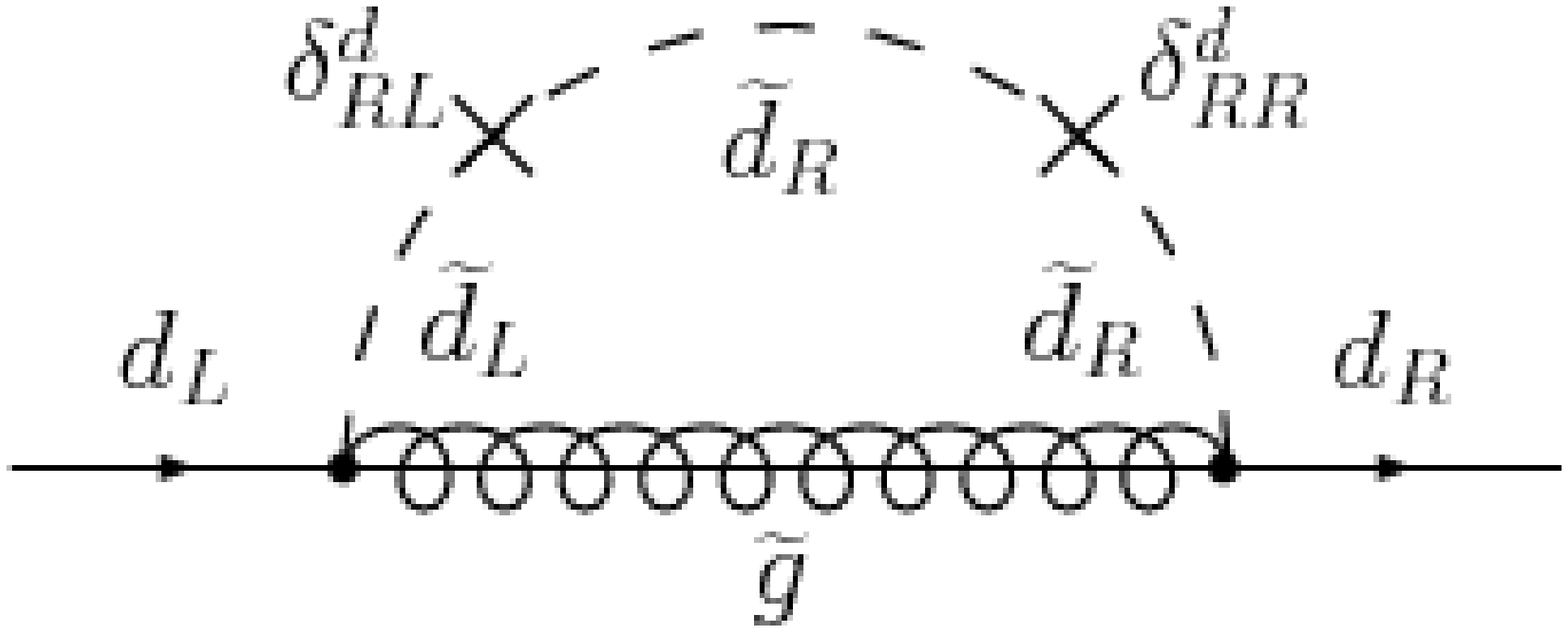}
\includegraphics[angle=0,width=0.40\textwidth]{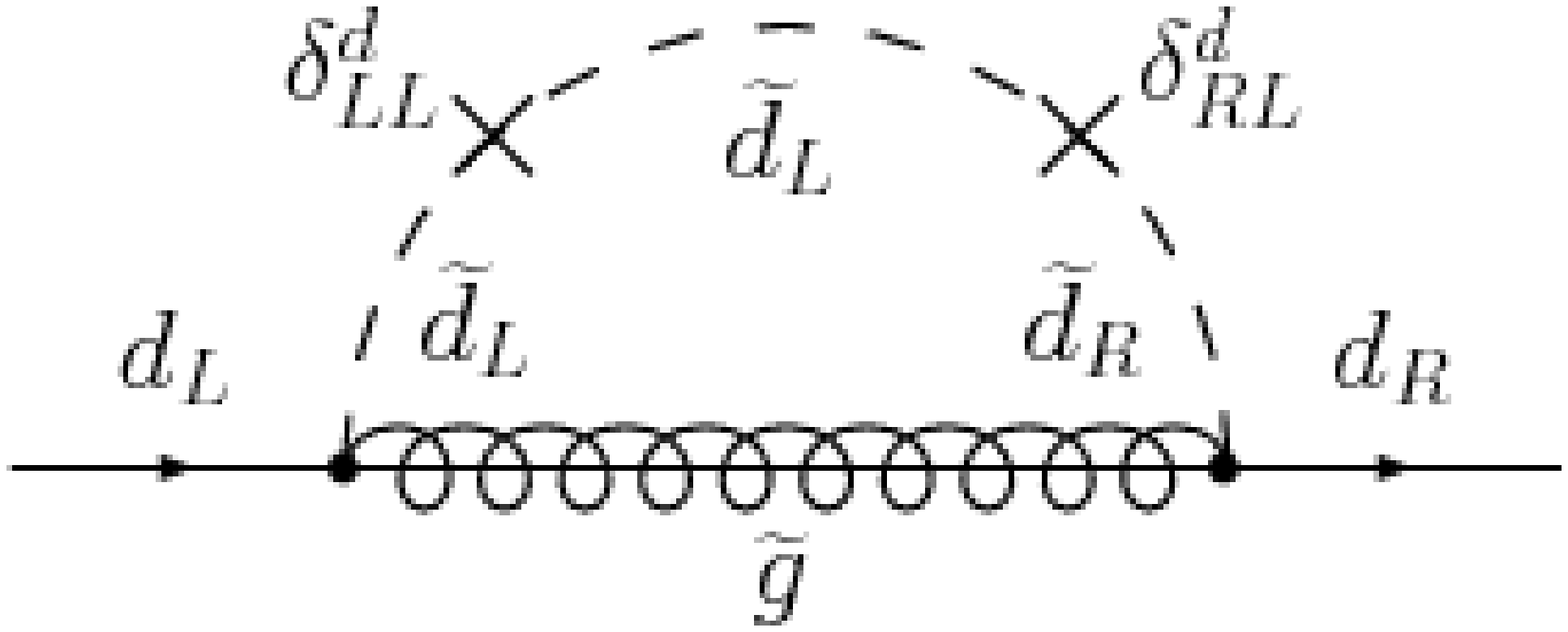}
\caption{GFM contributions to the off--diagonal elements of $\mdbare$.
\label{MIA:mdb:fig3}}
}
\begin{align}
\left(\mdbare\right)_{ij}&=
\frac{\epsilon_{RL}\xdrl\mgl}{\BLOfacj}\left(\drl\right)_{ij}
-\frac{\epsilon_{RR}\xdr\left(\mdphys\right)_{jj}\tanb}
{\left(\BLOfacj\right)^2}\left(\drr\right)_{ij}\nonumber\\
&-\frac{
\big[\epsilon_{LL}\xdl\left(\dll\right)_{ij}
         +\epsilon_Y Y_t^2 K_{ti}^{\ast}K_{tj}\big]\left(\mdphys\right)_{ii}\tanb}
{\left(\BLOfaci\right)\left(\BLOfacj\right)},
\label{MIA:mdb:offdiag}
\end{align} 
where $\epsilon_{RL}$ and $\epsilon_{LL}$ are given by
\begin{align}
\epsilon_{RL}&=-\frac{\alpha_s}{2\pi}C_2\left(3\right)
H_2\left(\xdr,\xdl\right),
&\epsilon_{LL}&=-\frac{\alpha_s}{2\pi}C_2(3)\frac{\mu}{\mgl}
H_3\left(\xdr,\xdr,\xdl\right)
\label{MIA:mdb:aux4}
\end{align}
and $\epsilon_{RR}$ can be obtained by making the substitution
$L \leftrightarrow R$ in the formula for $\epsilon_{LL}$. 
The loop functions $H_2$ and $H_3$ are defined
in appendix~\ref{LF:H}, the CKM matrix $K$ is defined in~\eqref{GA:CKM},
$\epsi$ and $\epsj$ are defined in~\eqref{MIA:mdb:aux1}, and
$\xdrl$ is given by
\begin{align}
\xdrl&=\frac{\sqrt{m_{d,LL}^2 m_{d,RR}^2}}{\mgl^2}.
\end{align}

It will be useful to see how the above
expressions behave in the limit of degenerate sparticle masses.
For instance, the various $\epsilon$--factors that appear in the
above formulae become
\begin{align}
\epsg=&\frac{\alpha_s}{3\pi}sgn\left({\mu}\right),
&\epsy=&\frac{1}{32\pi^2}sgn\left(\frac{A_t}{\mu}\right),
\label{MIA:lims1}
\end{align}
\begin{align}\epsilon_G=&\frac{\alpha_s}{18\pi}sgn\left(\mu\right),
&\epsilon_{RL}=&\frac{\alpha_s}{3\pi},
&\epsilon_{LL}=&-\frac{\alpha_s}{9\pi}sgn\left(\mu\right).
\label{MIA:lims2}
\end{align}
From~\eqref{MIA:lims1} it is easy to see that, in the phenomenologically
favoured region $\mu>0$, $A_t<0$, the chargino and gluino contributions
in~\eqref{MIA:mdb:aux1} for $i=3$ partially cancel. This can lead to a
reduction of BLO contributions compared to a case where only
gluino contributions are taken into account.

As we are chiefly concerned with flavour violation in the down
squark sector, we can safely omit the effects induced by LR, RL
and RR mixings amongst the up squarks. In other words
we assume that $m^2_{u,LR}$ and $m^2_{u,RR}$ are diagonal matrices.
However, the insertion $\ull$ is related by ${\rm SU}(2)$ symmetry
to $\dll$ and its effects on the bare mass matrix should be included. 
In the approximation used in this subsection however, the
contributions proportional to $\ull$, that arise solely from
higgsino exchange, are rather small, as they are suppressed
by factors of the Yukawa couplings of the first two generations.
We will see in section~\ref{ImpApp} however, that, once one
includes the effects induced by non--zero electroweak couplings,
additional contributions are possible.

%%%%%%%%%%%%%%%%%%%%%%%%%%%%%%%%%%%%%%%%%%%%%%%%%%%%%%%%%%%%%%%%%
\subsection{Corrections to Electroweak Vertices in the MIA}
\label{MIA:EW}
%%%%%%%%%%%%%%%%%%%%%%%%%%%%%%%%%%%%%%%%%%%%%%%%%%%%%%%%%%%%%%%%%

Now let us consider the effect of supersymmetric contributions to the
various electroweak vertices in the MIA. As stated in
section~\ref{GA:EW},
the CKM matrix that appears, for example, in the chargino
vertex~\nqs{GA:SU:Char}{GA:SU:CdR}
is related to the physical CKM matrix by the relation~\eqref{GA:EW:KM}.
At first order in the MIA, the vertex and self energy corrections
arising from gluino exchange
cancel due to $\EWSym$ gauge symmetry. The first corrections to $K$
therefore appear at second order,
through diagrams involving two $\EWSym$ breaking insertions on one
of the squark lines. The contributions to the vertex therefore
tend to be suppressed by factors of either $\mbphys$
or $\cotb$ and, whilst we take into account these effects in our
numerical analysis, to a good approximation we may set $K=\Keff$.
A similar result holds for the effective right handed coupling of the
$W$ boson~\eqref{GA:EW:WVer}.

Turning to the $Z$ boson vertex, once again we find that, to first
order in the MIA, the self energy and vertex corrections cancel
due to $\EWSym$ gauge symmetry
The first non--zero contribution arises from the diagram
shown in \fig{MIA:EW:Zverfig} involving two $\EWSym$ breaking
insertions.
\FIGURE[t!]{
\parbox{0.8\textwidth}{
\begin{center}
\includegraphics[angle=0,width=0.30\textwidth]{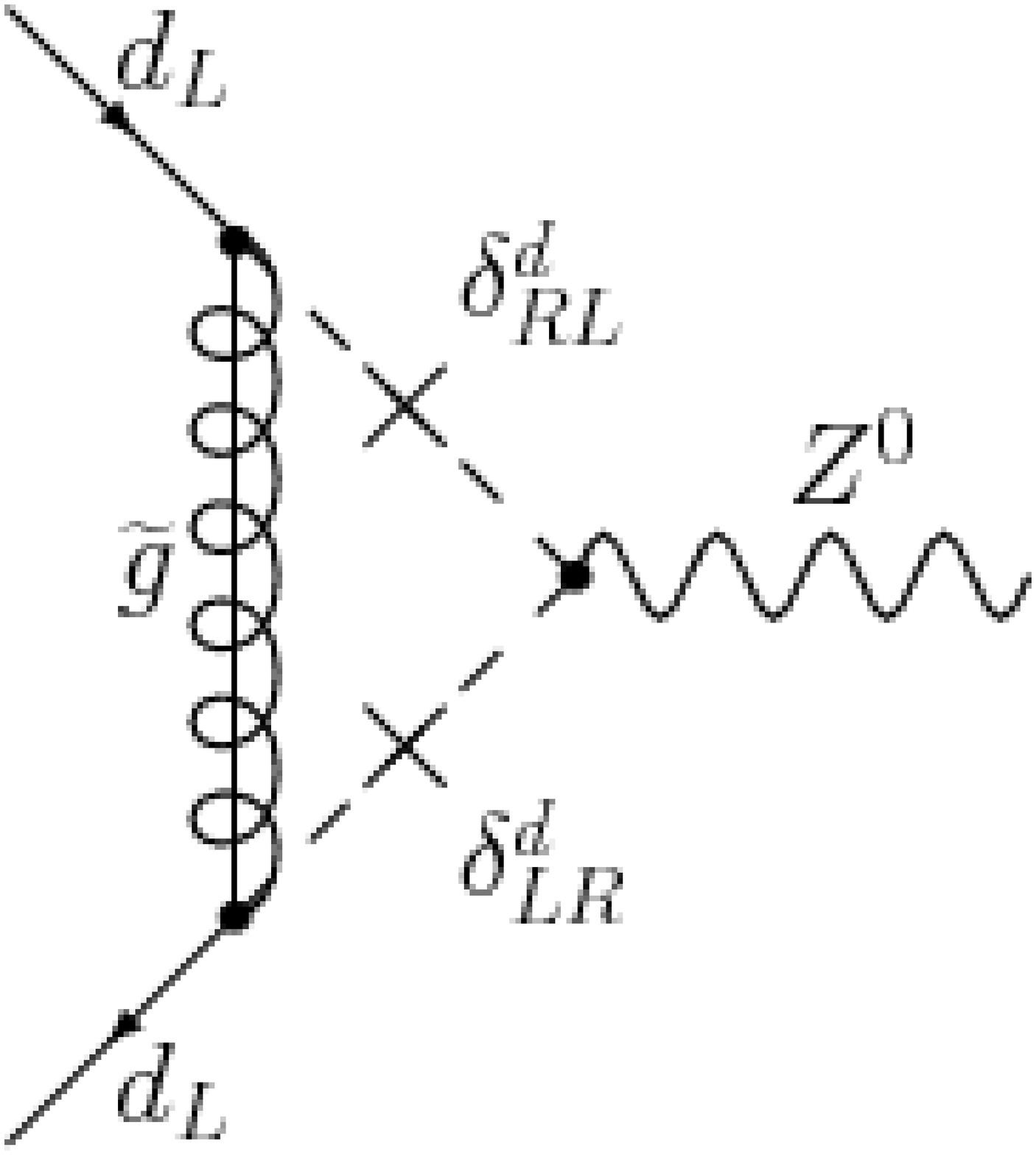}
\end{center}
}
\caption{The gluino correction to the Z boson vertex that arises at
second order in the MIA.}
\label{MIA:EW:Zverfig}
}
Evaluating the contributions to the effective vertex we find for
$\left(C_L^Z\right)_{23}$
\begin{align}
\left(C_L^Z\right)_{23}=-
\frac{g_2}{2\cos\theta_W}
\frac{\alpha_s C_2\left(3\right)}{2\pi}
\frac{f_Z\left(\xd\right)}{\mgl^2\left(\BLOfacg\right)}
\left[\left(m_{d,LR}\right)_{33}-\frac{m_b\mu\tanb}{\BLOfact}\right]
\left(\dlr\right)_{23},
\label{MIA:EW:ZVer}
\end{align}
where the function $f_Z$ is given in appendix~\ref{LF:bsm}.
The expression in square brackets in the above expression
represents the effect of the flavour diagonal RL insertion.
The off--diagonal elements of the bare mass matrix can also induce
terms proportional to $\dll$ and $\drr$, that can viably compete
with the corresponding contributions that arise at third order in
the MIA. Although the vertex~\eqref{MIA:EW:ZVer} is enhanced by $\tanb$,
we shall see later that the contributions to a given process due to this
vertex scale as $m_Z^2/M_{SUSY}^2$ and are typically rather small.

Now let us turn to the Higgs sector, where the effects induced by
supersymmetric contributions to the charged and neutral Higgs
couplings are known to be large~\cite{DGG:bsg,CGNW:bsg,Dedes:bdec}. These
corrections can, in turn, affect FCNC processes especially in
regions of parameter space where FCNC mediated solely by
sparticle exchange are suppressed by large sparticle masses.

The charged Higgs vertex receives corrections~\cite{CGNW:Htver,DGG:bsg,CGNW:bsg}
from both gluino and higgsino
exchange. To second order in the MIA, the effective charged Higgs
coupling is given by
\FIGURE[t!]{
\includegraphics[angle=0,width=0.30\textwidth]{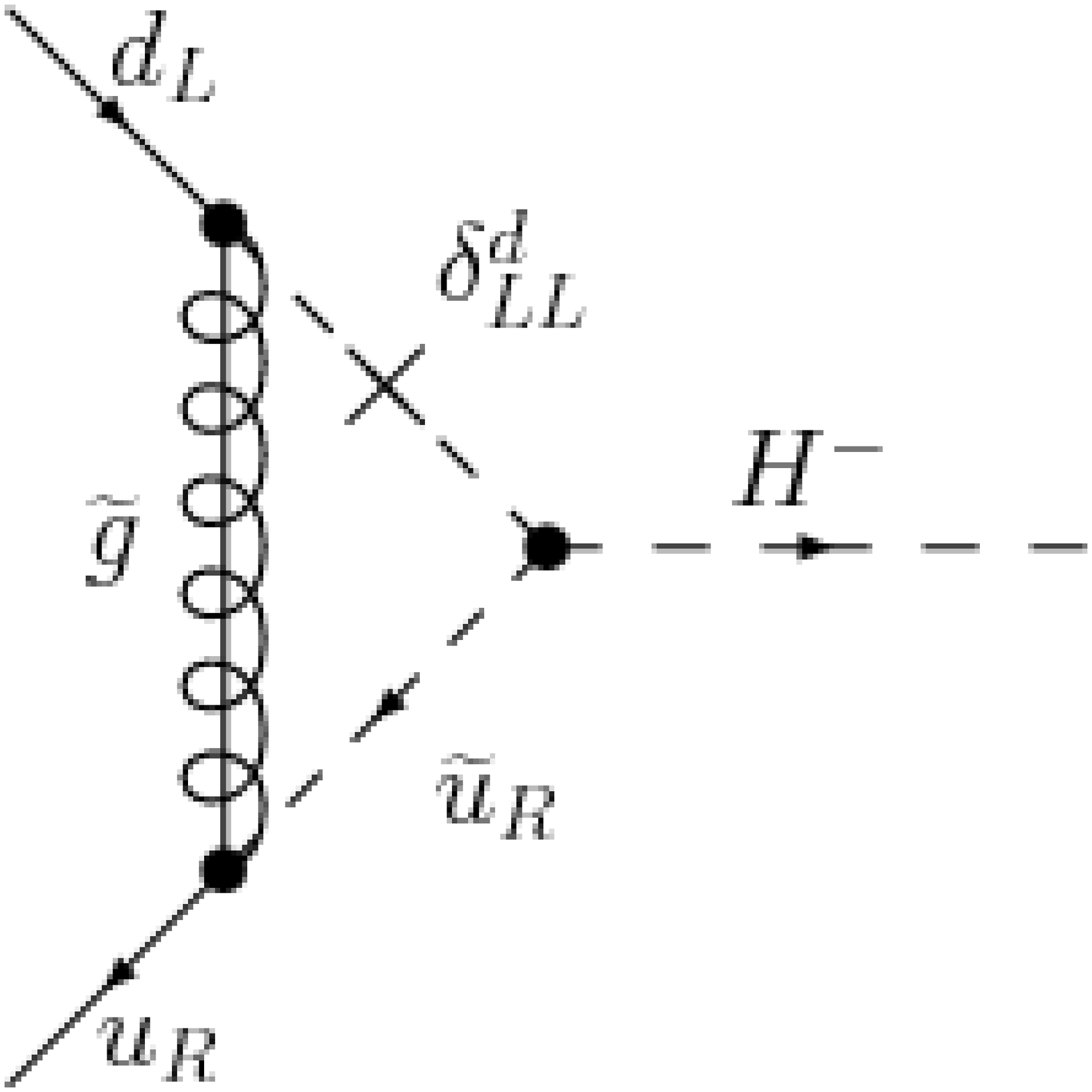}
\includegraphics[angle=0,width=0.30\textwidth]{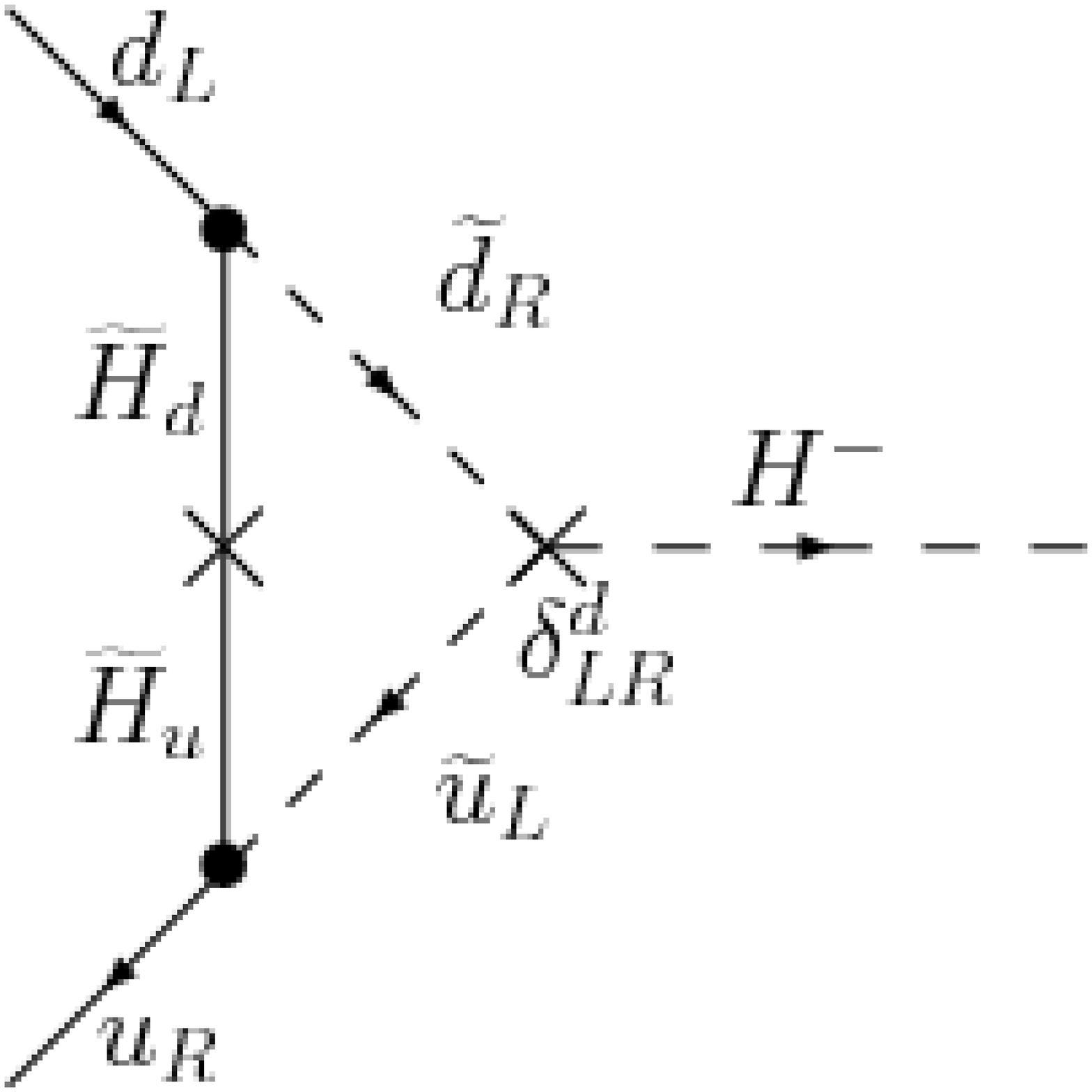}
\caption{The dominant GFM contributions to
the left--handed charged Higgs vertex,
arising from gluino and higgsino exchange.\label{MIA:EW:CHfig}}
}
\begin{align}
\left(C_L^{H^+}\right)_{ij}=&\frac{g_2}{\sqrt{2}m_W}\left(m_u\right)_{ii}\cotb
\left[K_{ij}
\left(1-\epsilon_s^{\prime}\tanb+\epsilon_Y^{\prime}(\ybpsckm)^2\delta_{3j}\tanb\right)
+\Lambda_{ij}^L\right],
\label{MIA:EW:CL}
\\
\left(C_R^{H^+}\right)_{ij}=&\frac{g_2}{\sqrt{2}m_W}\frac{\left(\mdphys\right)_{jj}}{\left(\BLOfaci\right)}\tanb\left(K_{ij}+\Lambda_{ij}^R\right),
\label{MIA:EW:CR}
\end{align}
where $i,j=1,2,3$ and $\ybpsckm=(\ydpsckm)_{33}$. The factors
$\epsilon_s^{\prime}$ and $\epsilon_Y^{\prime}$ are given by
\begin{align}
\epsilon_{s}^{\prime}&=-\frac{\alpha_s}{2\pi}C_2(3)\frac{\mu}{\mgl}
H_2\left(\xur,\xdl\right),
&\epsilon_{Y}^{\prime}&=-
\frac{1}{16\pi^2}\frac{\left(m_{d,LR}\right)_{33}}{\mu\left(\mdbare\right)_{33}}
H_2\left(\yul,\ydr\right).
\end{align}
The arguments of the loop functions $H_2$ can be obtained by
the appropriate generalisations of~\eqref{MIA:xydef}. Finally,
the $3\times 3$ matrices $\Lambda_{ij}^{L,R}$ denote the
additional off--diagonal contributions that arise in both
MFV and GFM models due to the off--diagonal elements of the
bare mass matrix and the GFM parameters. $\Lambda_{ij}^{L,R}$ may
be decomposed as follows
\begin{align}
\Lambda_{ij}^{L,R}=\Delta_{ij}^{L,R}+\gamma_{ij}^{L,R}.
\label{MIA:EW:Lam}
\end{align}
The MFV contributions $\Delta_{ij}^L$  to the vertex have been
highlighted in~\cite{AGIS:bdec,BCRS:bdec}. In the formalism
developed in section~\ref{GA}, they arise due to the presence
of the bare mass matrix in the neutralino vertex, and have
the following form
\begin{align}
\Delta_{ij}^L=&K_{ij}\frac{\epsilon_Y^{\prime}\epsilon_Y(\ybpsckm Y_t\tanb)^2}{\BLOfacj},
&(i,j)=(3,1),(3,2),\\
\Delta_{ij}^L=&0,
&{\rm otherwise}.
\label{MIA:EW:Del}
\end{align}
It should be noted that the additional terms found in~\cite{BCRS:bdec}
for $(i,j)=(1,3),(2,3)$ do not appear as we do not assume that
the trilinear soft terms are proportional to the bare Yukawa
coupling. If one adopts the parameterisation
described in appendix~\ref{dlrapp}, where such a relation is assumed,
it can be shown that one obtains an additional contribution to
$\Delta_{ij}^L$ in agreement with~\cite{BCRS:bdec}.

The GFM contributions to $\Lambda_{ij}^L$ arise from the two
diagrams shown in \fig{MIA:EW:CHfig}. Evaluating the
contributions yields
\begin{align}
\gamma_{ij}^L=&-K_{ii}\tanb
\Bigg[
\Bigg(\epsilon_{LL}^{\prime}-\frac{\epsilon_{LL}\epsilon_{Y}^{\prime}(\ybpsckm)^2\tan\beta}{\BLOfacj}\Bigg)\xdl\left(\dll\right)_{ji}
\nonumber\\
&+\frac{\mgl}{\mbphys}\frac{\epsilon_{RL}\epsilon_Y^{\prime}(\ybpsckm)^2\xdrl\left(\BLOfact\right)}{\left(\BLOfacj\right)}\left(\dlr\right)_{ji}
\nonumber\\
&+\epsilon_{RL}^{\prime}(\ybpsckm)(\ydpsckm)_{jj}\ydrl
\left(\drl\right)_{ji}
\Bigg],
&(i,j)=(3,1),(3,2),
\label{MIA:EW:GamL1}
\\
\gamma_{ij}^L=&-K_{ii}\tanb\left[\epsilon_{LL}^{\prime}\xdl\left(\dll\right)_{ij}
+\epsilon_{RL}^{\prime}(\ybpsckm)^2\ydrl\left(\dlr\right)_{ij}\right],
&(i,j)=(1,3),(2,3),
\label{MIA:EW:GamL2}
\end{align}
where $\epsilon_{LL}^{\prime}$ and $\epsilon_{RL}^{\prime}$ are
\begin{align}
\epsilon_{LL}^{\prime}&=-\frac{\alpha_s}{2\pi}C_2(3)\frac{\mu}{\mgl}
H_3\left(\xur,\xdl,\xdl\right),
&\epsilon_{RL}^{\prime}&=
-\frac{1}{16\pi^2}\frac{\mu}{\left(\mdbare\right)_{33}}
H_2\left(\yul,\ydr\right).
\label{MIA:EW:CHaux}
\end{align}
It should be noted that the third term in~\eqref{MIA:EW:GamL1}
is proportional the Yukawa coupling of the down or strange
quark. We include it however as the factors of $\cosb$ present in the
denominators of the Yukawa couplings (we remind the reader that
$Y_d\sim m_d/m_W\cosb$) can effectively
lead $\gamma^L_{ij}$ to vary as $\tan^3\beta$. This term can
become important if $\left(\dlr\right)_{32}=\drl$ is
large $\mathcal{O}\left(10^{-2}\right)$.

Now consider the right--handed coupling of the charged Higgs. 
In this case the dominant corrections to the vertex are due
to the self--energy correction $\Sigma_{mL}^d$.
The MFV contributions to the vertex are reflected by the
appearance of a factor of $\left(\BLOfaci\right)$ in the
denominator of~\eqref{MIA:EW:CR}, in agreement with~\cite{BCRS:bdec}.

In models with GFM it is possible to generate additional terms
of the form
\begin{align}
\gamma_{ij}^R=-K_{ii}\bigg[&\frac{\epsilon_{RR}\xdr\tanb}
{\left(\BLOfaci\right)}
\frac{\left(\mdphys\right)_{ii}}{\left(\mdphys\right)_{jj}}
\left(\drr\right)_{ij}
\nonumber\\
&+\epsilon_{RL}\xdrl\epsi\frac{\mgl}{\left(\mdphys\right)_{jj}}\left(\drl\right)_{ji}\bigg],
&(i,j)=(3,1),(3,2),
\label{MIA:EW:GamR1}
\\
\gamma_{ij}^R=-K_{ii}\bigg[&\frac{\epsilon_{LL}\xdl\tanb}
{\left(\BLOfacj\right)}\left(\dll\right)_{ij}
\nonumber\\
&+\epsilon_{RL}\xdrl\epsi
\frac{\mgl}{\left(\mdphys\right)_{jj}}\left(\dlr\right)_{ij}\bigg],
&(i,j)=(1,3),(2,3).
\label{MIA:EW:GamR2}
\end{align}
A particularly interesting consequence of~\eqref{MIA:EW:GamR1} is that
one can often avoid the factor of the strange quark
mass, that appears in the right handed vertex~\eqref{MIA:EW:CL} when
$i=3$ and $j=2$, via flavour violation in either the RL or RR sectors.

It is apparent from the above expressions that GFM contributions,
to both the left and right--handed vertices, can play the r\^ole of
the off--diagonal elements of the CKM matrix. The off--diagonal
BLO corrections to the charged Higgs vertex can therefore be
rather large in the GFM scenario.  Substantial enhancement or
suppression of charged Higgs contributions to FCNC are therefore
possible, even
in the limit where the squarks decouple from the theory.
In addition, $\tanb$ enhanced corrections affect the underlying
structure of the charged Higgs vertex, in both GFM and MFV, via the factors
of $\left(1+\epsi\tanb\right)$ that occur in the denominator in
\eqref{MIA:EW:CR} and the corrections $\epsilon_s^{\prime}$
and $\epsilon_Y^{\prime}$ that appear in~\eqref{MIA:EW:CL}.

The corrections to the charged Goldstone boson vertex \cite{DGG:bsg} prove
to be rather small, as the vertex is protected by ${\rm SU}(2)$ symmetry
and the self energy and vertex contributions approximately cancel.
These cancellations are required as, in a general $R_{\xi}$ gauge,
the corrected Goldstone boson vertex must act to cancel the $\xi$
dependence of the contributions originating from $W$ boson exchange.
The corrected vertex must therefore, in a similar manner to the
corrected $W$ boson vertex, be proportional to $\EWSym$ breaking
effects, even for GFM. 

Finally, let us consider the corrected neutral Higgs
vertices~\nqs{GA:EW:HVerL}{GA:EW:HVer}. The dominant
contributions originate from the self energy corrections
$\Sigma^d_{m\,L}$. To first order in the MIA the contributions
to the flavour diagonal elements of
the effective $A^0$ vertex become
\begin{align}
\left(C^{A^0}_L\right)_{ii}=&-\frac{i g_2}{2 m_W}\tanb\frac{\left(\mdphys\right)_{ii}}{\left(\BLOfaci\right)},
\label{MIA:EW:SLDiagA}
\end{align}
whilst the contributions to the effective $H^0$ and $h^0$ vertices are
\begin{align}
\left(C^{H^0}_L\right)_{ii}=&-\frac{g_2}{2\cos\beta m_W}\frac{\left(\mdphys\right)_{ii}}{\left(\BLOfaci\right)}\left(\cos\alpha+\epsilon_i\sin\alpha\right),
\\
\left(C^{h^0}_L\right)_{ii}=&+\frac{g_2}{2\cos\beta m_W}\frac{\left(\mdphys\right)_{ii}}{\left(\BLOfaci\right)}\left(\sin\alpha-\epsilon_i\cos\alpha\right).
\label{MIA:EW:SLDiagH}
\end{align}
At third order in the MIA, further corrections proportional to combinations
of $\dll$ and $\drr$ are generated in a similar manner to
\eqref{MIA:mdb:diagGFM} that can lead to large corrections to
the Yukawa couplings of the first two generations. Full expressions
can be found in~\cite{Demir:bdec}.

The off--diagonal elements of the coupling are generated by
MFV and GFM contributions and, in a similar manner to the charged
Higgs vertex, it is useful to perform the decomposition
\begin{align}
\left(C_{L,R}^{S^0}\right)_{ij}=
\left(C_{L,R}^{S^0}\right)^{\rm MFV}_{ij}+\left(C_{L,R}^{S^0}\right)^{\rm GFM}_{ij},
\label{MIA:EW:NHDecomp}
\end{align}
where $S^0=A^0,G^0,H^0,h^0$.
The dominant MFV contributions to the off--diagonal elements of the
coupling arise from higgsino exchange and are given
by~\cite{AGIS:bdec,BCRS:bdec},
\begin{align}
\left(C_L^{A^0}\right)^{\rm MFV}_{ij}=\frac{i g_2}{2 m_W}\frac{\left(\mdphys\right)_{ii}\epsilon_Y Y_t^2 K_{ti}^{\ast}K_{tj}\tan^2\beta}{\left(\BLOfaci\right)\left(\BLOfacj\right)}.
\label{MIA:EW:SLMFV}
\end{align}
The GFM contributions arise primarily from gluino exchange and yield
the additional contribution
\begin{align}
\left(C_L^{A^0}\right)^{\rm GFM}_{ij}=\frac{i g_2}{2 m_W}\tan^2\beta
\Bigg[
&\frac{\left(\mdphys\right)_{ii}\epsilon_{LL}\xdl}
{\left(\BLOfaci\right)
\left(\BLOfacj\right)}
\left(\dll\right)_{ij}
\nonumber\\
&+\frac{\epsilon_{RL}\epsj\mgl \xdrl}
{\left(\BLOfacj\right)}
\left(\drl\right)_{ij}
+\frac{\left(\mdphys\right)_{jj}\epsilon_{RR}\xdr}
{\left(\BLOfacj\right)^2}
\left(\drr\right)_{ij}
\Bigg].
\label{MIA:EW:SLGFM}
\end{align}
The off--diagonal couplings of the scalar Higgs bosons $H^0$ and $h^0$
may be obtained via the simple substitutions
\begin{align}
C_L^{H^0}=&\;i\sin\left(\alpha-\beta\right)C_L^{A^0},
&C_L^{h^0}=&\;i\cos\left(\alpha-\beta\right) C_L^{A^0},
\label{MIA:EW:SLConv}
\end{align}
whilst the right handed couplings can be obtained by taking
the Hermitian conjugate. 
Due to an accidental cancellation between the self--energy and
vertex corrections, the terms proportional
to $\drl$ and $\dlr$ vanish at LO. However, once BLO corrections are taken into
account, it is possible for these insertions to reappear through their
effects on the bare mass matrix $\mdbare$ \cite{FOR1:bdec}.

Once again, it should be noted that, due to ${\rm SU}(2)$ invariance,
the Goldstone boson vertex does not receive large corrections
even once GFM contributions are taken into account. As a result any
contributions to the corrected vertex are attributable solely
to $\EWSym$ breaking effects and are rather small.

%%%%%%%%%%%%%%%%%%%%%%%%%%%%%%%%%%%%%%%%%%%%%%%%%%%%%%%%%%%%%%%%%
\subsection{Additional Electroweak Effects}
\label{ImpApp}
%%%%%%%%%%%%%%%%%%%%%%%%%%%%%%%%%%%%%%%%%%%%%%%%%%%%%%%%%%%%%%%%%

As discussed at the beginning of this section
the results presented so far have been derived
in the limit where the electroweak gauge couplings
$g_1$ and $g_2$ are set equal to zero. The aim of this
subsection is to briefly discuss the dominant contributions
that arise once one proceeds beyond that approximation, and
to provide some simple substitutions such that these effects
can be taken into account.

First, let us consider the effect of such corrections on the bare
mass matrix. One of the most important corrections in this
case is due to the gaugino--higgsino mixing diagram shown
in \fig{MIA:ImpApp:fig1} that arises if the insertion $\ull$ is
non--zero~\cite{BRP:CKM}.
\FIGURE[t!]{
\parbox{0.8\textwidth}{
\begin{center}
\includegraphics[angle=0,width=0.40\textwidth]{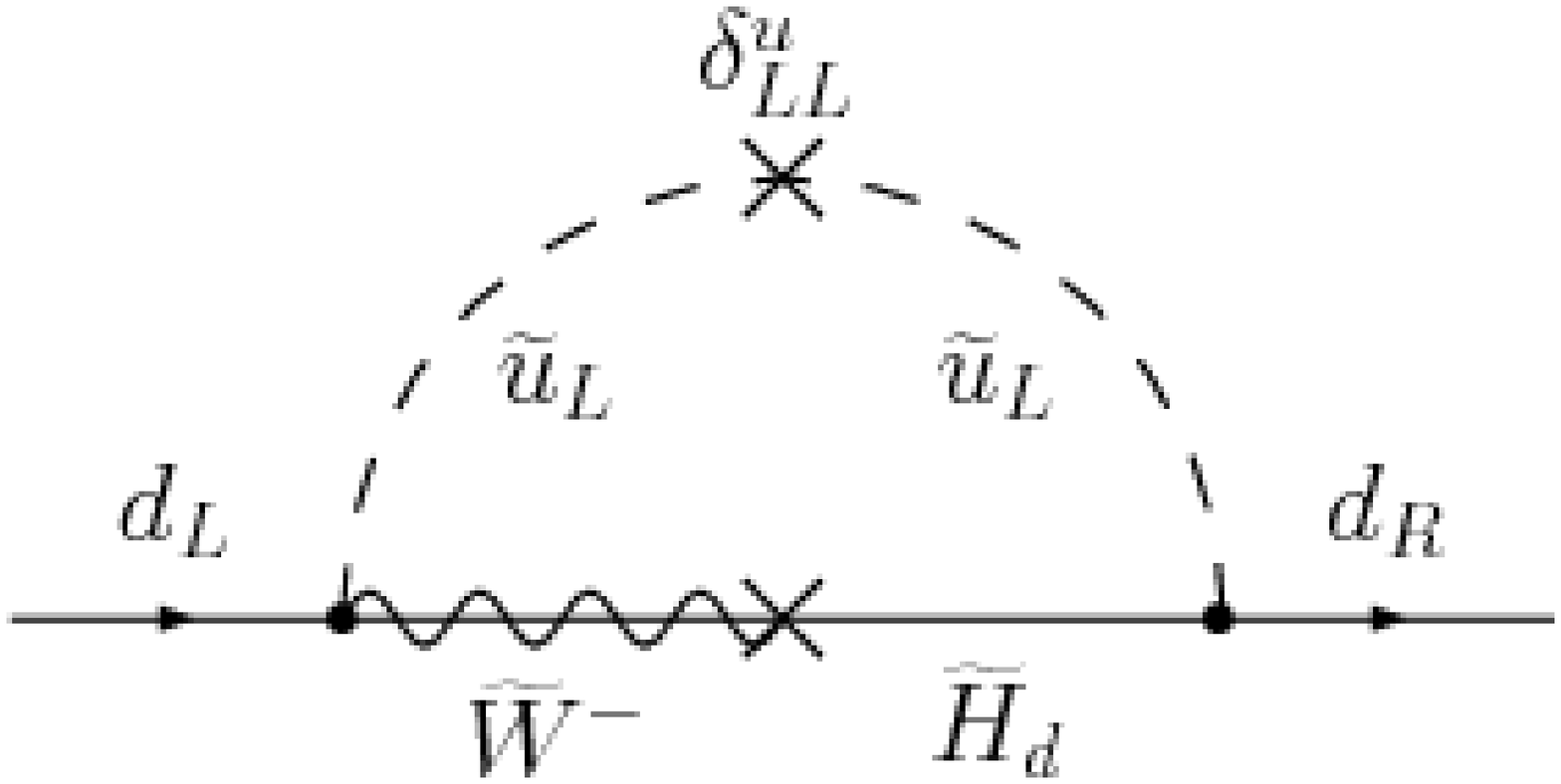}
\end{center}
}
\caption{Contributions to $\mdbare$ arising from the insertion $\ull$.
\label{MIA:ImpApp:fig1}}
}
The corrections induced by this diagram may be taken into
account by making the following substitution in~\eqref{MIA:mdb:offdiag}
\begin{align}
\epsilon_{LL}\xdl\to\epsilon_{LL}\xdl+\sum_{a=1}^{2}\epsilon_{\chi\,LL}^a\ydl^a,
\label{MIA:epsLLsubs}
\end{align}
where $\ydl^a=m_{d,LL}^2/m_{\chi^-_a}^2$ and $\epsilon^a_{\chi\,LL}$ is
given by
\begin{align}
\epsilon_{\chi\,LL}^a=
\frac{\alpha}{4\pi\sin^2\theta_W}\frac{V_{a1}^{\ast}m_{\chi^-_a}U_{a2}}{\sqrt{2}m_W\sinb}H_2\left(\ydl^a,\ydl^a\right).
\end{align}
$\alpha$ denotes the electromagnetic coupling constant. We have
made use of the relation~\eqref{GA:dllull} to express
the contribution in terms of flavour violation in the down
squark sector. In the phenomenologically interesting region
$\mu>0$ and $A_t<0$, $\epsilon^a_{\chi\,LL}$ interferes destructively
with the gluino contribution $\epsilon_{LL}$ and
acts to reduce the correction to $\mdbare$ that is proportional
to flavour violation in the LL sector.

Turning to the charged Higgs vertex, as discussed
in~\cite{BCRS:bdec}, large contributions to the left--handed
vertex arise from diagrams featuring gaugino and higgsino exchange.
They may be included by making the following substitution
in~\eqref{MIA:EW:CL}
\begin{align}
\epsgp\to\epsgp+\eps_{\chi}^{\prime}.
\end{align}
$\eps_{\chi}^{\prime}$ has the following form
\begin{align}
\eps_{\chi}^{\prime}=&-\frac{\alpha}{4\pi\sin^2\theta_W}
\sum_{a,\alpha}\frac{m_{\chi^0_\alpha}}{m_{\chi^-_a}}\overline{M}_{\alpha a}
\bigg\{-\frac{2}{3}V_{a2}^{\ast}N_{\alpha 1}^{\ast}\tan\theta_W
H_2\left(y_{\widetilde{u}_R}^a,w_{\alpha}^a\right)
\nonumber\\
&+\left[\frac{1}{2} V_{a2}^{\ast}\left(\frac{1}{3}N_{\alpha 1}^{\ast}\tan\theta_W-N_{\alpha 2}^{\ast}\right)
-\frac{1}{\sqrt{2}} V_{a1}^{\ast}N_{\alpha 4}^{\ast}\right]
H_2\left(y_{\widetilde{d}_L}^a,w_{\alpha}^a\right)
\bigg\},
\label{ImpApp:CHMFV}
\end{align}
where the quantity $\overline{M}_{\alpha a}$ is given by
\begin{align}
\overline{M}_{\alpha a}=U_{a2}\left(N_{\alpha 1}^{\ast}\tan\theta_W+N_{\alpha 2}^{\ast}\right)-\sqrt{2}U_{a1}N_{\alpha 3}^{\ast}.
\nonumber
\end{align}
Our results for $\epsilon_{\chi}^{\prime}$ agree with those originally
given in~\cite{BCRS:bdec}.  To include the additional effects induced
by GFM, one has to consider the diagrams shown in \fig{MIA:ImpApp:fig2}.
\FIGURE[t!]{
\includegraphics[angle=0,width=0.30\textwidth]{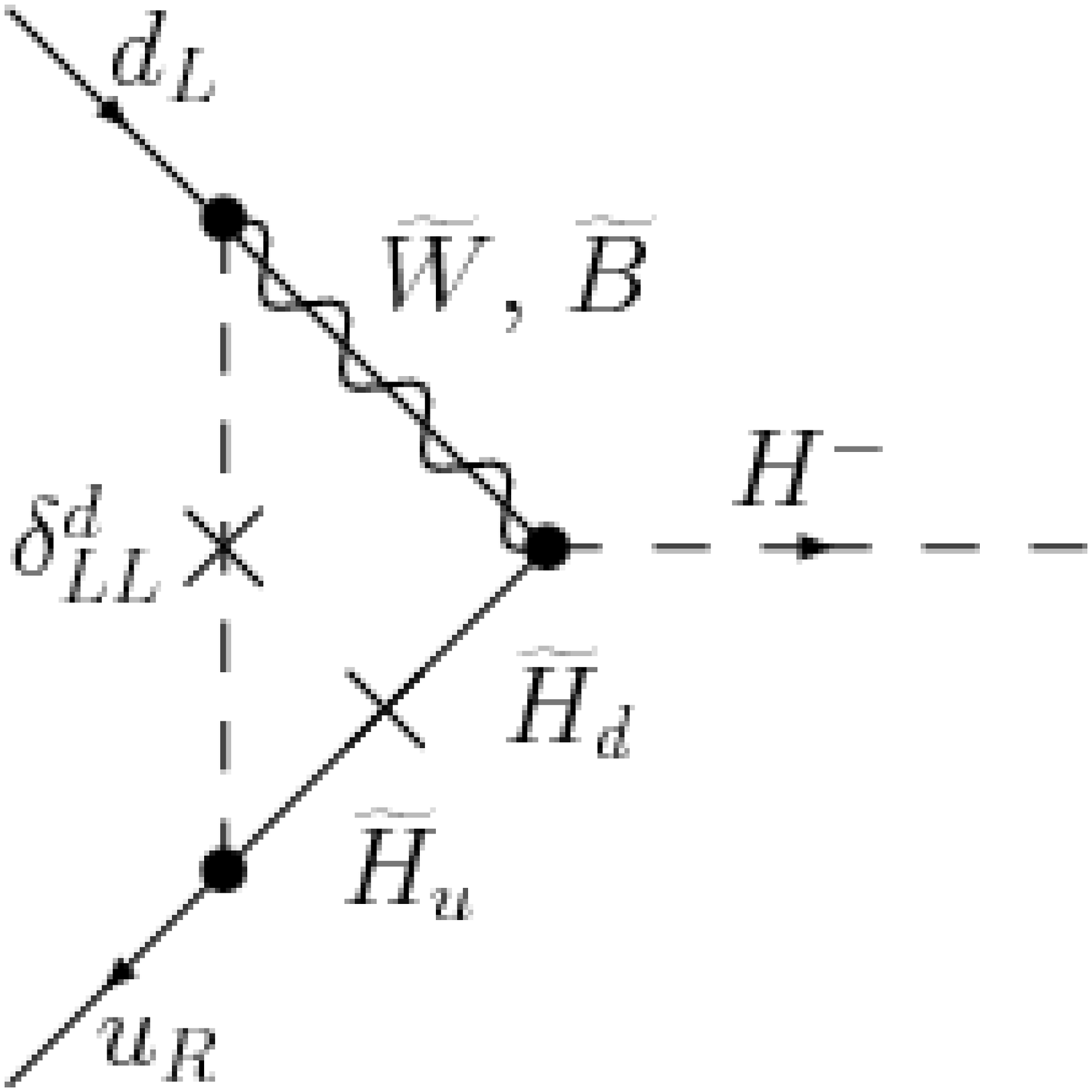}
\includegraphics[angle=0,width=0.30\textwidth]{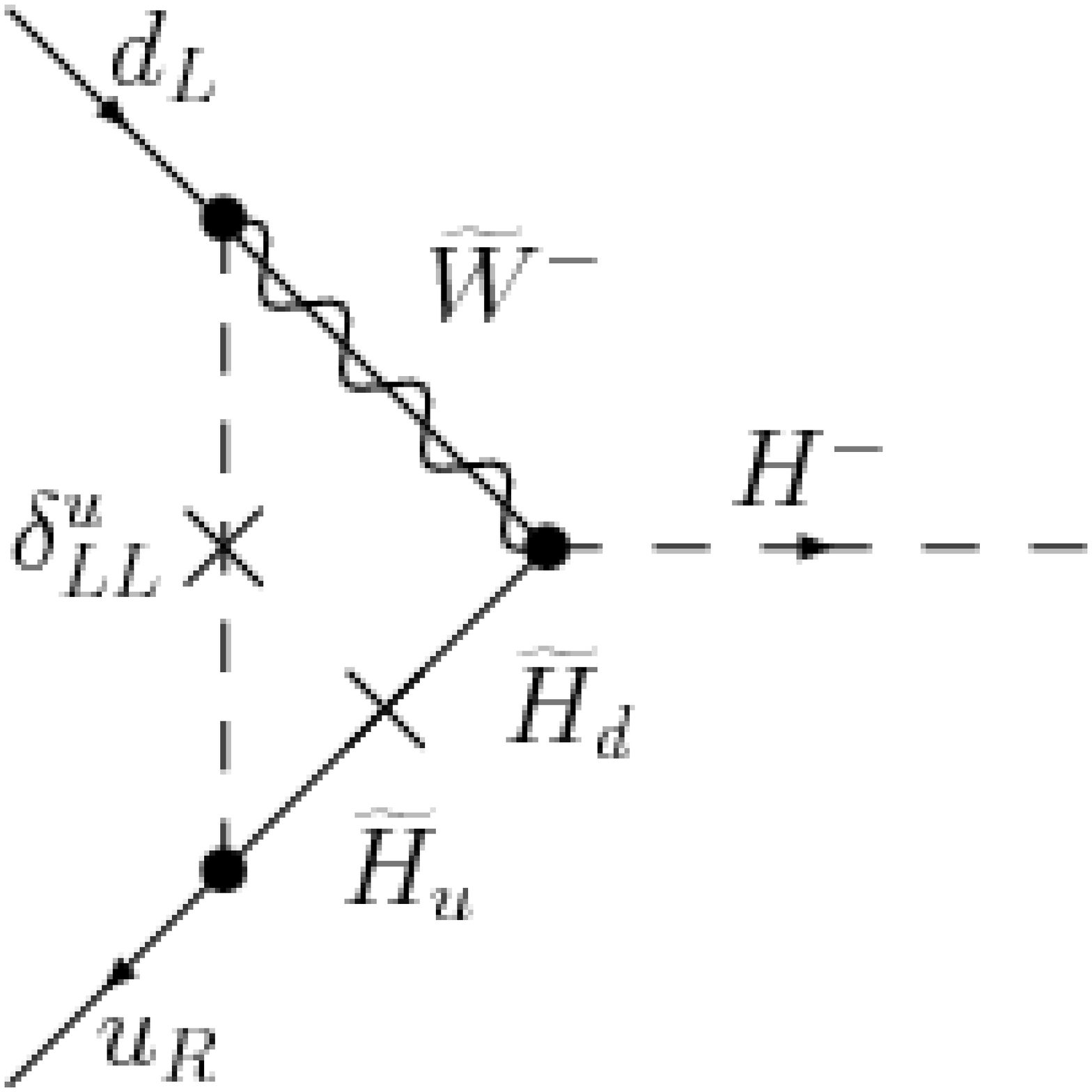}
\caption{Additional GFM contributions to
the left--handed charged Higgs vertex,
due to the electroweak effects considered in subsection~\ref{ImpApp}.
\label{MIA:ImpApp:fig2}}
}
Their effects may be included by making the following correction
to~\nqs{MIA:EW:GamL1}{MIA:EW:GamL2}
\begin{align}
\epsilon_{LL}^{\prime}\xdl\to\epsilon_{LL}^{\prime}\xdl
+\sum_{a}\ydl^a\epsilon^{a\;\prime}_{\chi\,LL},
\end{align}
where $\epsilon^{a\;\prime}_{\chi\;LL}$ is given by
\begin{align}
\eps_{\chi\,LL}^{a\;\prime}=&-\frac{\alpha}{4\pi\sin^2\theta_W}
\sum_{\alpha}\frac{m_{\chi^0_\alpha}}{m_{\chi^-_a}}\overline{M}_{\alpha a}
\nonumber\\
&\times\left[\frac{1}{2} V_{a2}^{\ast}\left(\frac{1}{3}N_{\alpha 1}^{\ast}\tan\theta_W-N_{\alpha 2}^{\ast}\right)
-\frac{1}{\sqrt{2}} V_{a1}^{\ast}N_{\alpha 4}^{\ast}\right]
H_3\left(\ydl^a,\ydl^a,w_{\alpha}^a\right).
\label{ImpApp:CHGFM}
\end{align}
Both the MFV~\eqref{ImpApp:CHMFV} and GFM~\eqref{ImpApp:CHGFM} corrections
typically interfere destructively with the dominant gluino contributions
and can lead to an appreciable reduction of BLO effects. 

Finally, let us consider the neutral Higgs vertex. As discussed
in the previous subsection, the dominant contributions
to the effective vertex arise from the self--energy corrections
$\Sigma^d_{mL}$. The effects induced by non--zero electroweak
couplings may therefore be included, in a similar manner to the
bare mass matrix, by making the substitution~\eqref{MIA:epsLLsubs}
in~\eqref{MIA:EW:SLGFM}.

%%%%%%%%%%%%%%%%%%%%%%%%%%%%%%%%%%%%%%%%%%%%%%%%%%%%%%%%%%%%%%%%%
\subsection{Other Methods}
\label{BLOeff:OM}
%%%%%%%%%%%%%%%%%%%%%%%%%%%%%%%%%%%%%%%%%%%%%%%%%%%%%%%%%%%%%%%%%

The method outlined in section~\ref{GA} takes into account both
$\tanb$ enhanced effects and those induced by non--minimal sources
of flavour violation. Other methods have been proposed in the
literature that can be modified to include the effects of
GFM and it shall be useful to briefly consider how two specific
examples compare with the method employed in this paper.

The first method, presented by Buras {\it et al}.~\cite{BCRS:bdec},
works in the bare SCKM basis. In this basis the Yukawa matrices
$\ydpsckm$ and $\yupsckm$ that appear in the superpotential are diagonal.
Calculating the self--energies in this basis gives the
physical quark masses
\begin{align}
m_d=D_R \left(v_d \ydbsckm+\delta\hat{m}^{(0)}\right) D_L^{\dag}
\label{BLOeff:OM:BCRS:bare}
\end{align}
where $\ydbsckm$ is the diagonalised Yukawa matrix, and
$\delta\hat{m}^{(0)}$ denotes the contributions of the self energy
corrections~\eqref{GA:deltamd} calculated in the bare SCKM basis.
$D_L$ and $D_R$ denote the unitary transformations performed on the squark
fields that transform between the bare and physical SCKM bases.
The bare mass matrix $\mdbare$ defined in~\eqref{GA:mdphys}
is related to these quantities by
\begin{align}
\mdbare=D_R\left(v_d\ydbsckm\right)D_L^{\dag}.
\label{BLOeff:OM:BCRS:barecon}
\end{align}
It is straightforward to relate the matrices $D_{L,R}$ to the
unitary matrices $V_{d_{L,R}}$ that appeared in section~\ref{GA}.
If, in analogy with the transformations~\nqs{GA:qtransd}{GA:qtransu},
one defines a transformation from the interaction basis to the
bare SCKM basis such that
\begin{align}
\ydbsckm=V_{d_R}^{(0)}\left(\ydint\right)V_{d_L}^{(0)\dag}.
\end{align}
$D_R$ and $D_L$ are then given by
\begin{align}
D_L&=V_{d_L}V_{d_L}^{(0)\dag},
&D_R&=V_{d_R}V_{d_R}^{(0)\dag}.
\end{align}
One may also define the bare CKM matrix in the SCKM basis
\begin{align}
K^{(0)}=
V_{u_L}^{(0)}V_{d_L}^{(0)\dag}
=U_L^{\dag} K D_L.
\label{BLOeff:OM:BCRS:bareCKM}
\end{align}
The r\^ole played by the off--diagonal elements of $\mdbare$ in
section \ref{GA} is taken by the unitary matrices
$D_L$, $D_R$ and the bare CKM matrix $K^{(0)}$. For instance,
the bare CKM matrix elements $K^{(0)}_{ts}$ and $K^{(0)}_{cb}$,
in the bare SCKM basis, are
related to the corresponding matrix elements K, defined in the physical
SCKM basis via the relation
\begin{align}
K_{ts}^{(0)}=&
\frac{\left(\BLOfact\right)}{\left(\BLOfacg\right)}K_{ts}
-\left[
\epsilon_{RL}\frac{\left(\BLOfact\right)}{\left(\BLOfacg\right)}\frac{\mgl}{m_b}\xdrl\dlr
-\frac{\epsilon_{LL}\tanb}{\left(\BLOfacg\right)}\xdl\dll\right]K_{tb},
\label{BLOeff:OM:BCRS:CKMts}
\\
K_{cb}^{(0)}=&
\frac{\left(\BLOfact\right)}{\left(\BLOfacg\right)}K_{cb}
+\left[
\epsilon_{RL}\frac{\left(\BLOfact\right)}{\left(\BLOfacg\right)}\frac{\mgl}{m_b}\xdrl\dlr
-\frac{\epsilon_{LL}\tanb}{\left(\BLOfacg\right)}\xdl\dll\right]K_{cs}.
\label{BLOeff:OM:BCRS:CKMcb}
\end{align}
Where we have used the shorthand $\dll=\left(\dll\right)_{23}$
and $\dlr=\left(\dlr\right)_{23}$.
Strictly speaking, the uncorrected CKM matrix $K$ should appear
in the above relations, however, as discussed in section~\ref{MIA:EW},
the vertex and self energy corrections are negligible and one may,
to a good approximation, set $K=\Keff$. An interesting consequence
of this formula is that the matrix element $K_{ts}^{(0)}$ obtained by
diagonalising the bare Yukawa couplings $Y_{d,u}^{(0)o}$ can be zero in
the presence of general flavour mixing~\cite{BRP:CKM,OR2:bsg}. We will discuss
the consequences of this in section~\ref{NRes:kzero}.

As the two methods are practically equivalent, choosing between
them
essentially becomes a choice as to which is more suitable for
the problem at hand. In MFV scenarios the method presented
in~\cite{BCRS:bdec} is generally  more convenient as it is only
necessary to calculate the diagonal parts to the vertex and
self--energy corrections induced by gluino exchange. For example,
when using the method described in section~\ref{GA} the
correct form of~\eqref{MIA:EW:SLMFV} is only obtained when one
considers the off--diagonal gluino contributions as well as the
higgsino exchange diagram.

In the GFM scenario the situation is rather different. As the
off--diagonal contributions to the electroweak vertices
and non--renormalizable operators induced by gluino exchange are
evaluated anyway, the method described in this paper
can become more preferable. In particular, the
flavour diagonal contributions, induced by the exchange of
the supersymmetric particles, to the various non--renormalizable
operators applicable to the process under investigation no longer
have to be calculated, as the r\^ole played by the matrices
$D_L$ and $D_R$ is replaced by the off--diagonal elements
of $\mdbare$.

The second method,
presented by Dedes and Pilaftsis~\cite{DP:bdec}, concerns itself
mainly with CP violation, however it is essentially applicable to both
MFV and GFM CP conserving scenarios as well. After translating
to the physical SCKM basis their expression for the bare mass matrix reads
\begin{align}
m_d=\mdbare R,
\label{OM:DP:bare}
\end{align}
where $R$ is a $3\times 3$ matrix.
It is then possible to express the various Higgs interactions via
an effective Lagrangian expressed in terms of the physical quark
masses and the inverse of $R$. In the context of MFV and GFM
scenarios with mixing only in the LL
sector this parameterisation is sufficient. However taking into account
all sources of flavour violation yields the more general form
\begin{align}
m_d=\mdbare R_L+R_R\mdbare+R_G.
\label{BLOeff:OM:DP:bareGFM}
\end{align}
In the MIA,  the $3\times 3 $ matrices $R_L$, $R_R$ and $R_G$
may be decomposed in the following way
\begin{align}
R_L&=\unitmatrix+\epsg\tanb+\sum_{u}\epsy K^{\dag} Y_u^2 K\tanb
+\epsilon_{LL}\tanb\;\xdl\dll,\nonumber\\
R_R&=+\epsilon_{RR}\tanb\;\xdr\drr,
\label{BLOeff:OM:DP:RMats}\\
R_G&=-\epsilon_{RL}\;\mgl \xdrl\drl.\nonumber
\end{align}
Obtaining a solution for $\mdbare$ is therefore rather more complicated
than simply finding the inverse of $R$. Considering each element
of $\mdbare$ in turn, however, it is possible to replicate the results
for $\mdbare$ presented in subsection~\ref{MIA:mdb}.

%%%%%%%%%%%%%%%%%%%%%%%%%%%%%%%%%%%%%%%%%%%%%%%%%%%%%%%%%%%%%%%%%
\section{{\boldmath $\bsg$} Beyond the LO}
\label{bsg}
%%%%%%%%%%%%%%%%%%%%%%%%%%%%%%%%%%%%%%%%%%%%%%%%%%%%%%%%%%%%%%%%%

Of all the FCNC processes involving transitions between the $b$ and
$s$ quarks, $\bsg$ is currently the best understood both experimentally
and theoretically. The data being taken by B--factories such as BaBar
and BELLE, is leading to an increasing degree of precision for the
measurement of the branching ratio of the decay. The current world
average is~\cite{HFAG}
\begin{align}
{\rm BR}\left(\bsg\right)_{\rm exp}=\left(3.39^{+0.30}_{-0.27}\right)\times 10^{-4}.
\label{bsg:exp}
\end{align}
This value takes into account the most recent BELLE~\cite{BELLE:bsg}
and BaBar results~\cite{BABAR:bsg}.

The SM prediction for the branching ratio is based on a NLO calculation
that was completed in Refs.~\cite{GM:bsg,BCMU:bsg,HLP:bsg}, resulting in the
prediction\footnote{This result includes the NNLO effect induced
by using the running charm quark mass rather than the
pole mass when calculating the charm quark contributions
to the decay~\cite{GM:bsg}. A more formal NNLO analysis of these
effects has been performed in~\cite{AGHHP:bsg}}
\begin{align}
{\rm BR}\left(\bsg\right)_{\rm SM}=\left(3.70\pm0.30\right)\times 10^{-4}.
\label{bsg:SM}
\end{align}

It has been pointed out recently~\cite{Neubert:bsg} that, if one
applies a realistic cut--off for the photon
energy (rather than $E_{\gamma}>1/20\;m_b$), a dependence on
two additional energy scales appears when calculating
the branching ratio for the decay.
The first scale ($\mu_i=\sqrt{m_b\Delta}$) is associated
with the energy of the final hadronic state $X_s$, whilst the second
is dependent on the energy range under investigation
($\mu_0=\Delta=m_b-2 E_{\gamma}$). The perturbative uncertainties
associated with these scales are rather large and can lead
to a significant increase in the error associated with the
branching ratio. However, in exchange, the final result can
be compared directly with those determined directly from experiment
rather than model dependent extrapolations to $E_{\gamma}>1/20\;m_b$.

We should also briefly mention that steps are now
being taken towards a NNLO calculation~\cite{NNLO:bsg,AGHHP:bsg},
that should increase the accuracy of the SM prediction to roughly
5\%.

The effective Hamiltonian relevant to $\Delta F=1$ processes
such as the decay $\bsg$ is
\begin{equation}
\mathcal{H}_{eff}=-\frac{4 G_F}{\sqrt{2}} \Keff_{ts}{}^{\ast} \Keff_{tb} \sum_{i=1}^{8}
\left[C_i\left(\mu\right)\mathcal{O}_i\left(\mu\right)+C_i^{\prime}\left(\mu\right)\mathcal{O}_i^{\prime}\left(\mu\right)\right].
\label{bsg:Ham}
\end{equation}
The operators most relevant to the decay are
\begin{align}
\mathcal{O}_7&=\frac{e}{16\pi^2}
m_b\left(\overline{s}_L\sigma^{\mu\nu}b_R\right) F_{\mu\nu},
&\mathcal{O}_8&=\frac{g_s}{16\pi^2}
m_b\left(\overline{s}_L\sigma^{\mu\nu} T^a b_R\right) G_{\mu\nu}^a.
\label{bsg:Ops}
\end{align}
(The six remaining operators can be found, for example, 
in~\cite{GM:bsg}). The primed operators can be obtained via the
simple substitution $L\leftrightarrow R$. The contributions to the
primed operators are negligible in the SM. However, in more general
models, such as the MSSM with general flavour mixing, their effects
can no longer be ignored. As the primed and unprimed operators
do not interfere with one another, any new physics contributions
to $C_7^{\prime}$ and $C_8^{\prime}$ enter quadratically and
therefore act to increase the value of the branching ratio.
New physics contributions to $C_7$ and $C_8$, on the other hand,
interfere directly with the SM contribution and can lead to far more
varied effects.

The good agreement (within $1\sigma$) of the SM prediction and the current
experimental results allows one to place increasingly stringent bounds on the
effects and mass scale of new physics contributions. In doing so it
is important to include the effects of new physics at a similar
precision to the SM result. NLO matching conditions have been completed
for several extensions of the SM, for example, the NLO matching
conditions relevant to the 2HDM were presented
in~\cite{BG:bsg,CDGG1:bsg} whilst a more general analysis was
presented in~\cite{BMU:bsg}. Turning to the MSSM, however,
things become rather more complicated. A complete NLO calculation
would involve the evaluation two loop diagrams involving both gluons
and gluinos. For MFV this task is already underway and, for example,
the NLO matching conditions for the charged Higgs contribution have
been discussed in~\cite{BGY:bsg}. Theoretical calculations
have, thus far, concentrated on particular cases.
The calculation presented in~\cite{CDGG2:bsg}, for example,
considers a realistic but specific region of MSSM parameter space
where the charginos and lightest stop are relatively light compared
to the rest of the sparticle spectrum and $\tanb$ is rather small.
These results were extended to the large $\tanb$ regime
in~\cite{DGG:bsg,CGNW:bsg}. The same papers also considered
the dominant effects that occur BLO for generic SUSY scenarios
taking into account effects enhanced by large logs and $\tanb$.
These results were subsequently extended to include 
CP violation~\cite{DO:bsg}, additional
CKM effects~\cite{AGIS:bdec} and $\EWSym$ breaking and
electroweak effects in~\cite{BCRS:bdec}.

In the GFM scenario the LO matching conditions have been
known for some time~\cite{BBMR:bsg,BGHW:bsg,BGH:bsg} and, in a similar manner to
the MFV calculation, the NLO matching conditions have been
derived in the limit where the gluino decouples and $\tanb$
is small~\cite{BMU:bsg}. An extension of the analysis given in~\cite{DGG:bsg}
to the GFM scenario was presented in~\cite{OR:bsg,OR2:bsg} where
it was found that BLO corrections can play a large
r\^ole and can lead to a significant relaxation of the limits
placed on GFM parameters compared to a LO analysis. 
The aim of this section is to present calculations in the MIA
for both the electroweak and SUSY contributions to $\bsg$, allowing
one to easily determine the dominant effects that occur once GFM
is taken into account compared to MFV calculations, as well as
presenting the calculations detailed in~\cite{OR2:bsg} in a
more transparent way. In doing so we therefore adopt the
approximations discussed at the beginning of
section~\ref{MIA}.

%%%%%%%%%%%%%%%%%%%%%%%%%%%%%%%%%%%%%%%%%%%%%%%%%%%%%%%%%%%%%%%%%
\subsection{BLO Corrections to Electroweak Contributions in the MIA}
\label{bsg:EW}
%%%%%%%%%%%%%%%%%%%%%%%%%%%%%%%%%%%%%%%%%%%%%%%%%%%%%%%%%%%%%%%%%

As emphasised in section~\ref{GA}, including $\tanb$ enhanced
corrections to the charged and neutral Higgs vertices can lead to
large effects. Let us first consider how the LO charged Higgs
contributions to the decay $\bsg$ are altered once these effects 
are taken into consideration. Using the corrected vertices
\nqs{MIA:EW:CL}{MIA:EW:CR} the dominant MFV contribution
in the large $\tanb$ regime is given
by~\cite{DGG:bsg,AGIS:bdec,BCRS:bdec}
\begin{align}
\left(\delta^{H^-}C_{7,8}\right)^{\rm MFV}&=
\frac{1}{\BLOfact}
\left(1-\epsilon_s^\prime\tanb+\frac{\epsilon_Y^{\prime}\epsilon_Y (\ybpsckm Y_t\tan\beta)^2}{\BLOfacg}\right)F_{7,8}^{(2)}\left(\frac{m_t^2}{m_{H^+}^2}\right).
\label{bsg:EW:CHMFV}
\end{align}
The loop function $F_{7,8}^{(2)}$ is given in appendix~\ref{LF:bsg}.
Note that~\eqref{bsg:EW:CHMFV} includes the LO contribution
in addition to the BLO corrections. 
Turning to the degenerate mass limit we see that $\epsilon_s^{\prime}$
depends on the sign of $\mu$. In the phenomenologically favoured region
$\mu>0$, for example, the BLO corrections
induced by gluino exchange typically reduce the branching ratio
compared to a simple LO calculation. The higher order contribution
proportional to $Y_b^{(0)2}$, first pointed out in~\cite{AGIS:bdec},
on the other hand is dependent on the sign of the trilinear soft terms
and can therefore interfere destructively or constructively with the
(dominant) gluino correction depending on the model at hand. 
It should be noted that~\eqref{bsg:EW:CHMFV} only serves as a rough
approximation of the BLO effects and that the additional effects
arising, for example, from gaugino mediated exchange and $\EWSym$
breaking can lead to deviations from this idealised result in
some regions of parameter space~\cite{BCRS:bdec}. The dominant
effects that arise from these corrections may be included by performing
the substitutions presented in subsection~\ref{ImpApp}.

The GFM contributions to the charged Higgs vertex, discussed in
section~\ref{MIA:EW}, give rise to additional BLO corrections
to $C_{7,8}$ given by\footnote{From now on we shall
adopt the conventional shorthand
$\delta_{XY}^d=\left(\delta_{XY}^d\right)_{23}$.} 
\begin{align}
\left(\delta^{H^-}C_{7,8}\right)^{\rm GFM}=
-\frac{K_{tb}^{\ast}}{K_{ts}^{\ast}}
\Bigg\{
&\left[\frac{\epsilon_{LL}^{\prime}\xdl\tanb}{\left(\BLOfact\right)}
-\frac{\epsilon_Y^{\prime}\epsilon_{LL}\xdl(\ybpsckm\tan\beta)^2}{\left(\BLOfacg\right)\left(\BLOfact\right)}
\right]\dll\nonumber\\
&+\frac{\mgl}{\mbphys}\frac{\epsilon_{RL}\epsilon_Y^{\prime}\xdrl(\ybpsckm)^2\tanb}{\BLOfacg}\dlr\nonumber\\
&+\epsilon_{RL}^{\prime}\ydrl\ybpsckm\yspsckm\tanb\;\drl
\Bigg\}
F_{7,8}^{(2)}\left(\frac{m_t^2}{m_{H^+}^2}\right).
\label{bsg:EW:CHGFM}
\end{align}
The factor of $K_{tb}^{\ast}/K_{ts}^{\ast}$ that appears in front
of~\eqref{bsg:EW:CHGFM} reflects the fact that the flavour change
is governed by the GFM parameters $\dll$ and $\dlr$, rather
than the CKM matrix. The additional GFM contributions interfere
directly with the MFV corrections to the LO result and, depending
on the sign of $\dll$ or $\dlr$, can easily lead to large reductions
or enhancements of the MFV result. In addition to these contributions
to $C_{7,8}$ it is also possible, for non--zero $\drr$ and $\drl$, to
induce corrections to the primed Wilson coefficients
\begin{align}
\left(\delta^{H^-}C_{7,8}^{\prime}\right)^{\rm GFM}=
-\bigg[&\frac{\mbphys^2}{m_t^2}\frac{\tan^2\beta}{\left(\BLOfact\right)^2}
F_{7,8}^{(1)}\left(\frac{m_t^2}{m_{H^+}^2}\right)
\nonumber\\
&+\frac{[1-(\epsilon_s^{\prime}+\epsilon_Y^{\prime}(\ybpsckm)^2)\tanb]}
{\left(\BLOfact\right)}
F_{7,8}^{(2)}\left(\frac{m_t^2}{m_{H^+}^2}\right)\bigg]
\nonumber\\
&\times
\frac{K_{tb}^{\ast}}{K_{ts}^{\ast}}\left[\frac{\epsilon_{RR}\xdr\tanb}{\left(\BLOfact\right)}\drr
+\frac{\mgl}{\mbphys}
\epsilon_{RL}\epst\xdrl\drl
\right]
.
\label{bsg:EW:CHGFMpr}
\end{align}
As the LO contributions to the primed coefficients are suppressed by
factors of $m_s/m_b$ the dominant behaviour, once BLO corrections
are taken into account, is determined solely by GFM effects. Note
that these GFM effects persist even if the squarks decouple from
the theory.

It has been pointed out in Refs.~\cite{AGIS:bdec,BCRS:bdec} that
the corrected neutral Higgs vertex can also induce contributions to
$C_{7,8}$ through the diagram where a neutral Higgs boson and a bottom
quark undergoing a chirality flip are exchanged. In the limit of
MFV using the corrected vertex~\eqref{MIA:EW:SLMFV} one
obtains~\cite{AGIS:bdec,BCRS:bdec}
\begin{align}
\left(\delta^{H^0}C_{7,8}\right)^{\rm MFV}=&
-\frac{1}{36}\frac{\mbphys^2}{m_A^2}\tan^3\beta
\frac{\epsilon_Y Y_t^2}{\left(\BLOfacg\right)\left(\BLOfact\right)^2}.
\label{bsg:EW:NHMFV}
\end{align}
The $\tan^3\beta$ dependence of the Wilson coefficient is characteristic
of the corrected Higgs vertex~\eqref{MIA:EW:SLMFV} and can
compensate for the suppression factor $m_b^2/m_A^2$.

Turning to the effects of GFM contributions, using~\eqref{MIA:EW:SLGFM}
it is possible to induce additional corrections to $C_{7,8}$
\begin{align}
\left(\delta^{H^0}C_{7,8}\right)^{\rm GFM}=
&-\frac{1}{36}\frac{\mbphys^2}{m_A^2 K_{ts}^{\ast}K_{tb}}\tan^3\beta
\Bigg[
\frac{\epsilon_{LL}\xdl}{\left(\BLOfacg\right)\left(\BLOfact\right)^2}\dll
\nonumber\\
&+\frac{\mgl}{\mbphys}
\frac{\epsilon_{RL}\xdrl\epsilon_s}{\left(\BLOfacg\right)\left(\BLOfact\right)}
\dlr
\Bigg].
\label{bsg:EW:NHGFM}
\end{align}
In a similar manner to~\eqref{bsg:EW:NHMFV}, the GFM contributions arising
from neutral Higgs exchange vary as $\tan^3\beta$. The primed coefficients
also receive contributions if $\drl$ or $\drr$ are non zero
\begin{align}
\left(\delta^{H^0}C_{7,8}^{\prime}\right)^{\rm GFM}=
&-\frac{1}{36}\frac{\mbphys^2}{m_A^2 K_{ts}^{\ast}K_{tb}}\tan^3\beta
\Bigg[
\frac{\epsilon_{RR}\xdr}{\left(\BLOfact\right)^3}\drr
\nonumber\\
&+\frac{\mgl}{\mbphys}
\frac{\epsilon_{RL}\epst\xdrl}{\left(\BLOfact\right)^2}
\drl
\Bigg].
\label{bsg:EW:NHGFMpr}
\end{align}
%%

%%%%%%%%%%%%%%%%%%%%%%%%%%%%%%%%%%%%%%%%%%%%%%%%%%%%%%%%%%%%%%%%%
\subsection{BLO Corrections to SUSY Contributions in the MIA}
\label{bsg:SU}
%%%%%%%%%%%%%%%%%%%%%%%%%%%%%%%%%%%%%%%%%%%%%%%%%%%%%%%%%%%%%%%%%

The supersymmetric contributions to the decay $\bsg$ can proceed through
a number of channels. In MFV, the only SUSY contributions arise
from diagrams involving chargino exchange. Once GFM effects are taken into
account, additional diagrams arising from FCNC mediated by gluinos
and neutralinos can occur and give rise to contributions to both the
unprimed and primed Wilson coefficients. As the gluino contributions
are enhanced by factors of $\alpha_s$ (compared to the MFV contributions),
these effects are rather large and
can play an important r\^ole for even small deviations from MFV.

All four insertions give rise to contributions to either $C_{7,8}$
or their primed counterparts and it will be useful, for our purposes,
to decompose the overall gluino mediated contribution to the decay
as follows
\begin{align}
\delta^{\widetilde{g}}C_{7,8}=
\left(\delta^{\widetilde{g}}C_{7,8}\right)^{\rm MFV}
+\left(\delta^{\widetilde{g}}C_{7,8}\right)^{\rm LL}
+\left(\delta^{\widetilde{g}}C_{7,8}\right)^{\rm LR}
+\left(\delta^{\widetilde{g}}C_{7,8}\right)^{\rm RL}
+\left(\delta^{\widetilde{g}}C_{7,8}\right)^{\rm RR}
\label{bsg:SU:GLdec}
\end{align}
The primed coefficients and other SUSY contributions may be defined
in a similar manner. The dominant BLO corrections to the gluino
contributions, shown in \fig{bsg:SU:glfig},
\FIGURE[t!]{
\includegraphics[angle=0,width=0.4\textwidth]{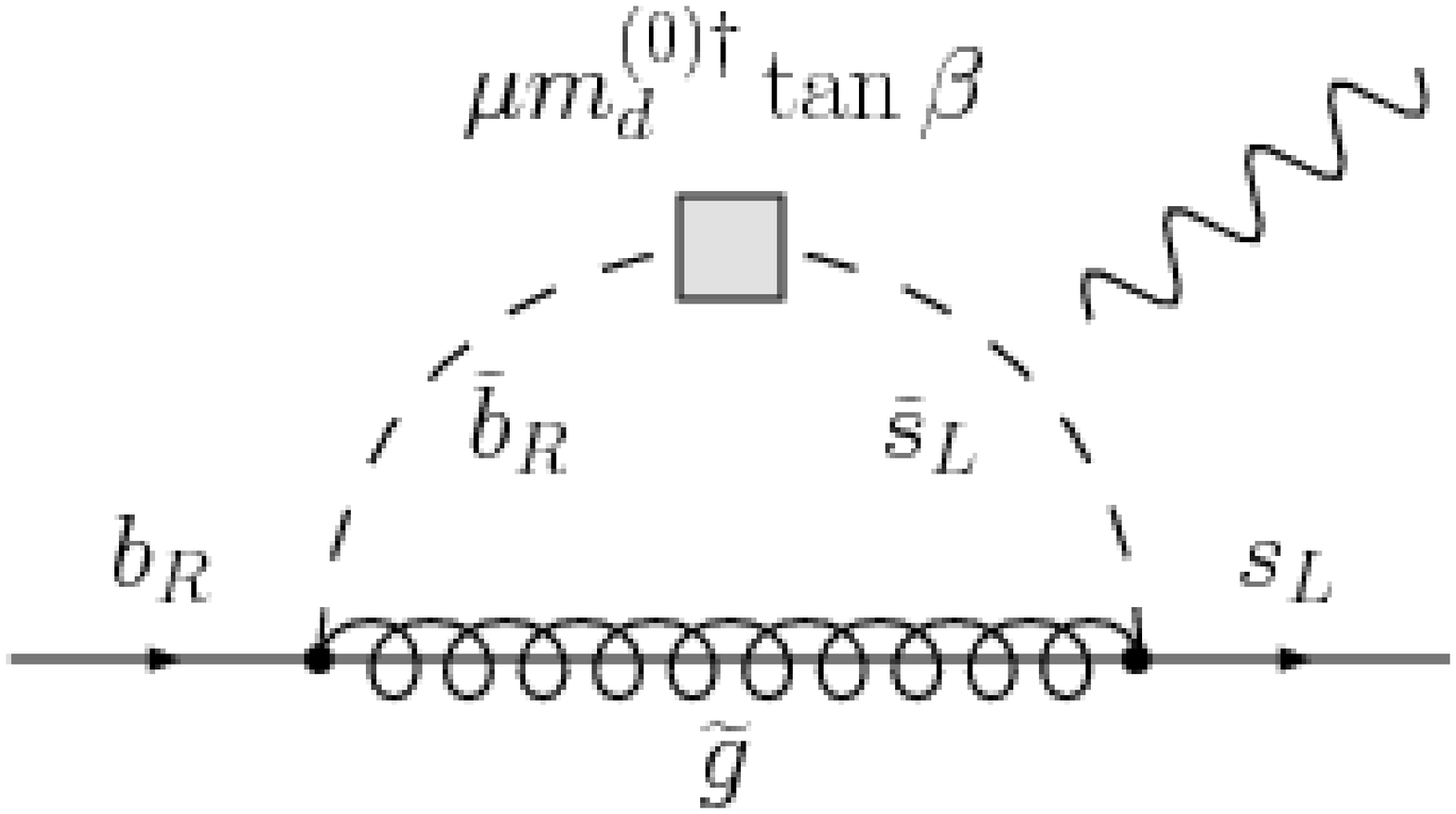}
\includegraphics[angle=0,width=0.4\textwidth]{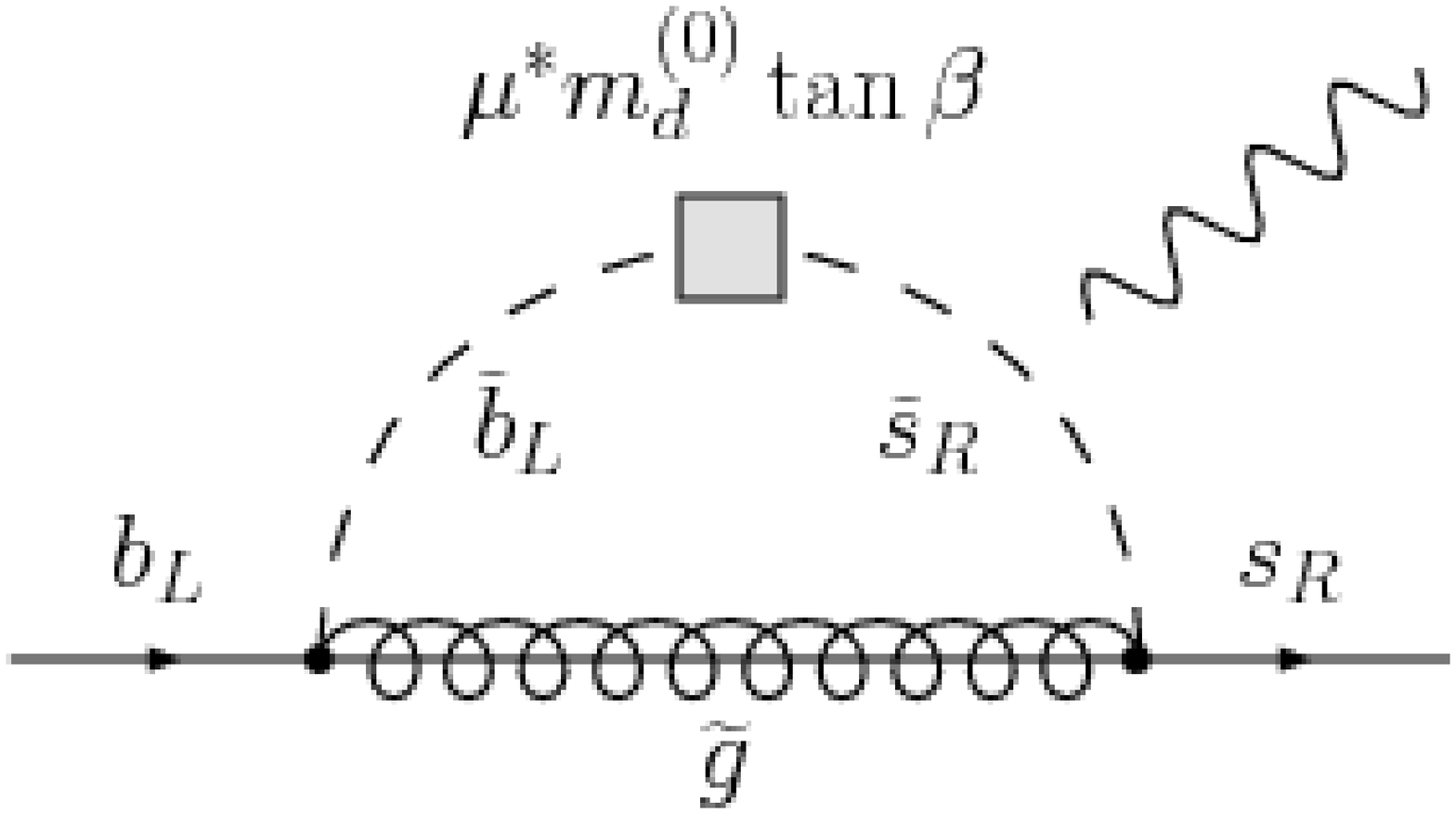}
\caption{BLO corrections to $C_7$ ($C_8$) and $C_7^{\prime}$
($C_8^{\prime}$) arising from gluino exchange, the photon
(gluon) line is attached in every possible manner.\label{bsg:SU:glfig}}
}
arise from the flavour violation
mediated by the off--diagonal elements of the bare mass matrix
and are proportional to $\mdbare\tanb$.

The MFV terms present in the bare mass matrix~\eqref{MIA:mdb:offdiag}
can lead to a correction to the gluino contribution of the
following form
\begin{align}
\left(\delta^{\widetilde{g}}C_{7,8}\right)^{\rm MFV}&=
-\frac{8}{3}
\frac{\alpha_s}{\alpha}
\frac{\sin^2\theta_W m_W^2}{\mgl^2}
\frac{\epsilon_{Y} Y_t^2\tan^2\beta}
{\left(\BLOfacg\right)\left(\BLOfact\right)}
\frac{\mu}{\mgl}
I_6^{[7,8]}\left(\xd\right).
\label{bsg:SU:gluMFV}
\end{align}
The loop functions $I_i^{[7,8]}\left(x\right)$ 
and $J_{i}^{[7,8]}\left(x\right)$, that appear throughout
this section, can be found in appendix~\ref{LF:bsg},
$\xd$ denotes the ratio
\begin{align}
\xd=\frac{\msq^2}{\mgl^2}
\label{bsg:SU:xddef}
\end{align}
where $\msq^2$ is a common mass of the quadratic soft
terms~\eqref{GA:SCKMst} (that is,
$\msq^2=\left(m_{d,LL}^2\right)_{ii}=\left(m_{d,RR}^2\right)_{ii}$).
Due to the $1/\mgl^3$ suppression of the amplitude it would be expected
that these effects are typically rather small when compared to the
additional BLO effects arising, for example, from the modified
charged Higgs vertex. However, the terms feature a dependence on
$\tan^2\beta$ in the numerator and
should be included if one wants to consider the effects of all $\tanb$
enhanced corrections. We should note that this correction
is entirely consistent with the definition of MFV presented
in~\cite{AGIS:bdec} and is a result of the transition between
the bare and physical super--CKM bases.

The GFM contributions to the Wilson coefficient $C_{7,8}$ arising
from gluino exchange are due mainly to the LL and the LR insertions. 
The contributions due to the RL and the RR insertions are
suppressed by factors of the strange quark mass and may be safely
ignored.

Contributions arising from the insertion $\dll$ are generated at
first and second order in the
MIA. At first order, the chirality flip is generated via the bottom
quark that appears in the operators~\eqref{bsg:Ops}. At second order, the
contribution arises from the diagram involving a diagonal LR
insertion, an LL insertion and a chirality flip on the gluino line.
This correction can play an important r\^ole for even moderate
$\tanb$, and for large $\tanb$ dominates the overall behaviour of the
contribution to $C_7$. If we ignore the effects generated by the
diagonal elements of the trilinear soft terms, we have
\begin{align}
\left(\delta^{\widetilde{g}}C_{7,8}\right)^{\rm LL}=&
\frac{8}{3 K_{ts}^{\ast}K_{tb}}
\frac{\alpha_s\sin^2\theta_W}{\alpha}
\left(\frac{m_W}{\mgl}\right)^2\xd\;
\Bigg\{
\bigg[
I_5^{[7,8]}\left(\xd\right)
\nonumber\\
-&\frac{\epsilon_{LL}\tan^2\beta}
{\left(\BLOfacg\right)\left(\BLOfact\right)}
\frac{\mu}{\mgl}
I_6^{[7,8]}\left(\xd\right)
\bigg]
+\frac{\mu}{2\mgl}\frac{\tanb}{\left(\BLOfact\right)}
\bigg[
J_6^{[7,8]}\left(\xd\right)
\nonumber\\
-&\frac{m_b^2\mu}{\mgl^3}\frac{\epsilon_{LL}\tan^2\beta}
{\left(\BLOfacg\right)\left(\BLOfact\right)}
J_5^{[7,8]}\left(\xd\right)
\bigg]
\Bigg\}\;\dll.
\label{bsg:SU:gluLL}
\end{align}
The first and second terms in square brackets that appear
in~\eqref{bsg:SU:gluLL} arise at the respective
order in the MIA. 
The chirally enhanced BLO term (that is proportional to
$I_6^{[7,8]}\left(\xd\right)$) occurs at first order
in the MIA, due to the off--diagonal elements of $\mdbare$.
This term tends to reduce the overall effect of the contribution that
arises at second order in the MIA (the term proportional to
$J_6^{[7,8]}\left(\xd\right)$) for $\mu>0$ (this is one of the
contributions to the focusing effect discussed in~\cite{OR:bsg,OR2:bsg}).
For $\mu<0$ on the
other hand, the two contributions interfere constructively and
increase the contribution to $C_{7,8}$ relative to a LO calculation.
The LO contribution proportional to $I_5^{[7,8]}\left(x\right)$ also
undergoes a similar reduction once BLO effects are taken into account.
However, in this case, the BLO correction is reduced by a factor of
$m_b^2/\mgl^2$.

For non--zero $\dlr$, the dominant contribution at LO arises
from the diagram involving an LR insertion and a chiral flip
on the gluino line. This contribution is therefore enhanced by
a factor of $\mgl/\mbphys$. Higher order contributions in the
MIA do not feature this enhancement and are typically rather
small. To second order in the MIA we have
\begin{align}
\left(\delta^{\widetilde{g}}C_{7,8}\right)^{\rm LR}&=
-\frac{8\sin^2\theta_W}{3 K_{ts}^{\ast}K_{tb}}
\frac{\alpha_s}{\alpha}
\left(\frac{m_W}{\mgl}\right)^2 \xd\;
\Bigg[
\frac{\mgl}{m_b}\frac{1}{\left(\BLOfacg\right)}
I_6^{[7,8]}\left(\xd\right)
\nonumber\\
&+\frac{\mu m_b}{2\mgl^2}
\frac{\tanb}{\left(\BLOfacg\right)\left(\BLOfact\right)}
J_5^{[7,8]}\left(\xd\right)
\Bigg]\;\dlr
\label{bsg:SU:gluLR}
\end{align}
Once again the first and second terms in the square bracket
arise at the respective order in the MIA.
From~\eqref{bsg:SU:gluLR} it can be seen that BLO effects can reduce,
or enhance, the dominant contribution due to the insertion $\dlr$ that
arises at first order in the MIA, depending on the sign of
$\epsilon_s$. In the phenomenologically
favoured scenario $\mu>0$, in particular, $\epsilon_s$ is positive
and BLO effects act to reduce the LO contribution to $C_{7,8}$.
The term that occurs at second order in the MIA tends to be
subdominant, compared with the chirally enhanced term that appears
at first order, but acts to reduce the contribution to $C_{7,8}$ further.

Turning to the primed sector, the corrections due to MFV, LL and LR
contributions are suppressed by factors of $m_s$ and are
rather small. We are therefore left with the contributions
arising from RL and RR insertions. 

The contribution due to the insertion $\drl$ to second order
in the MIA is given by
\begin{align}
\left(\delta^{\widetilde{g}}C_{7,8}^{\prime}\right)^{\rm RL}=&
-\frac{8\sin^2\theta_W}{3 K_{ts}^{\ast}K_{tb}}
\frac{\alpha_s}{\alpha}
\left(\frac{m_W}{\mgl}\right)^2\xd\;
\Bigg[
\frac{\mgl}{m_b}
\frac{\left(1+\epsilon_Y Y_t^2\tanb\right)}{\left(\BLOfact\right)}
I_6^{[7,8]}\left(\xd\right)
\nonumber
\\
&+\frac{\mu m_b}{2\mgl^2}
\frac{\tanb\left(1+\epsilon_Y Y_t^2\tanb\right)}{\left(\BLOfact\right)^2}
J_5^{[7,8]}\left(\xd\right)
\Bigg]\;\drl.
\label{bsg:SU:gluRL}
\end{align}
Comparing the above expression with~\eqref{bsg:SU:gluLR} we can
see that the form of the two are rather similar, the only differences
being the replacement of $\epsg$ with $\epst$ in the denominator of
\eqref{bsg:SU:gluLR} and multiplication by an overall factor of
$1+\epsilon_Y Y_t^2\tanb$. We therefore see that BLO corrections,
once again, act to reduce the contribution due to $\drl$ with respect
to a purely LO calculation if $\mu>0$.

Finally the contribution to $C_{7,8}^{\prime}$ arising from
non--zero $\drr$ has the form
\begin{align}
\left(\delta^{\widetilde{g}}C_{7,8}^{\prime}\right)^{\rm RR}&=
\frac{8\sin^2\theta_W}{3 K_{ts}^{\ast}K_{tb}}
\frac{\alpha_s}{\alpha}
\left(\frac{m_W}{\mgl}\right)^2 \xd\;
\Bigg\{
\left[
I_5^{[7,8]}\left(\xd\right)
-\frac{\epsilon_{RR}\tan^2\beta}
{\left(\BLOfact\right)^2}
\frac{\mu}{\mgl}
I_6^{[7,8]}\left(\xd\right)
\right]
\nonumber\\
+&\frac{\mu}{2\mgl}\frac{\tanb}{\left(\BLOfact\right)}
\left[
J_6^{[7,8]}\left(\xd\right)
-\frac{m_b^2\mu}{\mgl^3}\frac{\epsilon_{RR}\tan^2\beta}
{\left(\BLOfact\right)^2}
J_5^{[7,8]}\left(\xd\right)
\right]
\Bigg\}\;\drr
\label{bsg:SU:gluRR}
\end{align}
In a similar manner to~\eqref{bsg:SU:gluLL}, the chirally enhanced BLO
term arising at first order in the MIA, due to the off--diagonal
elements of the bare mass matrix, can once again affect the
dominant, chirally enhanced, LO contribution that arises at second
order in the MIA (the first and second order terms that
appear in the square brackets respectively).

We now turn to the chargino contributions to $C_{7,8}$ and
$C_{7,8}^{\prime}$, and use an analogous decomposition
to~\eqref{bsg:SU:GLdec}.
The dominant BLO corrections to chargino exchange arise
from the diagrams shown in \fig{bsg:SU:chfig}.
\FIGURE[t!]{
\parbox{1.0\textwidth}{
\begin{center}
\includegraphics[angle=0,width=0.4\textwidth]{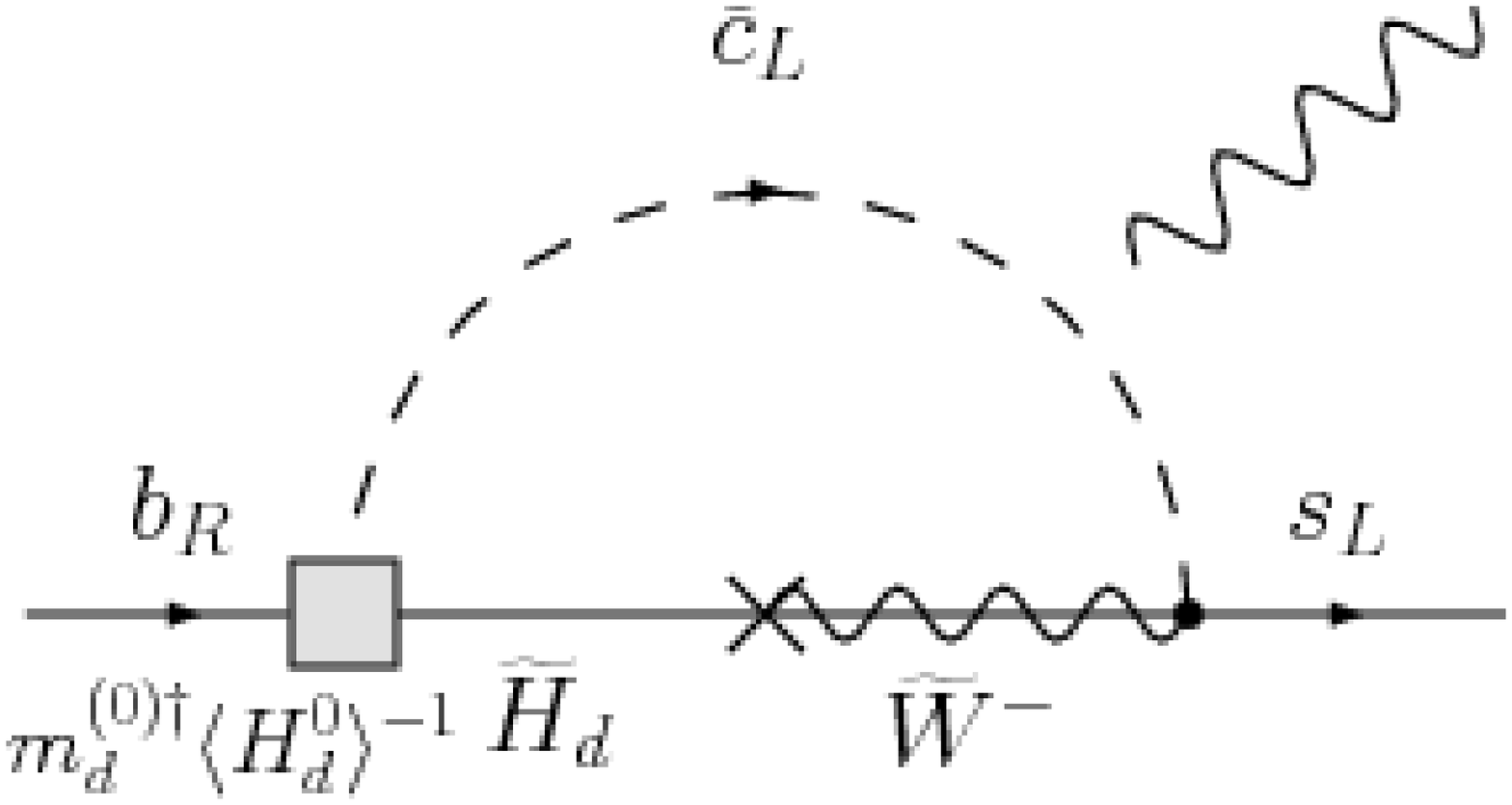}
\\
\includegraphics[angle=0,width=0.4\textwidth]{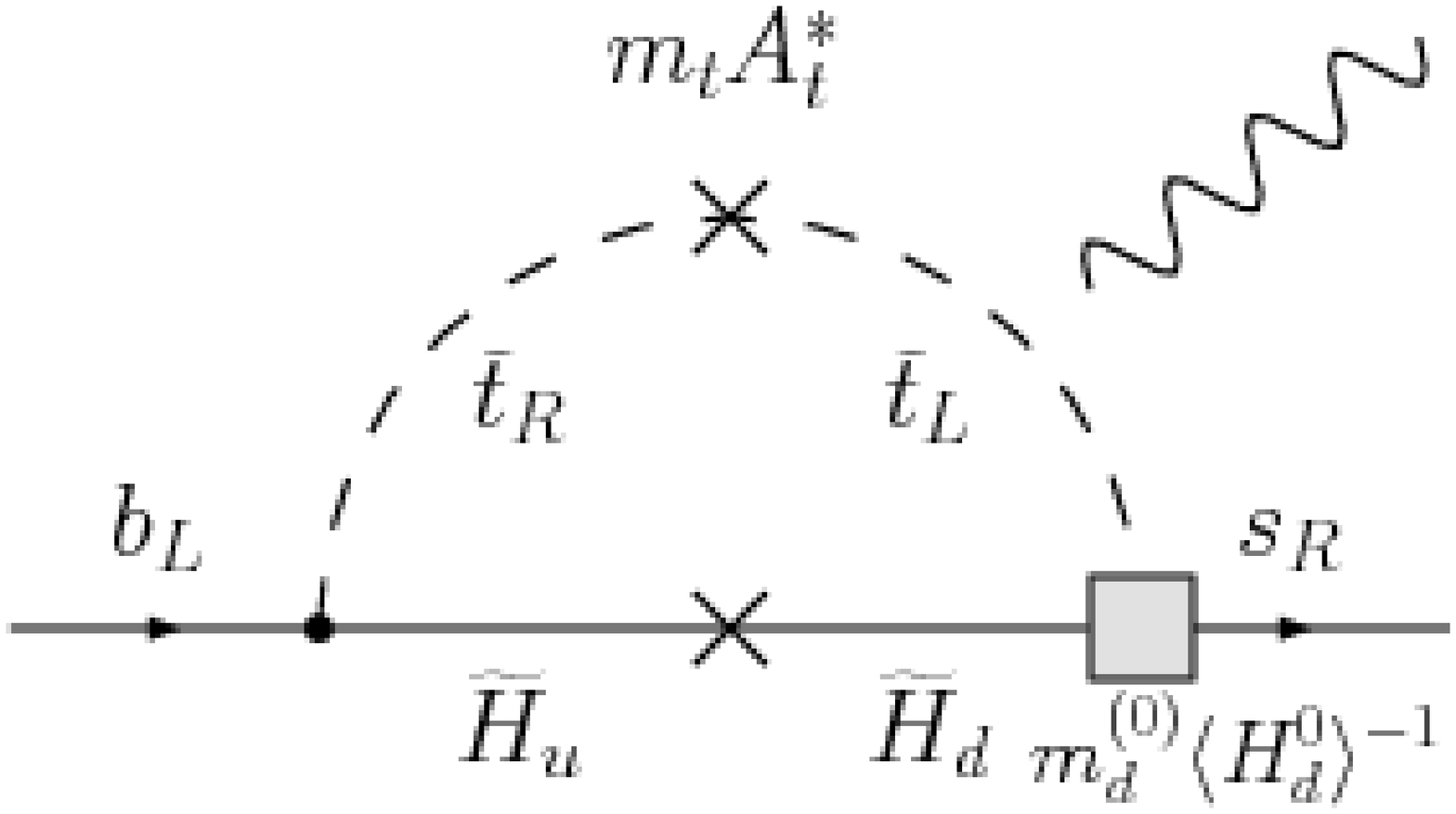}
\includegraphics[angle=0,width=0.4\textwidth]{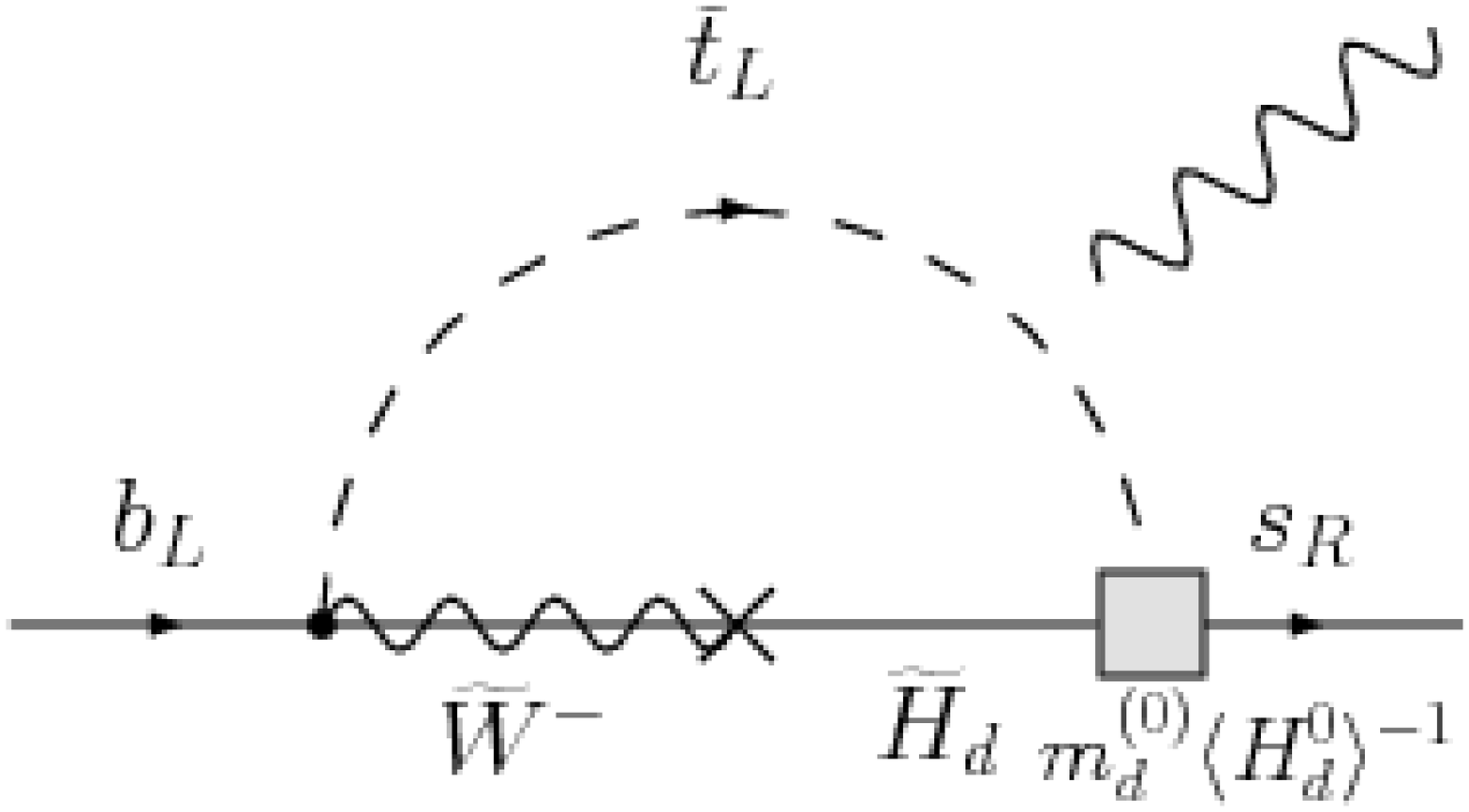}
\end{center}
}
\caption{BLO corrections to $C_7$ ($C_8$) and $C_7^{\prime}$ ($C_8^{\prime}$) arising from chargino exchange, the photon (gluon) line is
attached in every possible manner.\label{bsg:SU:chfig}}
}

The effect of BLO corrections
in the limit of MFV are well known~\cite{DGG:bsg,BCRS:bdec}.
In GFM models, however, it is possible that additional sources of
flavour violation in both the up and down squark sectors can significantly
alter the MFV result.

Contributions from flavour violation in the up squark sector
enter at LO and are therefore rather large. As we are chiefly
concerned with flavour violation in the down squark sector,
the only relevant source of flavour violation
in the up squark sector is the insertion $\ull$, which is related,
by ${\rm SU}\left(2\right)$ symmetry, to $\dll$~\eqref{GA:dllull}.
Flavour violation in
the down squark sector, on the other hand, only enters via BLO
effects induced by the off--diagonal elements of $\mdbare$. 

The contributions that arise due to LL insertions have the
form
\begin{align}
&\left(\delta^{\chi^-}C_{7,8}\right)^{\rm LL}=
\frac{K_{cs}^{\ast}}{K_{ts}^{\ast}}\sum_{a=1}^{2}
\bigg\{
\frac{m_W^2}{m_{\chi^-_a}^2}y_{a}
\bigg[
V_{a1}^{\ast}V_{a1} I_1^{[7,8]}\left(\yd^a\right)
\nonumber\\
&-\frac{m_{\chi^-_a}}{m_W}\frac{U_{a2}^{\ast} V_{a1}}{\sqrt{2}\cos\beta\left(\BLOfact\right)} I_2^{[7,8]}\left(\yd^a\right)
\bigg]\ull
\nonumber\\
&+\frac{K_{cs}}{K_{tb}}
\frac{m_W}{m_{\chi^-_a}}
\frac{U_{a2}^{\ast}V_{a1}\epsilon_{LL}\xd\tanb}{\sqrt{2}\cosb\left(\BLOfacg\right)\left(\BLOfact\right)}
\left(H_2^{[7,8]}\left(\yd^a\right)+\lambda^{[7,8]}\right)
\dll
\bigg\}.
\label{bsg:SU:chaLL}
\end{align}
Once again the loop functions $H_i^{[7,8]}$ and $I_i^{[7,8]}$ can be found in
appendix~\ref{LF:bsg}, the ratio $\yd^a$ is given by
\begin{align}
\yd^a=\frac{\msq^2}{m_{\chi^-_a}^2}.
\end{align}
The dominant contribution at LO for large $\tanb$ arises from the second
term in square brackets in~\eqref{bsg:SU:chaLL} that is enhanced
by factors of $m_{\chi^-_a}/m_W$ and $1/\cosb$.
This term, for $\mu>0$, has the
same sign as the gluino contribution~\eqref{bsg:SU:gluLL} and
acts to enhance the contributions due to flavour
violation in the LL sector. The BLO corrections to the Wilson coefficient
are reflected by a factor of $\left(\BLOfact\right)$ that appears
in the denominator of the second term in square brackets in~\eqref{bsg:SU:chaLL},
and the term that appears in the second line of~\eqref{bsg:SU:chaLL},
proportional to $\dll$. Both of these BLO corrections for $\mu>0$
act to decrease the LO contribution. For $\mu<0$, on the other
hand, both corrections act to increase the Wilson coefficient
relative to the LO result.

The LR insertions also contribute to the Wilson coefficient
\begin{align}
\left(\delta^{\chi^-}C_{7,8}\right)^{\rm LR}=&
-\frac{K_{cs}K_{cs}^{\ast}}{K_{tb}K_{ts}^{\ast}}
\sum_{a=1}^{2}
\frac{m_W}{\sqrt{2}m_{\chi^-_a}\cosb}V_{a1} U_{a2}^{\ast}
\frac{\mgl}{\mbphys}\frac{\epsilon_{RL}\xd}{\left(\BLOfacg\right)}
\nonumber\\
\times&\left(H_2^{[7,8]}\left(\yd^a\right)+\lambda^{[7,8]}\right)
\dlr.
\label{bsg:SU:chaLR}
\end{align}
In this case, GFM contributions only enter via BLO effects induced
by the off--diagonal elements of the bare mass matrix. For
$\mu>0$ and large $\tanb$, the contribution~\eqref{bsg:SU:chaLR}
has the opposite sign to the gluino contribution~\eqref{bsg:SU:gluLR} and
large cancellations are possible, which contribute significantly
to the focusing effect pointed out in~\cite{OR:bsg}.

Before proceeding with the results relevant to the primed sector
we should note that once again, the contributions to $C_{7,8}$
arising from RL and RR insertions are suppressed by factors of $m_s$.

In a similar manner to the corrections that arise from gluino
exchange, the only dominant contributions to the primed
coefficients arise from RL and RR insertions. The contribution
arising from RL insertions is given by
\begin{align}
\left(\delta^{\chi^-}C_{7,8}^{\prime}\right)^{\rm RL}=&
\sum_{a=1}^{2}\frac{K_{tb}^{\ast}}{K_{ts}^{\ast}m_{\chi^-_a}}
\Bigg[\frac{m_t^2 A_t}{2 m_{\chi^-_a}^2}V^{\ast}_{a2}U_{a2}\tanb I_2^{[7,8]}\left(\yd^a\right)
\nonumber\\
&-\frac{m_W}{\sqrt{2}\cosb}V^{\ast}_{a1}U_{a2}
\left(H_2^{[7,8]}\left(\yd^a\right)+\lambda^{[7,8]}\right)\Bigg]
\frac{\mgl}{\mbphys}\frac{\epsilon_{RL}\xd}{\left(\BLOfact\right)}\drl,
\label{bsg:SU:chaRL}
\end{align}
and the contribution arising from RR insertions is
\begin{align}
\left(\delta^{\chi^-}C_{7,8}^{\prime}\right)^{\rm RR}=&
-\sum_{a=1}^{2}\frac{K_{tb}^{\ast}}{K_{ts}^{\ast}m_{\chi^-_a}}
\Bigg[\frac{m_t^2 A_t}{2 m_{\chi^-_a}^2}V^{\ast}_{a2}U_{a2}\tanb I_2^{[7,8]}\left(\yd^a\right)
\nonumber\\
&-\frac{m_W}{\sqrt{2}\cosb}V^{\ast}_{a1}U_{a2}
\left(H_2^{[7,8]}\left(\yd^a\right)+\lambda^{[7,8]}\right)\Bigg]
\frac{\epsilon_{RR}\xd\tanb}{\left(\BLOfact\right)^2}\drr.
\label{bsg:SU:chaRR}
\end{align}
LO contributions to both coefficients arising from either MFV
or non--zero $\ull$ are suppressed by factors of $m_s/m_b$ and
are therefore rather small. At large $\tanb$, therefore, the BLO
effects dominate the behaviour of the chargino contributions
to the primed operators. We should also note that in a similar
manner to the case of the LR insertion both corrections 
(for $\mu>0$) have the opposite sign to the gluino
contributions~\nqs{bsg:SU:gluRL}{bsg:SU:gluRR}.

In the GFM scenario, neutralino contributions also play a r\^ole.
They can become especially important when, for example, the
gluino and chargino contributions partially cancel~\cite{BGH:bsg}.
However, as the
neutralino couplings~\nqs{SUSYver:NdL}{SUSYver:NdR} are
rather complicated, we shall refrain from presenting complete analytic
expressions for the coefficients in the
MIA. The overall effect of including BLO corrections is to
modify the Wilson coefficient in a similar manner to the gluino
contributions~\nqs{bsg:SU:gluLR}{bsg:SU:gluRR}. As an example,
the contribution arising from LR insertions to $C_{7,8}$ due to
bino exchange becomes
\begin{align}
\left(\delta^{\tilde{B}} C_7\right)^{\rm LR}=\frac{m_W^2\tan^2\theta_W}{9 K_{ts}^{\ast}K_{tb}\mbphys M_1}\frac{z_{\tilde{d}} I_4^{[7,8]}\left(z_{\tilde{d}}\right)}{\left(\BLOfacg\right)}\dlr.
\end{align}
The loop function $I_4^{[7,8]}$ can be found in appendix~\ref{LF:bsg}
whilst its argument is given by $z_{\tilde{d}}=\msq^2/M_1^2$.
The contributions induced by neutral gaugino--higgsino mixing, on the other,
hand are more complicated due to the appearance of the bare
quark mass matrix in the couplings~\nqs{SUSYver:NdL}{SUSYver:NdR}.

Finally, let us consider the effect of evolving these coefficients
from the SUSY matching scale $\mu_{SUSY}$ to the electroweak
scale $\mu_{W}$. Considering only the mixing between
$\mathcal{O}_7$ and $\mathcal{O}_8$ we have the LO
relation~\cite{MM:bsg}
\begin{align}
\delta C_7\left(\mu_W\right)=
\eta^{\frac{16}{21}}\delta C_7\left(\mu_{SUSY}\right)
+\frac{8}{3}\left(\eta^{\frac{2}{3}}-\eta^{\frac{16}{21}}\right)\delta C_8\left(\mu_{SUSY}\right).
\end{align}
The factors of $\eta$ in the above expression reflect the resummation
of leading logarithms and are given by
\begin{align}
\eta=\frac{\alpha_s\left(\mu_{SUSY}\right)}{\alpha_s\left(\mu_{W}\right)},
\end{align}
where $\alpha_s\left(\mu_{SUSY}\right)$ and $\alpha_s\left(\mu_{W}\right)$
should be evaluated with the QCD $\beta$ function relevant for six
active flavours.
If we retain only the first logarithm that appears when expanding
the factors $\eta$ we have
\begin{align}
\delta C_7\left(\mu_W\right)=\delta C_7\left(\mu_{SUSY}\right)
-\frac{4}{3\pi}\alpha_s\left(\mu_W\right)
\left[\delta C_7\left(\mu_{SUSY}\right)
-\frac{1}{3}\delta C_8\left(\mu_{SUSY}\right)\right]
\log\frac{\mu_{SUSY}^2}{\mu_{W}^2}.
\end{align}
From the above expression, it is apparent that the evolution of the
coefficients from $\mu_{SUSY}$ to $\mu_W$ acts to reduce
the overall SUSY contribution. In addition mixing with the
coefficient $C_8$ can also play a r\^ole. If $\delta C_8$ has the
opposite sign compared to $\delta C_7$, for example, further reductions
are possible.

Finally, let us provide a recipe for
implementing BLO corrections into existing LO gluino matching
conditions calculated in the MIA~\cite{EKRWW:bsg,LMSS:bsll}.
When performing LO calculations, one ignores the
corrections to the bare mass matrix, discussed in section~\ref{GA:FR},
and the $F$--terms are therefore assumed to be flavour diagonal.
Once one proceeds beyond the LO however, this assumption no longer
holds and additional off--diagonal elements in the squark mass
matrix appear, that are attributable to the factors of $\mdbare$
that appear in~\eqref{GA:Fterms}. In the LR sector, in particular,
the off--diagonal elements of the bare mass matrix are enhanced
by a factor of $\tanb$. The effect of these BLO corrections
can be included in existing LO expressions by making the
following substitutions
\begin{align}
\left(\dlr\right)_{ij}&\to\left(\dlr\right)_{ij}-\frac{\mu\tanb\left(\mdbare{}^{\dag}\right)_{ij}}{\sqrt{\left(m_{d,LL}^{2}\right)_{ii}\left(m^{2}_{d,RR}\right)_{jj}}},
\label{MIA:dLR1}
\\
\left(\drl\right)_{ij}&\to\left(\drl\right)_{ij}-\frac{\mu\tanb\left(\mdbare\right)_{ij}}{\sqrt{\left(m_{d,RR}^{2}\right)_{ii}\left(m^{2}_{d,LL}\right)_{jj}}}.
\label{MIA:dLR2}
%% \\
%% \left(\dll\right)_{ij}&\to\left(\dll\right)_{ij}+
%% \frac{\left(\mdbare{}^{\dag}\mdbare\right)_{ij}}
%% {\sqrt{\left(m_{d,LL}^{2}\right)_{ii}\left(m^{2}_{d,LL}\right)_{jj}}},
%% &\left(\drr\right)_{ij}&\to\left(\drr\right)_{ij}+
%% \frac{\left(\mdbare\mdbare{}^{\dag}\right)_{ij}}
%% {\sqrt{\left(m_{d,RR}^{2}\right)_{ii}\left(m^{2}_{d,RR}\right)_{jj}}}.
%% \label{MIA:dLL}
\end{align}
%%
%Typically the effects induced by the substitutions~\eqref{MIA:dLL}
%are $\mathcal{O}(m_b^2/m_{SUSY}^2)$ and may therefore be safely omitted.
%On the other hand, in the LR sector, the off--diagonal elements of the
%bare mass matrix are enhanced by a factor of $\tanb$ and can therefore
%play a rather large r\^ole.
Similar substitutions exist for the insertions $\dll$ and $\drr$,
however the effects are typically proportional to
$\mathcal{O}(m_b^2/m_{SUSY}^2)$ and may therefore be safely omitted.
In each of the substitutions given above, the off--diagonal elements of the
bare mass matrix are enhanced by a factor of $\tanb$ and can therefore
play a rather large r\^ole.
One should also note that the factor of
the down quark mass, that appears in flavour--diagonal LR mixings,
should also be replaced by the appropriate element of $\mdbare$.
Following this recipe, it is relatively easy to modify the LO
calculation presented in, for example,~\cite{EKRWW:bsg} and obtain results in
agreement with those presented above. We should note that, provided
one calculates $\mdbare$ to a similar precision, the
substitutions can be used to all orders in the MIA. One
may also use these substitutions in LO expressions for
the chargino and neutralino matching conditions, however
here one must also take into account the factors of the
bare mass matrix that appear in the chargino and neutralino
vertices. Finally, let us emphasise that the
substitutions~\nqs{MIA:dLR1}{MIA:dLR2} do not amount to a
redefinition of the $\delta$'s given in~\nqs{GA:dels1}{GA:dels2}
but merely represent the form of the BLO corrections to LO expressions.

%%%%%%%%%%%%%%%%%%%%%%%%%%%%%%%%%%%%%%%%%%%%%%%%%%%%%%%%%%%%%%%%%
\subsection{Full Calculation}
\label{bsg:Full}
%%%%%%%%%%%%%%%%%%%%%%%%%%%%%%%%%%%%%%%%%%%%%%%%%%%%%%%%%%%%%%%%%

With our results derived in the MIA in mind let us now outline
the steps required to implement BLO corrections in the general
framework outlined in section~\ref{GA} where the squark mass
matrices are diagonalised numerically. After performing the
iterative procedure described in subsection~\ref{num} one
may obtain the BLO charged Higgs and SM contributions by using
the matching conditions presented in~\cite{BG:bsg,CDGG1:bsg,BMU:bsg},
to account for the NLO gluon contribution, and using the
corrected vertices presented in section~\ref{GA} to evaluate the
LO matching conditions. As
discussed in subsection~\ref{MIA:EW} the corrections to the SM
contributions tend to be rather small and can generally be
ignored. The effect of the neutral Higgs contribution can also be
included by using the matching conditions presented
in~\cite{AGIS:bdec,BCRS:bdec}.
% SM + CH MC (45) and (46) of BMU:bsg
% NH MC (6.61) of BCRS:bdec

Turning to the supersymmetric contributions, BLO
corrections can be incorporated by using the supersymmetric vertices
detailed in section~\ref{GA:SU} and appendix~\ref{SUSYver} in the LO
matching conditions given in~\cite{BMU:bsg,OR2:bsg}. It should
be noted that one should use the unitary matrices $\Gamma_{d,u}$ that
are obtained at the end of the iterative procedure described in
subsection~\ref{num} when evaluating these contributions. 
One may also, with care, include the additional NLO effects that
appear if the gluino decouples by using the matching conditions
presented in~\cite{BMU:bsg}. Once one has evaluated all the
supersymmetric corrections they may be evolved from the SUSY
matching scale to the electroweak matching scale using the NLO
six flavour anomalous dimension matrix presented in~\cite{MM:bsg}.
% SUSY MC (E.13)--(E.37) in OR2:bsg
% Anom Dim. (4.12)--(4.15) in OR2:bsg

With the supersymmetric and electroweak contributions evaluated
at the scale $\mu_W$ it is finally possible to calculate the
branching ratio for $\bsg$ according to~\cite{GM:bsg}.
Let us note that this recipe is quite general 
and may be applied to any other process providing
the relevant matching conditions and anomalous dimension
matrices are available.

In summary, in this section we have included all the
relevant corrections that appear beyond the LO in the MIA. In
the electroweak sector, we have seen that the additional GFM contributions
to the charged and neutral Higgs vertices can lead to potentially
large modifications to the MFV results depending on the sign
of the insertion at hand.  The interplay and
cancellations between the various supersymmetric contributions,
has also been shown to be significant once one
proceeds BLO~\cite{OR:bsg} and leads to a focusing
effect in the phenomenologically interesting region $\mu>0$ and~$A_t<0$.
For the insertions $\dlr$, $\drl$ and $\drr$, in particular,
large cancellations can arise between the gluino and chargino
contributions to the decay. For the insertion $\dll$ on the other hand,
the cancellations play a more minor role, as a LO correction
to the chargino correction already exists (arising from the
insertion $\ull$) and tends to reduce
the effect of BLO corrections. Finally we have seen that the
RG evolution  of these corrections can lead to further reductions
to the supersymmetric contributions to the decay. 

%%%%%%%%%%%%%%%%%%%%%%%%%%%%%%%%%%%%%%%%%%%%%%%%%%%%%%%%%%%%%%%%%
\section{{\boldmath $\bsm$} Beyond the LO}
\label{bsm}
%%%%%%%%%%%%%%%%%%%%%%%%%%%%%%%%%%%%%%%%%%%%%%%%%%%%%%%%%%%%%%%%%

As B--factories do not run at the energy required to produce
large quantities of $B_s$ mesons, the best experimental constraint
on the rare decay $\bsm$ is provided by collider experiments. The
current 95\% confidence limits provided by the CDF~\cite{CDFpr:bsm} and
D{\O}~\cite{D0pr:bsm} experiments at the Tevatron
are\footnote{These results are preliminary, one can find the
most recent published results in~\cite{CDF:bsm,D0:bsm}.}
\begin{align}
&{\rm BR}\left(\bsm\right)_{\rm CDF}<2.0\times 10^{-7},\\
&{\rm BR}\left(\bsm\right)_{\rm D{\O}}<3.8\times 10^{-7}.
\end{align}
CDF and D{\O}~intend to further probe regions of up to 
$\mathcal{O}\left(10^{-8}\right)$. At the LHC, on the other hand,
branching ratios of up to $\mathcal{O}\left(10^{-9}\right)$ are
easily obtainable after
a few years of running at ATLAS, CMS and LHCb~\cite{LHC:bdec}. 

Theoretically, the decay $\bsm$ provides one of the cleanest
FCNC $\Delta F=1$ decay channels. It is described by the effective
Hamiltonian~\cite{BEKU2:bsm}
\begin{align}
\mathcal{H}_{eff}=-\frac{G_F\alpha}{\sqrt{2}\pi}\Keff_{ts}{}^{*} \Keff_{tb}\sum_{i}\left[C_i(\mu)\mathcal{O}_i(\mu)+C_i^{\prime}(\mu)\mathcal{O}_i^{\prime}(\mu)\right]
\label{bsm:Ham}
\end{align}
where the operators $\mathcal{O}_i$ are given by
\begin{align}
\mathcal{O}_{10}&=\left(\bar{s}_{\alpha}\gamma^{\mu}P_Lb_{\alpha}\right)\left(\bar{l}\gamma_{\mu}\gamma_5l\right),
&\mathcal{O}_{10}^{\prime}&=\left(\bar{s}_{\alpha}\gamma^{\mu}P_Rb_{\alpha}\right)\left(\bar{l}\gamma_{\mu}\gamma_5l\right),\nonumber\\
\mathcal{O}_{S}&=m_b\left(\bar{s}_{\alpha}P_Rb_{\alpha}\right)\left(\bar{l}l\right),
&\mathcal{O}_{S}^{\prime}&=m_s\left(\bar{s}_{\alpha}P_Lb_{\alpha}\right)\left(\bar{l}l\right),\nonumber\\
\mathcal{O}_{P}&=m_b\left(\bar{s}_{\alpha}P_Rb_{\alpha}\right)\left(\bar{l}\gamma_5l\right),
&\mathcal{O}_{P}^{\prime}&=m_s\left(\bar{s}_{\alpha}P_Lb_{\alpha}\right)\left(\bar{l}\gamma_5l\right).\label{bsm:Ops}
\end{align}
As the anomalous dimensions of all three operators are zero, the RG running
is trivial and the overall branching ratio for the process $l=\mu$ is
given by
\begin{align}
{\rm BR}\left(\bsm\right)=&\frac{G_F^2\alpha^2 m_{B_s}^2 f_{B_s}^2\tau_{B_s}}{64 \pi^3}\lvert \Keff_{ts}{}^{\ast}\Keff_{tb}\rvert^2\sqrt{1-4\hat{m}_{\mu}^2}
\nonumber\\
&\times\left[\left(1-4\hat{m}_{\mu}^2\right)\lvert F_S\rvert^2+\lvert F_P+2\hat{m}_{\mu}^2 F_{10}\rvert^2\right],
\label{bsm:br}
\end{align}
where $\hat{m}_{\mu}=m_{\mu}/m_{B_s}$ and the dimensionless quantities
$F_i$ are given by
\begin{align}
F_{S,P}&=m_{B_s}\left[\frac{C_{S,P}m_b-C_{S,P}^{\prime}m_s}{m_b+m_s}\right],
&F_{10}=C_{10}-C_{10}^{\prime}.
\nonumber
\end{align}
It should be noted from the above expression that the Wilson coefficient
of the operator $\mathcal{O}_{10}$ is helicity suppressed by a factor
of $\hat{m}_{\mu}^2$ as the $B_s$ meson has spin zero.
The SM contributions are only proportional to $\mathcal{O}_{10}$
as the Higgs mediated contributions to $\mathcal{O}_{S,P}$ can be
safely neglected. The SM contributions to $C_{10}$ have been evaluated
to NLO~\cite{NLO:bsm} resulting in the branching ratio~\cite{Buras:bsm}
\begin{align}
{\rm BR}\left(\bsm\right)_{\rm SM}=\left(3.46\pm1.5\right)\times 10^{-9}.
\label{bsm:smval}
\end{align}
The large uncertainty is mainly attributable to the hadronic matrix
element $f_{B_s}$ that can be determined from either lattice or
QCD sum rule calculations. A representative value 
for $f_{B_s}$ is\footnote{The current unquenched lattice calculations
for $f_{B_s}$ vary from $215\mev$~\cite{JLQCD:lat} up to
$260\mev$~\cite{WDGLS:lat} (for details of the errors associated
with these values we refer the reader to the original papers). As
the branching ratio is proportional to the square of
$f_{B_s}$ we decide to take a rather conservative estimate
for the magnitude of $f_{B_s}$ as recommended
in~\cite{Hashimoto:lat}.}
\begin{align}
f_{B_s}=\left(230\pm30\right)\mev.
\label{bsm:fbs}
\end{align}

In scenarios beyond the SM, particularly SUSY with large $\tanb$,
the contributions to $C_S$ and $C_P$ arising from neutral Higgs
penguins can become large and dominate
the underlying behaviour of the branching ratio. Studies in the
MSSM have focussed on both MFV~\cite{BK:bsm,BEKU1:bsm,IR1:bsm}
and GFM scenarios~\cite{CS:bsm,IR2:bsm,FOR1:bdec}
where the corrections induced by the corrected neutral Higgs
vertex~\nqs{MIA:EW:SLMFV}{MIA:EW:SLGFM} lead
the branching ratio for the decay to vary
as $\tan^6\beta$ (for a review see~\cite{Dedes:bdec}). At large
$\tanb$ it is therefore quite possible
for ${\rm BR}\left(\bsm\right)$ to be enhanced by a few a orders
of magnitude compared to the SM value, providing an ideal
signal for physics beyond the SM. The aim of this section
is to present analytic expressions for the contributions 
to $C_S$ and $C_P$ in the MIA that include the BLO
effects discussed in section~\ref{GA}. We also discuss
briefly the effect BLO contributions have on the subdominant
contributions that arise from box diagrams mediated by neutralinos
and charginos. Finally we discuss the application of the recipe
given at the end of the previous section to $\bsm$.

%%%%%%%%%%%%%%%%%%%%%%%%%%%%%%%%%%%%%%%%%%%%%%%%%%%%%%%%%%%%%%%%%
\subsection{BLO Corrections to Electroweak Contributions in the MIA}
\label{bsm:EW}
%%%%%%%%%%%%%%%%%%%%%%%%%%%%%%%%%%%%%%%%%%%%%%%%%%%%%%%%%%%%%%%%%

As above for $\bsg$, here we will present MIA calculations for
the contributions that arise due to the effective vertices presented
in section~\ref{GA:EW}.

Corrections to the effective $Z$ vertex~\ref{MIA:EW:ZVer}
lead to contributions to $C_{10}$ proportional to $\dlr$
\begin{align}
\left(\delta^{Z}C_{10}\right)^{\rm GFM}=&\frac{1}{3 K_{tb} K_{ts}^*}\frac{\alpha_s}{\alpha}
\frac{\mu m_b}{\mgl^2}
\frac{\tanb}{\left(\BLOfacg\right)\left(\BLOfact\right)}f_Z\left(\xd\right)\;\dlr
\label{bsm:EW:Z}
\end{align}
whilst there is a similar contribution proportional to $\drl$
to $C_{10}^{\prime}$. Terms proportional to $\dll$ may be generated
at third order in the MIA, which undergo BLO corrections from terms
that appear when using the bare mass matrix~\eqref{MIA:mdb:offdiag}.
As stated in the previous section however, $C_{10}$
and $C_{10}^{\prime}$ are both helicity suppressed by factors of
$m_{\mu}^2$ and their contribution to the overall branching ratio
is therefore typically limited to the low $\tanb$ regime.

The charged Higgs contributions to $C_{10}$ and $C_{10}^{\prime}$ 
arise from $Z$ penguins and box diagrams. The contribution
to $C_{10}$ is suppressed by a factor of $\cot^2\beta$ and,
whilst the contribution to $C_{10}^{\prime}$ is enhanced by a factor of
$\tan^2\beta$, it is suppressed by a factor of $m_s$.
Including BLO effects can alleviate these suppression factors.
However, as the Wilson coefficients are suppressed by a factor of
$m_{\mu}^2$ the overall effect is rather small.

The LO contributions to $C_{S,P}$ induced by neutral Higgs
penguins and box diagrams have been calculated in~\cite{BEKU1:bsm}.
For completeness we present them here
\begin{align}
\delta^{H^-}C_{S,P}^{(0)}&=\pm\frac{m_{\mu}\tan^2\beta}{4m_W^2\sin^2\theta_W}\frac{y\log y}{1-y},
&\delta^{H^-}C_{S,P}^{(0)\;\prime}&=\pm\left(\delta^{H^-}C_{S,P}^{(0)}\right)
\end{align}
\label{bsm:EW:CHLO}
where $y=m_t^2/m_{H^+}^2$.

BLO effects can be included by using the couplings
presented in subsection~\ref{MIA:EW} when calculating the matching
conditions. The largest correction induced by using these couplings
is attributable to the factor of $\left(\BLOfact\right)$ that
accompanies the right handed coupling of the charged Higgs.
This factor typically acts to reduce the charged Higgs
contribution relative to the LO prediction.
The GFM corrections to the vertex can act to replace
the factors of $K_{ts}^{\ast}K_{tb}$ that characterise
flavour change in the MFV contribution with the flavour violating
insertions~\nqs{GA:dels1}{GA:dels2}.

The contributions that arise due to the corrected neutral Higgs vertex
proceed via the penguin diagram shown in \fig{bsm:EW:NHfig}. 
\FIGURE[t!]{
\includegraphics[angle=0,width=0.30\textwidth]{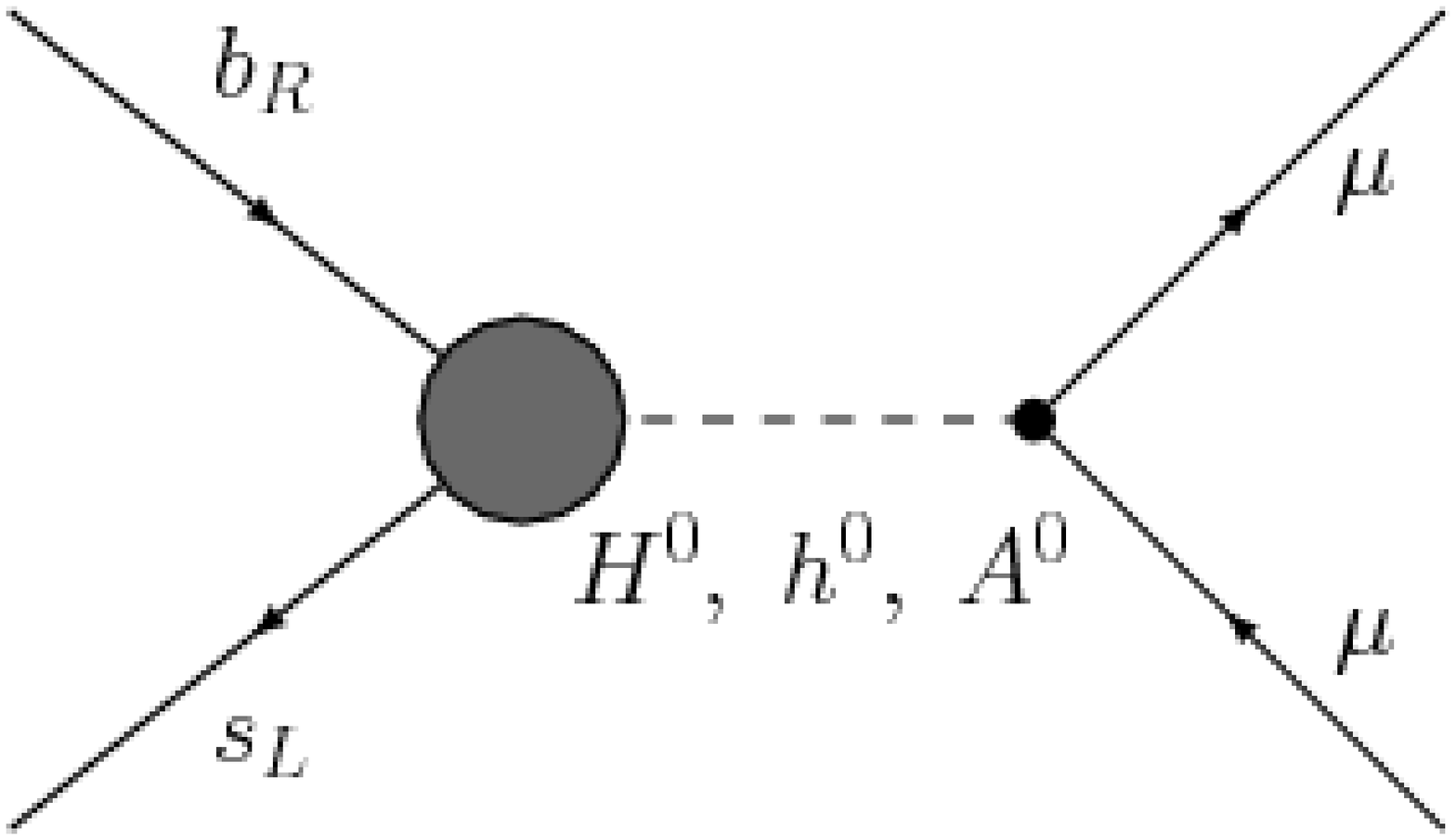}
\includegraphics[angle=0,width=0.30\textwidth]{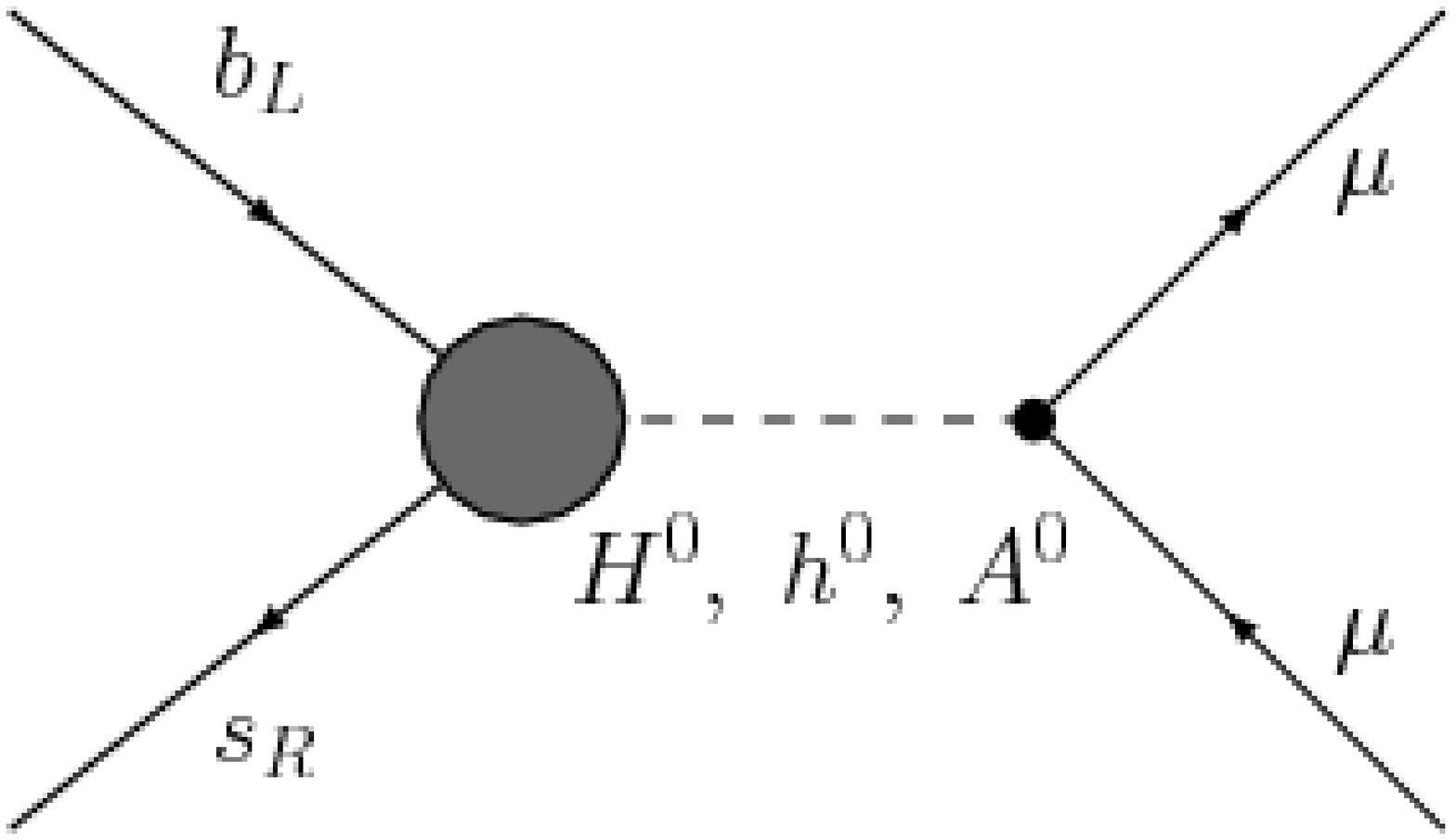}
\caption{The neutral Higgs penguin contributions to $C_{S,P}$ (on the
left) and $C_{S,P}^{\prime}$ (on the right).\label{bsm:EW:NHfig}}
}
Using~\eqref{MIA:EW:SLMFV} it is relatively easy to obtain the
dominant contribution arising from chargino exchange in the limit
of MFV~\cite{BK:bsm,IR1:bsm}
\begin{align}
\left(\delta^{H^0}C_{S,P}\right)^{\rm MFV}=&\pm\frac{m_{\mu}m_{t}^2 A_t\tan^3\beta}{4 m_W^2 m_A^2\mu\sin^2\theta_W}\frac{H_2\left(\yur,\yul\right)}{\left(\BLOfact\right)\left(\BLOfacg\right)}.
\label{bsm:EW:NHMFV}
\end{align}
$m_A$ denotes the mass of the pseudoscalar Higgs, whilst we decompose
the MFV and GFM contributions in a similar
manner to~\eqref{MIA:EW:NHDecomp}.
The most striking aspect of this contribution stems from the
factor of $\tan^3\beta$ that appears in the numerator of~\eqref{bsm:EW:NHMFV}
coupled
with a relatively weak dependence on the underlying SUSY mass scale.
It is therefore possible in SUSY models with large $\tanb$ that large
contributions to $C_{S,P}$ can occur even if the sparticle masses
are $\mathcal{O}\left(1\tev\right)$. (Provided, of course, that
the Higgs sector does not decouple, too.) The BLO contributions
in~\eqref{bsm:EW:NHMFV} are contained in the factors of $\left(\BLOfact\right)$
and $\left(\BLOfacg\right)$
that appear in the denominator. In the limit of degenerate sparticle
masses, for $\mu>0$, these corrections tend to reduce the neutral
Higgs contribution to $C_{S,P}$ compared with the LO limit
$\epsy,~\epsg\to 0$. 

The GFM corrections to the effective neutral Higgs vertex
\eqref{MIA:EW:SLGFM} contribute to $C_{S,P}$~\cite{FOR1:bdec}
\begin{align}
\left(\delta^{H^0}C_{S,P}\right)^{\rm GFM}=
&\pm\frac{4\alpha_s}{3\alpha}\frac{\mu
  m_{\mu}} {m_A^2} \frac{\tan^3\beta}{K_{tb}K_{ts}^{\ast}}
\Bigg[
\frac{\epsilon_{RL}}{\left(\BLOfacg\right)}
\frac{\xdrl}{m_b}
H_2(\xdr,\xdl)\dlr\nonumber\\
+&\frac{1}{\left(\BLOfacg\right)\left(\BLOfact\right)}
\frac{\xdl}{\mgl}
H_3(\xdr,\xdl,\xdl)\dll
\Bigg],
\label{bsm:EW:NHGFM}  
\end{align}
and the primed coefficients
\begin{align}
\left(\delta^{H^0}C_{S,P}^{\prime}\right)^{\rm GFM}=
& \frac{4\alpha_s}{3\alpha}
\frac{\mu m_{\mu}}{m_A^2}
\frac{\tan^3\beta}{K_{tb}K_{ts}^{\ast}}
\Bigg[
\frac{\left(\epsilon_{RL}+\frac{\mgl}{\mu}\epsilon_Y
    Y_t^2\right)}{\left(\BLOfact\right)} 
\frac{\xdrl}{m_s} 
H_2(\xdl,\xdr)\drl\nonumber\\
&+\frac{1}{\left(\BLOfact\right)^2}
\frac{\xdr}{\mgl}\frac{m_b}{m_s}
H_3(\xdl,\xdr,\xdr)\drr
\Bigg].
\label{bsm:EW:NHGFMpr}
\end{align}
The contributions arising from the insertions $\dll$ and $\drr$
are modified in a similar manner to the MFV contribution
\eqref{bsm:EW:NHMFV} and for $\mu>0$ undergo the familiar reduction
once BLO effects are taken into account. It should be noted that
once one proceeds beyond the approximation of setting electroweak
couplings to zero, and uses the substitutions gathered in
subsection~\ref{ImpApp} an additional contribution, proportional to
the insertion, $\dll$ arises
\begin{align}
\left(\delta^{H^0} C_{S,P}\right)^{\rm EW}=\mp\sum_{a=1}^2
\frac{m_{\mu}V_{a1}^{\ast}m_{\chi^-_a}U_{a2}}{2\sqrt{2}m_W m_A^2}\frac{\tan^3\beta}{K_{tb}K_{ts}^{\ast}\sinb\sin^2\theta_W}
\ydl^a H_2\left(\yul^a,\yul^a\right)\dll
\label{bsm:EW:NHGFMEW}
\end{align}
This LO correction tends to interfere destructively with the
dominant gluino contribution given in~\eqref{bsm:EW:NHGFM} and
is typically the largest contribution attributable to the
insertion $\dll$ once one proceeds beyond
the approximation described in section~\ref{MIA}.

Turning to the insertions $\dlr$ and $\drl$, their
appearance is a strictly BLO effect and can lead to
large deviations from LO results where the contributions
arising from the insertions accidentally cancel as we have shown
in~\cite{FOR1:bdec}.

%%%%%%%%%%%%%%%%%%%%%%%%%%%%%%%%%%%%%%%%%%%%%%%%%%%%%%%%%%%%%%%%%
\subsection{BLO Corrections to SUSY Contributions}
\label{bsm:SU}
%%%%%%%%%%%%%%%%%%%%%%%%%%%%%%%%%%%%%%%%%%%%%%%%%%%%%%%%%%%%%%%%%

As the gluino does not couple to the leptonic sector the supersymmetric
contributions to $\bsm$ take place via the box diagrams mediated
by chargino and neutralino exchange shown in \fig{bsm:EW:boxfig}.
\FIGURE[t!]{
\includegraphics[angle=0,width=0.30\textwidth]{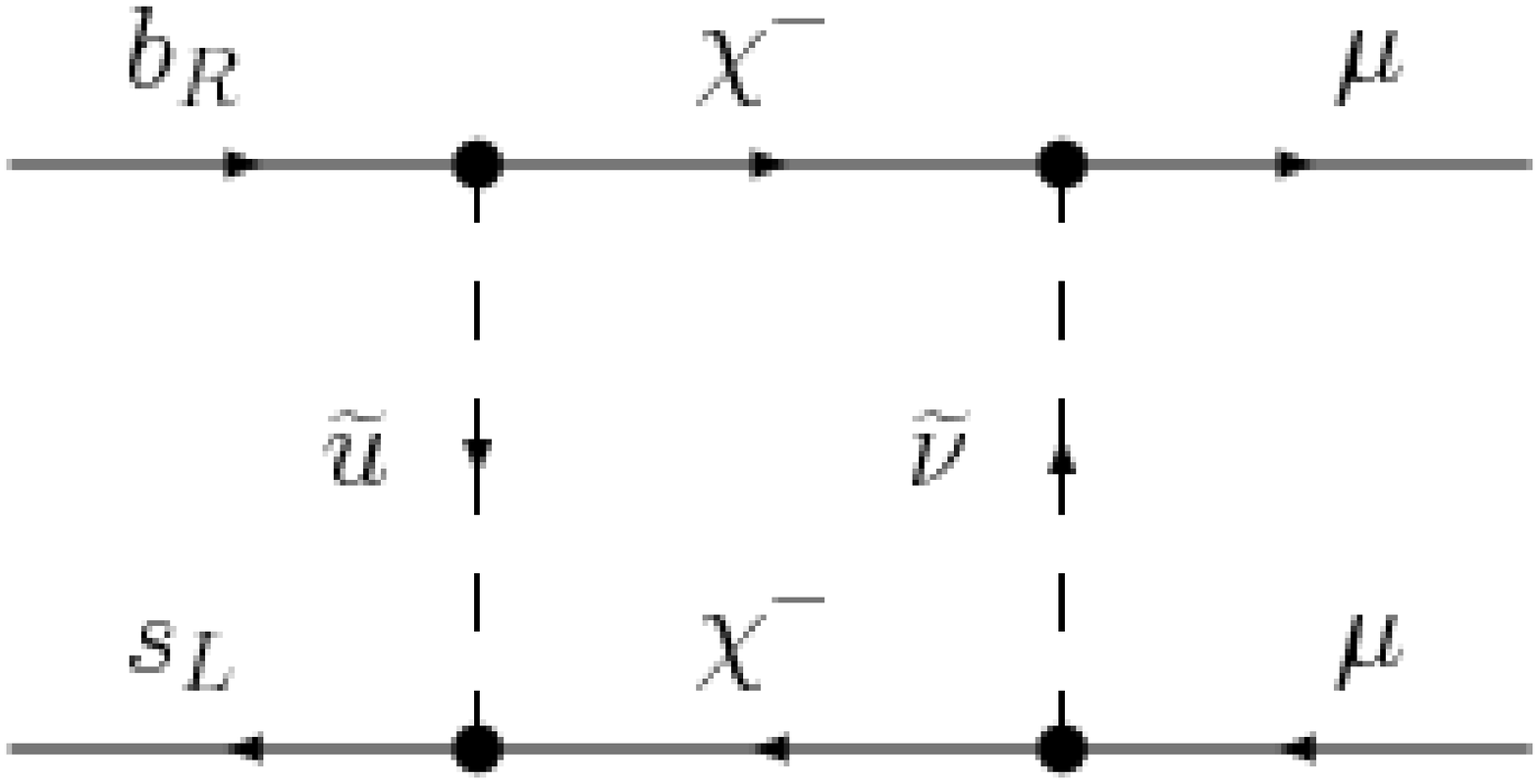}
\includegraphics[angle=0,width=0.30\textwidth]{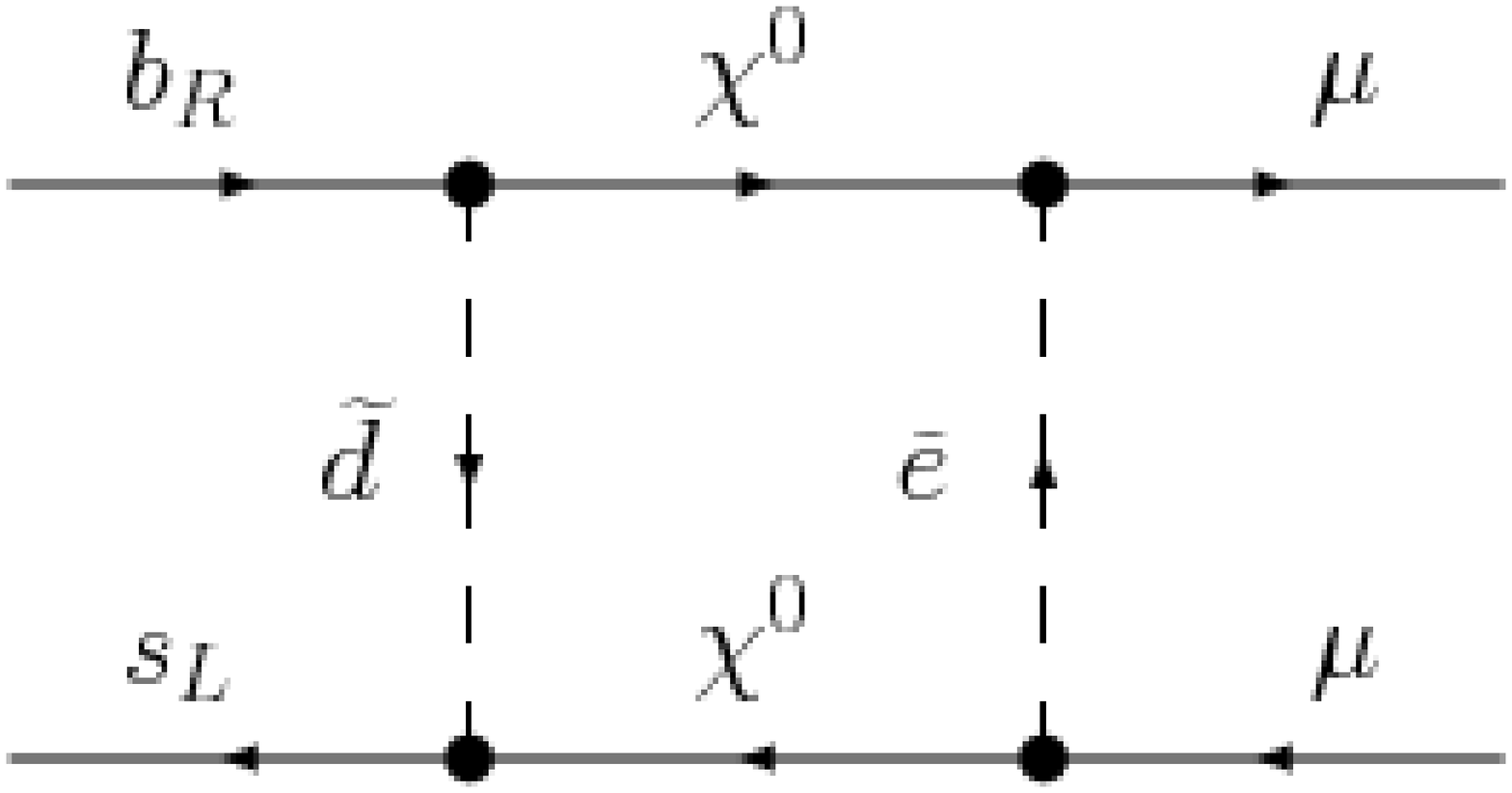}
\includegraphics[angle=0,width=0.30\textwidth]{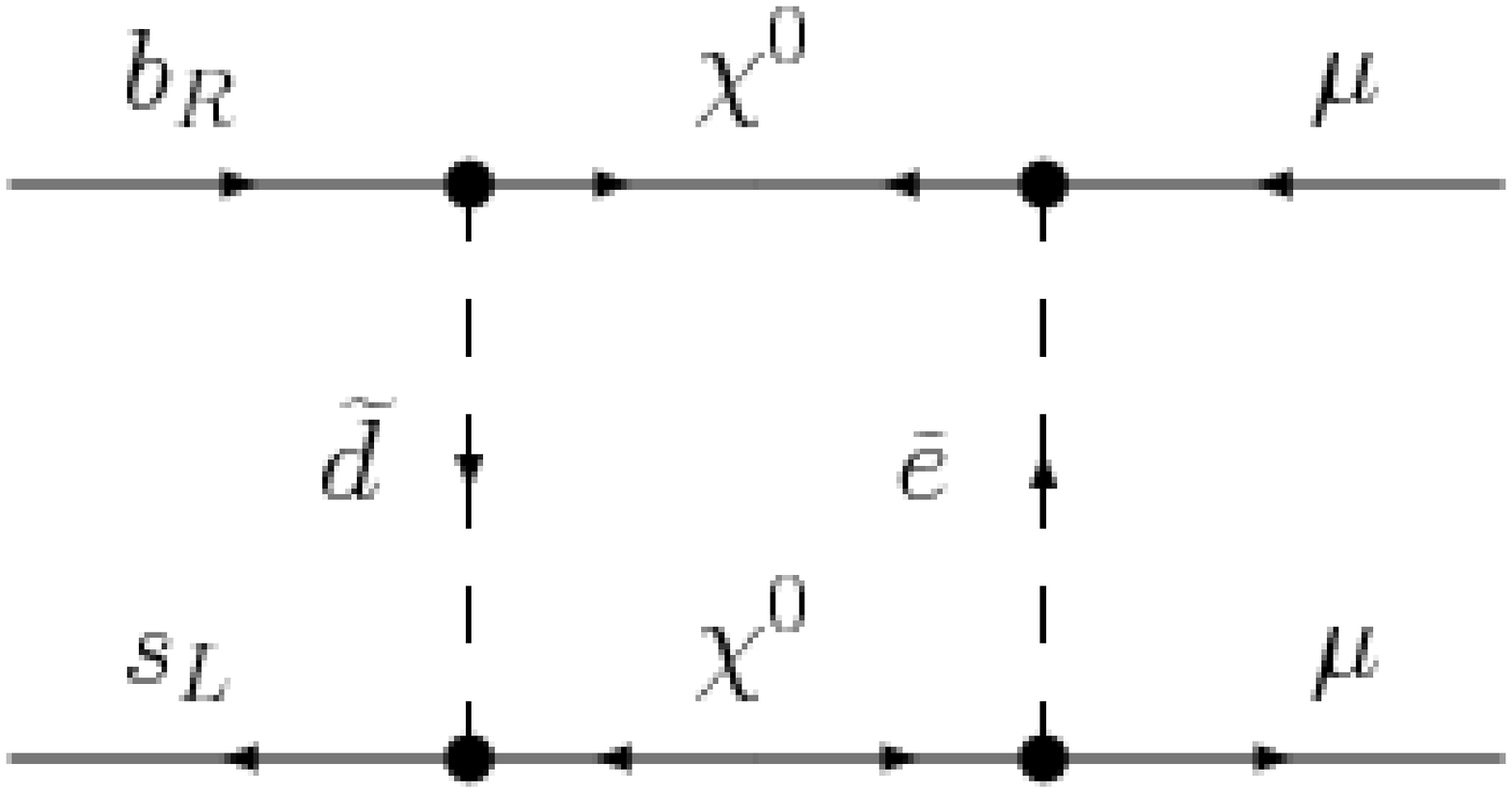}
\caption{The contributions to $\bsm$ arising from box diagrams
mediated by charginos and neutralinos.\label{bsm:EW:boxfig}}
}
Including BLO effects when calculating these contributions
introduces a dependence on the bare mass matrix, through the
vertices~\nqs{GA:SU:CdL}{GA:SU:CdR} and
\nqs{SUSYver:NdL}{SUSYver:NdR}. Sources of flavour violation can
therefore enter through either the chargino or the neutralino
contributions. However these contributions tend to scale
as $\tan^2\beta$ and, coupled with the underlying dependence
on $1/M_{SUSY}^2$, rather than $1/m_A^2$, are rather small when
compared to the effects induced by the neutral Higgs penguins
discussed in the previous subsection.

%%%%%%%%%%%%%%%%%%%%%%%%%%%%%%%%%%%%%%%%%%%%%%%%%%%%%%%%%%%%%%%%%
\subsection{Full Calculation}
\label{bsm:Full}
%%%%%%%%%%%%%%%%%%%%%%%%%%%%%%%%%%%%%%%%%%%%%%%%%%%%%%%%%%%%%%%%%

During our numerical analysis we shall 
follow the recipe described in subsection~\ref{bsg:Full}
and use the expressions gathered in appendix~\ref{RadCor}
to evaluate the corrections to the various effective
vertices. We therefore include higher order terms in the MIA
as well as any (subdominant) $\EWSym$ breaking effects.
Concerning the SM and charged Higgs contributions to the
decay we use the matching conditions gathered in~\cite{BBKU:bsm}
to evaluate the NLO gluon contribution. The contributions that
arise from SUSY boxes are given in~\cite{HLYZ:bsm,BEKU1:bsm}.
% e.g. 5.11 in BEKU1:bsm for chargino box.

In conclusion, in this section we have discussed how the BLO
effects discussed in section~\ref{GA} alter LO contributions
to $\bsm$. In the electroweak sector, these corrections typically manifest
themselves as factors of either $\left(\BLOfact\right)$
or $\left(\BLOfacg\right)$ that act to reduce both MFV and
GFM contributions to the decay. In addition, we have seen
that new flavour structures, absent at LO, can appear once
BLO effects are taken into account, leading to potentially
large deviations from LO calculations.

%%%%%%%%%%%%%%%%%%%%%%%%%%%%%%%%%%%%%%%%%%%%%%%%%%%%%%%%%%%%%%%%%
\section{{\boldmath $\bbb$} Mixing Beyond the LO}
\label{bbb}
%%%%%%%%%%%%%%%%%%%%%%%%%%%%%%%%%%%%%%%%%%%%%%%%%%%%%%%%%%%%%%%%%

The final process that we will consider concerns mixing in the
$B_s$ meson system. In a similar manner to the neutral kaon and
$B_d$ systems mixing can occur between the $B_s$ and $\bar{B}_s$
mesons via $\Delta F=2$ loop diagrams. In contrast to the
neutral kaon and $B_d$ systems, however, the mass difference
$\delmbs$ between the physical states formed from the two
mesons has so far remained unobserved. The best bound provided
by experiment is currently~\cite{HFAG}
\begin{align}
\delmbs^{\rm exp}>14.5 \ps^{-1}.
\label{bbb:exp}
\end{align}
In future, the experiments at the Tevatron intend to increase this limit
by 20--30\%~\cite{Tev:exp} whilst even after a year of low luminosity running
ATLAS, CMS and LHCb intend to place limits of $30{\rm ps}^{-1}$,
$26{\rm ps}^{-1}$ and $48{\rm ps}^{-1}$~\cite{LHC:bdec}
respectively on $\delmbs$.
Comparing these limits
with the NLO Standard Model prediction~\cite{BJW:bbm,BCJL:bbm}
\begin{align}
\delmbs^{\rm SM}=\left(18.0\pm 3.7\right) \ps^{-1},
\label{bbb:SM}
\end{align}
it can be seen that the full range of values allowed by the SM can be
probed in a relatively short time after data taking has commenced
at the LHC.

The effective Hamiltonian that most generally describes $\bbb$ mixing
effects is given by~\cite{BMU:bbm,BJU:bbm}
\begin{equation}
\mathcal{H}_{eff}=
\frac{G_F^2}{16\pi^2}m_W^2\left(\Keff_{tb}{}^*\Keff_{ts}\right)^2
\sum_{i}C_i\left(\mu\right)\mathcal{O}_i\left(\mu\right).
\label{bbb:Ham}
\end{equation}
In the SM the only non-negligible contribution is proportional to the
operator
\begin{align}
\mathcal{O}^{VLL}&=
(\bar{b}^{\alpha}\gamma_{\mu}P_L s^{\alpha})(\bar{b}^{\beta}\gamma^{\mu}P_L s^{\beta}).
\label{bbb:SMOps}
\end{align}
However, in the presence of any source of new physics it is possible to
induce additional contributions to the operators
\begin{align}
\mathcal{O}^{LR}_{1}&=(\bar{b}^{\alpha}\gamma_{\mu}P_L s^{\alpha})(\bar{b}^{\beta}\gamma^{\mu}P_R s^{\beta}),
&\mathcal{O}^{LR}_{2}&=(\bar{b}^{\alpha}P_L s^{\alpha})(\bar{b}^{\beta}P_R s^{\beta}),\label{bbb:Ops1}\\
\mathcal{O}^{SLL}_{1}&=(\bar{b}^{\alpha}P_L s^{\alpha})(\bar{b}^{\beta}P_L s^{\beta}),
&\mathcal{O}^{SLL}_{2}&=(\bar{b}^{\alpha}\sigma_{\mu\nu}P_L s^{\alpha})(\bar{b}^{\beta}\sigma^{\mu\nu}P_L s^{\beta}),\label{bbb:Ops2}
\end{align}
as well as the parity flipped operators $\mathcal{O}^{VRR}$ and
$\mathcal{O}^{SRR}_{i}$ that can be obtained by substituting $P_L$
with $P_R$ in~\eqref{bbb:SMOps} and~\eqref{bbb:Ops2}.
The mass difference $\delmbs$ may then be evaluated by taking the matrix
element
\begin{align}
\delmbs=2\lvert\langle\bar{B}^0_s\lvert\mathcal{H}_{eff}\rvert B^0_s\rangle\rvert,
\label{bbb:delmb}
\end{align}
where $\langle\bar{B}^0_s\lvert\mathcal{H}_{eff}\rvert B^0_s\rangle$ is given
by
\begin{align}
\langle\bar{B}^0_s\lvert\mathcal{H}_{eff}\rvert B^0_s\rangle=&
\frac{G_F^2}{48\pi^2}m_W^2 m_{B_s} f_{B_s}
\left(\Keff_{tb}{}^{\ast}\Keff_{ts}\right)^2
\sum_{i}P_i C_i\left(\mu_W\right),
\label{bbb:matel}
\end{align}
$m_{B_s}$ denotes the mass of the $B_s$ meson, whilst
$f_{B_s}$ is given in~\eqref{bsm:fbs}.
The coefficients $P_i$ contain the effects due to RG running between
$\mu_t$ and $\mu_b$ as well as the relevant hadronic matrix element
for the operator in question. Using the $\overline{{\rm MS}}$ lattice
calculation~\cite{lattice:bbm} the coefficients $P_i$ have the form
\begin{align}
P^{VLL}_1=&0.73,
&P^{LR}_1=&-1.97,
&P^{LR}_2=&2.50,
&P^{SLL}_1=&-1.02,
&P^{SLL}_2=&-1.97.
\label{bbb:pcoef}
\end{align}
where we have taken $\mu_b=4.25\gev$ and $\mu_W=m_t(m_t)$. The coefficients
$P^{VRR}_1$, \etc,~may be obtained by simply exchanging $L$ and $R$.
One interesting aspect of~\eqref{bbb:pcoef} is that QCD effects
act to enhance the contributions arising from the scalar operators
relative to the SM operator $C^{VLL}$.

The new physics contributions to neutral meson mixing have been
discussed extensively in the literature. The NLO charged Higgs
contributions to $C^{VLL}$, for instance, have been known
for some time now~\cite{UKJS:bbm}. Turning to the MSSM,
the NLO matching conditions have, in a similar
manner to the decay $\bsg$, only been derived for some special
cases in the MFV limit~\cite{KS:bbm}. Most analyses have
therefore focussed on using LO matching conditions.
For example, in~\cite{Ciuchini:bbm} the LO gluino matching
conditions, relevant for mixing in the kaon system, were used
alongside the NLO anomalous dimension matrix and lattice
matrix elements to place limits on the insertions
governing non--minimal flavour violation between down and
strange squarks. This analysis has since been extended
to the $B_d$~\cite{It:bbm} and $B_s$~\cite{CFMS:bdec} meson
systems. In the $\bbb$ mixing
system, in particular, large contributions to $\delmbs$ of
up to $120\ps^{-1}$ have been found to arise. In the large
$\tanb$ regime another possibility for large contributions
arises from the inclusion of the effects induced by
the neutral Higgs penguin. Although strictly speaking an
NLO contribution the corrections arising from flavour violation mediated
by two neutral Higgs penguins have been shown to vary
as $\tan^4\beta$. Such corrections have been analysed
in the context of MFV~\cite{HPT:bbm,CS:bsm,BCRS:bbm,IR1:bsm,BCRS:bdec} and
GFM~\cite{HPT:bbm,CS:bsm,IR2:bsm,FOR1:bdec} and can similarly
induce rather large corrections to $\delmbs$. 

The aim of this section is to discuss how the inclusion
of BLO corrections affect the electroweak
and supersymmetric contributions to the Wilson coefficients
associated with the operators~\nqs{bbb:SMOps}{bbb:Ops2}. In
particular, we shall discuss, in detail, the
contributions that arise from double Higgs penguins
in the GFM scenario and the dominant BLO corrections to
the existing LO gluino matching conditions. When presenting our
analytic expressions we shall use the MIA and follow the
approximation outlined at the beginning of section~\ref{MIA}.
We therefore only include the effects induced by gluino and higgsino
exchange and ignore additional corrections induced by electroweak
gaugino exchange and $\EWSym$ breaking. One may include the
additional contributions that appear once one proceeds
beyond this approximation by using the substitutions gathered
in subsection~\ref{ImpApp}. Finally, we shall briefly discuss
the application of the recipe given in subsection~\ref{bsg:Full} to
$\bbb$ mixing.

%%%%%%%%%%%%%%%%%%%%%%%%%%%%%%%%%%%%%%%%%%%%%%%%%%%%%%%%%%%%%%%%%
\subsection{BLO Corrections to Electroweak Contributions in the MIA}
\label{bbb:EW}
%%%%%%%%%%%%%%%%%%%%%%%%%%%%%%%%%%%%%%%%%%%%%%%%%%%%%%%%%%%%%%%%%

%%
\FIGURE[t!]{
\parbox{0.8\textwidth}{
\begin{center}
\includegraphics[angle=0,width=0.40\textwidth]{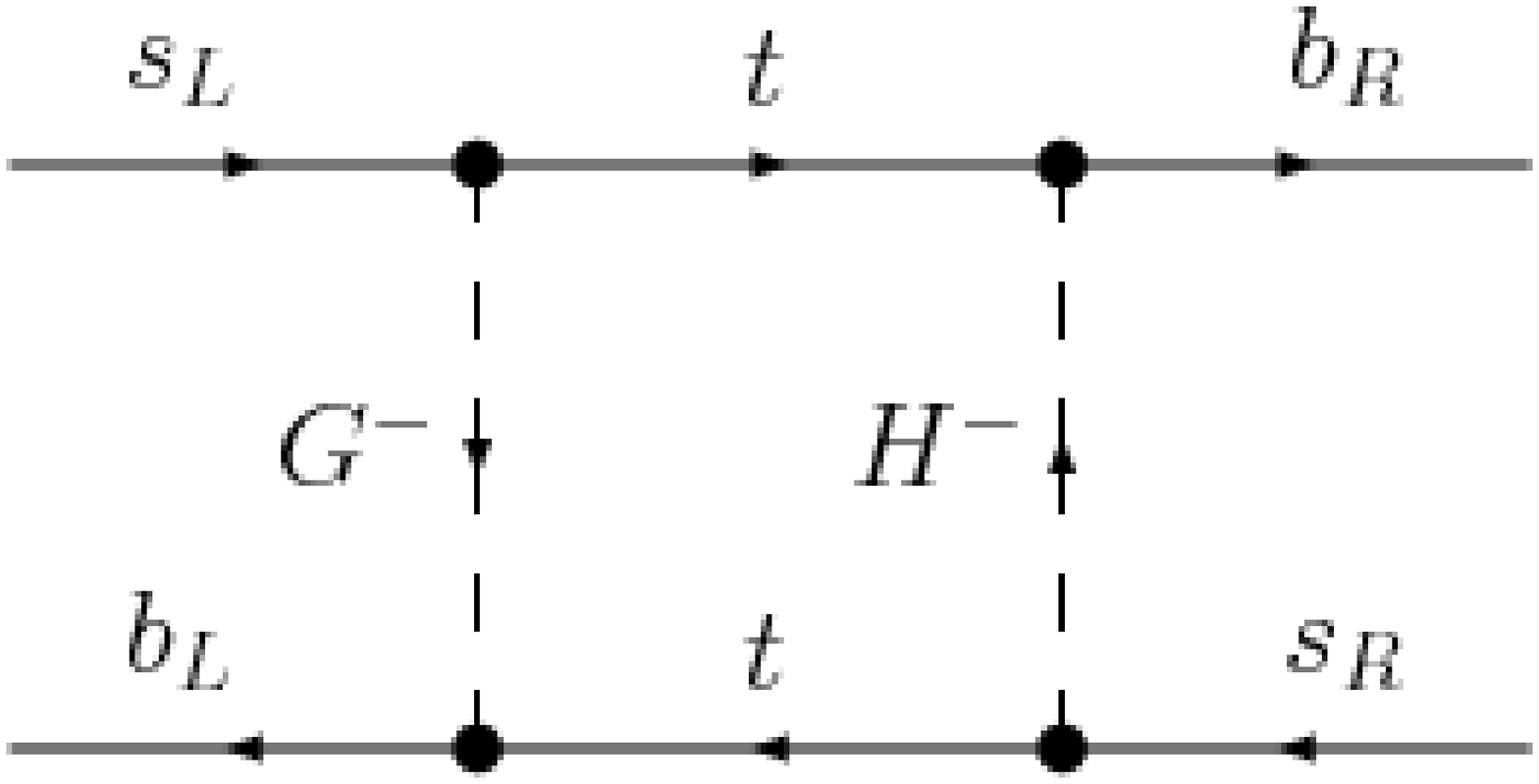}
\end{center}
}
\caption{The $\tan^2\beta$ enhanced contribution to $C^{LR}_2$ arising
from charged Higgs and Goldstone boson exchange\label{bbb:EW:CHfig}.}
}
Charged Higgs exchange leads to contributions to all of the operators
given in~\nqs{bbb:SMOps}{bbb:Ops2} (with the exception of $\mathcal{O}^{SLL}_{2}$
and its parity flipped counterpart).
BLO effects can be easily included by using the matching conditions
given in~\cite{BCRS:bdec} and the corrected vertices given in
section~\ref{MIA:EW}. As an example, the contribution to
$C^{LR}_2$ arising from the diagram shown in \fig{bbb:EW:CHfig} is given by
\begin{align}
\delta^{H^-}C^{LR}_2=\frac{8 m_t^4\mbphys\msphys\tan^2\beta}{m_{H^+}^4
  m_W^2\left(\BLOfact\right)^2}\left(1+\frac{\Lambda^R_{32}}{K_{ts}}\right)H_3
\left(\frac{m_W^2}{m_{H^+}^2},\frac{m_t^2}{m_{H^+}^2},\frac{m_t^2}{m_{H^+}^2}\right).
\label{bbb:CHCLR}
\end{align}
(The other contributions undergo similar corrections.) The factor
of $\left(\BLOfact\right)$ that appears in the denominator
acts, once again, to reduce the contributions relative to a LO
calculation. The GFM corrections included in the term
proportional to $\Lambda^{R}_{32}$, defined in~\eqref{MIA:EW:Lam},
can remove the dependence
on the strange quark mass that appears in~\eqref{bbb:CHCLR} and
lead to potentially large corrections to the MFV result.

The corrected neutral Higgs vertex can give rise to large
corrections to $\delmbs$ in the large $\tanb$ regime via
the double Higgs penguin diagrams shown in \fig{bbb:EW:NHfig}~\cite{HPT:bbm}.
The corresponding matching conditions are
\FIGURE[t!]{
\includegraphics[angle=0,width=0.25\textwidth]{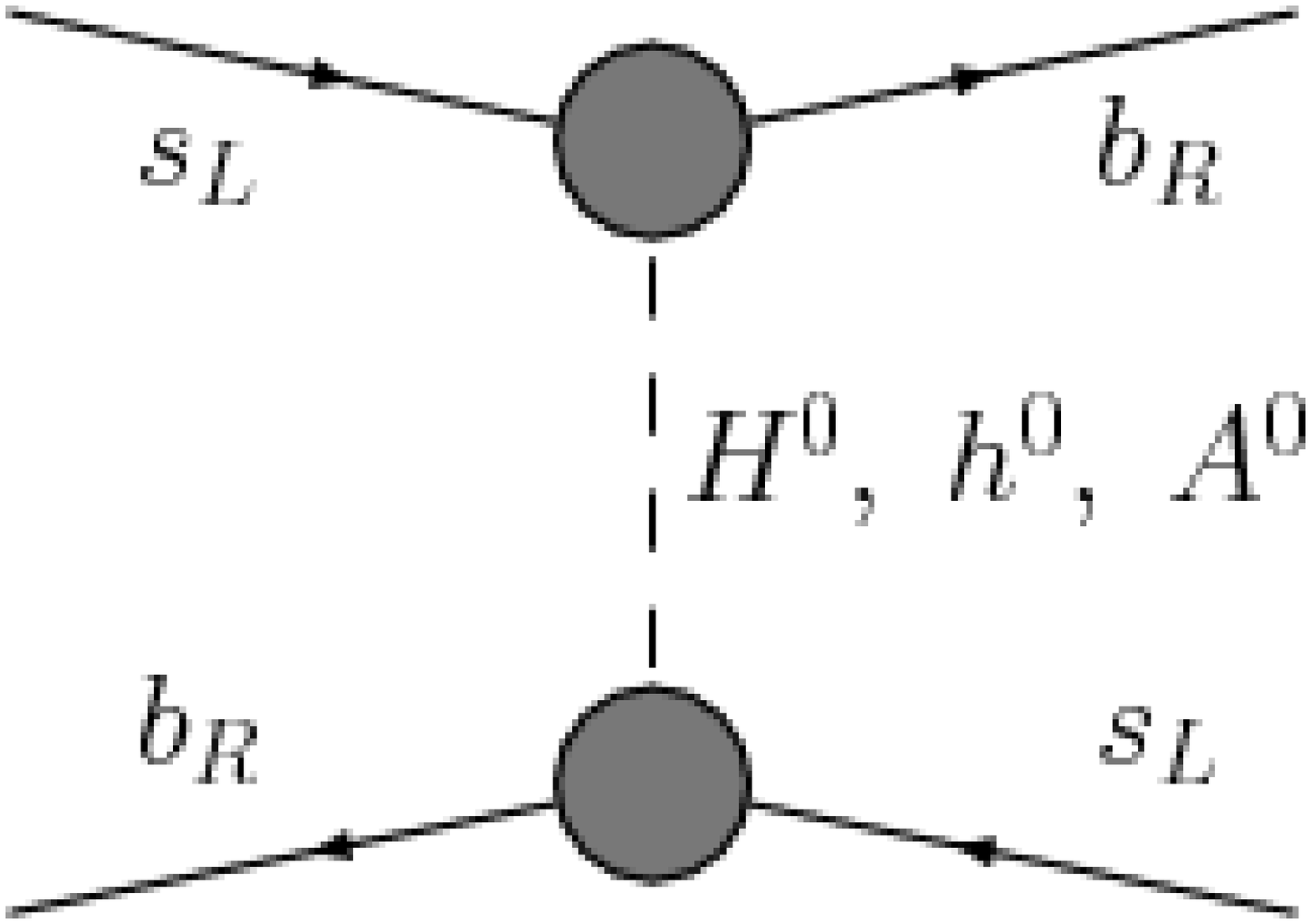}
\includegraphics[angle=0,width=0.25\textwidth]{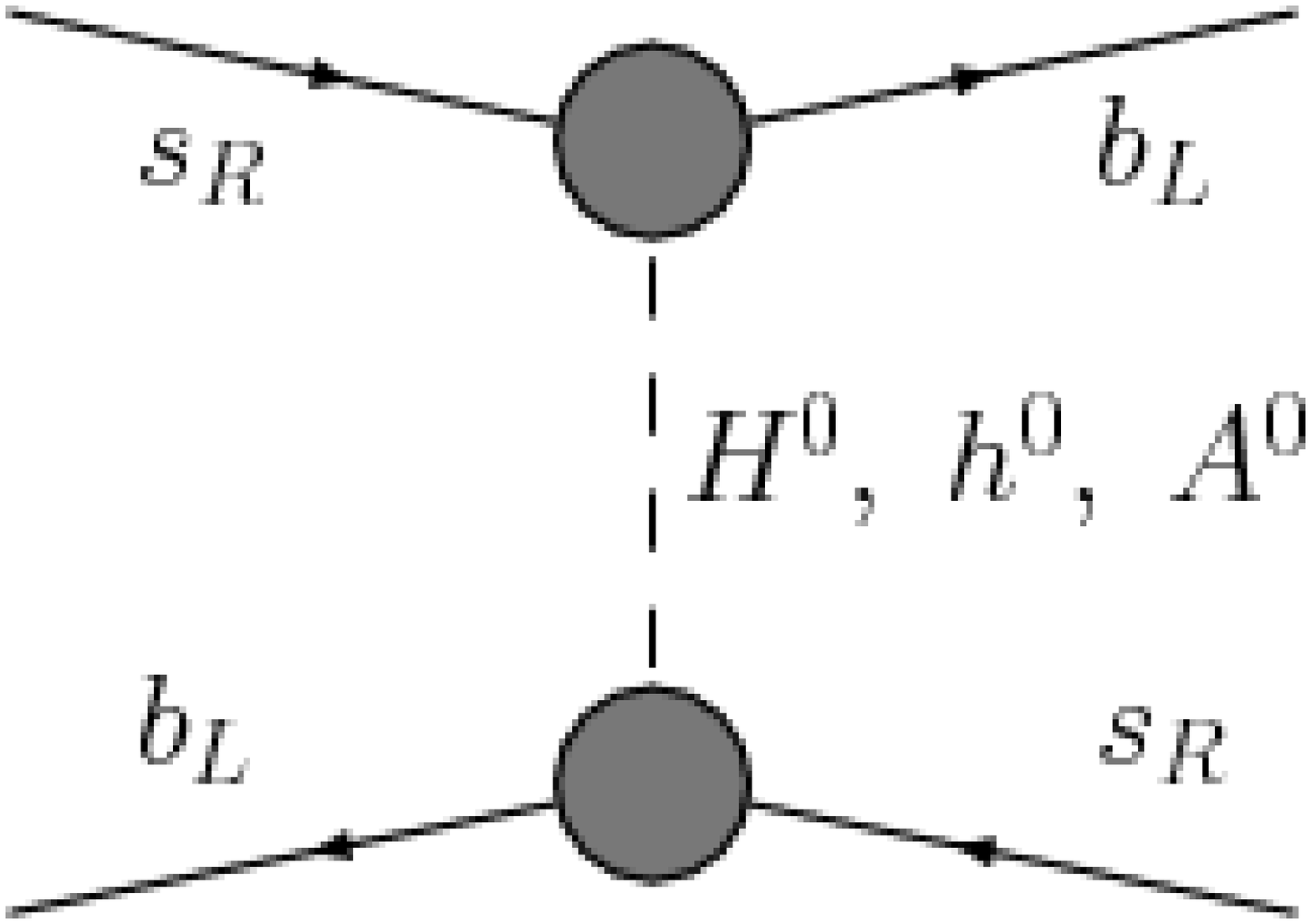}
\includegraphics[angle=0,width=0.25\textwidth]{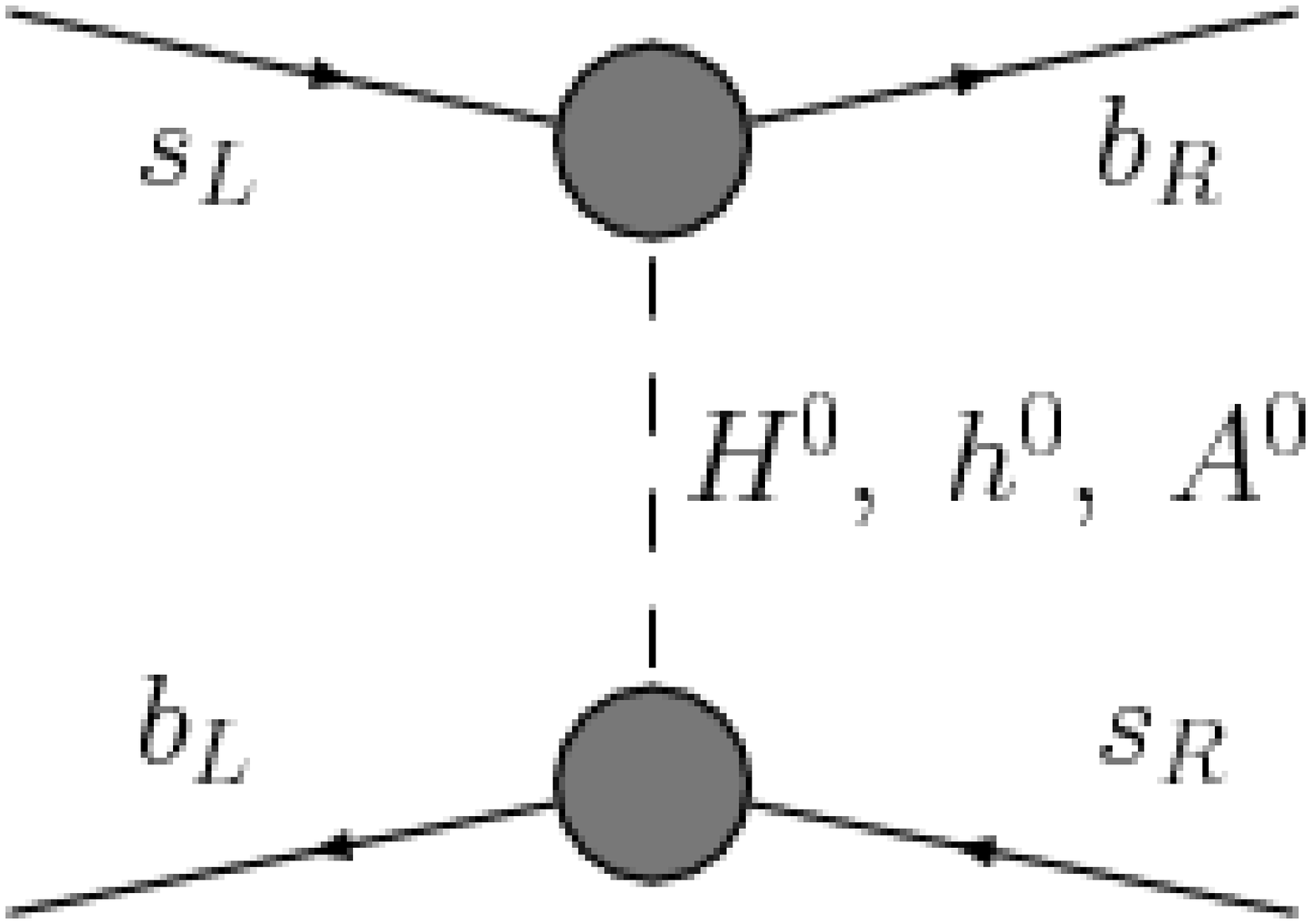}
\caption{The contributions to $\bbb$ mixing arising from
neutral Higgs penguins.\label{bbb:EW:NHfig}}
}
\begin{align}
C^{SLL}_1=&-\frac{16\pi^2}{2G_F^2 m_W^2 \left(K_{tb}^{\ast}K_{ts}\right)^2}
\sum_{S^0=H^0,h^0,A^0}\frac{\left(C^{S^0}_L\right)_{32}\left(C^{S^0}_L\right)_{32}}{m_{S^0}^2}, 
\label{bbb:EW:NHCSLL}
\\
C^{LR}_2=&-\frac{16\pi^2}{G_F^2 m_W^2 \left(K_{tb}^{\ast}K_{ts}\right)^2}
\sum_{S^0=H^0,h^0,A^0}\frac{\left(C^{S^0}_L\right)_{32}\left(C^{S^0}_R\right)_{32}}{m_{S^0}^2}.
\label{bbb:EW:NHCLR}
\end{align}
The contribution to the parity flipped operator $\mathcal{O}^{SRR}_1$ can
be obtained via the substitution $L\leftrightarrow R$ in~\eqref{bbb:EW:NHCSLL}. Using the
corrected neutral Higgs vertex~\eqref{MIA:EW:SLMFV} in the limit
of MFV it is easy to obtain the results for $C^{SLL}_1$~\cite{BCRS:bdec}
\begin{align}
\left(\delta^{H^0}C_{1}^{SLL}\right)^{\rm MFV}=-\frac{G_{F}\mbphys^2 m_t^4}{2\sqrt{2}\pi^2 m_W^2}
\frac{\epsilon_Y^2\left(16\pi^2\right)^2\tan^4\beta}{\left(\BLOfacg\right)^2\left(\BLOfact\right)^2}\mathcal{F}^-
\label{bbb:EW:NHCSLLMFV}
\end{align}
and $C^{LR}_2$
\begin{align}
\left(\delta^{H^0}C_{2}^{LR}\right)^{\rm MFV}=-\frac{G_{F}\mbphys\msphys m_t^4}{\sqrt{2}\pi^2 m_W^2}
\frac{\epsilon_Y^2\left(16\pi^2\right)^2\tan^4\beta}{\left(\BLOfacg\right)^2\left(\BLOfact\right)^2}\mathcal{F}^+.
\label{bbb:EW:NHCLRMFV}
\end{align}
Here we decompose the MFV and GFM contributions to the Wilson
coefficients in similar manner to~\eqref{bsg:SU:GLdec}.
We have adopted the notation
\begin{equation}
\mathcal{F}^{\pm}=\frac{\sin^2\left(\alpha-\beta\right)}{m_{H^0}^2}
+\frac{\cos^2\left(\alpha-\beta\right)}{m_{h^0}^2}
\pm\frac{1}{m_{A^0}^2}
\label{bbb:EW:Higgs}
\end{equation}
to represent the interference between the scalar and pseudoscalar
contributions. At large $\tanb$,~\eqref{bbb:EW:Higgs} becomes (for $m_A^2>m_Z^2$)
\begin{equation}
\mathcal{F}^{\pm}=\frac{1}{m_{H^0}^2}\pm\frac{1}{m_{A^0}^2},
\label{bbb:EW:HiggsLTB}
\end{equation}
and the scalar and pseudoscalar contributions to $C^{SLL}_1$
approximately cancel, whilst the contributions to $C^{LR}_2$
interfere constructively. As has been pointed out in~\cite{BCRS:bbm,BCRS:bdec}, the double penguin contribution
to $C^{LR}_2$ acts to reduce the value of $\delmbs$,
bringing it closer to the current experimental limit~\eqref{bbb:exp}.
The factor of $m_s$ that appears~\eqref{bbb:EW:NHCLRMFV} suppresses
the contribution somewhat, however it can still be of the order of 
50\% of the Standard Model contribution in certain regions of parameter
space even once the current limits on $\bsm$ are taken into
account~\cite{BCRS:bdec}.

Including the effects of GFM when evaluating the contributions
to the Wilson coefficients $C^{SLL}_1$ and $C^{LR}_2$ can
lead to far more varied effects. The full expressions for
$C^{LR}_2$ and $C^{SLL}_1$ can be obtained by using~\nqs{MIA:EW:NHDecomp}{MIA:EW:SLGFM}, however,
as they are rather complicated, let us highlight the phenomenologically
most interesting terms that appear. If the insertion $\dll$ is
non--zero the contribution to $C^{LR}_2$ becomes
\begin{align}
\left(\delta^{H^0}C^{LR}_2\right)^{\rm LL}=
-\frac{\mbphys\msphys}{2\pi^2 g_2^2 \left(K_{tb}^{\ast}K_{ts}\right)^2}
\frac{\epsilon_{LL}^2 \xdl^2\left(16\pi^2\right)^2\tan^4\beta}{\left(\BLOfacg\right)^2\left(\BLOfact\right)^2}
\mathcal{F}^+
\left(\dll\right)^2.
\label{bbb:EW:NHCLRGFMdll}
\end{align}
In a similar manner to the MFV correction~\eqref{bbb:EW:NHCLRMFV}, the
contribution acts to reduce the value of $\delmbs$. Once
again, however, $C_2^{LR}$ is suppressed by a factor of the
strange quark mass that tends to limit the size of the
correction to $\delmbs$. The insertion $\dlr$ also contributes
to $C_2^{LR}$ via the diagram where one penguin is mediated by
gluino exchange and the other is mediated by chargino exchange
\begin{align}
\left(\delta^{H^0}C^{LR}_2\right)^{\rm LR}= -\frac{\msphys\mgl m_t^2}{4\pi^2 m_W^2 K_{tb}^{\ast}K_{ts}}
\frac{\epsilon_{RL}\epsg\epsilon_Y \xdrl\left(16\pi^2\right)^2\tan^4\beta}{\left(\BLOfacg\right)^2\left(\BLOfact\right)}
\mathcal{F}^+
\dlr.
\label{bbb:EW:NHCLRGFMdlr}
\end{align}
In this case the contribution can increase or decrease the value
of $\delmbs$ depending on the sign of $\dlr$. However the factor
of $m_s$ that appears in~\eqref{bbb:EW:NHCLRGFMdlr} means that these
effects are, once again, rather small.

It is possible to avoid factors of the strange quark mass
that appear in the above expressions for $C_2^{LR}$ by
considering scenarios where the right handed Higgs
coupling that appears in~\eqref{bbb:EW:NHCLR} does not feature
such a suppression. This occurs
if either $\drl$ or $\drr$ are non--zero. If we consider
the diagram where one Higgs penguin is mediated by chargino
exchange and the other by gluino exchange we have, for non--zero
$\drr$~\cite{FOR1:bdec}
\begin{align}
\left(\delta^{H^0}C^{LR}_2\right)^{\rm RR}= -\frac{\mbphys^2 m_t^2}{4\pi^2 m_W^2 K_{tb}^{\ast}K_{ts}}
\frac{\epsilon_{RR}\epsilon_Y \xdr\left(16\pi^2\right)^2\tan^4\beta}{\left(\BLOfacg\right)\left(\BLOfact\right)^3}
\mathcal{F}^+
\drr.
\label{bbb:EW:NHCLRGFMdrr}
\end{align}
A similar contribution is possible for the insertion $\drl$
\begin{align}
\left(\delta^{H^0}C^{LR}_2\right)^{\rm RL}= -\frac{\mbphys\mgl m_t^2}{4\pi^2 m_W^2 K_{tb}^{\ast}K_{ts}}
\frac{\epsilon_{RL}\epst\epsilon_Y \xdrl\left(16\pi^2\right)^2\tan^4\beta}{\left(\BLOfacg\right)\left(\BLOfact\right)^2}
\mathcal{F}^+
\drl.
\label{bbb:EW:NHCLRGFMdrl}
\end{align}
As one penguin is mediated by chargino
exchange,~\nqs{bbb:EW:NHCLRGFMdrr}{bbb:EW:NHCLRGFMdrl}
feature only a linear dependence on $\drr$ and $\drl$. Large positive
or negative contributions to $\delmbs$ are therefore possible,
depending on the sign of the insertions.

If $\dll$ or $\dlr$ are non--zero, in addition to either
$\drl$ or $\drr$ the
diagram involving two gluino mediated penguins becomes viable.
For example, if $\dll$ and $\drr$ are non--zero we have the contribution
\begin{align}
\left(\delta^{H^0}C^{LR}_2\right)^{\rm LL+RR}=-\frac{\mbphys^2}{2\pi^2 g_2^2 \left(K_{tb}^{\ast}K_{ts}\right)^2}
\frac{\epsilon_{RR}\epsilon_{LL}\xdl\xdr\left(16\pi^2\right)^2\tan^4\beta}{\left(\BLOfacg\right)\left(\BLOfact\right)^3}
\mathcal{F}^+
\dll\drr.
\label{bbb:EW:NHCLRGFMdrrdll}
\end{align}
Similar contributions arise for the remaining combinations
LL+RL, LR+RR and LR+RL.
The $\tan^4\beta$ dependence of~\eqref{bbb:EW:NHCLRGFMdrrdll}, coupled
with its dependence on the strong coupling constant present in the factors
of $\epsilon_{LL}$ and $\epsilon_{RR}$, can lead to
large corrections to $\delmbs$ in the large $\tanb$ regime.
We should briefly mention here how this result compares
to that presented in~\cite{IR2:bsm}. As discussed in~\cite{FOR1:bdec}
the authors of~\cite{IR2:bsm} omit the
effects that arise when considering the GFM contributions to
the bare CKM matrix~\eqref{BLOeff:OM:BCRS:bareCKM}. Once one takes into account 
such contributions, one of the factors of $\left(\BLOfact\right)$
that appears in Eq.~(5.8) of~\cite{IR2:bsm} is replaced
by a factor of $\left(\BLOfacg\right)$. Taking this correction
into account, our results agree.

A common feature of all of these corrections lie in the factors
of $\left(\BLOfact\right)$ and $\left(\BLOfacg\right)$ that
appear in the denominators of all double Higgs penguin contributions.
These factors represent the resummation of $\tanb$ enhanced
effects and act to reduce the contributions for $\mu>0,~A_t<0$
compared to calculations where resummation is not taken into account.

%%%%%%%%%%%%%%%%%%%%%%%%%%%%%%%%%%%%%%%%%%%%%%%%%%%%%%%%%%%%%%%%%
\subsection{BLO Corrections to SUSY Contributions in
the MIA}
\label{bbb:SU}
%%%%%%%%%%%%%%%%%%%%%%%%%%%%%%%%%%%%%%%%%%%%%%%%%%%%%%%%%%%%%%%%%

The supersymmetric contributions
to $\bbb$ mixing proceed via box diagrams mediated by
gluino, chargino and neutralino exchange.  Unlike
the decay $\bsg$ however, sizeable effects due to these contributions
are often limited to regions where the squarks and gluinos
are relatively light $\mathcal{O}\left(500\gev\right)$.

As $\bbb$ mixing is a $\Delta F=2$ process, the gluino and
neutralino contributions, at LO, feature combinations of two
insertions. The LO gluino matching conditions have been discussed,
in the context of the MIA, in~\cite{It:bbm}. It is relatively
easy to modify them to take into account the BLO effects discussed
in section~\ref{GA} by using the recipe discussed at the
end of subsection~\ref{bsg:SU}.
The gluino contribution to the operator $C_2^{LR}$ is
given by (to second order in the MIA)
\begin{align}
&\left(\delta^{\widetilde{g}}C_2^{LR}\right)^{\rm GFM}=\frac{16\alpha_s^2\sin^4\theta_W}{9\alpha^2\left(K_{tb}^{\ast}K_{ts}\right)^2}
\frac{m_W^2}{\mgl^2}
\Bigg\{
\frac{11\left(1+\epsyII\tanb\right)f_{\widetilde{g}}^{[2]}\left(\xd\right)}{\left(\BLOfacg\right)\left(\BLOfact\right)}
\dlr\drl
\nonumber\\
&+\left[-42f_{\widetilde{g}}^{[1]}\left(\xd\right)
+\left(6+11\frac{\mu^2m_b^2}{\mgl^2}
\frac{\epsilon_{LL}\epsilon_{RR}\tan^4\beta}{\left(\BLOfacg\right)\left(\BLOfact\right)^3}\right)f_{\widetilde{g}}^{[2]}\left(\xd\right)
\right]\dll\drr
\Bigg\}.
\label{bbb:SU:glu}
\end{align}
The functions $f_{\widetilde{g}}^{[1,2]}\left(\xd\right)$
are given in appendix~\ref{LF:bbm} and $\xd$ is defined
in~\eqref{bsg:SU:xddef}. Once again, we see that the contributions
arising from the insertions
$\dlr$ and $\drl$ are modified in the same manner as the supersymmetric
contributions to $\bsg$ (namely~\eqref{bsg:SU:gluLR} and~\eqref{bsg:SU:gluRL}) that
appear in section~\ref{bsg:SU}. The inclusion
of BLO effects therefore tends to reduce the overall contribution to
$C_2^{LR}$ arising from these insertions. The BLO term proportional
to $\dll\drr$, on the other hand, tends to interfere with terms
that arise at fourth order in the MIA and on the whole tends
to be rather small. In
addition, the combinations $\dll\drl$ and $\dlr\drr$ also appear
once BLO effects are taken into account. Due to their $\tan^2\beta$
dependence, they can viably compete with the corresponding terms
that arise at higher orders in the MIA. Contributions arising
from MFV corrections to the bare mass matrix can also appear and
play the r\^ole of $\dll$ and $\dlr$ insertions. (Contributions
where the MFV corrections play the r\^ole of $\drl$ and $\drr$
insertions are suppressed by factors of $m_s$.)

Turning to the contributions arising from chargino box
diagrams, in MFV the dominant behaviour at LO arises from
contributions to the Wilson coefficient $C^{VLL}$. At large
values of $\tanb$, however, large corrections to
$C_{1,2}^{SLL}$ are possible that
interfere destructively with $C^{VLL}$ and can reduce
the contribution that arises from chargino box diagrams
to $\delmbs$~\cite{BCRS:bdec}.
The inclusion of BLO effects in MFV, however, tends to reduce this
cancellation somewhat; whilst the Wilson coefficient $C^{VLL}$
remains virtually unaffected by BLO corrections, the BLO corrections
to $C_{1,2}^{SLL}$ introduce factors of $\left(\BLOfact\right)$
that, in the phenomenologically favoured
region $\mu>0$ and $A_t<0$, act to reduce the contribution
to the Wilson coefficient. These effects therefore reduce the
cancellations that occur between the two Wilson
coefficients, and can lead to an enhancement
of the contributions that arise from chargino box
diagrams~\cite{BCRS:bdec}.

In the GFM scenario, the occurrence of the bare mass
matrix in the chargino vertex~\ref{GA:SU:Char} can play the r\^ole
of one (or both) of the factors of $K_{ts}$ that mediate the
flavour change in the leading order matching conditions.
A dependence on the flavour violating parameters $\delta^{d}_{XY}$
can therefore appear once one proceeds beyond the LO.
In particular, the
factor of $m_s$, that features in the matching condition
for $C_2^{LR}$ in MFV, can be bypassed in the GFM scenario
if either $\drl$ or $\drr$ are non--zero. A similar effect
occurs for the Wilson coefficients $C_{1,2}^{SLL}$, where
non--zero $\dll$ or $\dlr$ can lead to additional contributions
to the coefficient. It is therefore possible to enhance,
or decrease, the cancellations that occur between
$C^{VLL}$ and the remaining Wilson coefficients depending
on the sign of the insertions $\delta^{d}_{XY}$.

The appearance of the bare mass matrix in the neutralino
vertex also introduces additional BLO corrections that
can modify the LO contributions. In a similar manner
to the contributions to the decay $\bsg$, these
corrections typically manifest themselves as factors
of $(\BLOfact)$ or $(\BLOfacg)$ and tend to reduce the
contributions compared to a LO analysis.

Let us finally comment on the RG running of the SUSY
corrections. The six flavour anomalous dimension matrix
required to evolve the Wilson coefficients from the SUSY
matching scale to the electroweak scale was given in~\cite{BJU:bbm}.
If we consider the running
of the coefficient $C_{2}^{LR}$ we have
\begin{align}
\delta C_2^{LR}\left(\mu_W\right)=\delta C_2^{LR}\left(\mu_{SUSY}\right)
+\frac{\alpha_{s}\left(\mu_W\right)}{\pi}
\left[2\delta C_2^{LR}\left(\mu_{SUSY}\right)
-\frac{3}{2}\delta C_1^{LR}\left(\mu_{SUSY}\right)\right]\log\frac{\mu_{SUSY}^2}{\mu_{W}^2}.
\label{bbb:SU:RG}
\end{align}
In contrast to the decay $\bsg$, the RG evolution of the
Wilson coefficients, from $\mu_{SUSY}$ to $\mu_{W}$, acts
to increase the coefficients with respect to an analysis
where RG running is ignored. A similar effect exists
in the SLL sector. In the VLL sector, however, RG running
acts to decrease the Wilson coefficients in a similar manner
to $\bsg$. These effects can therefore act to enhance the
cancellations between contributions to the VLL sector
and the remaining operators.

%%%%%%%%%%%%%%%%%%%%%%%%%%%%%%%%%%%%%%%%%%%%%%%%%%%%%%%%%%%%%%%%%
\subsection{Full Calculation}
\label{bbb:Full}
%%%%%%%%%%%%%%%%%%%%%%%%%%%%%%%%%%%%%%%%%%%%%%%%%%%%%%%%%%%%%%%%%

Let us now discuss the calculation we perform in our numerical
analysis. Following the method discussed at the end of section~\ref{bsg},
we evaluate the SM matching conditions to NLO using the
matching conditions originally given in~\cite{BJW:bbm}.
For the charged Higgs contribution, we use the corrected vertices
given in appendix~\ref{RadCor} when evaluating the LO matching
conditions given in~\cite{BCRS:bdec}. When evaluating the SUSY
contributions we use the LO matching conditions collected in~\cite{BGOO2:drr}
(transformed into our operator basis). They are then subsequently
evolved from the SUSY matching scale to the EW scale using
the NLO anomalous dimension matrix given in~\cite{BJU:bbm}.
% Chargino + CH : Appendix A.4 of BCRS:bdec
% Other SUSY Boxes Appendix C BGOO2:drr
% Anomalous Dims. Appendix C BJU:bbm

In summary, in this section we have discussed the effects 
of all the dominant BLO corrections to $\bbb$ mixing. In particular,
we have considered all the contributions that arise once
one takes into account the double Higgs penguin
contribution. In the region where $\tanb$
and the sparticle masses are large, for example, these contributions can
dominate the corrections to $\delmbs$ that arise from new physics,
due to the non--decoupling property of the corrections to the
neutral Higgs vertex. Turning to the supersymmetric contributions
to the process, we see that, although they play a more minor r\^ole
compared to the neutral Higgs contributions in the large
$\tanb$ regime, the overall effect of BLO corrections is to
reduce the contributions arising from gluino exchange. We have
also seen that, in contrast to the decay $\bsg$, RG evolution
between $\mu_{SUSY}$ and $\mu_{W}$ can act to enhance the
contributions of the coefficient $C_2^{LR}$ in particular.

%%%%%%%%%%%%%%%%%%%%%%%%%%%%%%%%%%%%%%%%%%%%%%%%%%%%%%%%%%%%%%%%%
\section{Numerical Results}
\label{NRes}
%%%%%%%%%%%%%%%%%%%%%%%%%%%%%%%%%%%%%%%%%%%%%%%%%%%%%%%%%%%%%%%%%

Before proceeding with our numerical results let us first define
our parameterisation for the soft sector. We treat the
soft terms~\eqref{GA:SCKMst}, defined in the
physical SCKM basis, as input. For the
diagonal elements we set
\begin{align}
\left(m_{d,LL}^2\right)_{ii}&=\msq^2\;\delta_{ii},
&\left(m_{d,RR}^2\right)_{ii}&=\msq^2\;\delta_{ii},
&\left(m_{d,LR}^2\right)_{ii}&=A_d\left(m_d\right)_{ii},
\end{align}
The off--diagonal elements are related to the parameters
$\delta_{XY}^d$ via the relations defined in
\nqs{GA:dels1}{GA:dels2}.
The soft terms in the up--sector are defined analogously.
As inputs for the Higgs sector we take 
$m_A$,
$\mu$ and $\tanb$ and use {\it FeynHiggs 2.2.8}~\cite{FH} to determine
the remaining parameters. For the majority of this section
we will only vary one $\delta_{XY}^d$ at a time unless
stated otherwise. Finally the gaugino soft terms $M_1$ and $M_2$
are related to the gluino mass $\mgl$ via the usual unification relation.

Let us briefly specify the abbreviations used to denote the
various approximations used in this section:
\begin{itemize}
\item LO: a calculation that does not feature the resummation
procedure described in section~\ref{GA}.  We do, however,
include the LO effects that arise from RG evolution from
$\mu_{SUSY}$ to $\mu_W$ (in contrast with~\cite{OR2:bsg});

\item $\widetilde{g}$--BLO: the approximation
used in~\cite{OR2:bsg} where only gluino contributions were
taken into account in the resummation procedure;

\item BLO:  the results obtained using the
full expressions included in appendix~\ref{RadCor}.

\end{itemize}

%lr Let us briefly specify the abbreviations used to denote the
%lr various approximations used in this section. By LO we refer
%lr to a calculation that does not feature the resummation
%lr procedure described in section~\ref{GA}.  We do, however,
%lr include the LO effects that arise from RG evolution from
%lr $\mu_{SUSY}$ to $\mu_W$ (in contrast with~\cite{OR2:bsg}).
%lr The abbreviation $\widetilde{g}$--BLO denotes the approximation
%lr used in~\cite{OR2:bsg} where only gluino contributions are
%lr taken into account in the resummation procedure. Finally
%lr BLO is used to denote the results obtained using the
%lr full expressions included in appendix~\ref{RadCor}.

%%%%%%%%%%%%%%%%%%%%%%%%%%%%%%%%%%%%%%%%%%%%%%%%%%%%%%%%%%%%%%%%%
\subsection{$\bsg$}
\label{NRes:bsg}
%%%%%%%%%%%%%%%%%%%%%%%%%%%%%%%%%%%%%%%%%%%%%%%%%%%%%%%%%%%%%%%%%

A detailed analysis of the BLO effects relevant to $\bsg$ was
presented in~\cite{OR2:bsg} and we shall therefore tend
to focus on the additional effects induced by the inclusion
of electroweak contributions. We shall also compare
the approximate expressions, gathered in section~\ref{bsg},
with our complete calculation.

Before focusing on the GFM scenario,
\FIGURE[t!]{
    \includegraphics[angle=0,width=0.45\textwidth]{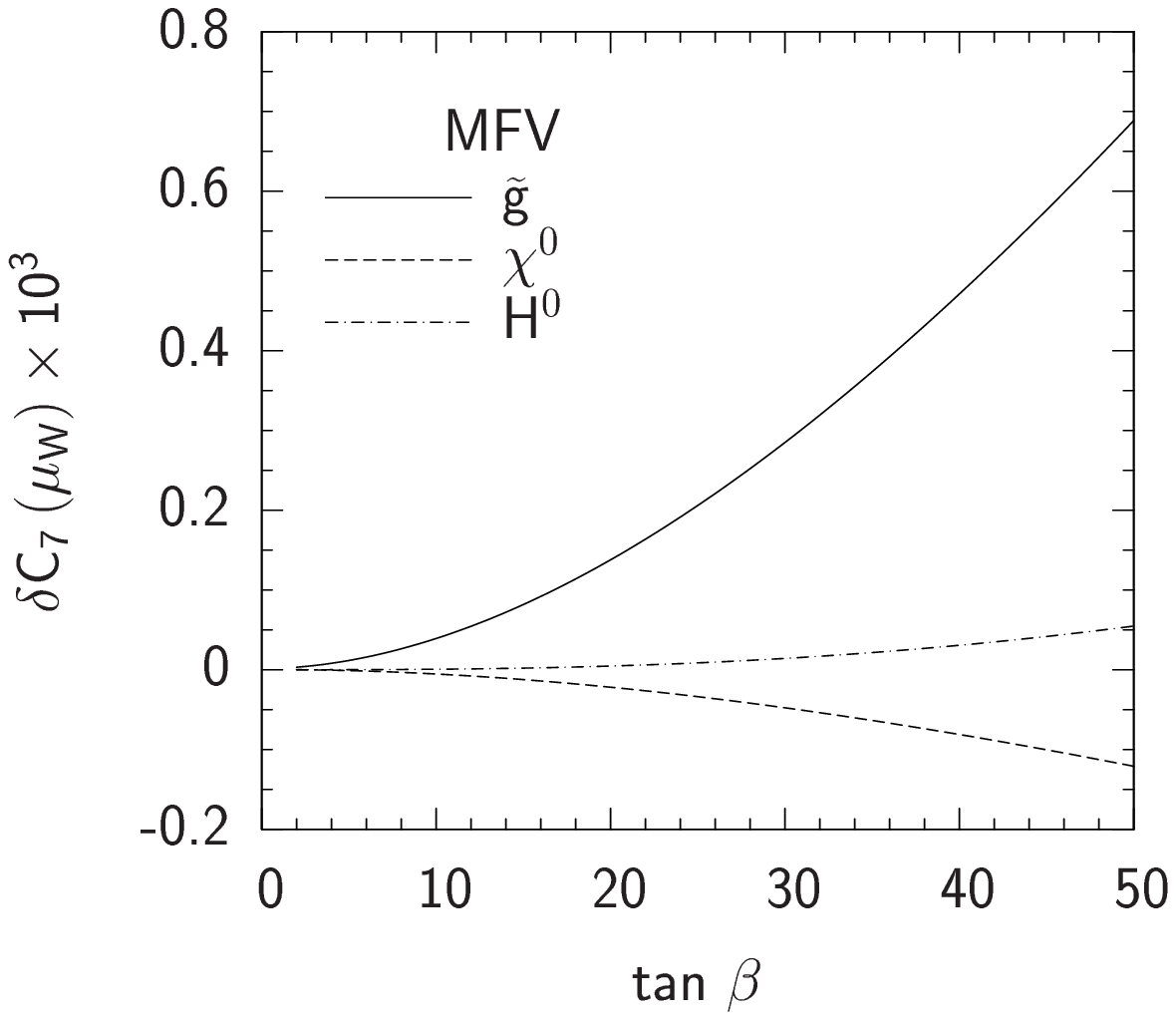}
    \includegraphics[angle=0,width=0.45\textwidth]{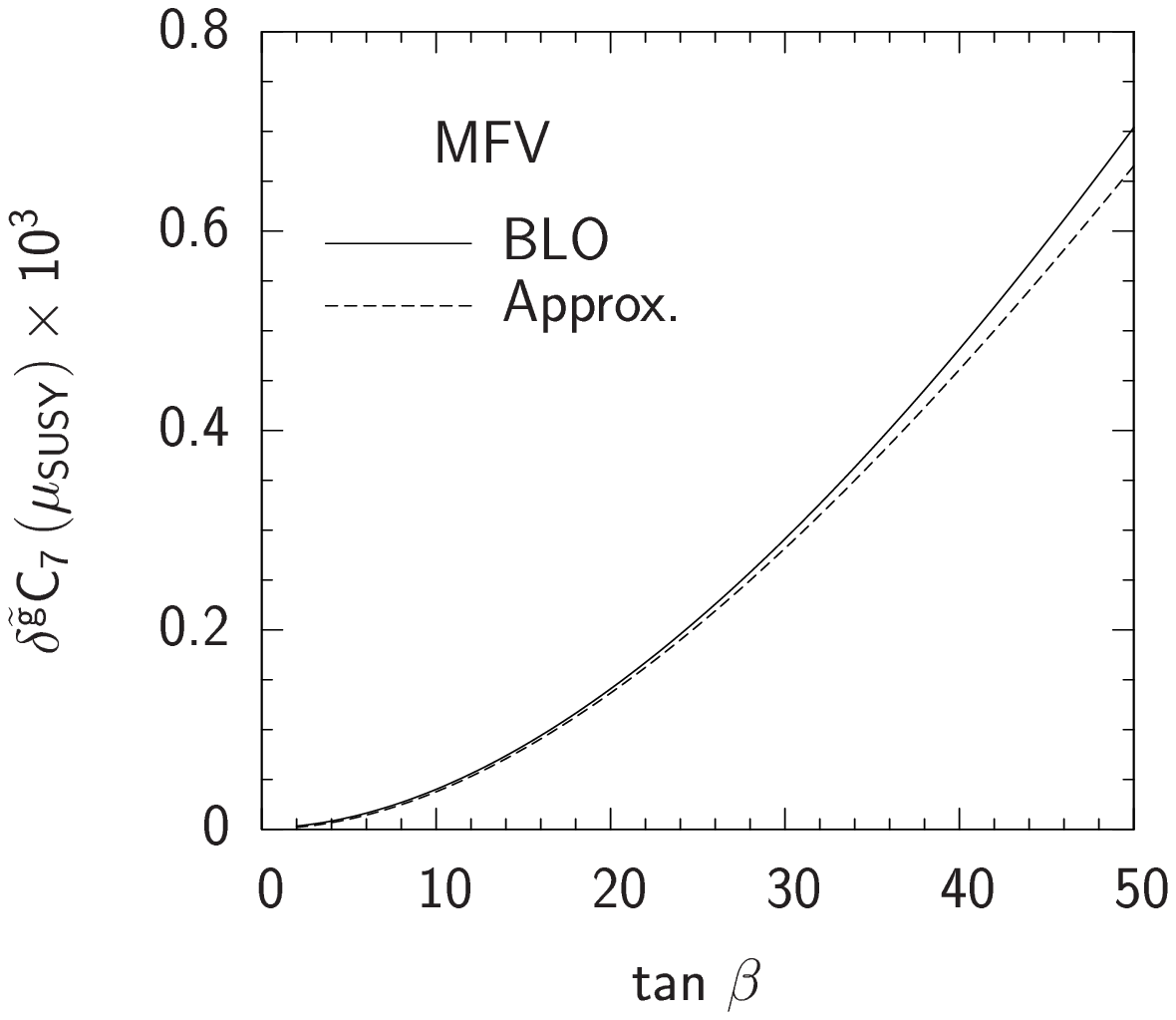}
  \caption{The panel on the left depicts corrections to the Wilson
    coefficient $C_7$ induced by gluino, neutralino and neutral
    Higgs exchange for MFV. The panel on the right illustrates
    the behaviour of the approximate result~\eqref{bsg:SU:gluMFV}
    compared with the full numerical calculation.
    The soft sector is parameterised as follows:
    $\msq=1\tev$, $\mgl=\sqrt{2}\,\msq$, $A_t=-500\gev$,
    $m_{A}=500\gev$, $\mu=500\gev$. \label{bsgres:MFVwc}}
}
let us briefly consider
the effects induced by the bare mass matrix and corrected
Higgs vertices in the case of MFV. The panel on the left
of \fig{bsgres:MFVwc} illustrates
the contributions mediated by gluinos, neutralinos and
neutral Higgs. As is evident from the graph, all three
contributions are rather small and, compared with the
corrections to $C_7$ induced by charged Higgs or chargino exchange,
are of the order of a few percent. However, it should be noted that the
gluino and neutralino contributions are both larger than
the neutral Higgs contribution in this particular region of
parameter space.
The panel on the right of \fig{bsgres:MFVwc} depicts the approximate
expression~\eqref{bsg:SU:gluMFV} alongside the result taken from our
numerical analysis. As is evident,~\eqref{bsg:SU:gluMFV}
describes the behaviour of the full numerical result rather well
for this choice of parameters and only differs by
about 5\% from the result of a full numerical calculation at
large $\tanb$.

Now let us turn to the GFM scenario.
\FIGURE[t!]{
  \begin{tabular}{c c}
    \includegraphics[angle=0,width=0.45\textwidth]{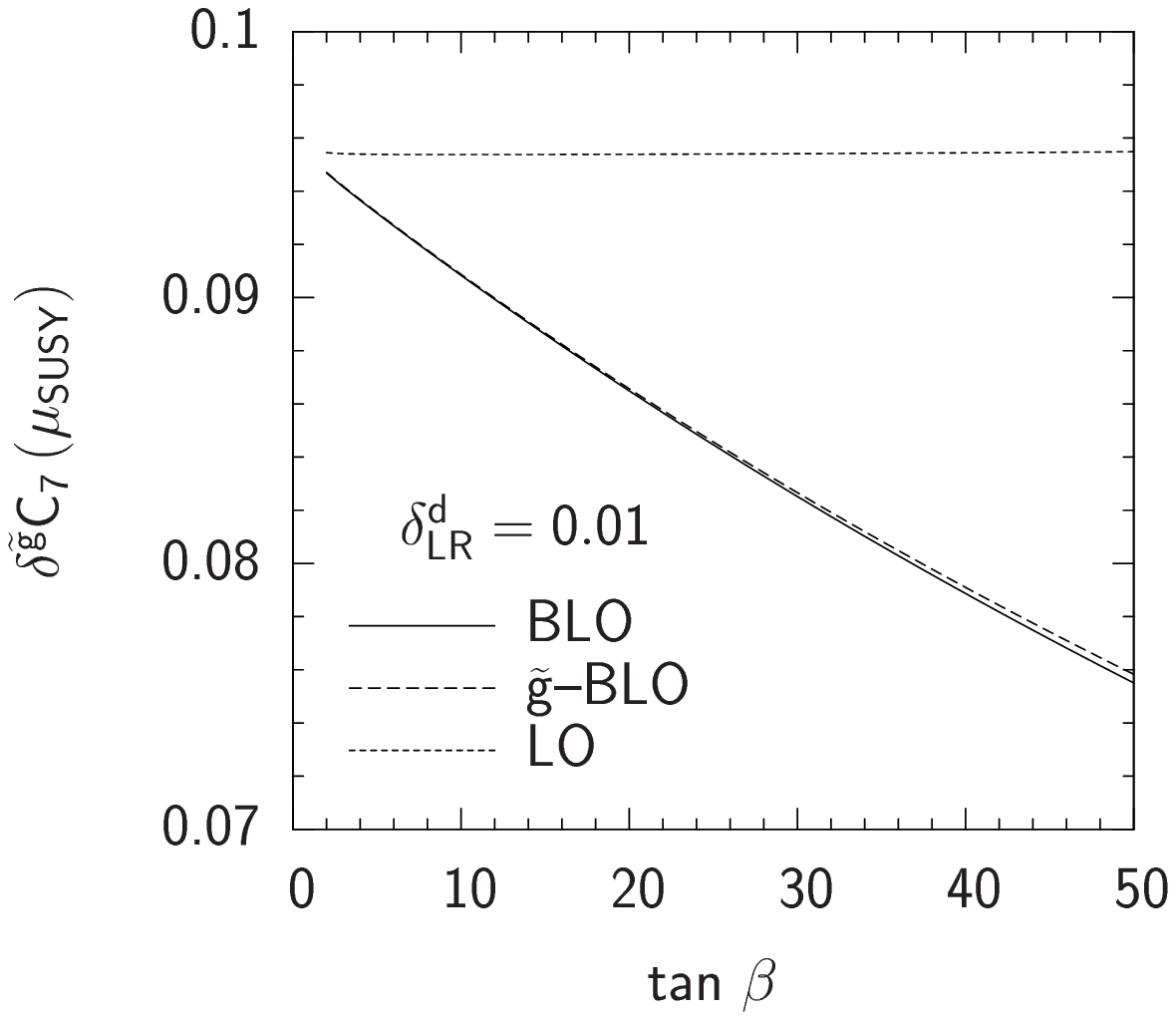}
    & \includegraphics[angle=0,width=0.45\textwidth]{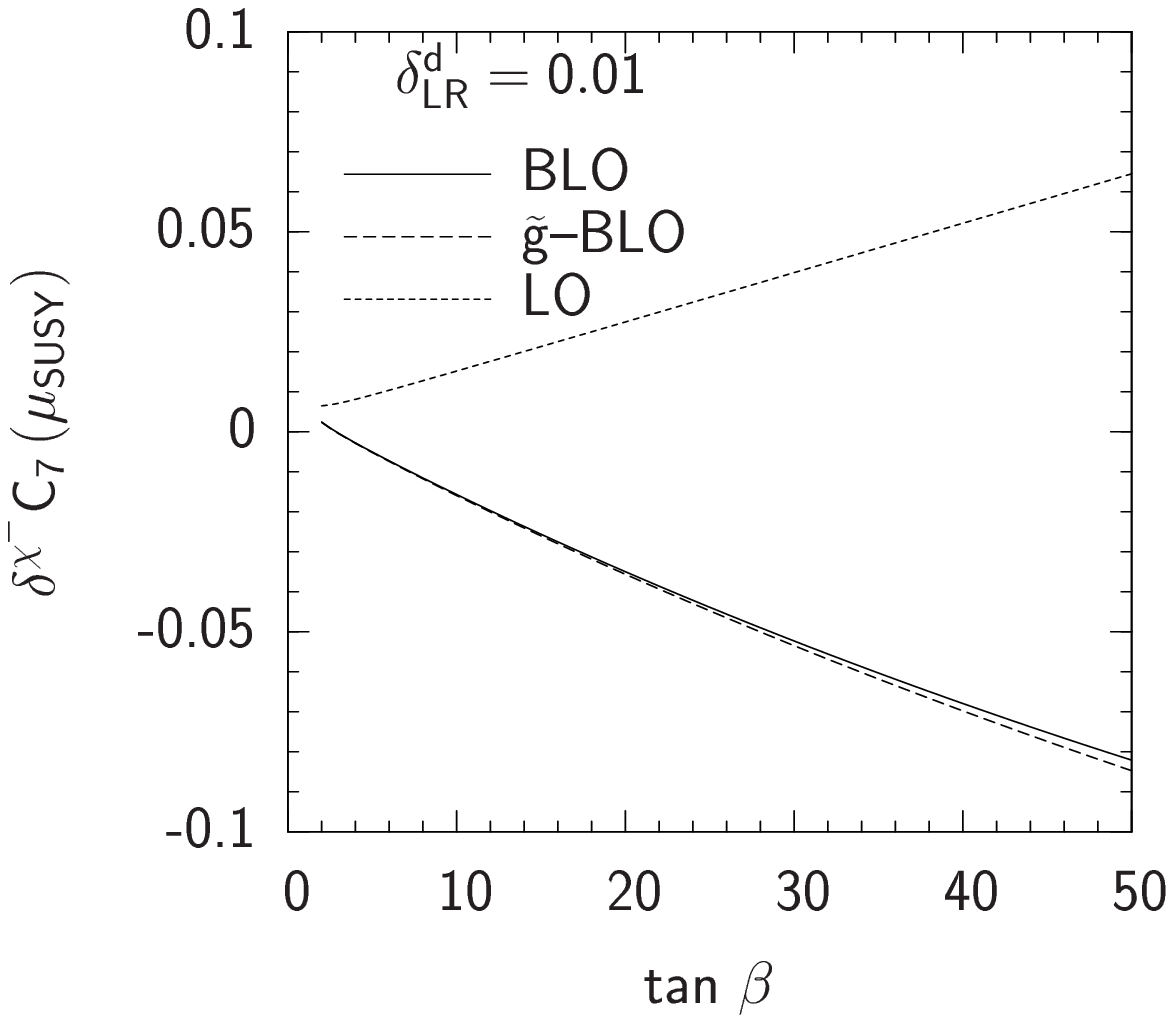}
    \\ \includegraphics[angle=0,width=0.45\textwidth]{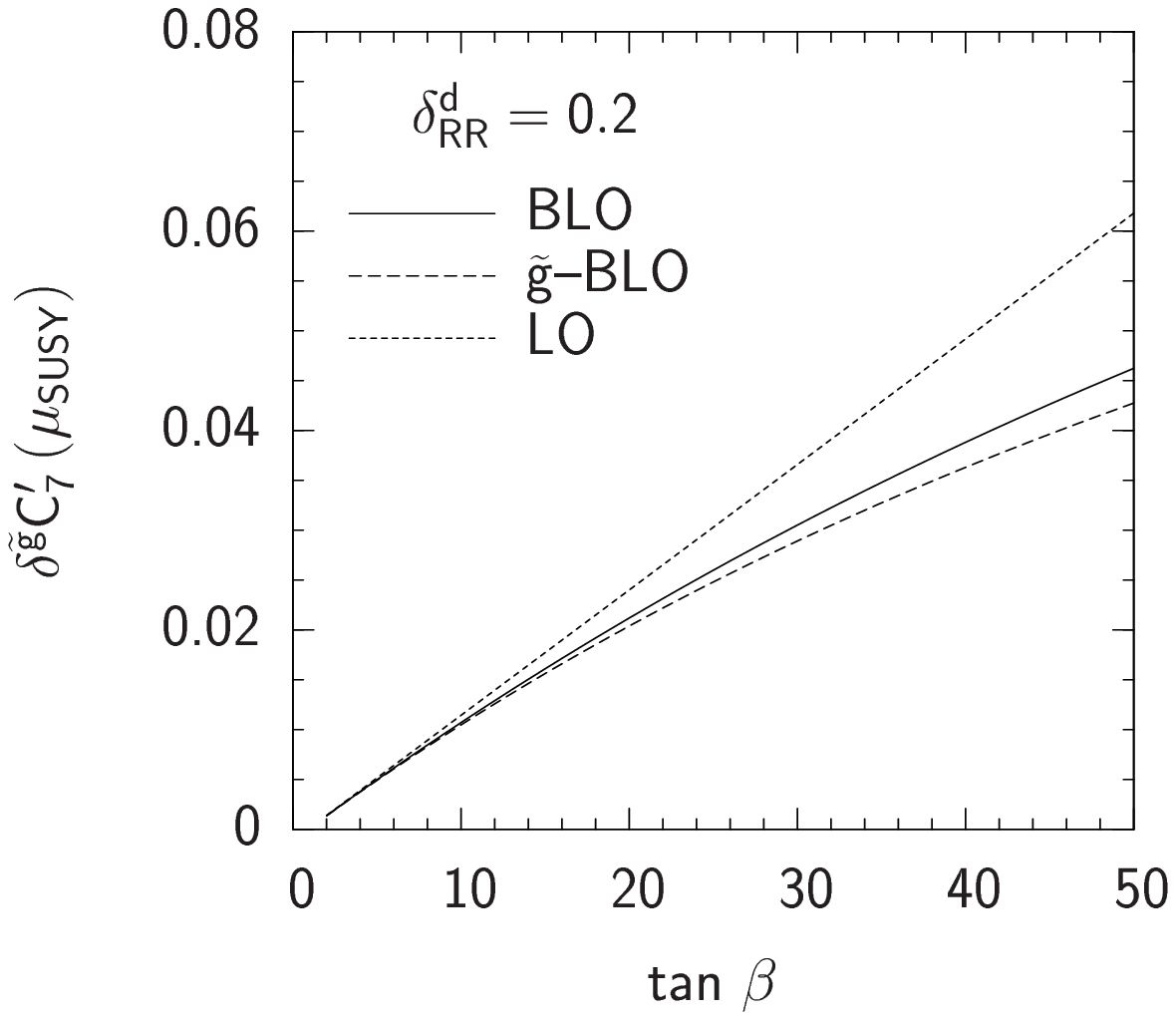}
    & \includegraphics[angle=0,width=0.45\textwidth]{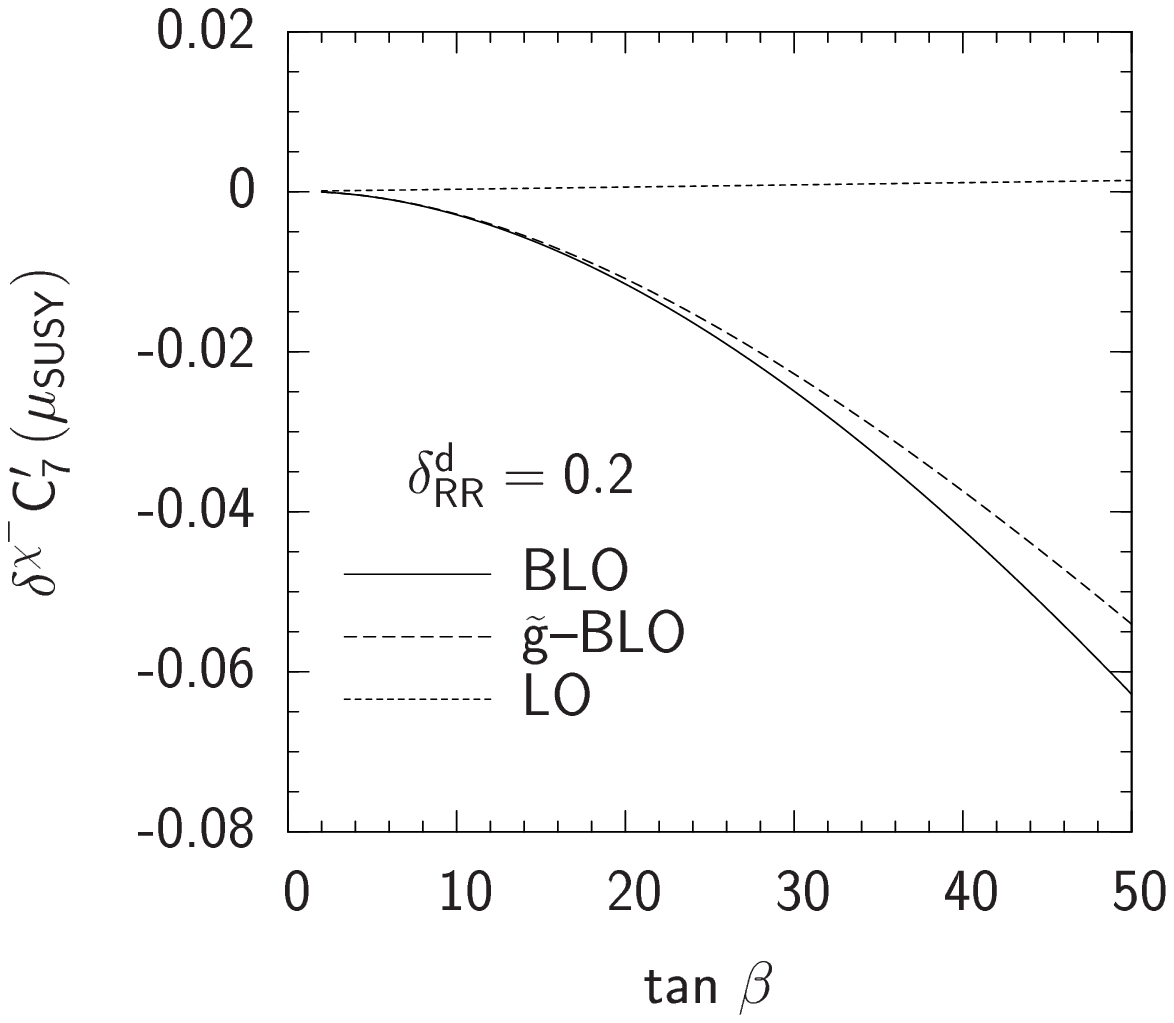}
  \end{tabular}
  \caption{The corrections to the Wilson coefficient $C_7$ (evaluated
    at the scale $\mu_{SUSY}$) induced by gluino exchange (on the
    left) and chargino exchange (on the right). 
    $\dlr=0.01$ and $\drr=0.2$ in the top and bottom panels,
    respectively.
    The soft sector is described by the same parameters as
    \fig{bsgres:MFVwc}. 
    As stated at the beginning of section~\ref{NRes},
    LO is used to denote a calculation
    that does not take into account the BLO effects discussed
    in section~\ref{GA}. $\widetilde{g}$--BLO denotes the
    approximation used in~\cite{OR:bsg} where only gluino
    contributions are taken into account in the resummation
    procedure. BLO is used to denote the results obtained
    using the expressions collected in appendix~\ref{RadCor}.
    \label{bsgres:glchwc}}
}
As stated in sections~\ref{bsg:EW}
and~\ref{bsg:SU}, it is possible, once BLO effects are taken into
account, that large cancellations can occur between the various
supersymmetric contributions. The top two plots in
\fig{bsgres:glchwc}, for example, illustrate
the cancellations that occur between the chargino
and gluino contributions for non--zero $\dlr$.
The dominant BLO chargino
contributions to $C_7$ stem from the appearance of the bare mass
matrix in the modified chargino vertex
\nqs{GA:SU:CdL}{GA:SU:CdR}. The inclusion
of electroweak effects, as indicated by the absence of a
term dependent on $\epsy$ in the terms proportional
to $\dlr$ in~\eqref{bsg:SU:gluLR} and~\eqref{bsg:SU:chaLR}, is rather small.
The lower two plots in \fig{bsgres:glchwc} show the contributions
in the primed sector for non--zero $\drr$. Here we see a similar
cancellation between the
BLO effects arising from the chargino contribution and
gluino contribution to $C_7^{\prime}$. The effect of including
electroweak effects is, however, larger than the case of the
LR insertion due to the appearance of  $\left(\BLOfact\right)$ 
(rather than $\left(\BLOfacg\right)$) in the
denominators of~\eqref{bsg:SU:gluRR} and~\eqref{bsg:SU:chaRR}.
However, one can see from the two graphs that
the increase in the gluino contribution tends to be compensated
by a similar correction to the chargino contribution.
As such the overall effect on the branching ratio is rather small.

Let us briefly discuss how well the MIA expressions presented
in section~\ref{bsg} describe the results of our numerical
analysis. The two panels in~\fig{bsgres:glchwcanal}
\FIGURE[t!]{
  \begin{tabular}{c c}
    \includegraphics[angle=0,width=0.45\textwidth]{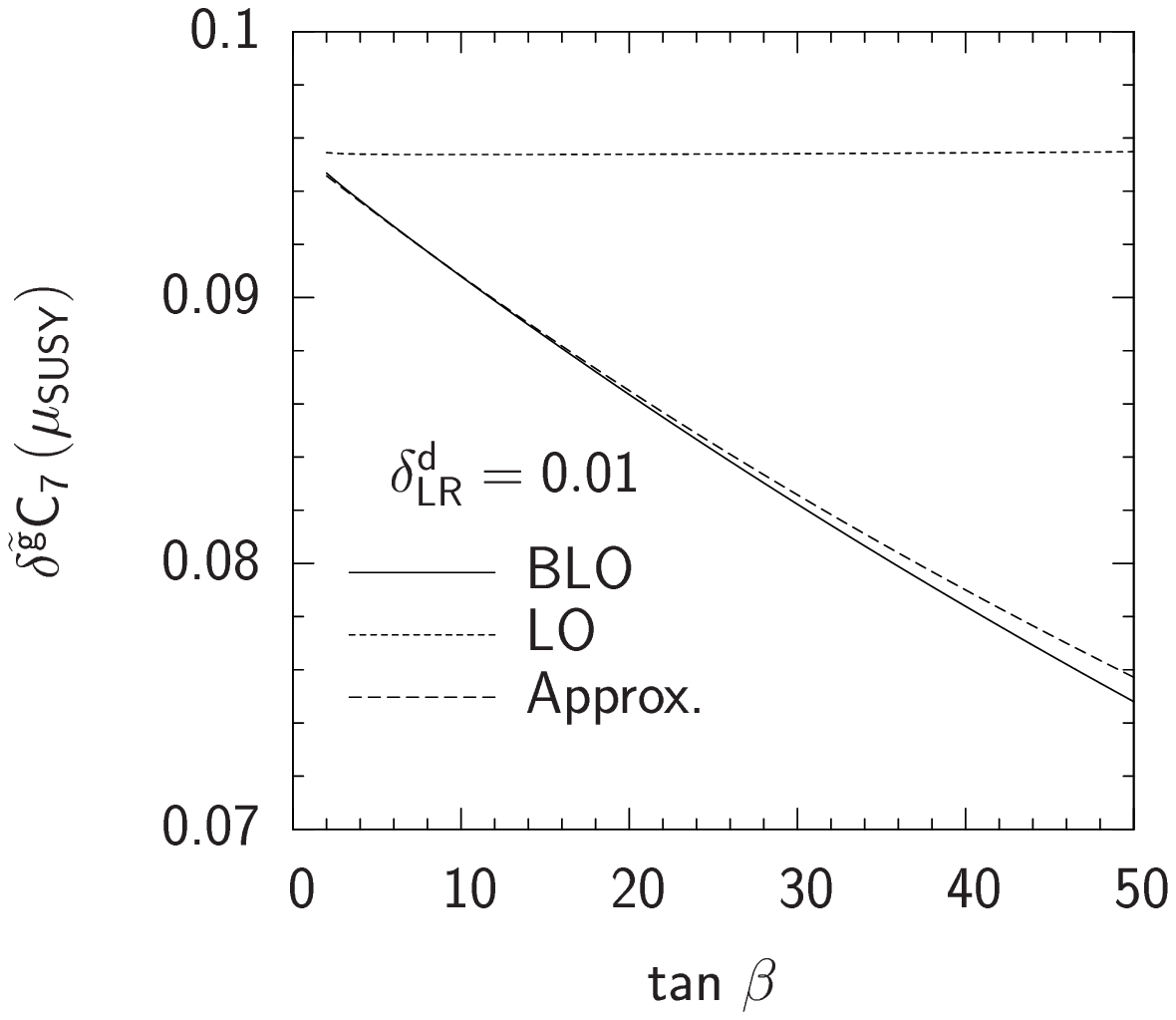}
    & \includegraphics[angle=0,width=0.45\textwidth]{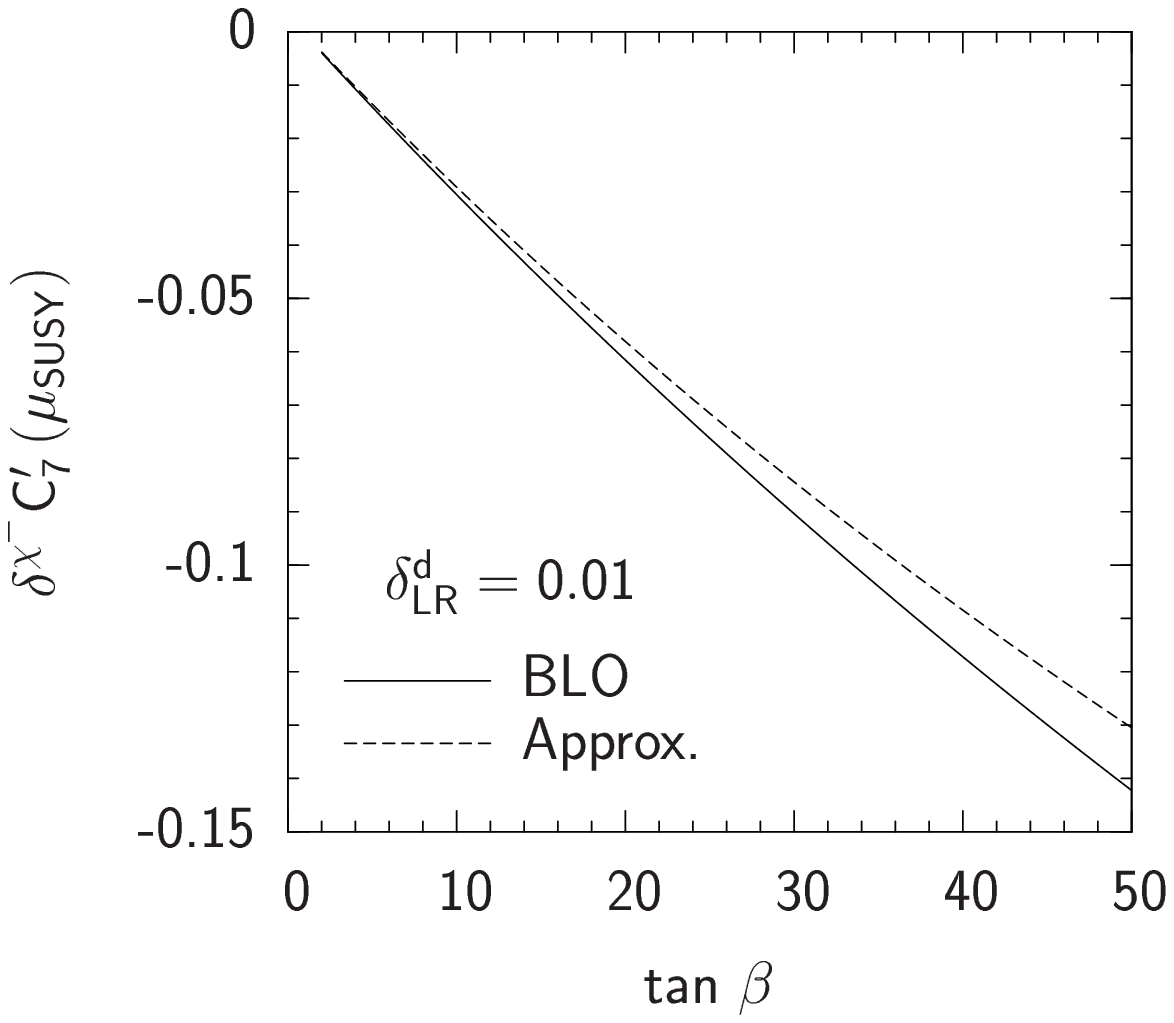}
  \end{tabular}
  \caption{Plots comparing of the results of our full numerical calculation
    with the analytic results, derived in the MIA,
    given in section~\ref{bsg}. The
    gluino contribution is shown in the panel on the left,
    whilst the chargino contribution is shown in the 
    panel on the right. $\dlr=0.01$ in both figures whilst
    the remaining parameters are described in the caption
    of \fig{bsgres:MFVwc}. The MFV
    contribution to the Wilson coefficient in question
    has been removed in both panels.\label{bsgres:glchwcanal}}
}
show the approximate expressions
for the gluino~\eqref{bsg:SU:gluLR} and chargino~\eqref{bsg:SU:chaLR}
contributions, alongside the results of our full numerical analysis,
performed in the mass eigenstate formalism where flavour violation
is communicated via the unitary matrices~\eqref{GA:GammaDecomp}. 
As is evident from the graph, in this region of parameter space
at least, the agreement between the approximate expression
and that of the full calculation is rather good (within 10\%).
We have checked that expressions for the chargino and gluino contributions
that arise from the remaining insertions also tend to agree
within a similar accuracy.

Now let us consider the effect of including electroweak corrections
when computing the corrected charged Higgs
vertex~\nqs{MIA:EW:CL}{MIA:EW:CR}.
\FIGURE[t!]{
  \begin{tabular}{c c}
    \includegraphics[angle=0,width=0.45\textwidth]{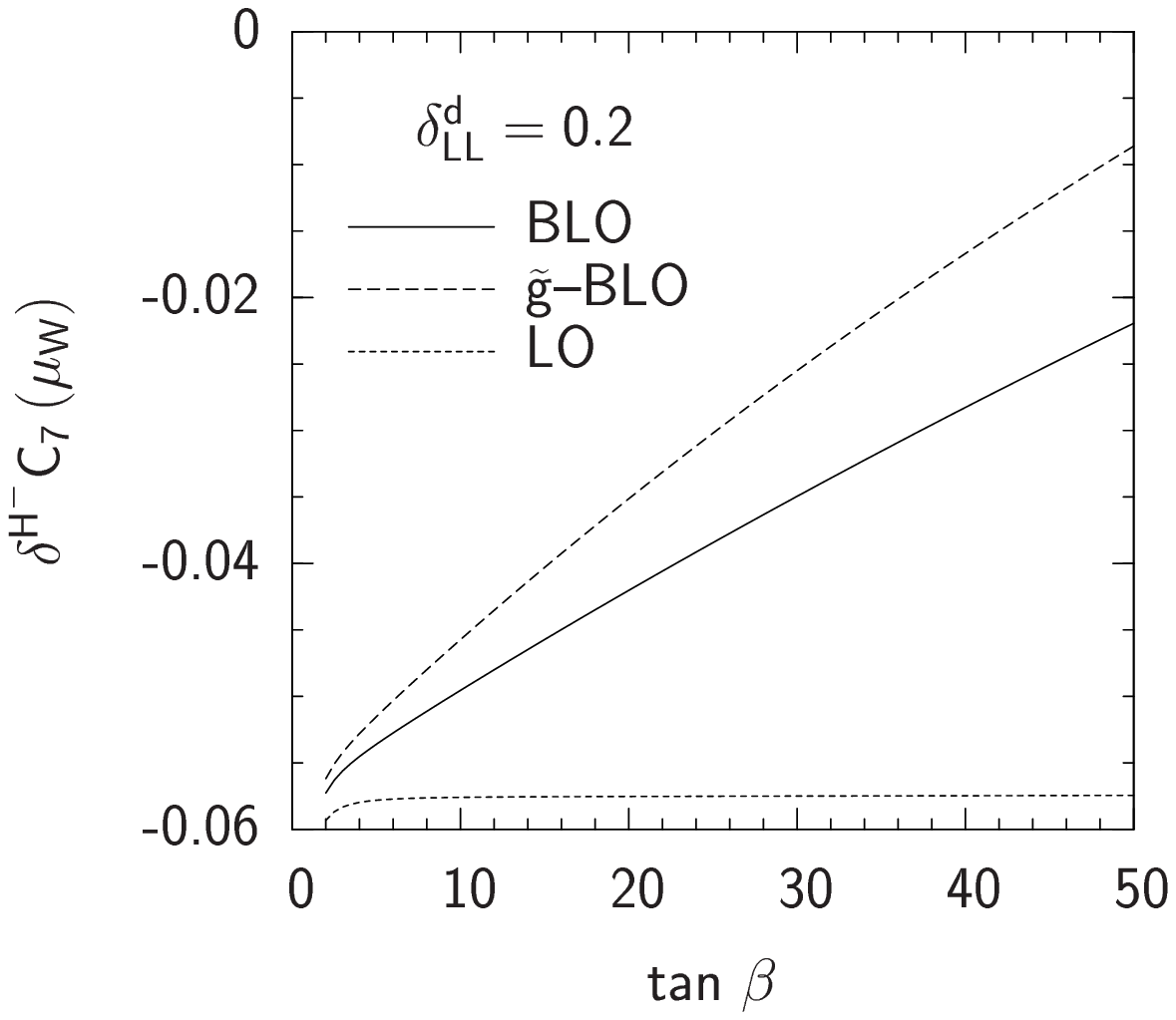}
    & \includegraphics[angle=0,width=0.45\textwidth]{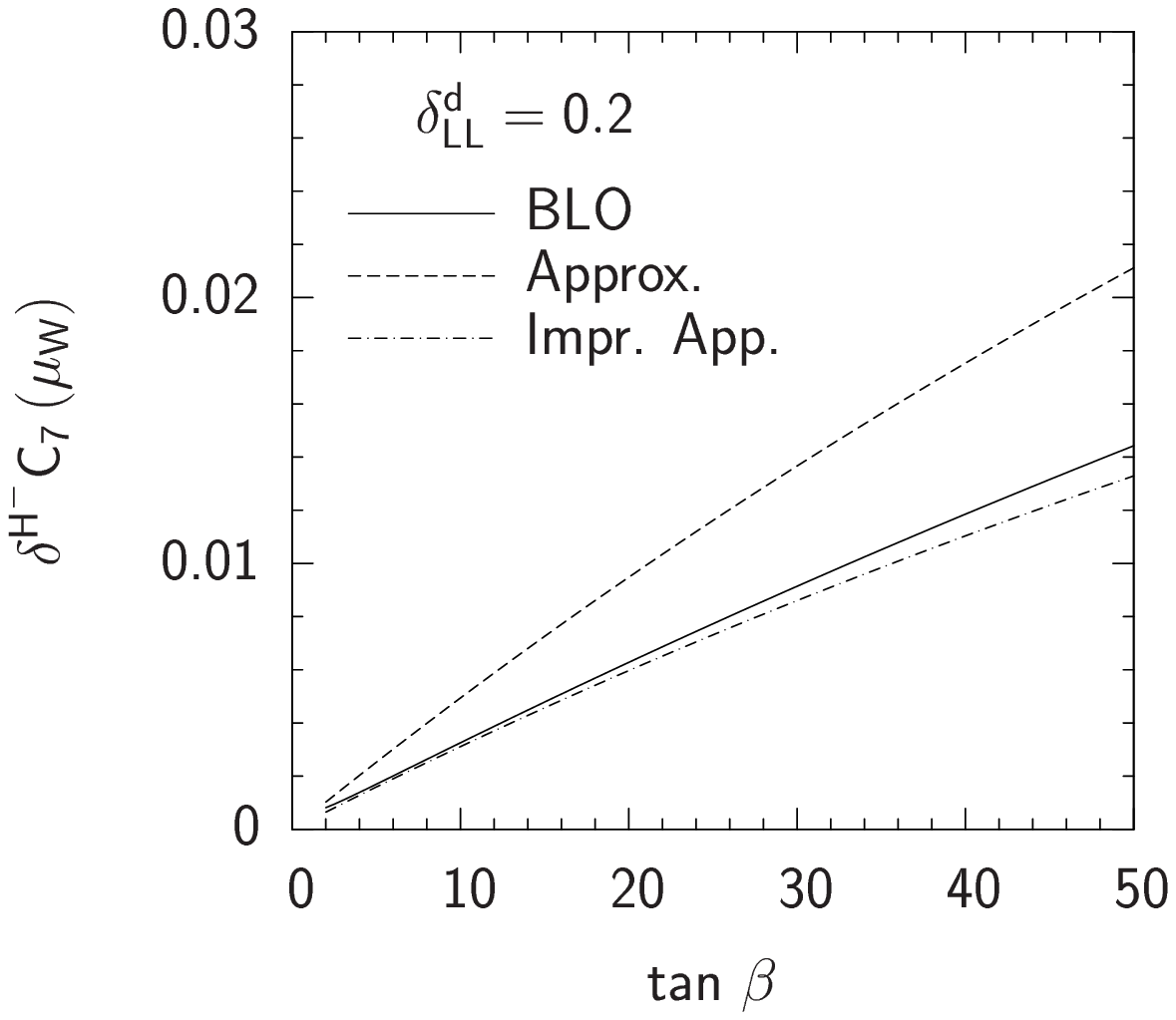}
    \\
        \includegraphics[angle=0,width=0.45\textwidth]{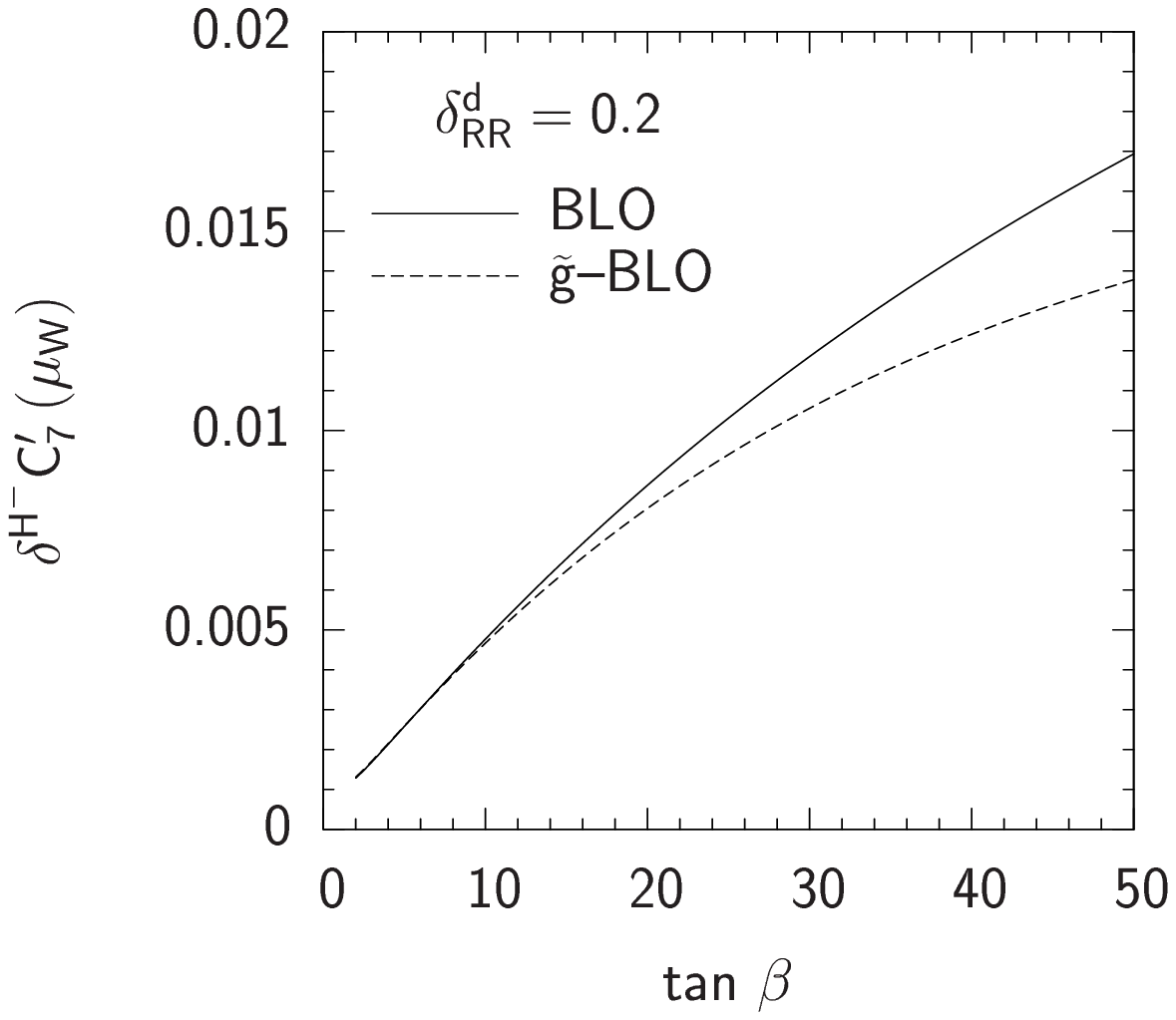}
    & \includegraphics[angle=0,width=0.45\textwidth]{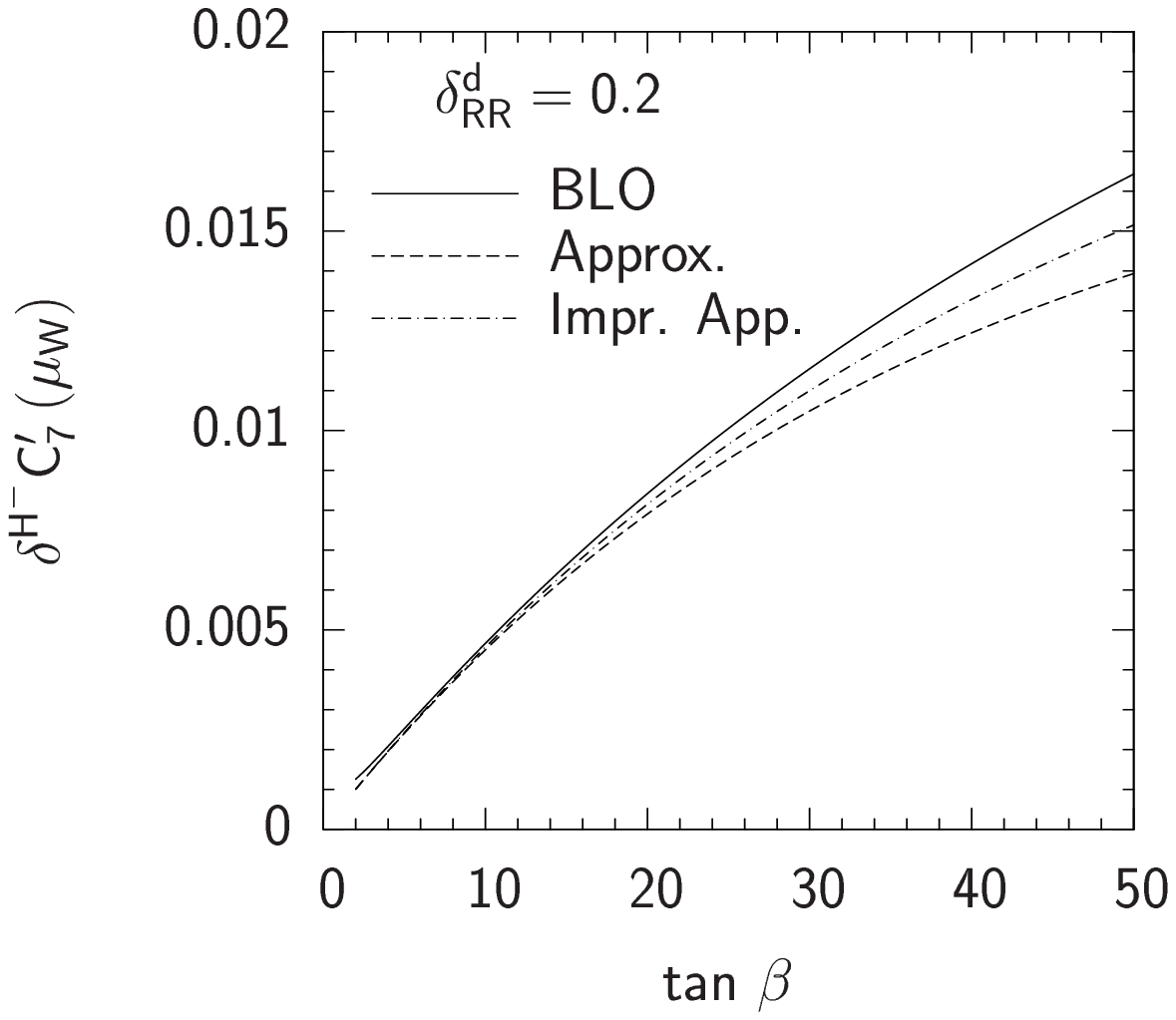}
  \end{tabular}
  \caption{
    The corrections to the Wilson coefficient $C_7$ arising from
    charged Higgs exchange. In the top two panels $\dll=0.2$
    whilst in bottom panels $\drr=0.2$. LO, BLO and $\widetilde{g}$
    are defined in the caption of \fig{bsgres:glchwc}. In the
    panels to the right our analytic
    results are compared with those of our complete numerical
    analysis. The MFV contributions
    to $C_7$ and $C_7^{\prime}$
    arising from charged Higgs exchange have been removed in the
    panels to the right. The abbreviation ``Approx.'' denotes the
    results gathered in section~\ref{bsg:EW}, whilst ``Impr. App.''
    is used to denote the same results supplemented by the substitutions
    gathered in subsection~\ref{ImpApp}.
    The soft sector is described by the same
    parameters as \fig{bsgres:MFVwc}.\label{bsgres:chwc}}
}
The panel on the top--left of \fig{bsgres:chwc}, illustrates the
effect of including BLO corrections when calculating the charged
Higgs mediated contribution to $C_7$, for non--zero $\dll$.
If $\dll$ is positive, the GFM corrections to the charged
Higgs vertex tend to interfere constructively with the MFV
contribution, reducing the charged Higgs contribution
to $C_7$ compared to a LO analysis.
On the other hand, if $\dll$ is negative,
the MFV and GFM corrections to the LO contribution interfere
destructively and can therefore lead to an enhancement of
the charged Higgs contribution to the decay.
As $A_b$ is set equal to zero, the bulk of the difference
induced by the inclusion of EW contributions in the figure results from
the correction to the right handed coupling~\eqref{MIA:EW:CR} and
the corrections that arise from the additional gaugino
mediated electroweak effects discussed in subsection~\ref{ImpApp}.
This is confirmed in the top--right panel that illustrates the
behaviour of the approximate expression~\eqref{bsg:EW:CHGFM} compared
with the results of
our full numerical calculation. As is evident from the figure,
the agreement between the curves is rather poor and the
approximate result where the electroweak couplings $g_1$ and $g_2$
are ignored provides a 30\% overestimate of the beyond leading order
effects. The origin of this discrepancy stems from graphs
featuring gaugino exchange. This result is not specific
to the GFM scenario and such effects have been discussed
before, in the context of MFV in~\cite{BCRS:bdec}. Once one
takes into account these effects, by applying the
substitutions found in subsection~\ref{ImpApp}, the agreement
between the numerical and approximate results improves dramatically.
As is evident from the line depicting the improved approximation
in the top--right panel of~\fig{bsgres:chwc}.

The lower two panels in \fig{bsgres:chwc} depict the contributions
to the primed coefficients arising from BLO corrections
to charged Higgs exchange for $\drr=0.2$. In a similar
manner to the insertion $\dll$, the majority of the difference
between 
the effects considered 
in~\cite{OR2:bsg} (the line labeled $\widetilde{g}$--BLO) and
the complete calculation
presented in this analysis, stems from the destructive
interference between the $\epsilon_s$ and $\epsilon_Y$ terms
that appear in the denominator of~\eqref{MIA:EW:CR} and
the additional electroweak corrections to the $(3,3)$ element
of the left--handed charged Higgs vertex. Both of these
effects act to increase the charged Higgs contribution
attributable to RR insertions. This situation is also
evident in the lower right panel, where the inclusion
of the additional electroweak corrections described
in subsection~\ref{ImpApp} roughly doubles the accuracy
of the approximate expression.

The contributions due to new physics for each insertion
are shown in \fig{bsgres:wctanbsum}.
\FIGURE[t!]{
  \begin{tabular}{c c}
    \includegraphics[angle=0,width=0.45\textwidth]{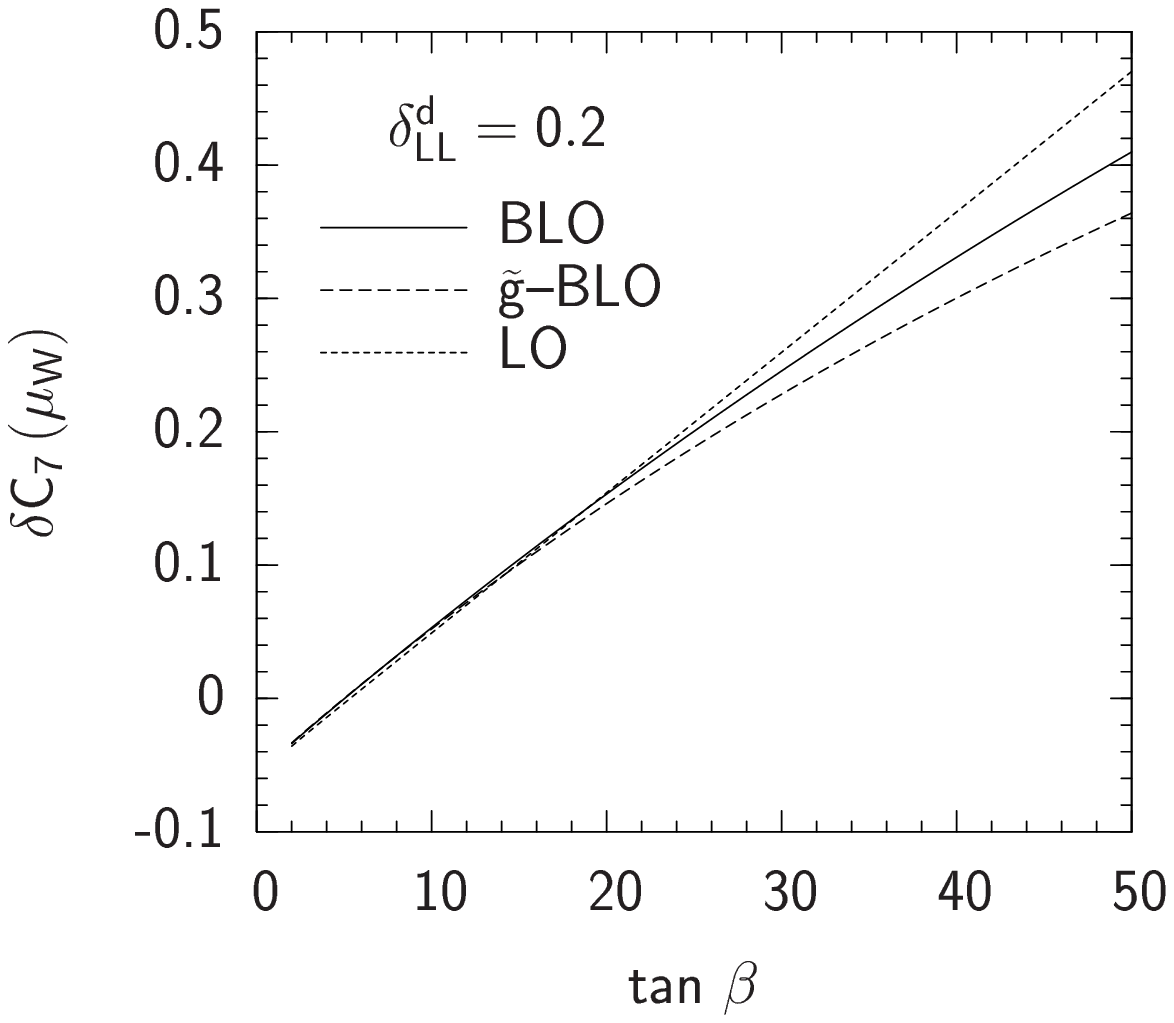}
    & \includegraphics[angle=0,width=0.45\textwidth]{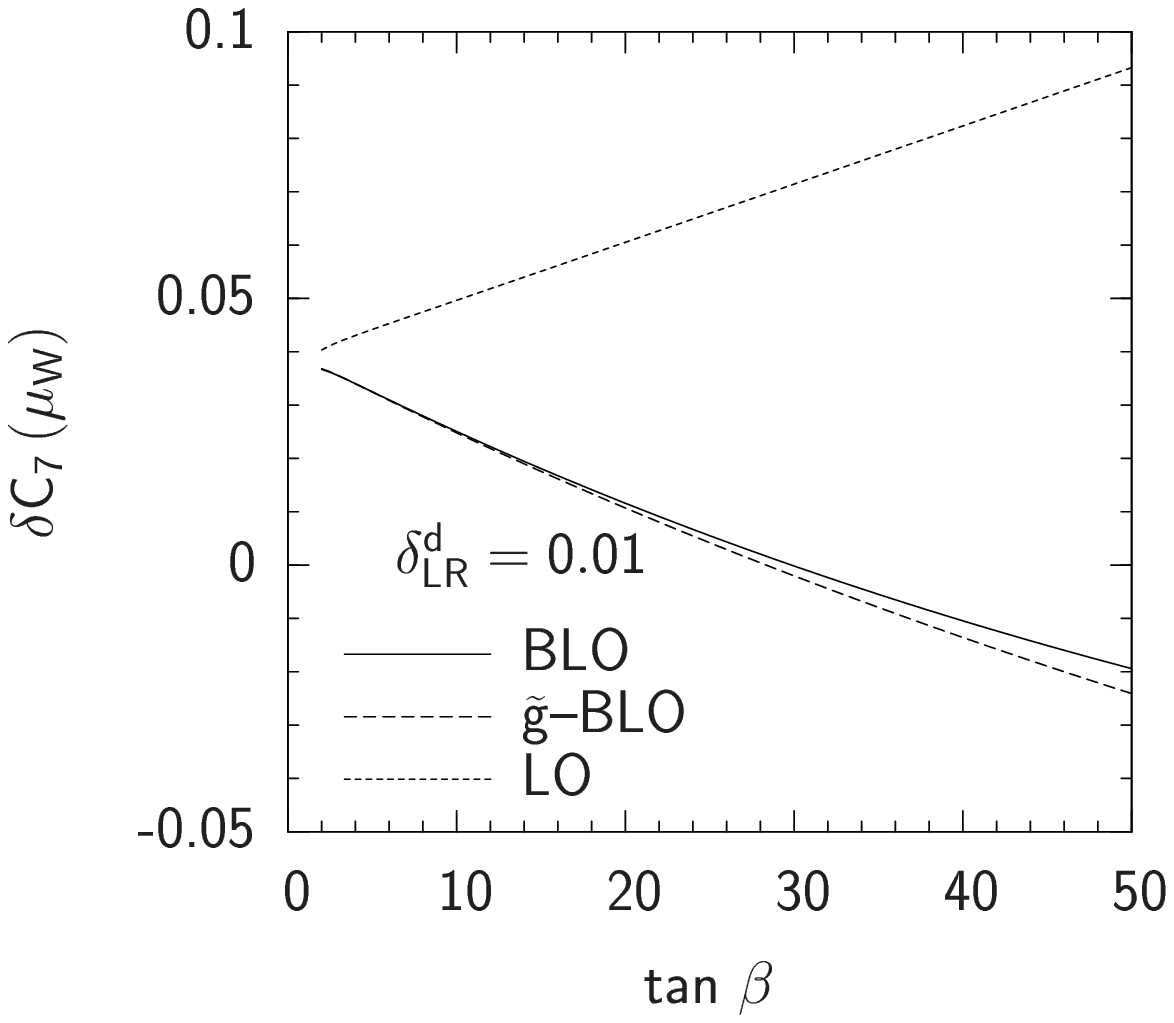}\\
    \includegraphics[angle=0,width=0.45\textwidth]{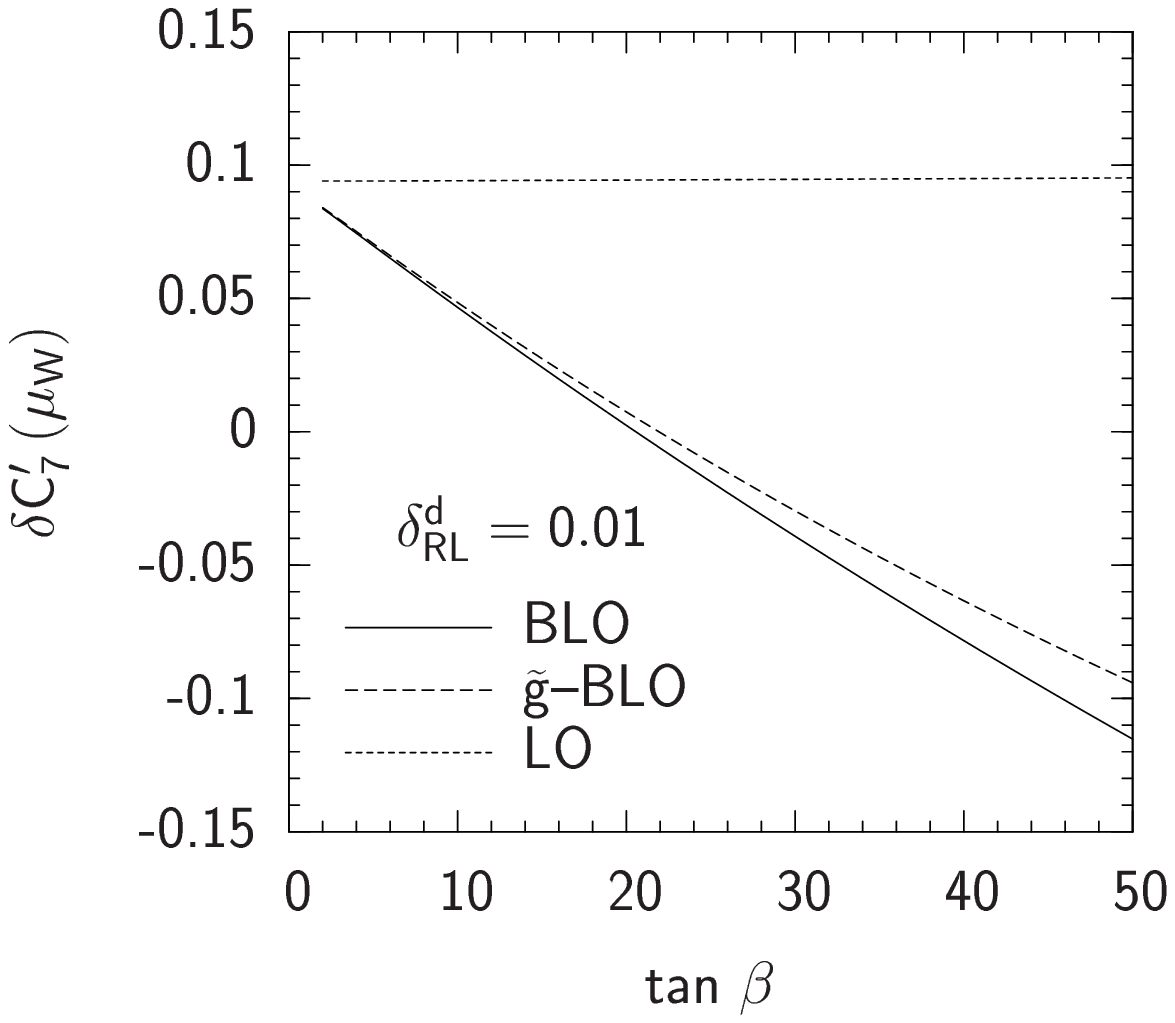}
    & \includegraphics[angle=0,width=0.45\textwidth]{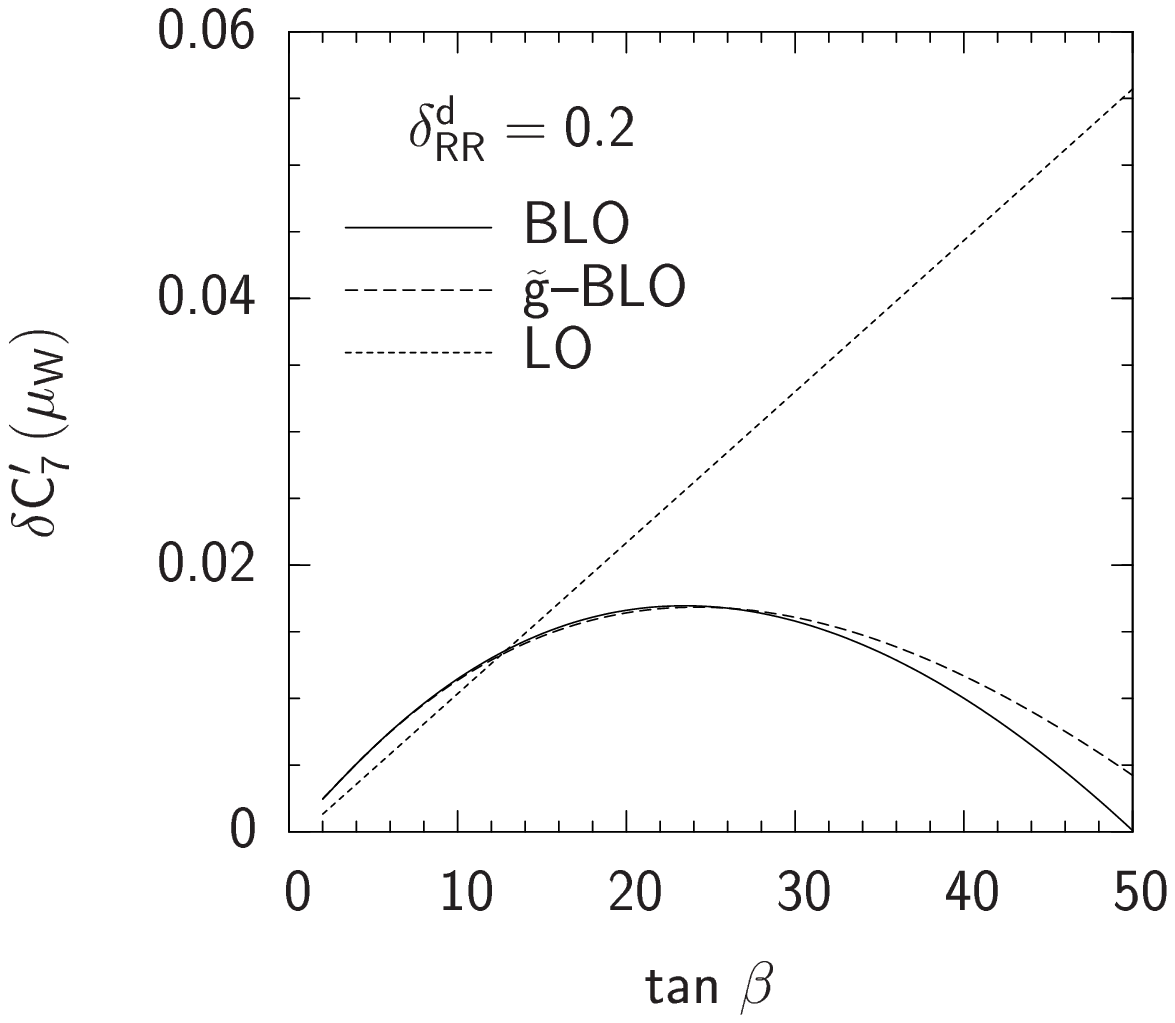}
  \end{tabular}
  \caption{The corrections to the Wilson coefficients $C_7$
    (the top     two panels) and $C_7^{\prime}$ (the lower two panels)
    due to contributions beyond the SM. The soft sector
    is described by the same parameters as \fig{bsgres:MFVwc}.
    \label{bsgres:wctanbsum}}
}
As is evident from the plots for the LR, RL and RR insertions,
the effect of BLO corrections tends to be rather large. For
example, in the case of the LR insertion (the top--right panel in
\fig{bsgres:wctanbsum}), the reduction of the gluino contribution with
increasing $\tanb$, coupled with its destructive interference
with the chargino contribution, can dramatically alter the
behaviour of the Wilson coefficient for even moderate $\tanb$.
Similar effects occur for the RL and RR insertions. For the
LL insertion the difference between the LO and BLO calculations
shown in the top--left panel of \fig{bsgres:wctanbsum} is rather
slight. This is because a $\tanb$ enhanced correction,
proportional to $\dll$,
appears at LO in the chargino Wilson coefficients.
It is therefore rather difficult for the BLO corrections to the
chargino contribution to play as large a r\^ole as they
do for the other three insertions.

Finally, let us consider the combined effect these
contributions have on the branching ratio. \fig{bsgres:dxycnt}
\FIGURE[t!]{
  \begin{tabular}{c c}
    \includegraphics[angle=0,width=0.45\textwidth]{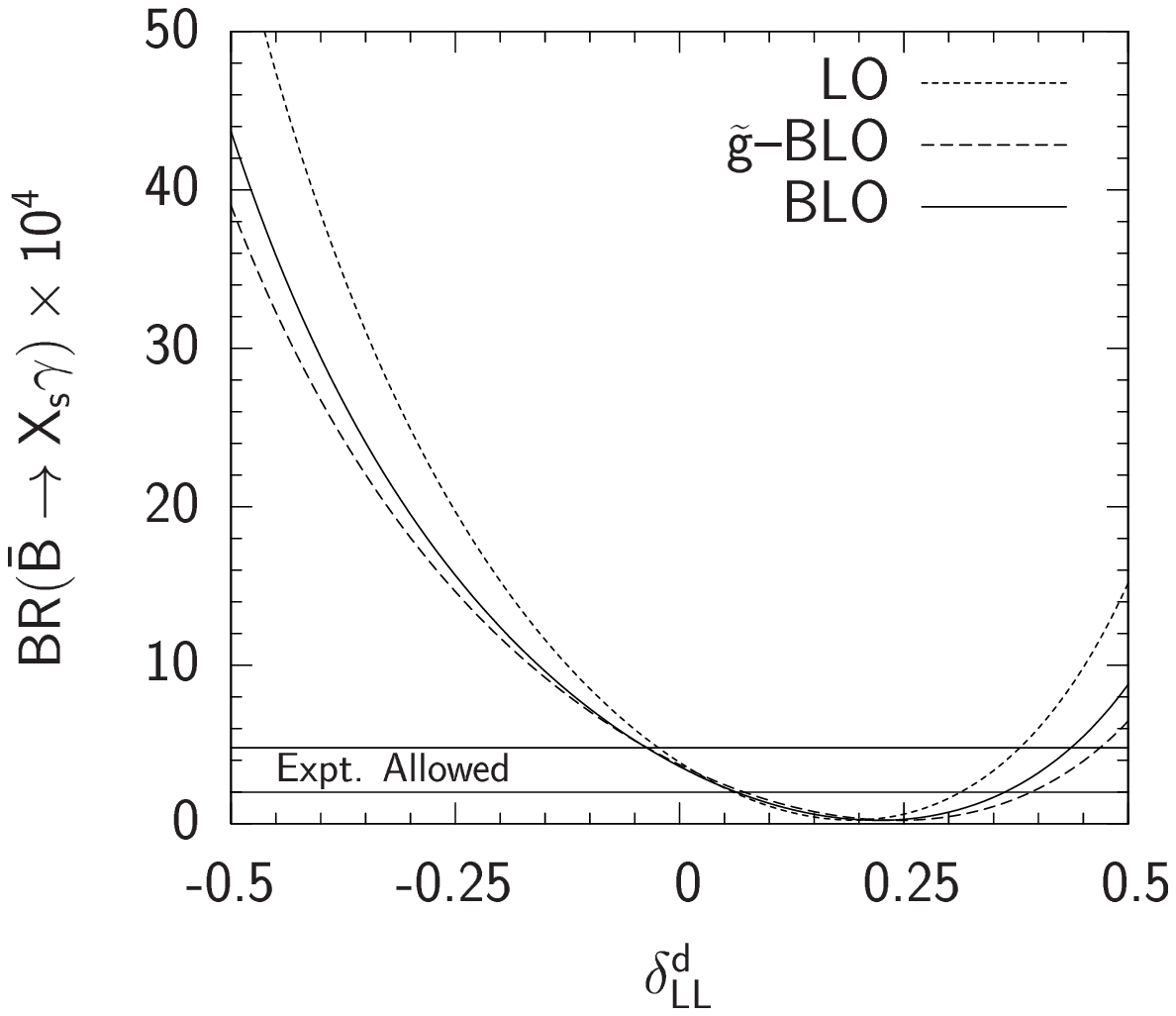}
    & \includegraphics[angle=0,width=0.45\textwidth]{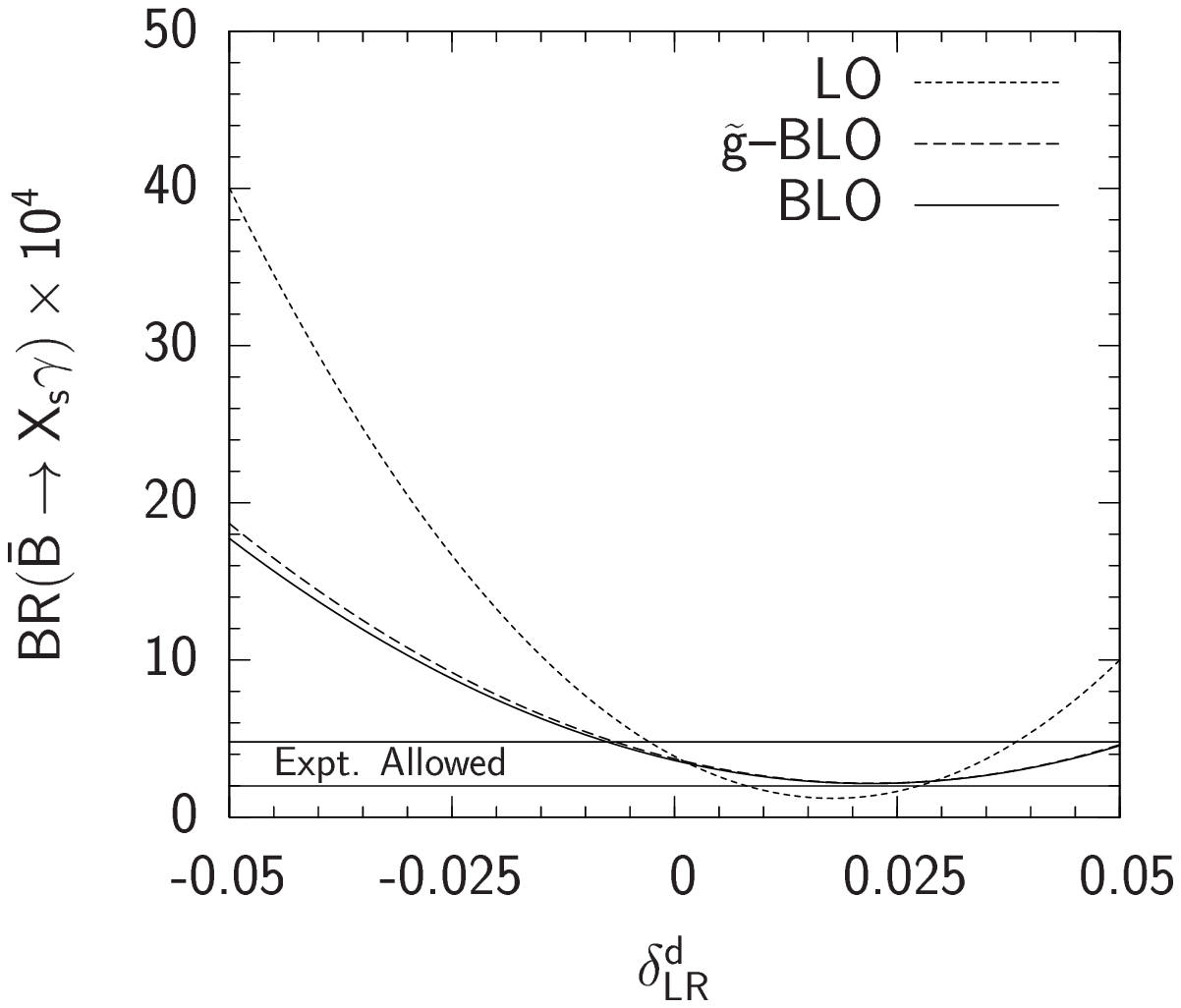}\\
    \includegraphics[angle=0,width=0.45\textwidth]{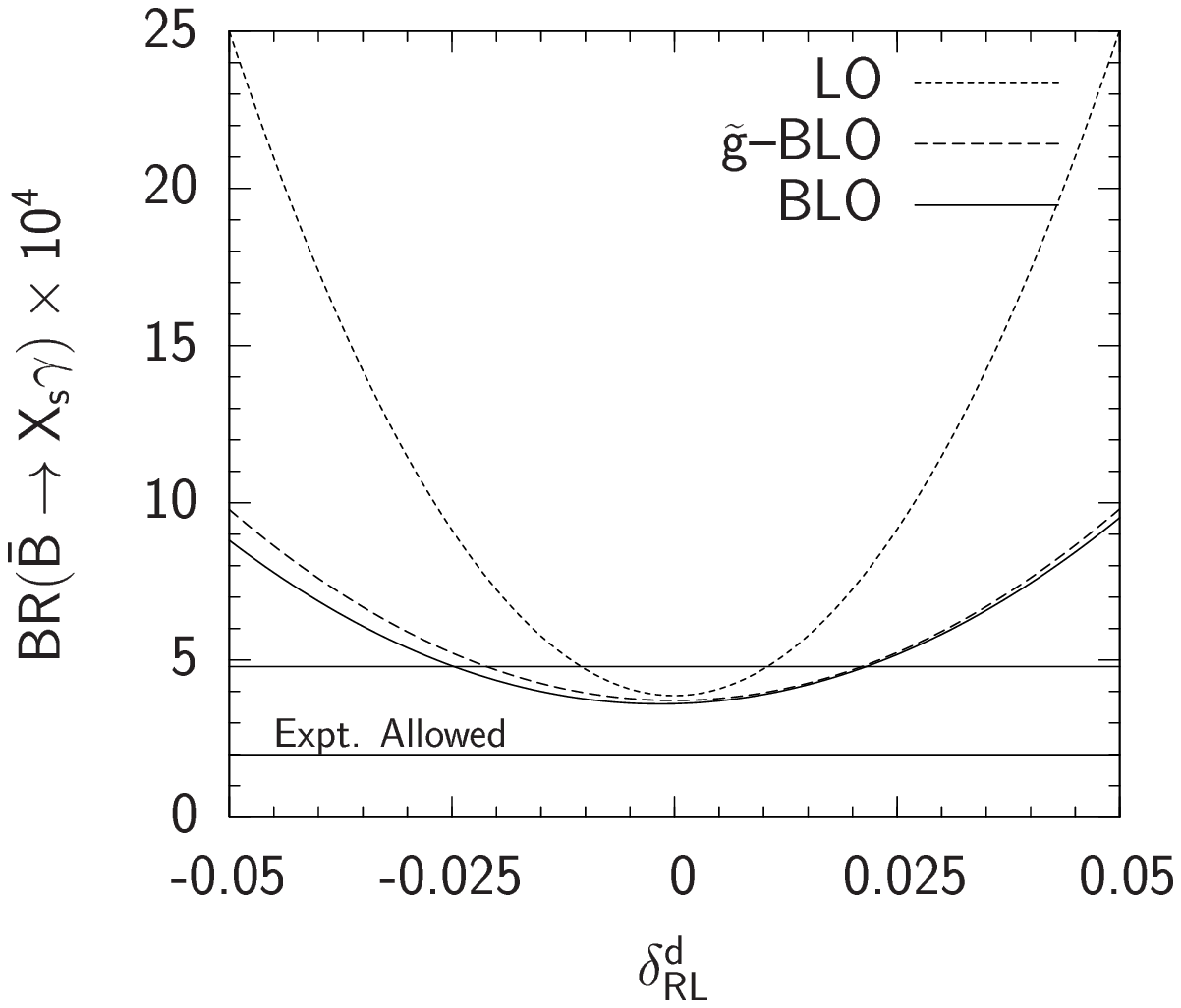}
    & \includegraphics[angle=0,width=0.45\textwidth]{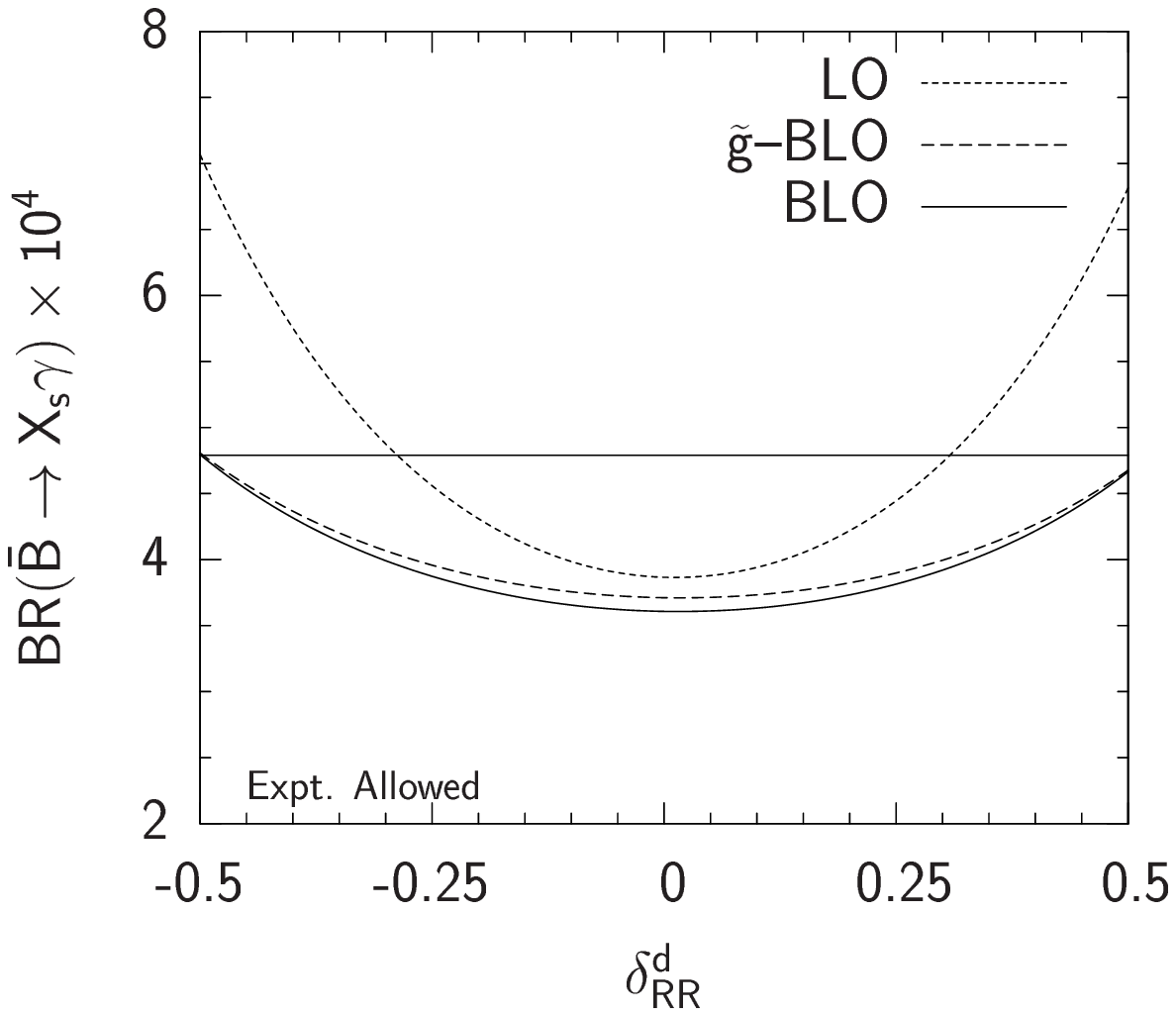}
  \end{tabular}
  \caption{The variation of BR$(\bsg)$ with
    the flavour violating parameters $\dxy$. The soft sector
    is parameterised as follows $\msq=1\tev$, $\mgl=\msq\,/\sqrt{2}$, 
    $A_u=-500\gev$, $m_A=500\gev$, $\mu=500\gev$ and $\tanb=50$.
    A broad region in agreement with the current experimental
    limit is shown in all four panels.\label{bsgres:dxycnt}}
}
illustrates the variation of the branching ratio
with the flavour violating parameters $\dxy$. As is
evident from the figure, BLO effects can be significant
for all four insertions. Contributions due to flavour
violation in the LR and RL sectors, in particular,
can undergo reductions by up to a factor of two compared
to a LO analysis. Turning to the inclusion of electroweak
effects, we see that although such corrections can effect
the Wilson coefficients by up to 30\%, the overall
difference between the approximation used in~\cite{OR2:bsg}
and the full calculation used in this analysis is
rather small. This is primarily because the origin
of the large discrepancy, between the LO and BLO calculations,
is mainly due to the cancellation between the gluino and chargino
contributions, that arise for each insertion. As both
the chargino and gluino contributions undergo similar
corrections, once electroweak effects are taken into account,
the overall effect on the branching ratio tends to be rather minor.

\fig{bsgres:tanb} depicts the variation of the branching 
ratio with $\tanb$.
\FIGURE[t!]{
  \begin{tabular}{c c}
    \includegraphics[angle=0,width=0.45\textwidth]{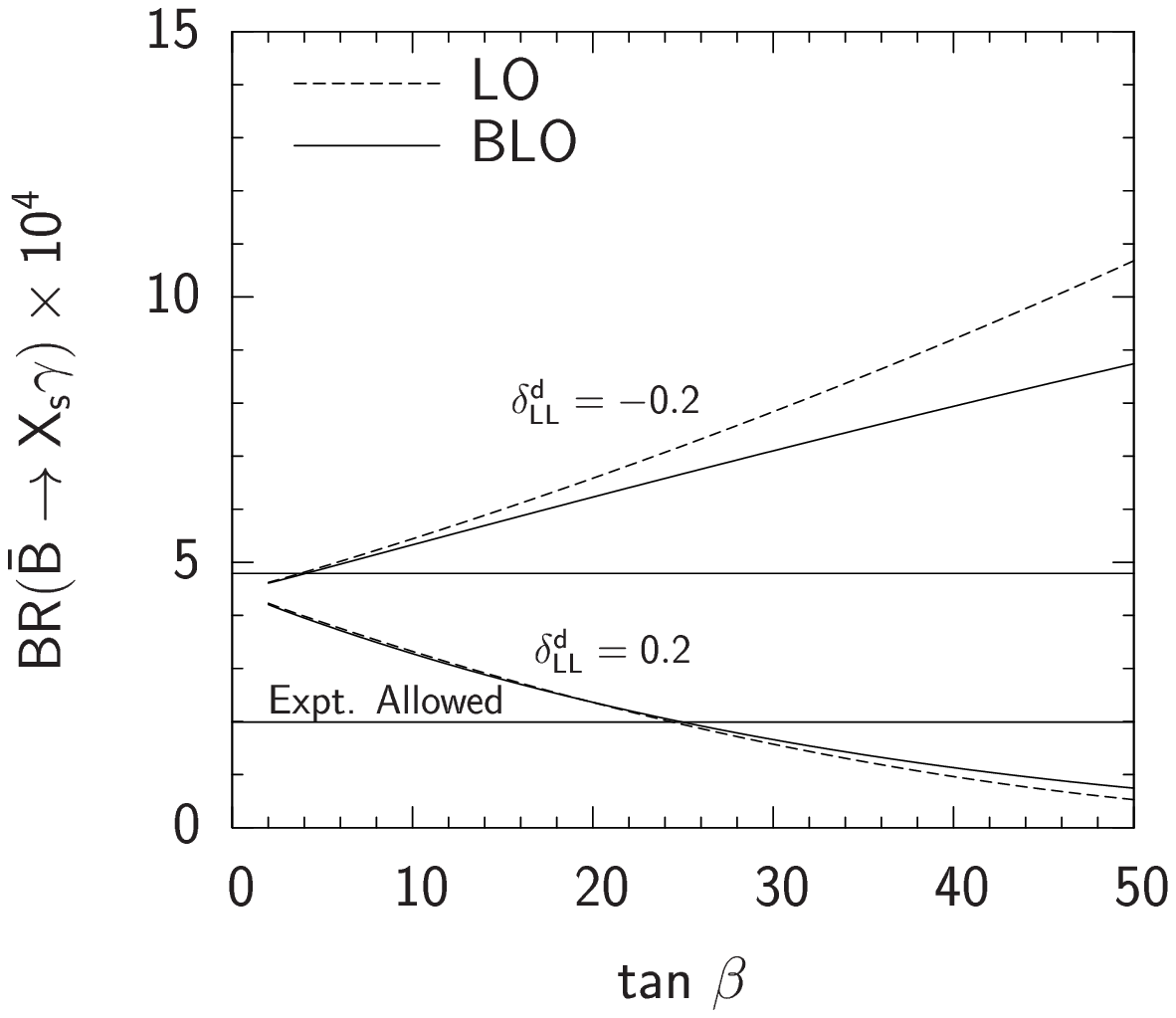}
    & \includegraphics[angle=0,width=0.45\textwidth]{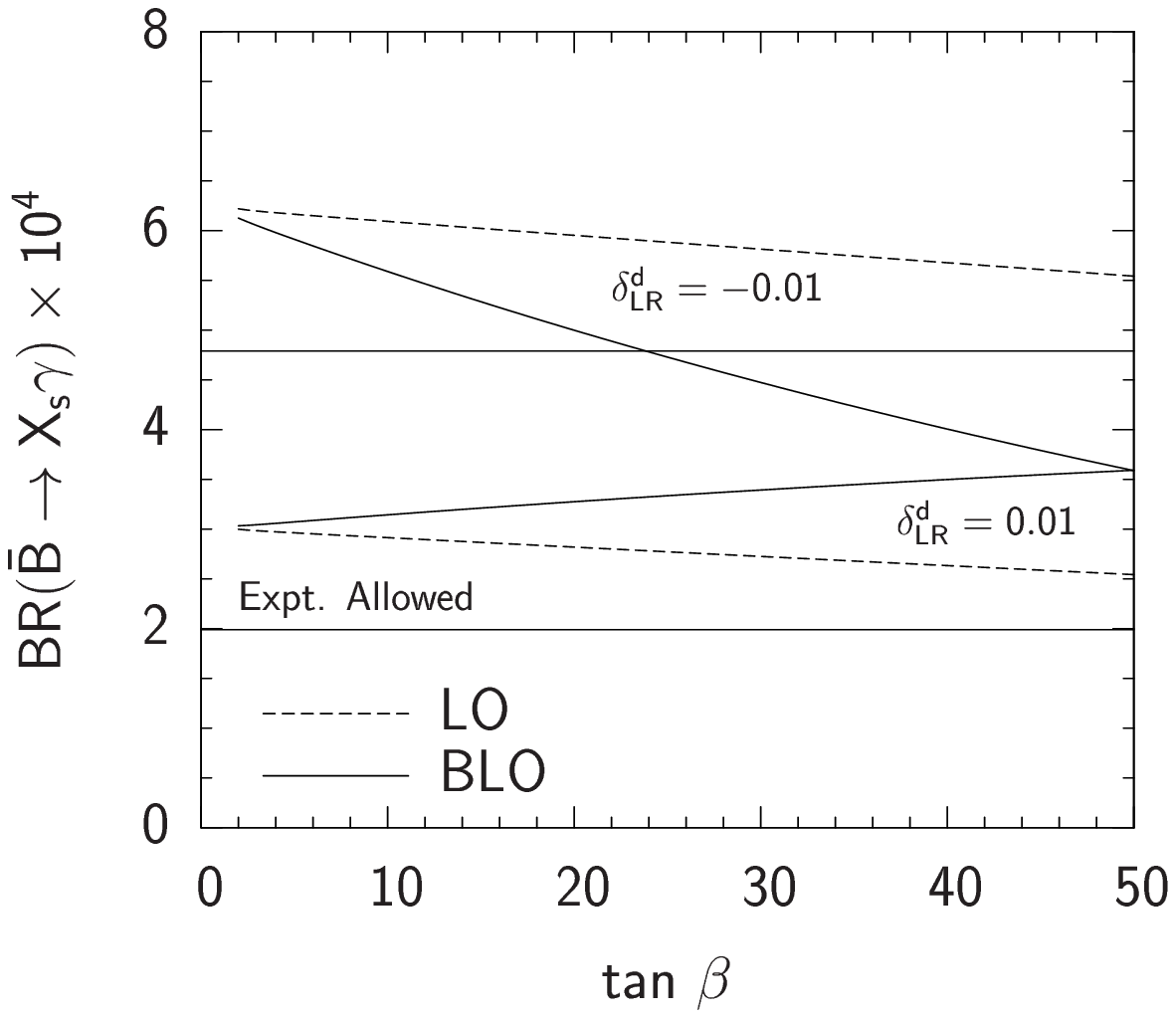}\\
    \includegraphics[angle=0,width=0.45\textwidth]{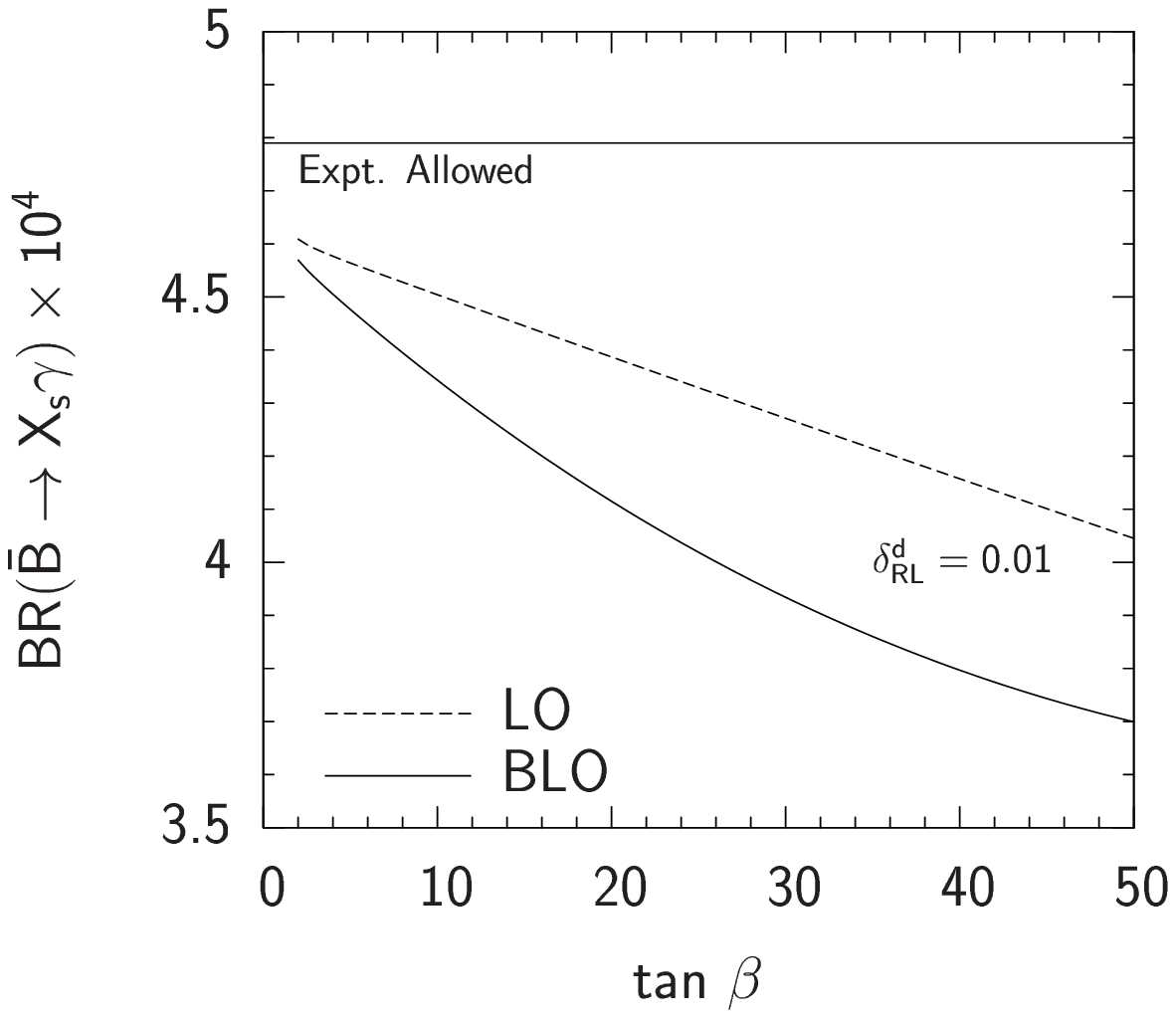}
    & \includegraphics[angle=0,width=0.45\textwidth]{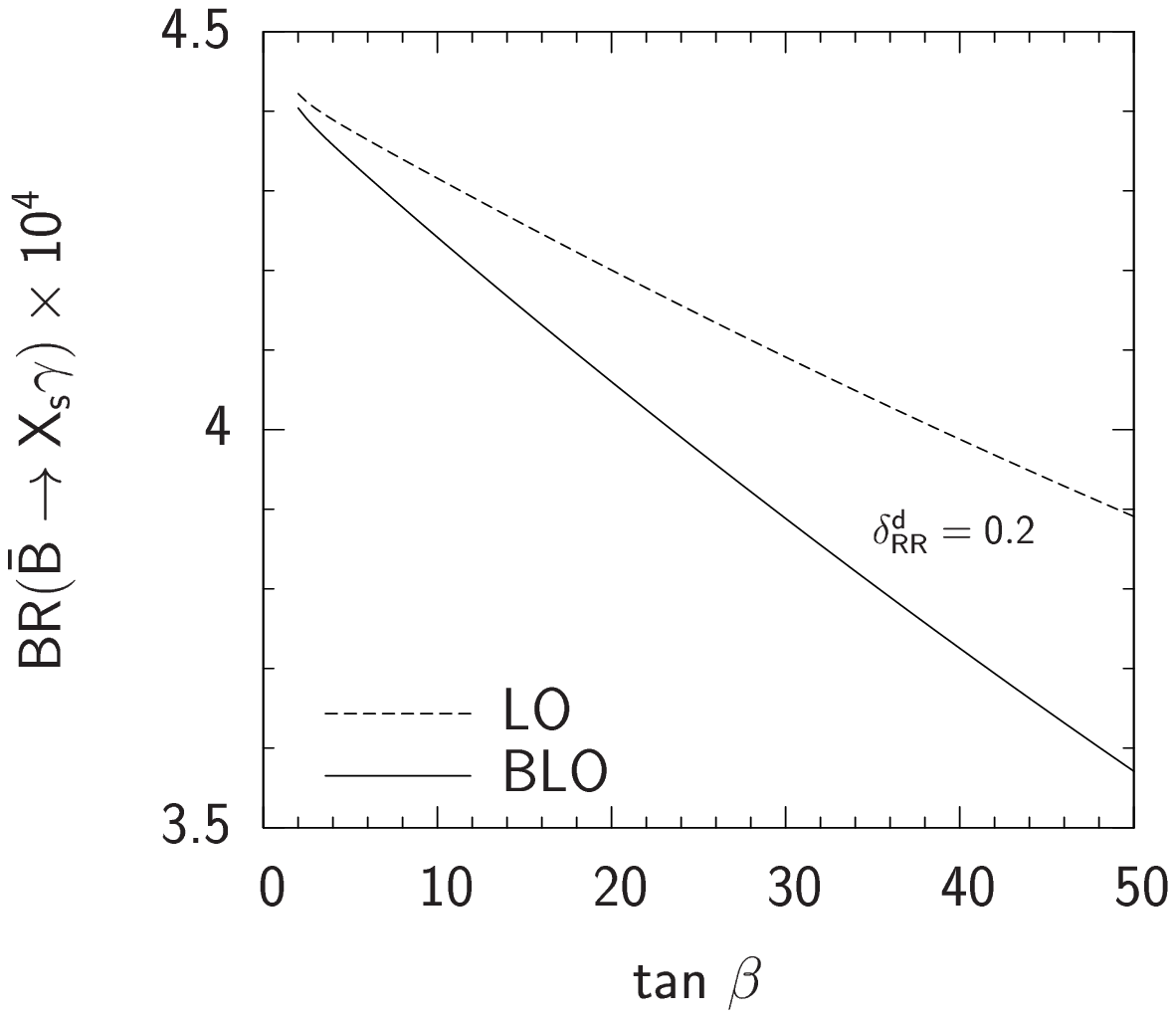}
  \end{tabular}
  \caption{The variation of BR$(\bsg)$ with
    the $\tanb$. The soft sector is parameterised as
    follows $\msq=1\tev$, $\mgl=\sqrt{2}\msq$, 
    $A_u=-500\gev$, $m_A=500\gev$ and $\mu=500\gev$.
    A broad region in agreement with the current experimental
    limit is also shown when applicable.\label{bsgres:tanb}}
}
All four panels exhibit the focusing effect
first described 
in~\cite{OR:bsg,OR2:bsg} where BLO corrections
act to reduce the LO result such that the SM result~\eqref{bsg:SM}
is preferred. Once again we see that for the insertion $\dll$
BLO corrections are rather small due to the presence of the
LO correction to the chargino contribution in~\eqref{bsg:SU:chaLL}. For
the remaining insertions we see that the BLO effects described
in this paper can significantly alter LO corrections for
$\tanb$ as low as 20.

In summary, we have seen that whilst electroweak corrections
can effect individual Wilson coefficients by up to 20\%, the
overall effect of such contributions is rather small. In particular,
the focusing effect described
in~\cite{OR:bsg,OR2:bsg} remains
even once one includes all the electroweak corrections
described in this paper. This is unsurprising as the focusing
effect mainly arises from the combined effect of two contributions.
The first arises from renormalization group evolution of the
Wilson coefficients evaluated at the SUSY scale to the electroweak
scale, which is naturally independent of electroweak corrections.
The second is due to cancellations between the gluino and chargino
contributions to the decay. We have seen however that electroweak
corrections typically alter each contribution in a similar manner,
and an increase in the gluino contribution, attributable to
the destructive interference of the gluino and chargino contributions
to the bare mass matrix, is often accompanied by a similar 
increase in the chargino contribution. Finally, we have seen
that the approximate expressions gathered in section~\ref{bsg}
tend to describe the overall behaviour of the supersymmetric
contributions rather well. The contributions arising from
the charged Higgs exchange however, often have to be modified according
to the improved approximation described in subsection~\ref{ImpApp}
to obtain a sufficient level of accuracy. 

%%%%%%%%%%%%%%%%%%%%%%%%%%%%%%%%%%%%%%%%%%%%%%%%%%%%%%%%%%%%%%%%%
\subsection{$\bsm$}
\label{NRes:bsm}
%%%%%%%%%%%%%%%%%%%%%%%%%%%%%%%%%%%%%%%%%%%%%%%%%%%%%%%%%%%%%%%%%

As discussed in section~\ref{bsm}, large corrections to the decay
$\bsm$ are possible in the large $\tanb$ regime due to 
the contributions of scalar and pseudoscalar operators that
are proportional to $\tan^3\beta$. \fig{bsmres:wctanb}
\FIGURE[t!]{
  \begin{tabular}{c c}
    \includegraphics[angle=0,width=0.45\textwidth]{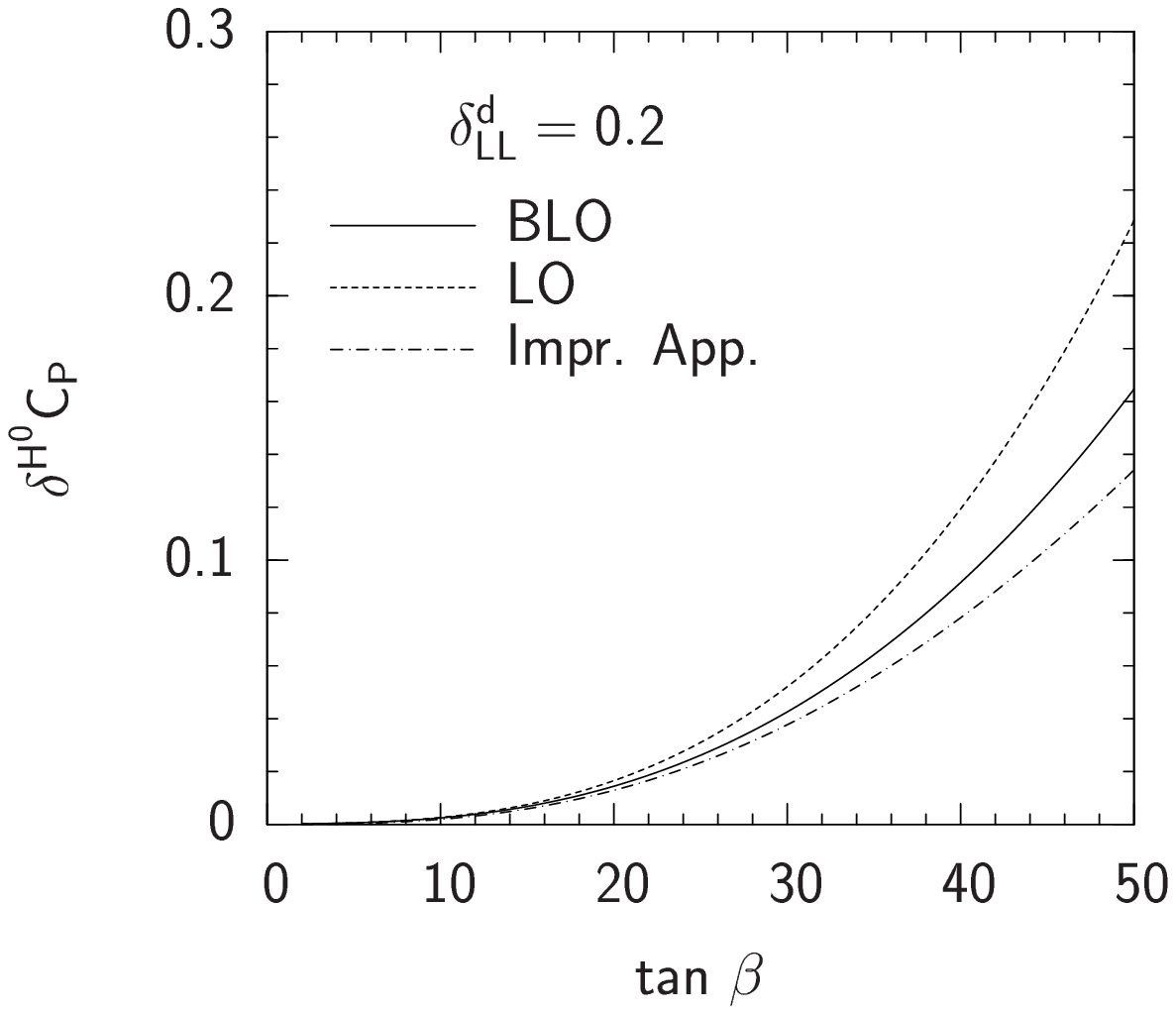}
    & \includegraphics[angle=0,width=0.45\textwidth]{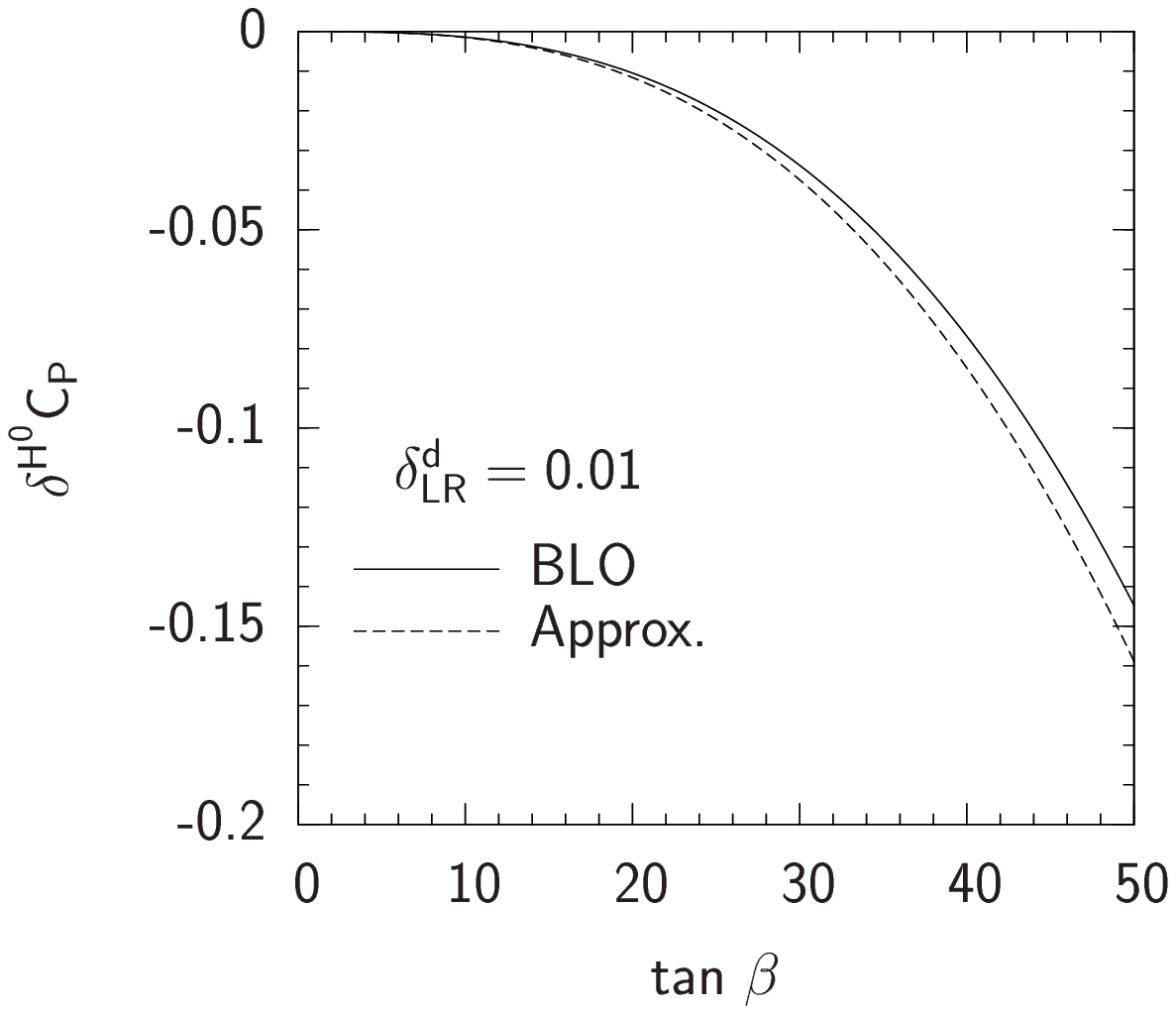}\\
    \includegraphics[angle=0,width=0.45\textwidth]{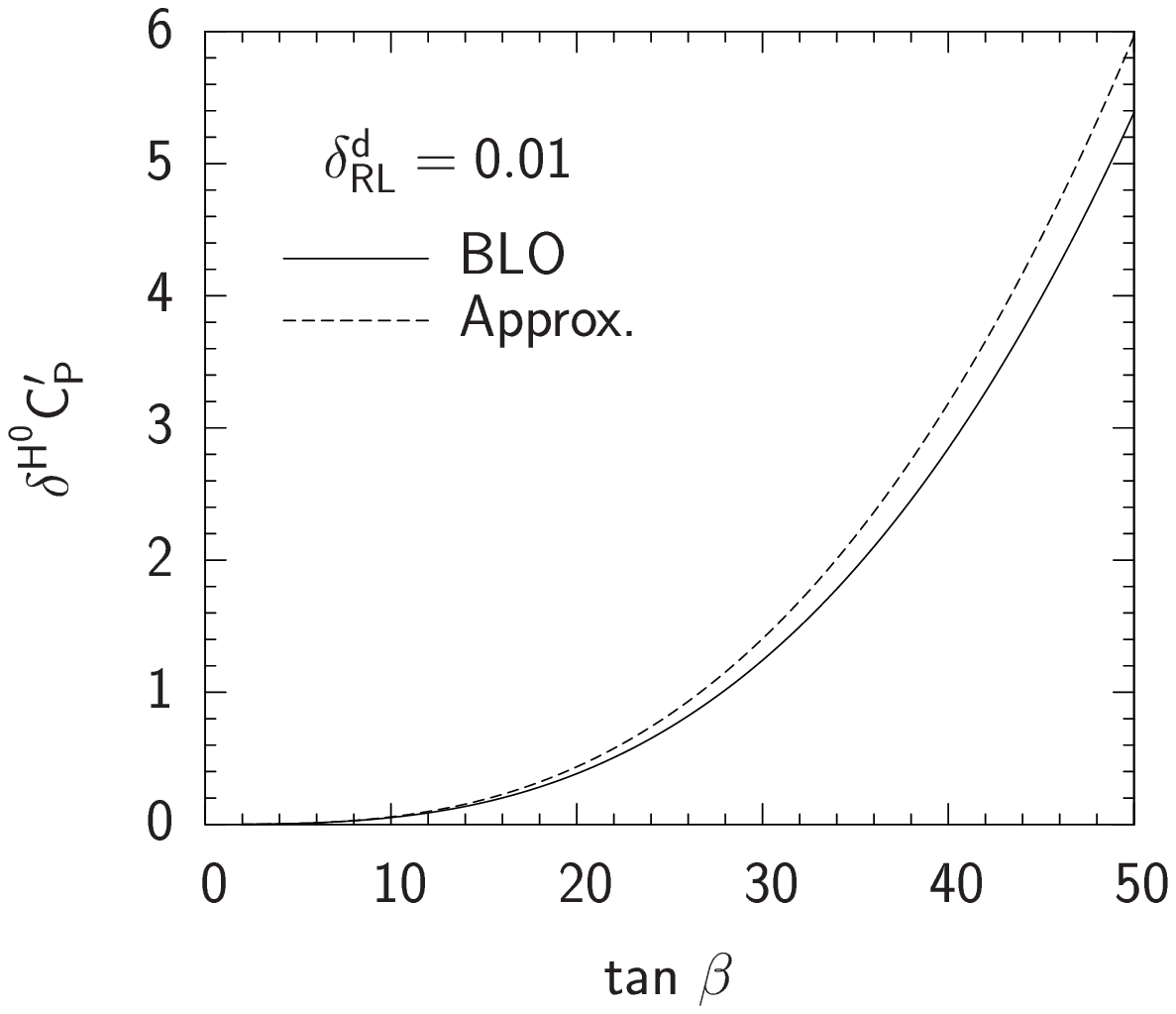}
    & \includegraphics[angle=0,width=0.45\textwidth]{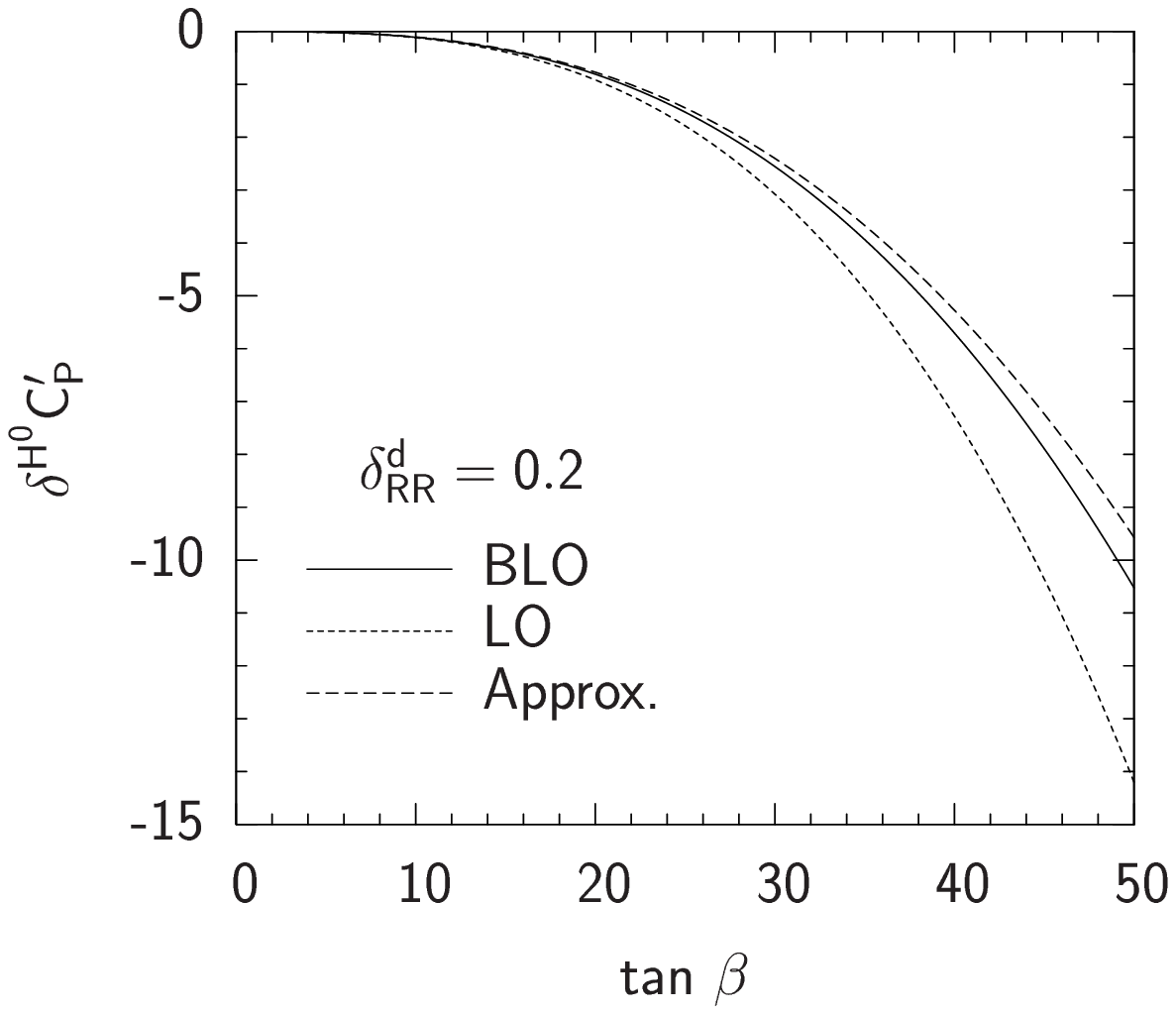}
  \end{tabular}
  \caption{
        The $\tanb$ dependence of the Wilson coefficients
    $C_P$ and $C_P^{\prime}$. In the top two panels $\dll=0.2$ and $\dlr=0.01$,
    respectively, whilst in the lower two $\drl=0.01$ and $\drr=0.2$.
    In all four panels the MFV contributions to $C_P^{(\prime)}$
    have been removed.
    The soft sector is parameterised as follows $\msq=1\tev$,
    $\mgl=\sqrt{2}\msq$, $A_u=-500\gev$,
    $m_A=500\gev$, $\mu=500\gev$. The abbreviation ``Approx.'' 
    is used to denote the contributions arising from
    formulae~\nqs{bsm:EW:NHGFM}{bsm:EW:NHGFMpr} whilst
    ``Imp. Appr.'' is used to denote the calculation
    that includes the additional electroweak
    contribution~\eqref{bsm:EW:NHGFMEW}.
    \label{bsmres:wctanb}
    }
}
depicts the dependence of our analytic and numerical results for
the Wilson coefficients $C_P$ and $C_P^{\prime}$
on $\tanb$ for various choices of flavour violating parameters.
The top--left panel in the figure shows the contribution due
to the insertion $\dll$. Let us point out that when
one only uses expression~\eqref{bsm:EW:NHGFM} to calculate
the contribution to $C_P$, one obtains a value roughly 20\% larger than the
result obtained from a full numerical analysis. Once one takes
into account the additional LO correction induced by gaugino--higgsino
mixing~\eqref{bsm:EW:NHGFMEW} (denoted by Impr. App. in~\fig{bsmres:wctanb})
the agreement improves to roughly 10\%. The remaining
sources of discrepancy are mainly due to wavefunction corrections
to the bare mass matrix, which can be as large as 10\%, and
the inevitable limitations associated with the MIA.
The contributions due to the insertions $\dlr$ and $\drl$
shown in the top--right and bottom--left figures respectively
are absent at LO as the effects of the insertions cancel~\cite{FOR1:bdec}.
However, once BLO effects are taken into account, a dependence
on the insertions is reintroduced due to their appearance in the
bare mass matrix~\eqref{MIA:mdb:offdiag}. (The $\dlr$ dependence
of~\eqref{MIA:mdb:offdiag} becomes apparent once one recalls
that $\left(\dlr\right)_{23}=\left(\drl\right)_{32}$.)
Comparing the approximate and numerical results for the BLO
corrections we see that they typically agree with one another
very well. (Unlike the $\dll$ insertion, contributions
due to wavefunction corrections are typically rather
small as they appear at second order in the MIA and
tend to be suppressed by factors of $m_b^2$.) Finally,
the bottom--right panel shows the contributions that
arise for non--zero $\drr$ where, once again, the analytic
and numerical results agree with one another rather well.

With these results in mind let us now consider the overall
effect of such corrections on the branching ratio of the
decay.
\fig{bsmres:dxydep} depicts the dependence the branching ratio on the
various sources of flavour violation in the squark sector.
The two graphs depicting the variation with $\dll$ and $\drr$ show
the characteristic suppression associated with the BLO factors of
$\left(\BLOfact\right)$ and $\left(\BLOfacg\right)$ that appear in the
denominators of the Wilson coefficients~\nqs{bsm:EW:NHMFV}{bsm:EW:NHGFMpr}.
This suppression can loosen the bounds placed on these insertions
by the $\bsm$ constraint. The panel depicting the variation
with $\dll$ displays a larger dependence on the the gluino mass than
the panel featuring the insertion $\drr$ as the gluino mass not only
features in the gluino contribution, but also in the corrections that
arise once one includes gaugino--higgsino mixing. (We remind the reader
that we assume that the mass of the wino and the gluino are related.)
Let us briefly comment however, that in contrast to $\bsg$, the
differences between the BLO and LO results for MFV and GFM
calculations tend to be rather similar.
This is because, now, the only dominant contribution to the decay is
via the neutral Higgs penguin and BLO effects tend to be limited
to the factors of $(\BLOfacg)$ and $(\BLOfact)$ that accompany
the LO matching conditions. Turning to
the $\dlr$ and $\drl$ insertions, large deviations from a
purely leading order calculation are possible due to the
reappearance of the insertion in the Wilson coefficients
\nqs{bsm:EW:NHGFM}{bsm:EW:NHGFMpr} once one proceeds beyond the
LO~\cite{FOR1:bdec}.
\FIGURE[t!]{
  \begin{tabular}{c c}
    \includegraphics[angle=0,width=0.45\textwidth]{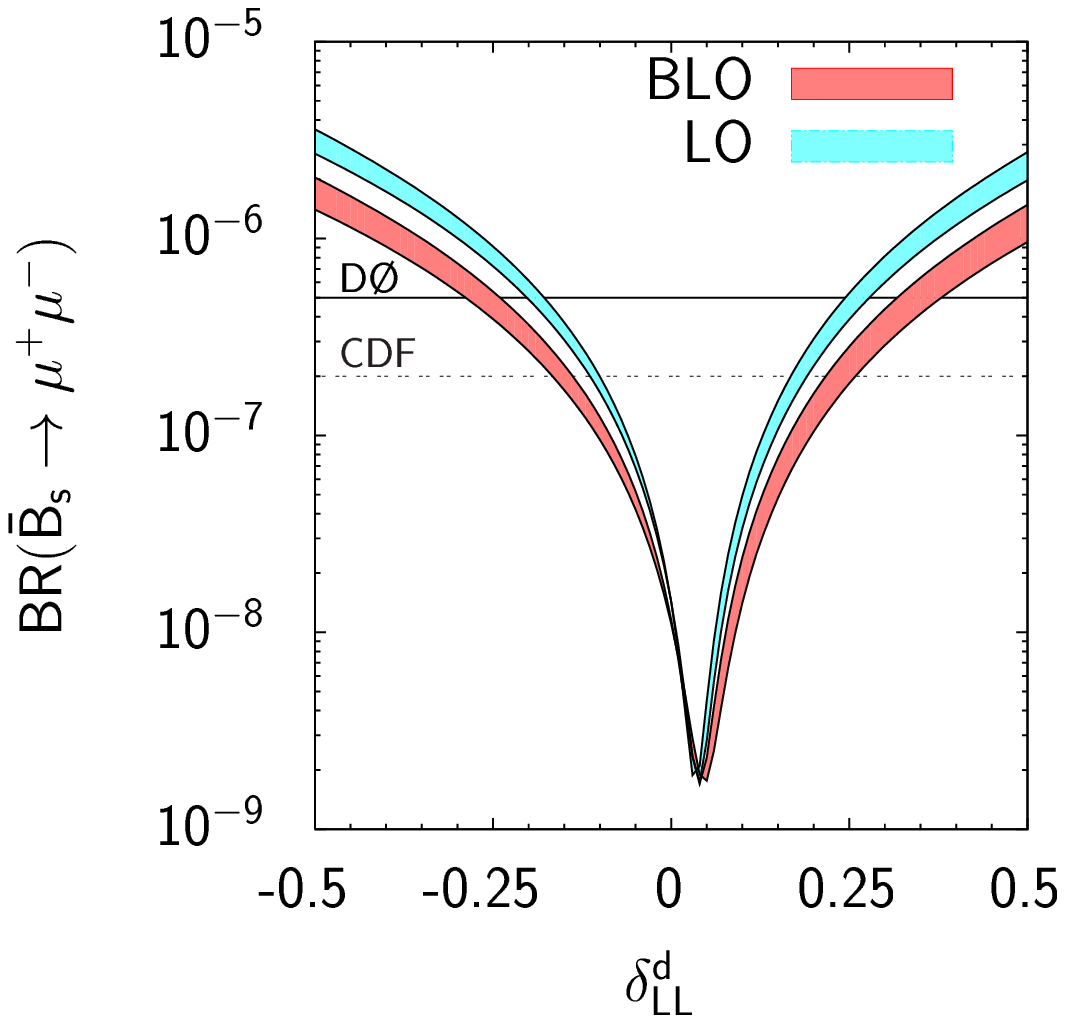}
    & \includegraphics[angle=0,width=0.45\textwidth]{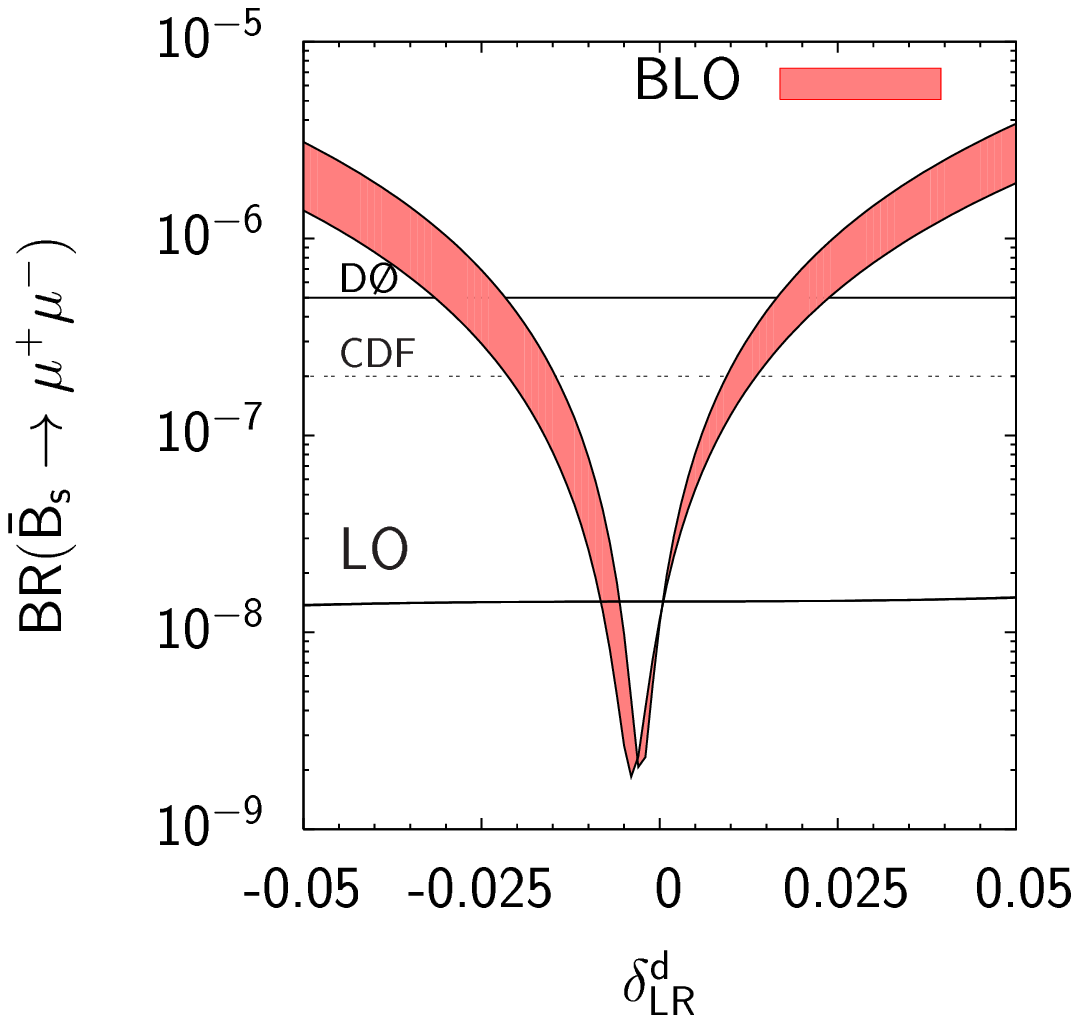}\\
    \includegraphics[angle=0,width=0.45\textwidth]{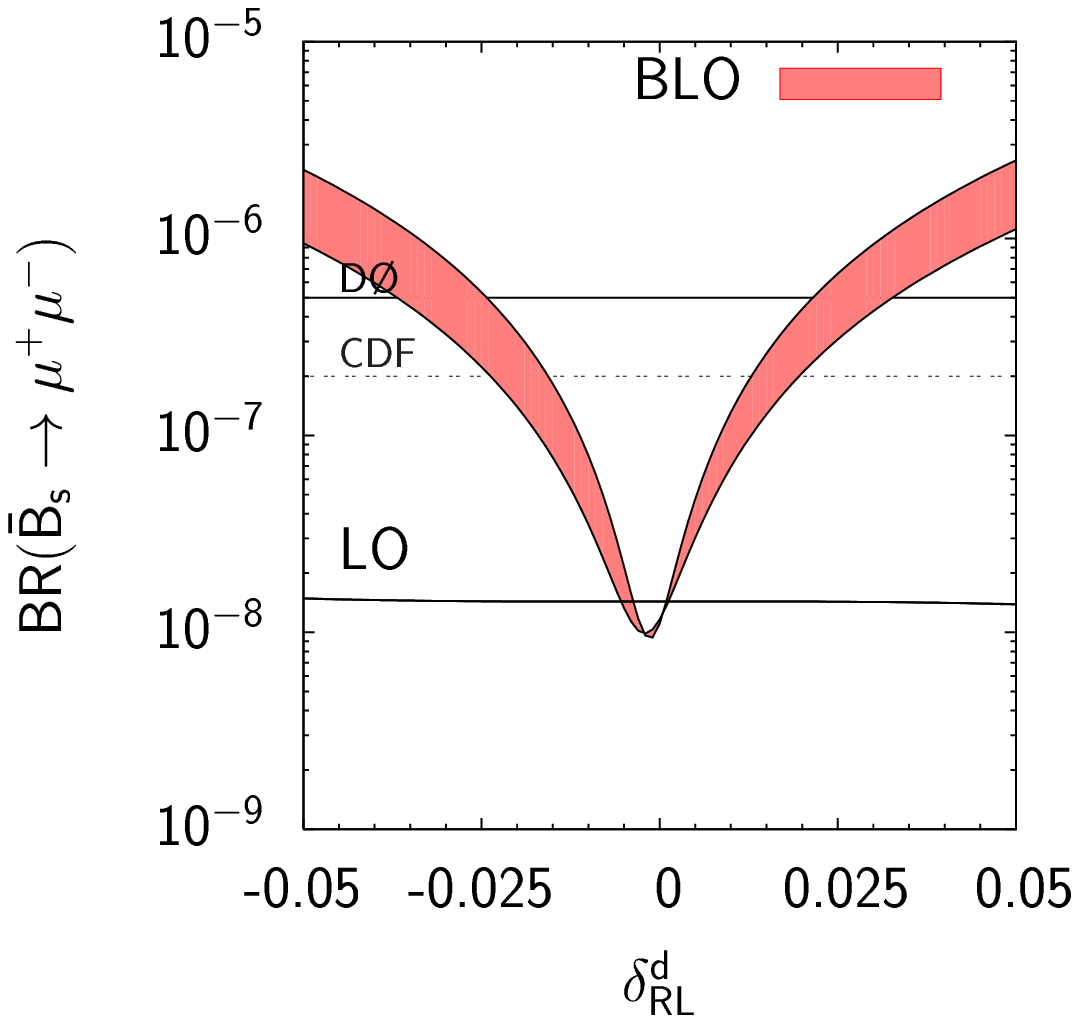}
    & \includegraphics[angle=0,width=0.45\textwidth]{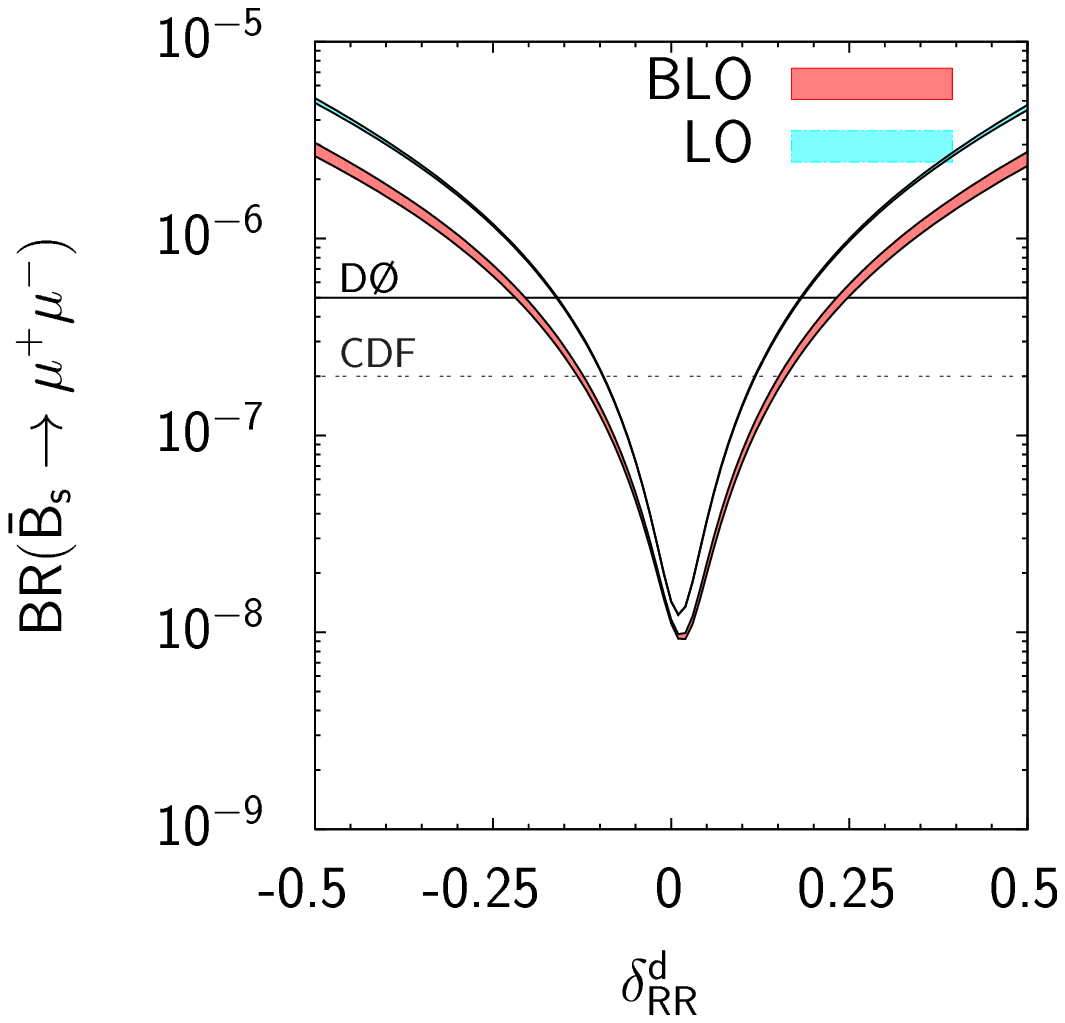}
  \end{tabular}
  \caption{The dependence of BR($\bsm$) on $\delta_{XY}^d$ for
    $\msq=1\tev$, $A_u=-500\gev$, 
    $m_A=500\gev$, $\mu=500\gev$ and $\tanb=50$. The light blue (light grey)
    and red (dark grey) bands depict the effect of varying $\mgl$ between
    $\msq/\sqrt{2}$ and $\sqrt{2}\msq$. The published D\O~and
    preliminary CDF limits for the decay are shown.\label{bsmres:dxydep}}
}
Finally, the asymmetric
nature of the $\dll$ and $\dlr$ curves arises as these contributions
interfere directly with the MFV contribution and it is therefore possible
to induce quite large cancellations. For $\drr$ and $\drl$ on the other
hand, direct interference with the MFV contributions is generally
not possible.  Cancellations with the Wilson coefficient $C_A$
when calculating the branching ratio~\eqref{bsm:br} can occur, however,
and lead the minima
of the curves to deviate slightly from MFV. In general however, in
a similar manner to $\bsg$, the overall effect of these insertions
is to increase the branching ratio with respect to the MFV result,
independent of the sign of the insertion.

The dependence of the branching ratio on $\tanb$ and the
pseudoscalar mass $m_A$ is shown in \fig{bsmres:tanbdep}.
\FIGURE[t!]{
  \begin{tabular}{c c}
    \includegraphics[angle=0,width=0.45\textwidth]{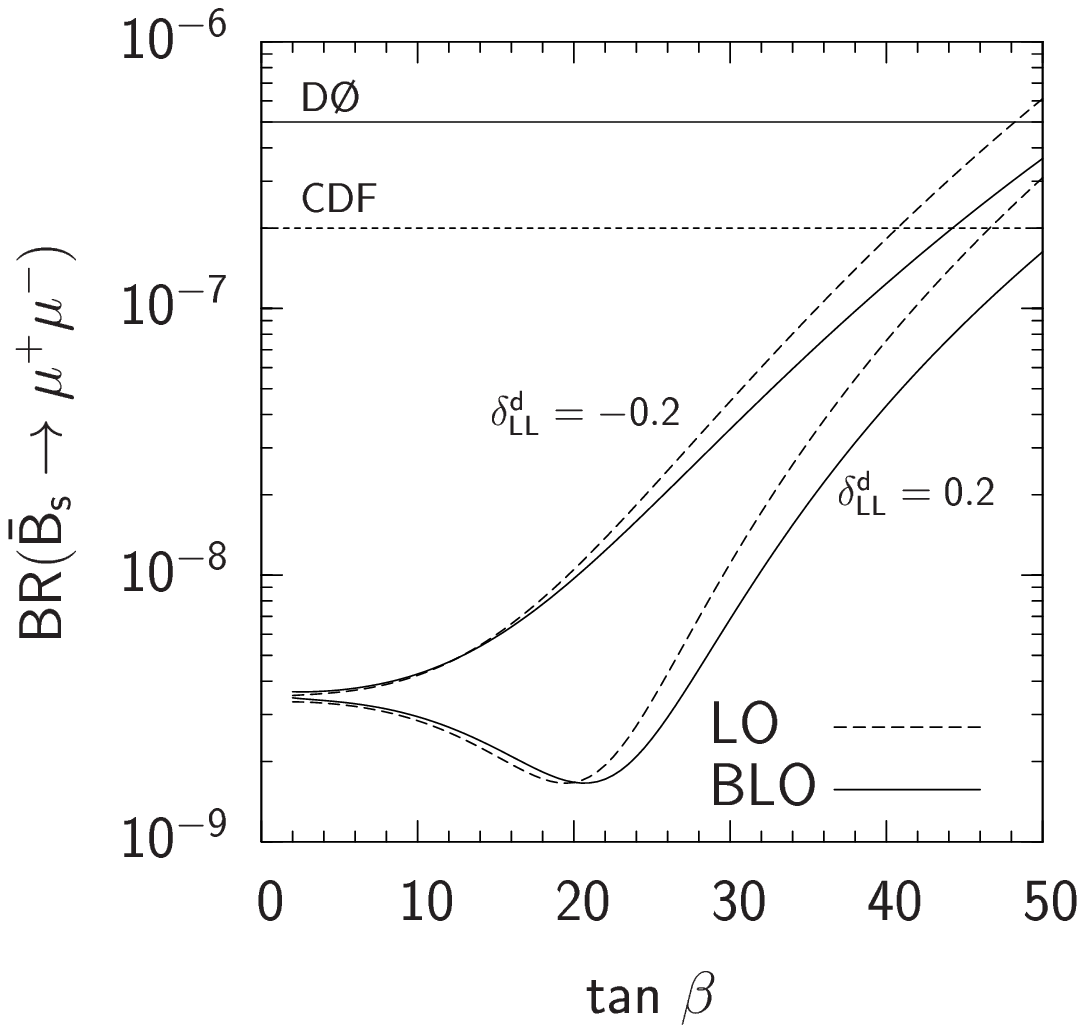}
    & \includegraphics[angle=0,width=0.45\textwidth]{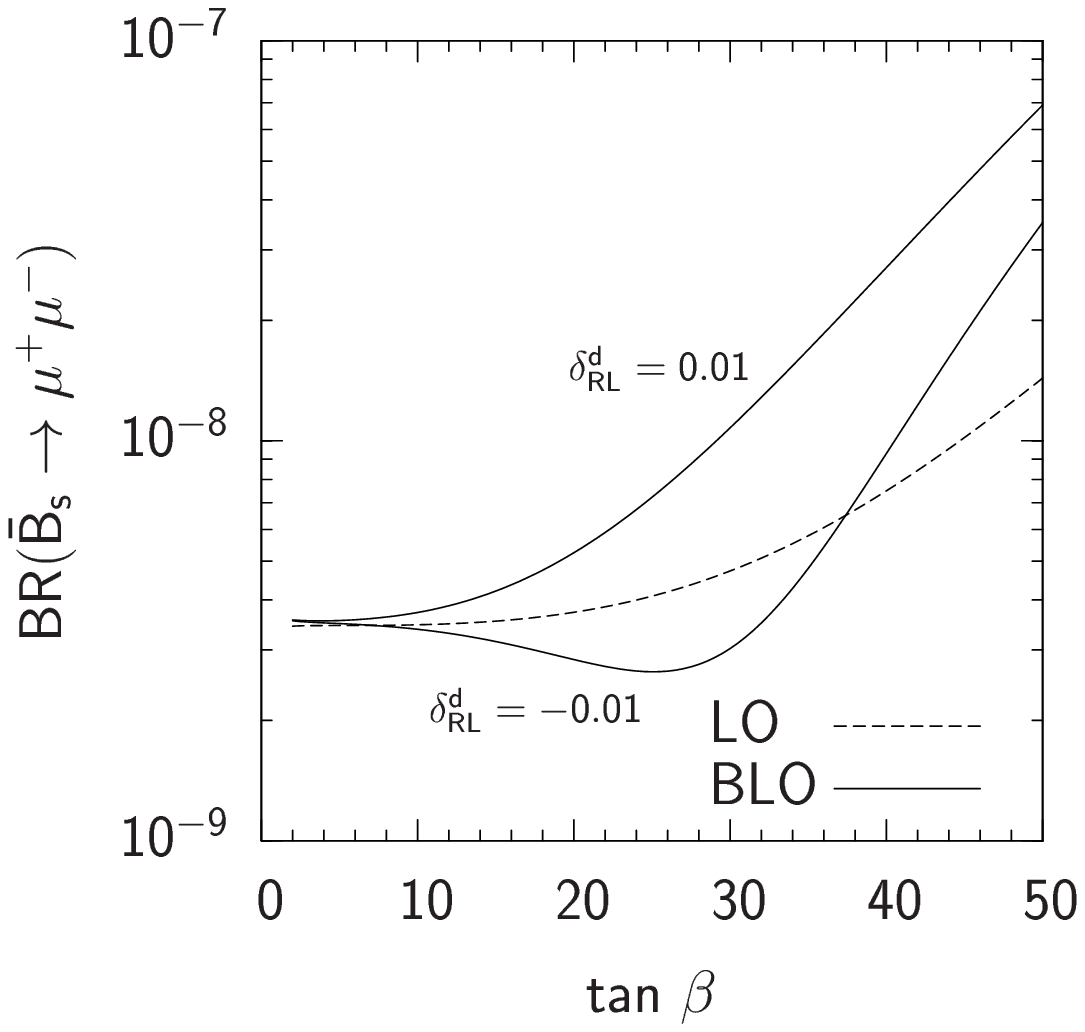}\\
    \includegraphics[angle=0,width=0.45\textwidth]{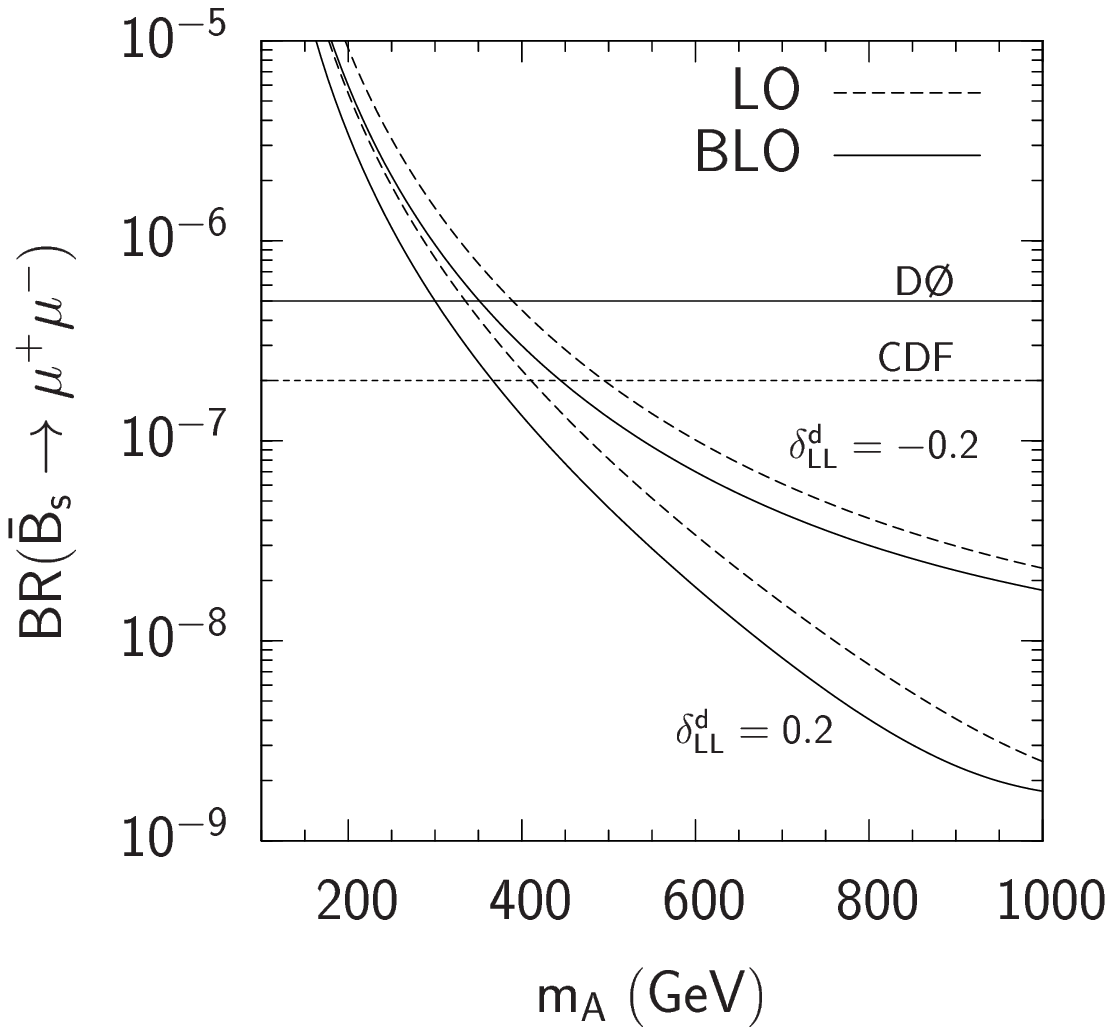}
    & \includegraphics[angle=0,width=0.45\textwidth]{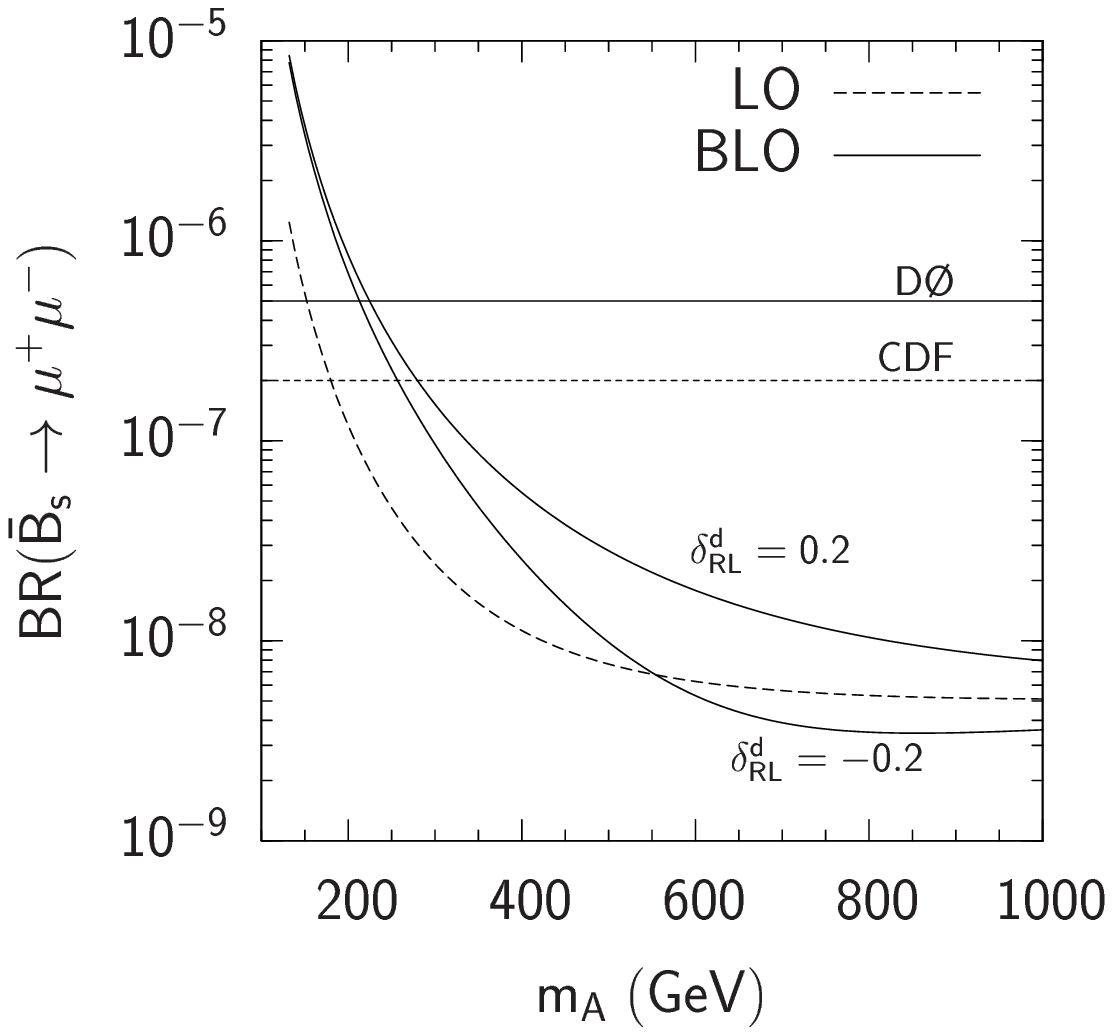}
  \end{tabular}
  \caption{The dependence of BR($\bsm$) on $\tanb$
    (the top two panels) and $m_A$ (the bottom two panels).
    The soft sector is parameterised as follows: $\msq=\sqrt{2}\,\mgl=1\tev$,
    $A_u=-500\gev$, $\mu=500\gev$. In the top two panels $m_A=500\gev$,
    in the lower two panels $\tanb=40$. 
    The published D\O~and the preliminary CDF bounds
    are shown when appropriate. Only one LO curve is shown
    in the panels on the right as the dependence on the
    insertion vanishes at LO.
    \label{bsmres:tanbdep}}
}
As discussed in section~\ref{bsm}
the scalar and pseudoscalar contributions to the decay
can lead the branching ratio to vary as $\tan^6\beta$. 
As is evident from the top two panels, values of BR$\left(\bsm\right)$
approaching (or even exceeding) the current experimental limit
are possible, even for $\tev$ scale sparticle masses, if $\tanb$ is
large $\sim 40$. A strong dependence on $m_A$ is also apparent
in the lower two panels. Both figures illustrate the
reductions associated with BLO calculations for the LL
insertion and the new effects that appear beyond the leading
order for the RL insertion.

In summary, we have seen that the approximate formulae gathered
in section~\ref{bsm} seem to describe the results of our numerical
analysis rather well (within 10\%) especially once one considers
the approximations involved in their derivation. For the LL
insertion the additional electroweak corrections, described
in subsection~\ref{ImpApp}, typically act to reduce the correction
to $\bsm$ by up to 20\% compared with calculations performed in
the limit of vanishing electroweak couplings. This reduction
coupled with the resummation of $\tanb$ enhanced effects can
relax the contribution due to $\dll$ by roughly 60\% compared
with a na\"ive LO analysis in which only the effects of the
gluino contribution to the neutral Higgs vertex are taken into account.

%%%%%%%%%%%%%%%%%%%%%%%%%%%%%%%%%%%%%%%%%%%%%%%%%%%%%%%%%%%%%%%%%
\subsection{$\bbb$ Mixing}
\label{NRes:bbm}
%%%%%%%%%%%%%%%%%%%%%%%%%%%%%%%%%%%%%%%%%%%%%%%%%%%%%%%%%%%%%%%%%

Turning now to $\bbb$ mixing, here we shall not compare the
expressions gathered in section~\ref{bbb} to those of our
numerical analysis as our approximations for
the effective Higgs vertex have been discussed in the
previous subsection.

As discussed in section~\ref{bbb}, contributions
to $\delmbs$ in the GFM scenario stem from
box diagrams mediated by SUSY particles and charged Higgs
and, in the large $\tanb$ regime, from double penguin diagrams.
\FIGURE[t!]{
  \begin{tabular}{c c}
    \includegraphics[angle=0,width=0.45\textwidth]{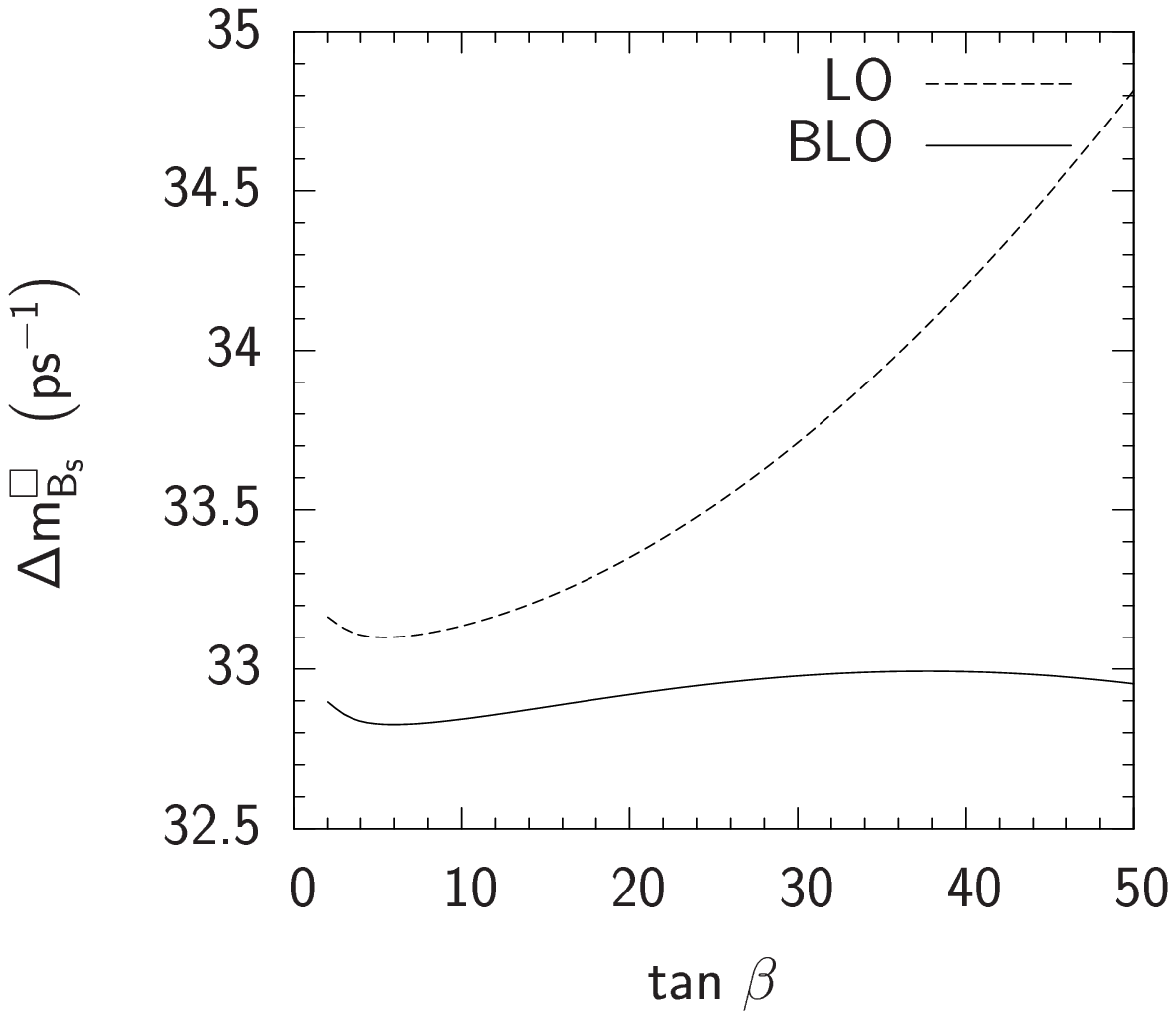}
    & \includegraphics[angle=0,width=0.45\textwidth]{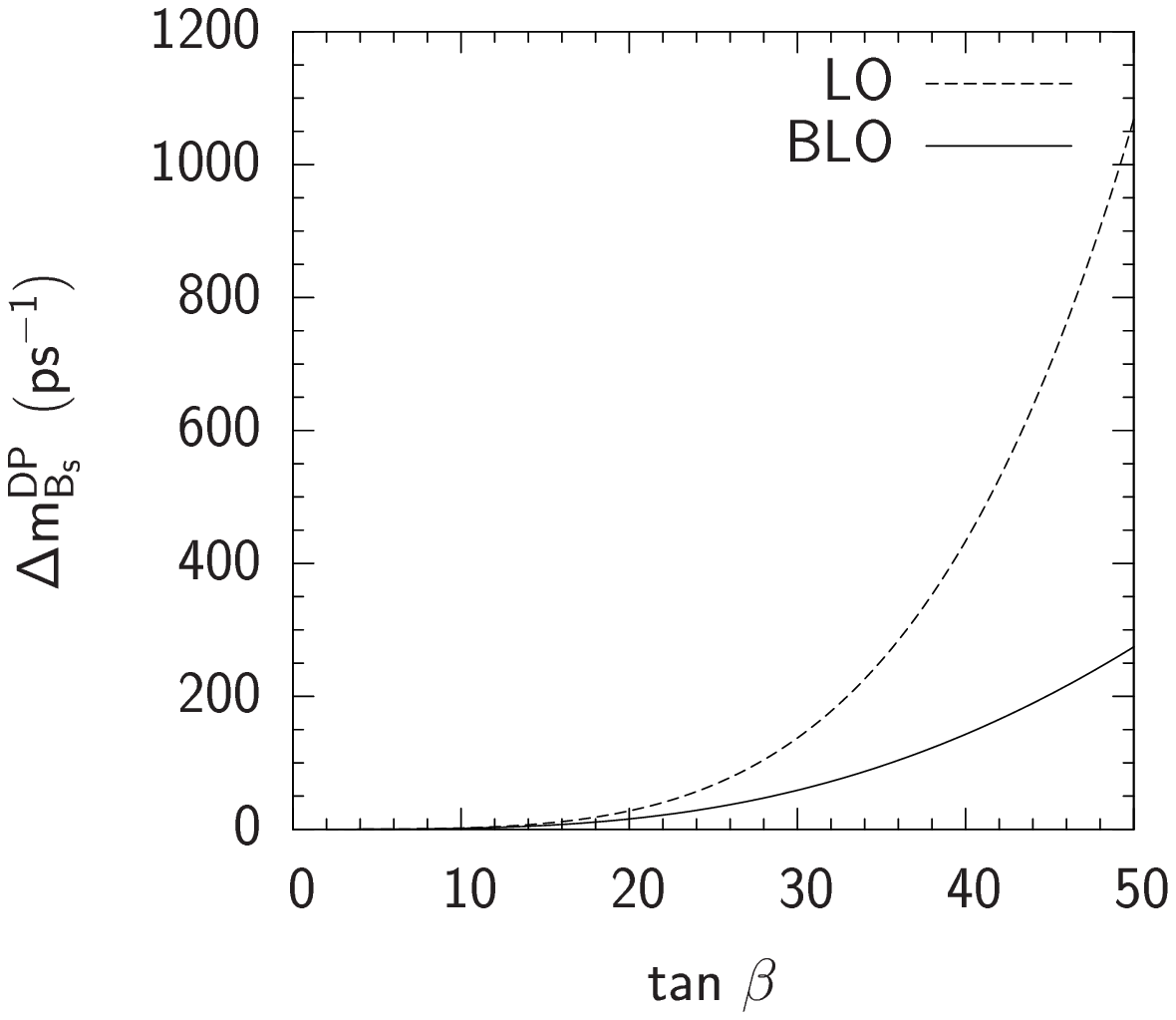}
  \end{tabular}
  \caption{Contributions to $\delmbs$ \vs\ $\tanb$. The combined effect
    of box diagrams mediated by SUSY particles is shown in the left
    panel whilst the right panel shows the double penguin contribution.
    The soft sector is parameterised as follows: $\msq=500\gev$,
    $A_t=-\msq$, $\mgl=\sqrt{2}\msq$, $\mu=\msq$ and $m_A=300\gev$. 
    The flavour violating parameters are $\dll=\drr=0.1$.
    \label{dmbres:tanbdep}}
}
The panel on the left of \fig{dmbres:tanbdep} shows the
$\tanb$ dependence of the contributions to $\delmbs$ arising
from box diagrams mediated by SUSY particles. The main difference
between the two curves at low $\tanb$, originates from the use
of the NLO anomalous dimension matrix to run the BLO calculation
from the SUSY matching scale to the electroweak matching scale.
At large $\tanb$ the interference between the dominant gluino contribution
and the BLO corrections to the chargino and neutralino contributions
acts to reduce the overall contribution to $\delmbs$ further. As is
apparent from plot however, the overall correction tends to be
only of the order of five to ten percent.

The double penguin contributions are depicted on the panel on the right of
\fig{dmbres:tanbdep}. Here we see a rather more dramatic difference
between LO and BLO calculations and at large $\tanb$ it is possible
that BLO corrections can lead to a reduction of LO effects by up
to a factor of three. 

Let us briefly comment on which values of $\tanb$ should
analyses, that typically only feature LO matching conditions
for the gluino contributions to $\bbb$ mixing, include
the double Higgs penguin contribution. As is evident from
\fig{dmbres:tanbdep}, the double penguin contribution completely
dominates the behaviour of $\delmbs$ at large $\tanb$ and can
become as important as the LO result for $\tanb$ as low as 
20. We have checked that a similar situation arises for
non--zero LR and RL insertions. 

With these results in mind let us consider the combined effect
of all the beyond Standard Model corrections in the large
$\tanb$ regime.
\FIGURE[t!]{
  \begin{tabular}{c c}
    \includegraphics[angle=0,width=0.45\textwidth]{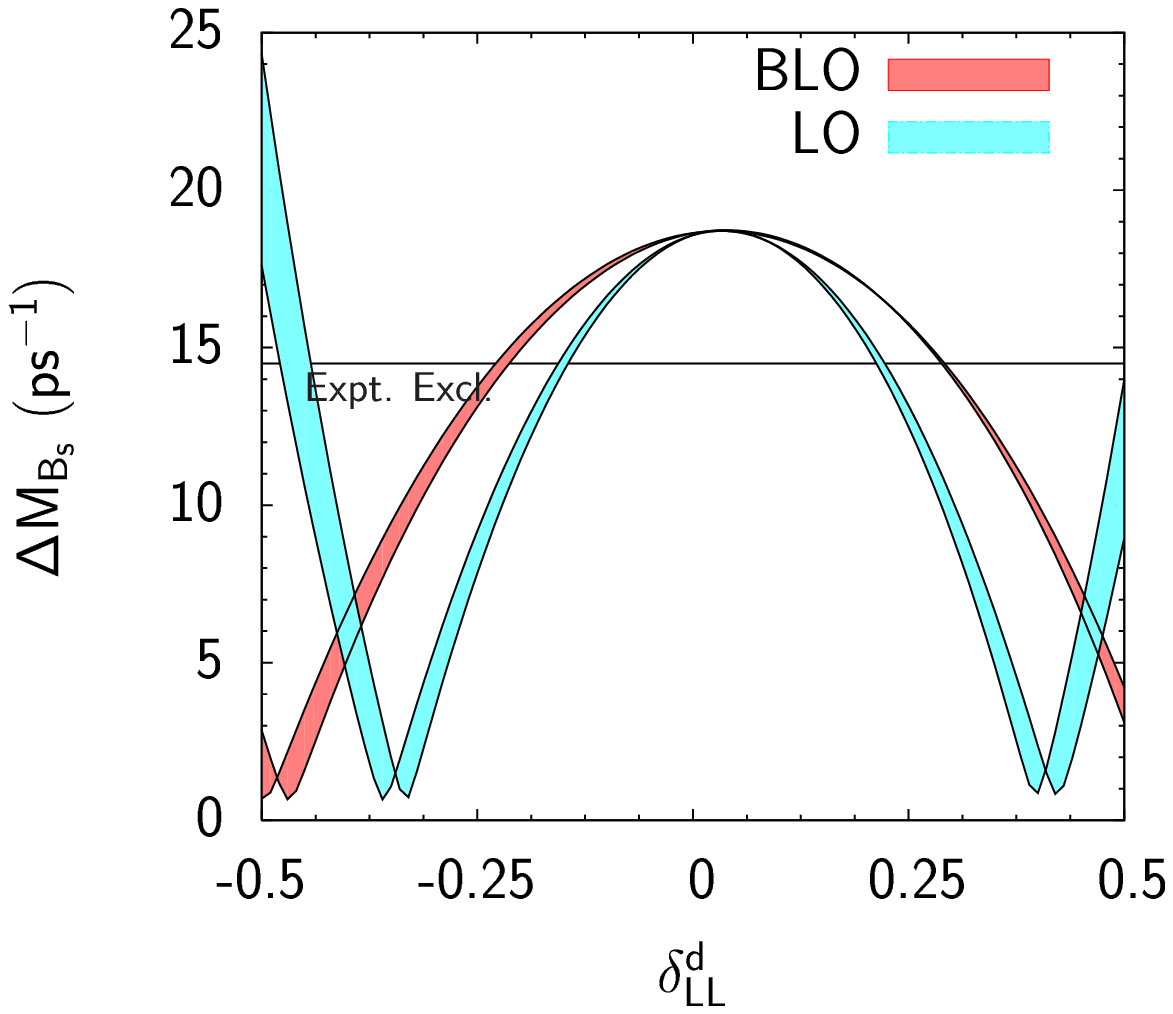}
    & \includegraphics[angle=0,width=0.45\textwidth]{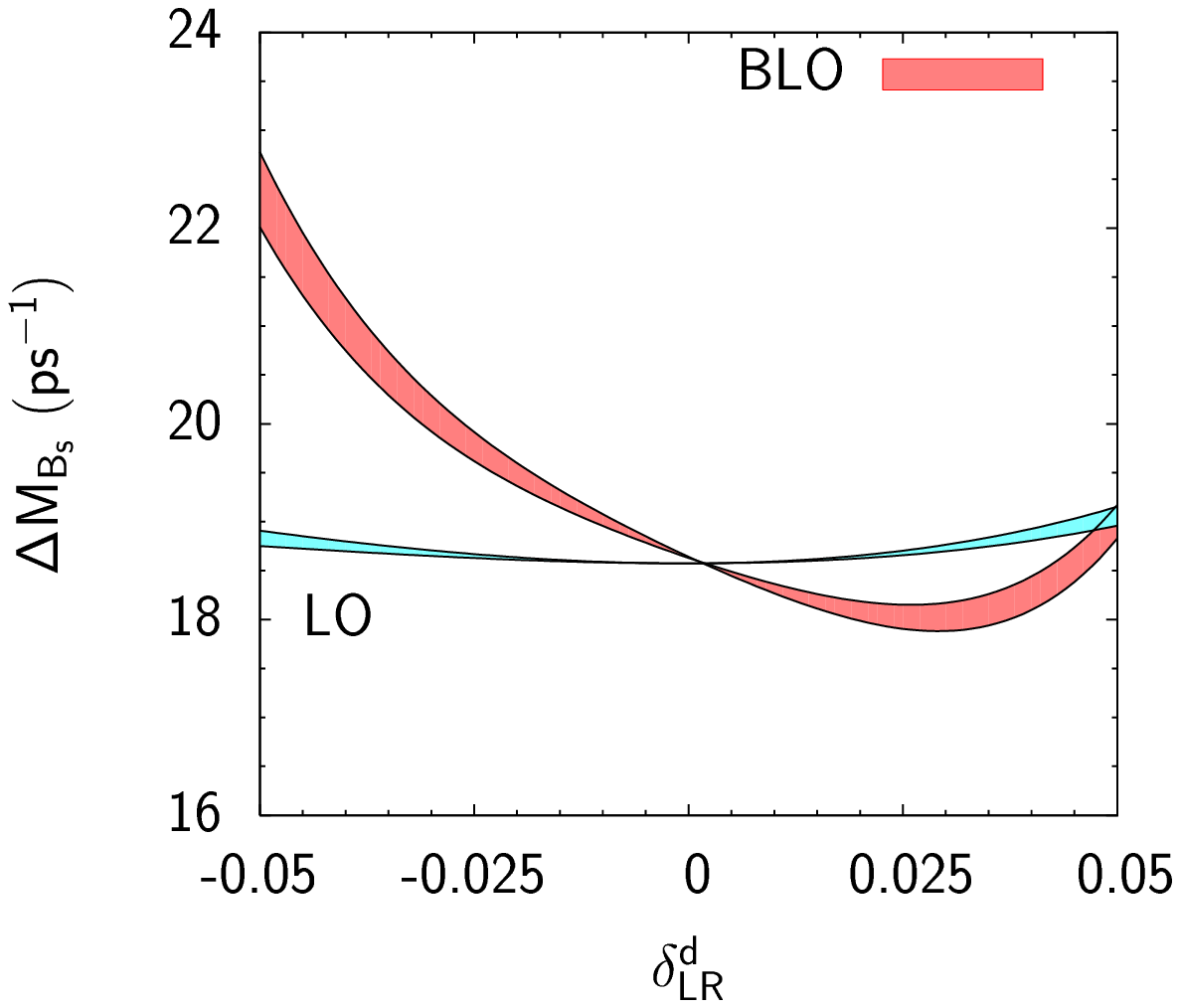}\\
    \includegraphics[angle=0,width=0.45\textwidth]{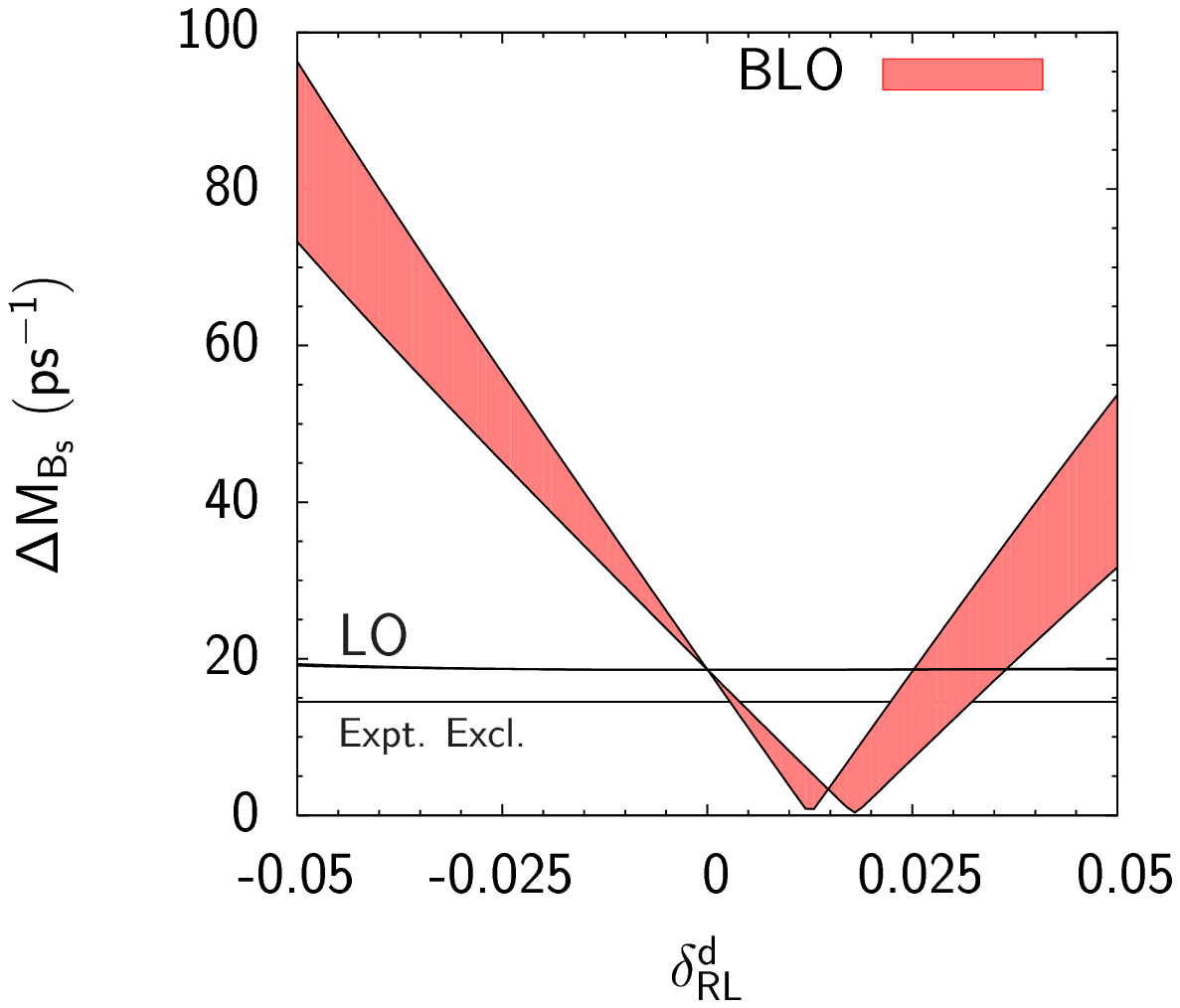}
    & \includegraphics[angle=0,width=0.45\textwidth]{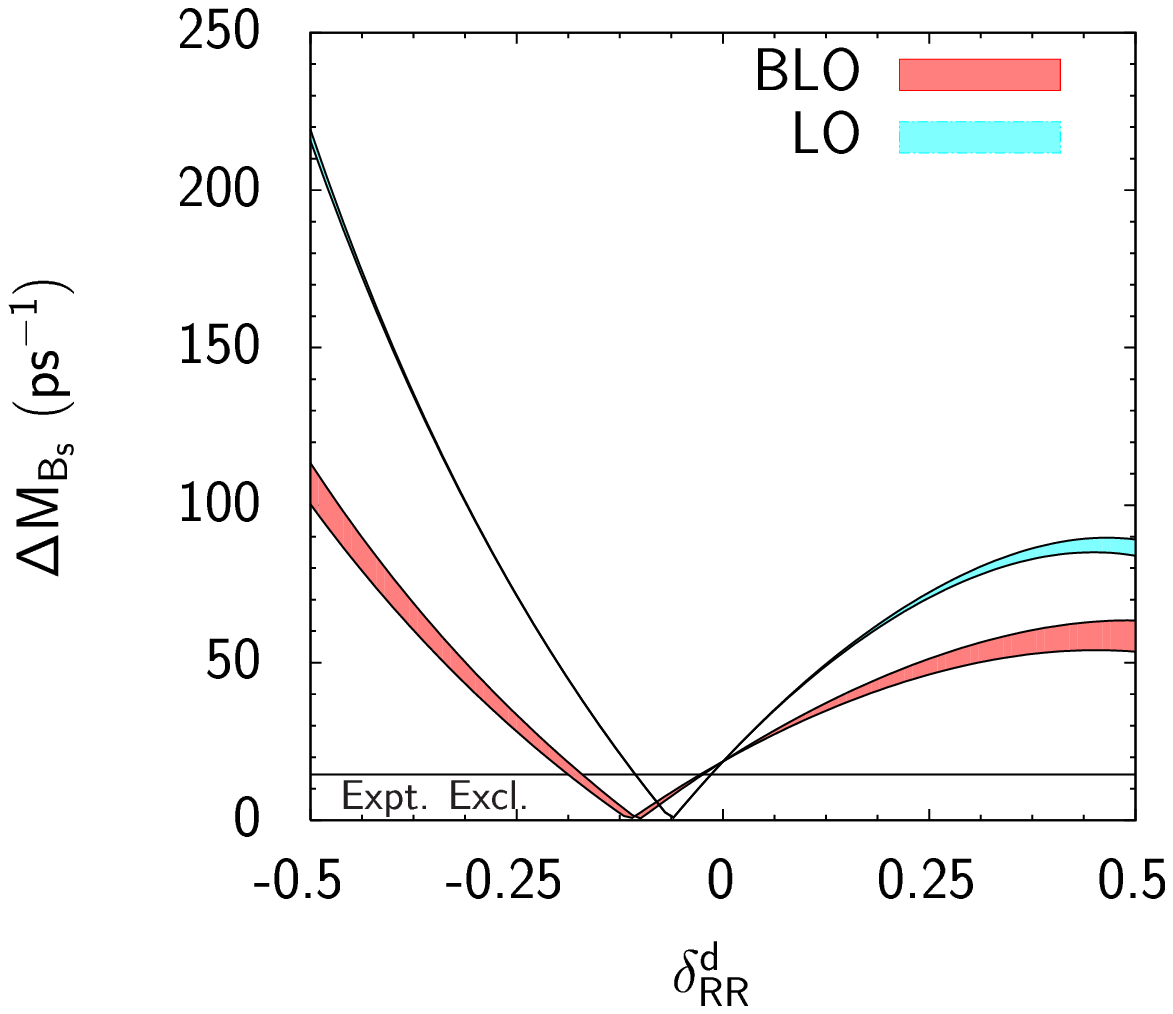}
  \end{tabular}
  \caption{The dependence of $\delmbs$ on $\delta_{XY}^d$ for the same parameters
    as \fig{bsmres:dxydep}.\label{dmbres:dxydep}}
}
The dependence of $\delmbs$ on each of the
flavour violating parameters is illustrated in
\fig{dmbres:dxydep}. In the top left panel corresponding to the
insertion $\dll$ one can see a largely quadratic dependence on the
insertion, in agreement
with the analytic result~\eqref{bbb:EW:NHCLRGFMdll}. The graph is not
centred on $\dll=0$ as, in a similar manner to the $\bsm$ graph
in \fig{bsmres:dxydep}, the MFV and GFM contributions
to the neutral Higgs vertex approximately cancel for $\dll\sim 0.08$. 
Turning to the top right panel, depicting the dependence
of $\delmbs$ on $\dlr$, we can see that the overall effect on $\delmbs$
is rather slight, for small $\dlr$ the effect is mainly linear
as $\dlr$ only contributes to the left--handed vertex that appears
in~\eqref{bbb:EW:NHCLR}. It is therefore necessary for the
other penguin to be mediated by chargino exchange~\eqref{bbb:EW:NHCLRGFMdlr}.
For larger values of $\dlr$, contributions to
$C_1^{SLL}$, as well as SUSY box diagrams, lead to a quadratic
dependence on the insertion to emerge, however, once again,
the corrections are rather small.

The bottom two panels depict
the larger effects induced in the $\drl$ and $\drr$ sectors. The linear
dependence of the contributions is due once again to one Higgs penguin
being mediated by chargino exchange and the other by gluino exchange. The
only alteration to this behaviour arises for very large $\drr$ ($\sim 0.4$)
where the gluino mediated contributions to the left--handed Higgs
coupling, that are suppressed by a factor of $m_s$, can become important
and interfere with the chargino contribution.
Finally, let us once again point out the large differences between
the LO and BLO calculations featured in all four plots.
For $\dll$ and $\drr$ we
see the characteristic suppression of LO effects that arise
from the factors of $\left(\BLOfact\right)$ and $\left(\BLOfacg\right)$
that appear
in~\nqs{bbb:EW:NHCSLLMFV}{bbb:EW:NHCLRGFMdrrdll}. These
typically lead to reductions proportional to factors of two
or three, if $\mu>0$. In a similar manner to the decay $\bsm$, a dependence
on the insertions $\dlr$ and $\drl$, which is absent at LO, 
reappears when BLO corrections are taken into
account~\cite{FOR1:bdec}.

Before ending this section let us briefly discuss the
correlation between $\bsm$ and $\delmbs$ at large $\tanb$.
As was pointed out earlier in this subsection, the double Higgs
penguin tends to completely dominate the contributions that
arise from new physics in the large $\tanb$ regime. It is
therefore natural to expect a degree of correlation with the
decay $\bsm$, that is also dependent on the neutral Higgs penguin
when $\tanb$ is large. Such a situation is illustrated by
the scatter plots shown in \fig{dmbres:scat}.
\FIGURE[t!]{
  \begin{tabular}{c c}
    \includegraphics[angle=0,width=0.45\textwidth]{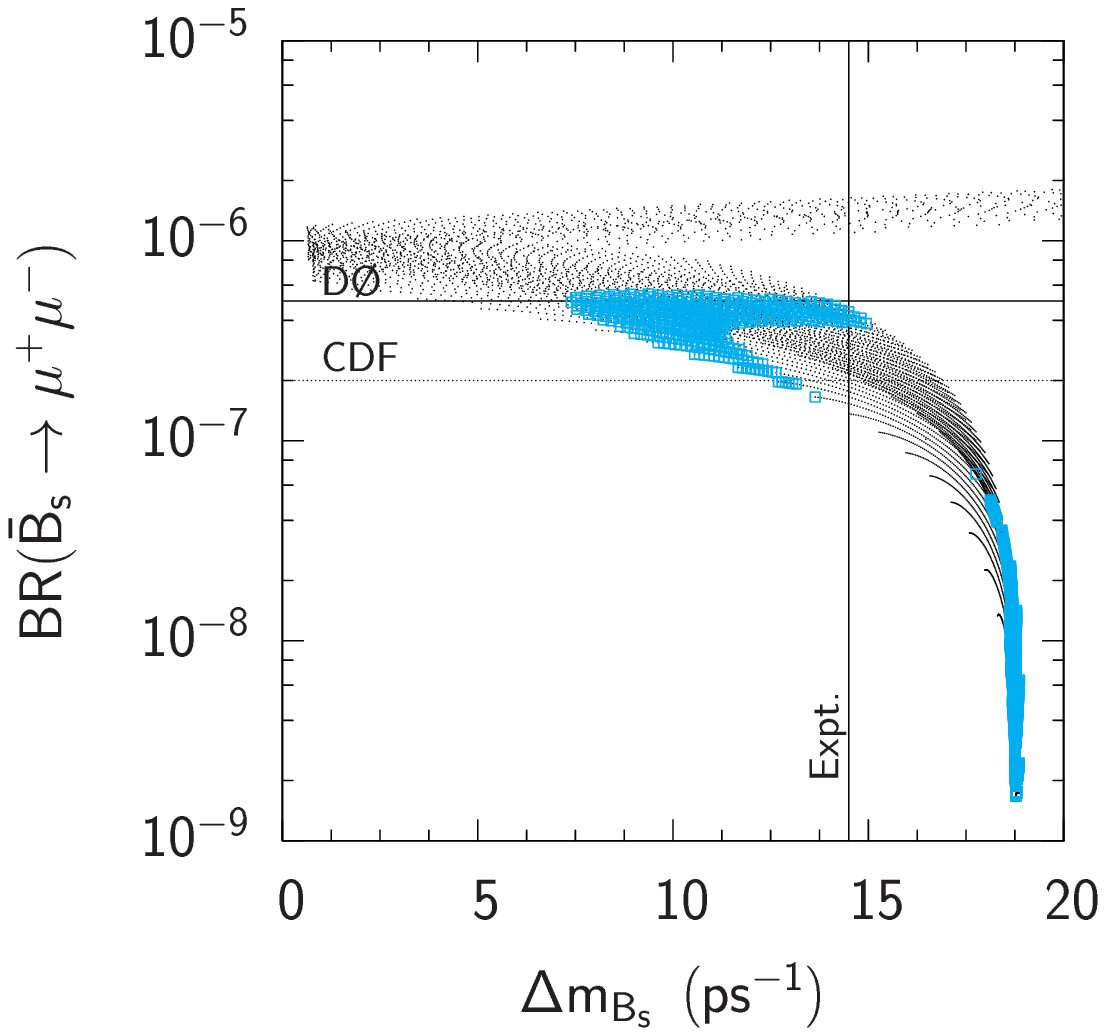}
    & \includegraphics[angle=0,width=0.45\textwidth]{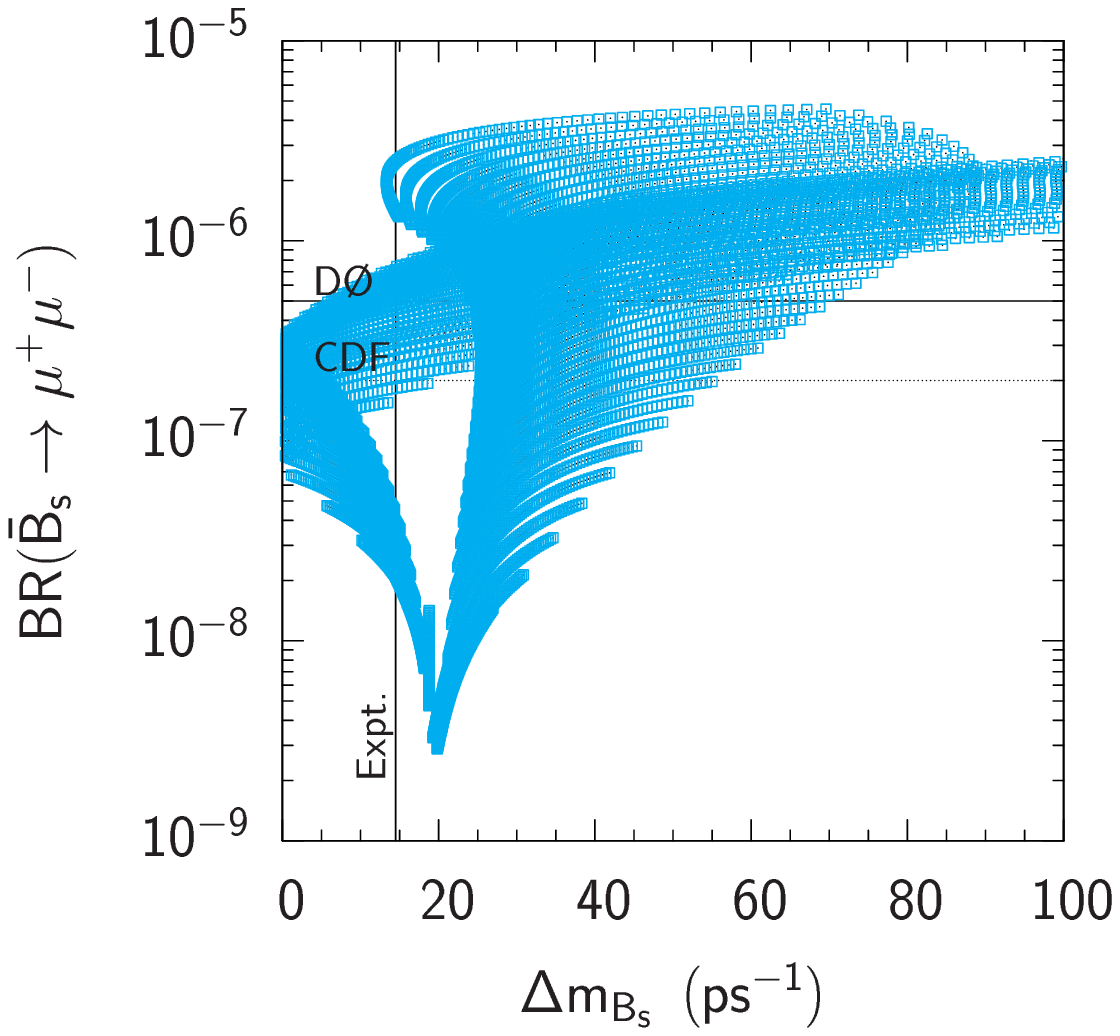}
  \end{tabular}
  \caption{Scatter plots showing the correlation between $\delmbs$
    and BR$(\bsm)$. In the panel to the left $\dll$ is varied over
    the range $[-0.8,0.8]$ whilst in the panel to the right the
    insertion $\drr$ is varied over the same range of values.
    In both plots $\msq$ is varied over the range $[500,1500]\gev$.
    The remaining parameters describing the SUSY sector
    are as follows: $\mu=500\gev$, $m_A=500\gev$, $A_u=-500\gev$
    and $\mgl=1\tev$, $\tanb=40$. The published D\O~and the
    preliminary CDF bounds for the decay $\bsm$ are shown,
    as well as the lower limit for $\delmbs$. Points compatible
    with $\bsg$ are highlighted by light blue (light grey) squares.
    \label{dmbres:scat}}
}
Here we see in both
panels that a large correlation exists between each process.
The panel on the left (where only the LL insertion is varied) features
only one branch as the double Higgs penguin in this case tends
to only lead to a reduction in $\delmbs$. Varying the RR
insertion can lead to reductions or enhancements of $\delmbs$
thanks to the linear dependence on the insertion exhibited in
the matching condition given in~\eqref{bbb:EW:NHCLRGFMdrr}. This
effect leads to two distinct branches being visible in the right
panel of \fig{dmbres:scat}. 

%%%%%%%%%%%%%%%%%%%%%%%%%%%%%%%%%%%%%%%%%%%%%%%%%%%%%%%%%%%%%%%%%
\subsection{Radiative Corrections to the CKM Matrix}
\label{NRes:kzero}
%%%%%%%%%%%%%%%%%%%%%%%%%%%%%%%%%%%%%%%%%%%%%%%%%%%%%%%%%%%%%%%%%

It was mentioned briefly in subsection~\ref{BLOeff:OM} that
the rotation from the bare to the physical SCKM
basis can induce large contributions to the bare CKM
matrix $K^{(0)}$~\cite{BRP:CKM}. 
One particularly interesting consequence
of this is that the CKM matrix elements $K_{ts}$ and $K_{cb}$
could be generated radiatively via GFM effects~\cite{OR2:bsg}. Such
a situation is illustrated in \fig{NRes:Comb:bareCKM}.
\FIGURE[t!]{
    \includegraphics[angle=0,width=0.45\textwidth]{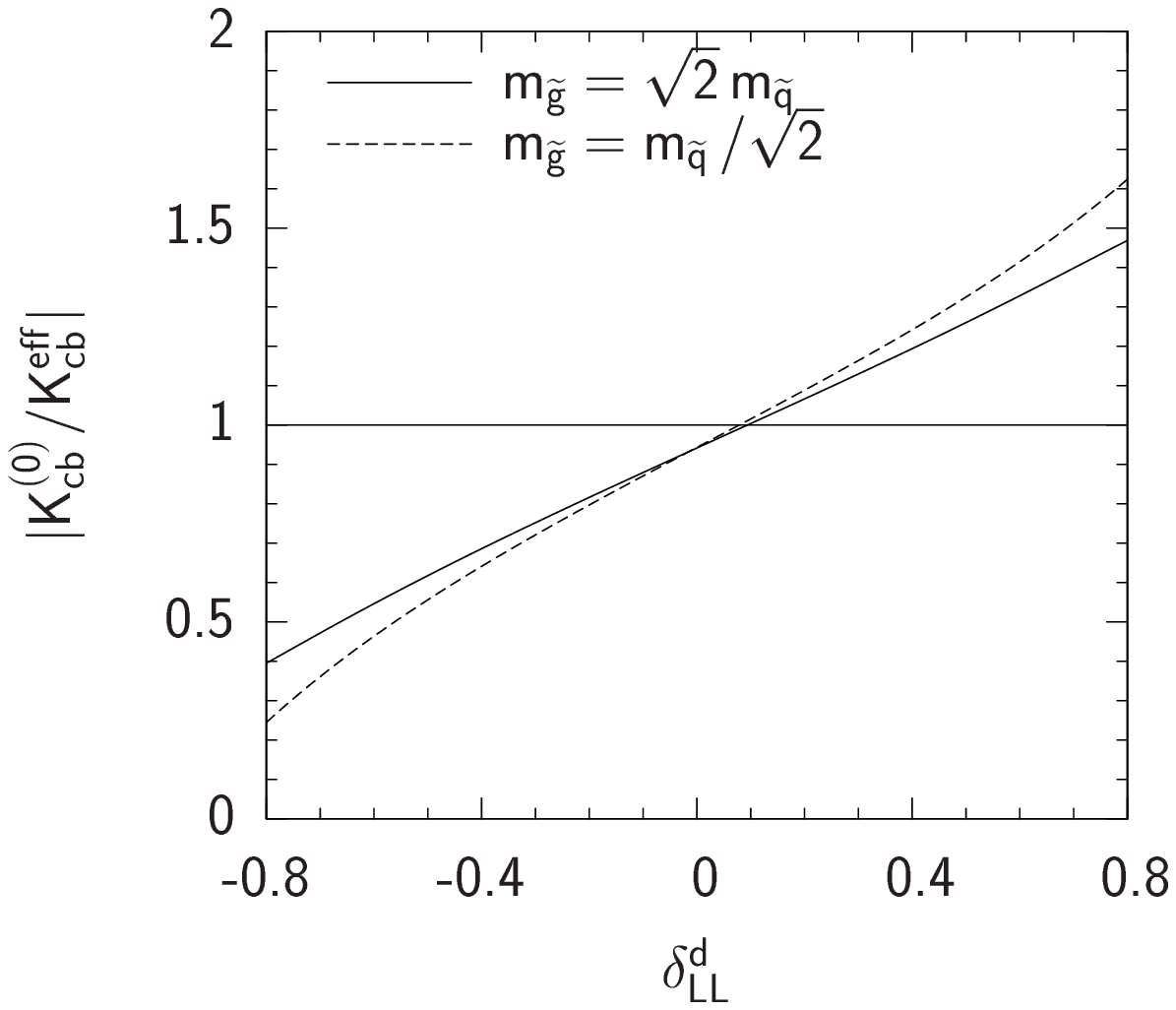}
    \includegraphics[angle=0,width=0.45\textwidth]{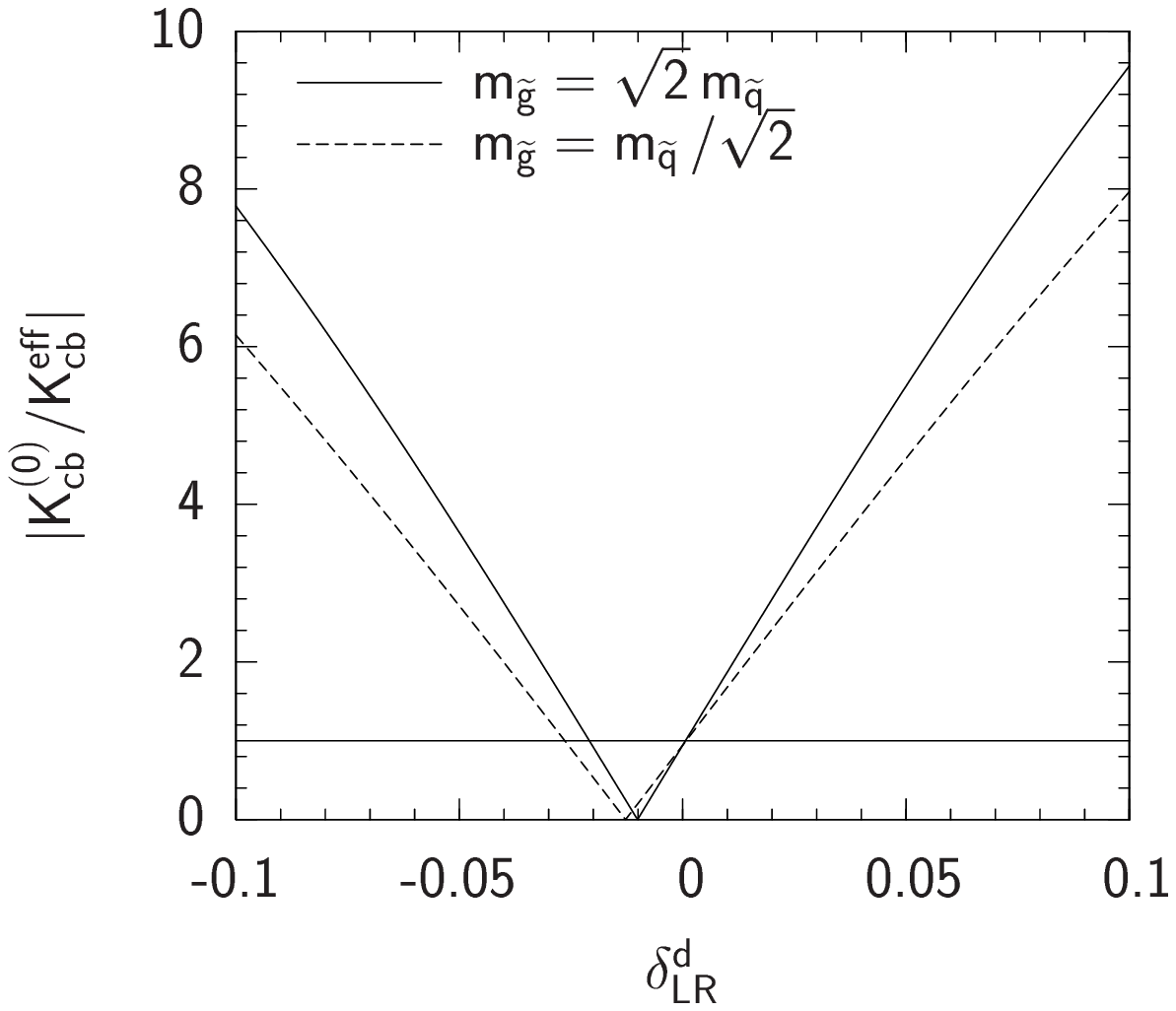}
  \caption{The ratio $K_{cb}^{(0)}/\Keff_{cb}$~\vs\ $\dll$ (the
    panel on the left) and $\dlr$ (the panel on the right).
    The soft sector is parameterised as follows: $\tanb=40$,
    $\msq=1\tev$, $\mu=500\gev$, $m_A=500\gev$ and 
    $A_t=-1\tev$. Solid and dashed lines denote $\mgl=\sqrt{2}\,\mgl$
    and $\mgl=\msq\,/\sqrt{2}$ respectively.\label{NRes:Comb:bareCKM}}
}
From both plots it is evident that $K_{cb}^{(0)}$
(and similarly $K_{ts}^{(0)}$) can receive significant
corrections due to the presence of GFM in the squark
sector. In the LR sector in particular, radiative generation
of the entire matrix
element can typically occur for $\dlr\sim -0.01$.
One could imagine the scenario where
corrections to the remaining matrix elements might
induce similar effects and lead the CKM matrix to
be fully diagonal before SUSY threshold corrections
are taken into account. 

From the top--right panel in \fig{bsmres:dxydep} it is also apparent
that, due to a cancellation between the chargino and gluino
corrections to neutral Higgs vertex, the branching ratio of
the decay $\bsm$ also tends to approach SM like values
for negative $\dlr$. The minima of the two curves however tend not to
coincide unless $\epsilon_Y Y_t^2\sim -\epsilon_s$ (this
typically only occurs if $A_t$ is rather large and negative).

%%%%%%%%%%%%%%%%%%%%%%%%%%%%%%%%%%%%%%%%%%%%%%%%%%%%%%%%%%%%%%%%%
\section{Summary}
\label{Conc}
%%%%%%%%%%%%%%%%%%%%%%%%%%%%%%%%%%%%%%%%%%%%%%%%%%%%%%%%%%%%%%%%%

We have presented here the first complete analysis that includes
the resummation of all
$\tanb$ enhanced BLO effects in SUSY with GFM that contribute to the
processes $\bsg$, $\bsm$ and $\bbb$ mixing. We have derived analytic
expressions applicable in general and in the MIA. As such, they match
the precision of similar calculations performed in the context of
MFV. We have provided a recipe for including BLO effects into LO
expressions. 

We have found that BLO effects in GFM can be large. In particular, we
have more fully analysed the focusing effect that was initially
pointed out in~\cite{OR:bsg,OR2:bsg} in the case of $\bsg$ and
next shown to 
also exist in the other two processes~\cite{FOR1:bdec}.  In the
phenomenologically interesting case of large $\tanb$ and $\mu>0$ the
focusing effect often leads to a significant relaxation of
experimental bounds on the soft mass mixings.

Finally, we have  examined radiative corrections to the CKM
entries. These can be large, or even dominant, due to large LR
mixings. We have pointed out a possible, although apparently
accidental, correlation with the processes analysed here.

The method presented here is rather general and can readily be
extended to include CP violating contributions, here
BLO corrections can carry additional phases and lead to potentially
large deviations from LO calculations. One might also
consider the process $\bsll$, where the BLO corrections to the Wilson coefficients
$C_7$ and $C_8$ (and their primed counterparts) may also play
a large r\^ole. Another possible
application would be to include the $\bbbd$ mixing system where,  at
large $\tanb$, one can expect strict bounds the insertions
$\left(\drr\right)_{13}$ and $\left(\drl\right)_{13}$
due to contributions to $\delmbd$ unsuppressed by $m_d$.
% Limits on $\bdm$ supplied by B--factories and hadron colliders
% will also be useful when constraining the remaining insertions.

The formalism presented here can be applied to data from present and
future B--factories and hadron colliders
in constraining mass
insertions and, eventually, in extracting  information on an emerging
pattern of flavour violation in the squark sector~\cite{FOR3}.
\\

\acknowledgments
We would like to thank D.~Demir, G.~Giudice and A.~Masiero
for helpful comments. J.F. would also like to thank
the HEP/PA group at Sheffield for use of the
HEPgrid cluster on which most of the numerical results of this
paper were prepared. J.F. has been supported by the research
fellowship MIUR PRIN 2004 -- ``Physics Beyond the Standard Model'' and
a PPARC Ph.D. studentship. K.O. has been supported by the grant--in--aid
for scientific research on priority areas
(No. 441):  ``Progress in elementary particle physics of the 21st
century through discoveries of Higgs boson and supersymmetry'' (No. 16081209)
and the Korean government grant KRF PBRG 2002-070-C00022.
The Feynman diagrams in this paper were prepared using
{\it Jaxodraw}~\cite{Jaxodraw}.

%%%%%%%%%%%%%%%%%%%%%%%%%%%%%%%%%%%%%%%%%%%%%%%%%%%%%%%%%%%%%%%%%
\begin{appendix}

%%%%%%%%%%%%%%%%%%%%%%%%%%%%%%%%%%%%%%%%%%%%%%%%%%%%%%%%%%%%%%%%%
\section{Loop Functions}
%%%%%%%%%%%%%%%%%%%%%%%%%%%%%%%%%%%%%%%%%%%%%%%%%%%%%%%%%%%%%%%%%

%%%%%%%%%%%%%%%%%%%%%%%%%%%%%%%%%%%%%%%%%%%%%%%%%%%%%%%%%%%%%%%%%
\subsection{The Functions $H_i$}
\label{LF:H}
%%%%%%%%%%%%%%%%%%%%%%%%%%%%%%%%%%%%%%%%%%%%%%%%%%%%%%%%%%%%%%%%%

The loop functions $H_i$ that appear throughout the text
are given by
\begin{align}
H_2(x_1,x_2)=&\frac{x_1\log x_1}{\left(1-x_1\right)\left(x_1-x_2\right)}
+\frac{x_2\log x_2}{\left(1-x_2\right)\left(x_2-x_1\right)},\\
H_3(x_1,x_2,x_3)=&\frac{H_2(x_1,x_2)-H_2(x_1,x_3)}{x_2-x_3},\\
H_4(x_1,x_2,x_3,x_4)=&\frac{H_3(x_1,x_2,x_3)-H_3(x_1,x_2,x_4)}{x_3-x_4}.
\label{LF:funcs}
\end{align}
In the limit of degenerate arguments the functions become
\begin{align}
H_2\left(1,1\right)&=-\frac{1}{2},
&H_3\left(1,1,1\right)&=\frac{1}{6},
&H_4\left(1,1,1,1\right)&=-\frac{1}{12}.
\label{LF:lims}
\end{align}

%%%%%%%%%%%%%%%%%%%%%%%%%%%%%%%%%%%%%%%%%%%%%%%%%%%%%%%%%%%%%%%%%
\subsection{$\bsg$}
\label{LF:bsg}
%%%%%%%%%%%%%%%%%%%%%%%%%%%%%%%%%%%%%%%%%%%%%%%%%%%%%%%%%%%%%%%%%

The loop functions $F_{7,8}^{(2)}\left(x\right)$ that appear in
the charged Higgs matching conditions~\nqs{bsg:EW:CHMFV}{bsg:EW:CHGFMpr}
are
\begin{align}
F_{7}^{(2)}\left(x\right)&=\frac{x\left(3-5x\right)}{12\left(x-1\right)^2}
+\frac{x\left(3x-2\right)}{6\left(x-1\right)^3}\log x,
\nonumber\\
F_{8}^{(2)}\left(x\right)&=\frac{x\left(3-x\right)}{4\left(x-1\right)^2}
-\frac{x}{2\left(x-1\right)^3}\log x.
\nonumber
\end{align}

The functions $H_i^{[7,8]}\left(x\right)$ that appear in
section~\ref{bsg:SU} are given by the following expressions. The
chargino contribution features the functions
\begin{align}
H_1^{[7]}\left(x\right)&=
\frac{-3x^2+2x}{6\left(1-x\right)^4}\log x
+\frac{-8x^2-5x+7}{36\left(1-x\right)^3},
\nonumber
\\
H_1^{[8]}\left(x\right)&=
\frac{x}{2\left(1-x\right)^4}\log x
+\frac{-x^2+5x+2}{12\left(1-x\right)^3},
\nonumber
\end{align}
\begin{align}
H_2^{[7]}\left(x\right)&=
\frac{-3x^2+2x}{3\left(1-x\right)^3}\log x
+\frac{-5x^2+3x}{6\left(1-x\right)^2},
\nonumber
\\
H_2^{[8]}\left(x\right)&=
\frac{x}{\left(1-x\right)^3}\log x
+\frac{-x^2+3x}{2\left(1-x\right)^2}.
\nonumber
\end{align}
The functions relevant to the neutralino contribution are given by
\begin{align}
H_3^{[7]}\left(x\right)&=-\frac{1}{3}H_1^{[8]}\left(x\right),
&H_3^{[8]}\left(x\right)&=H_1^{[8]}\left(x\right),
\nonumber
\\
H_4^{[7]}\left(x\right)&=
-\frac{1}{3}\left(H_2^{[8]}\left(x\right)+\frac{1}{2}\right),
&H_4^{[8]}\left(x\right)&=
H_2^{[8]}\left(x\right)+\frac{1}{2}.
\nonumber
\end{align}
The functions relevant to the gluino contribution are
\begin{align}
H_5^{[7]}\left(x\right)&=
-\frac{1}{3}H_1^{[8]}\left(x\right)
&H_5^{[8]}\left(x\right)&=
\frac{9x^2-x}{16\left(1-x\right)^4}\log x
+\frac{19x^2+40x-11}{96\left(1-x\right)^3},
\nonumber
\\
H_6^{[7]}\left(x\right)&=
-\frac{1}{3}\left(H_2^{[8]}\left(x\right)+\frac{1}{2}\right),
&H_6^{[8]}\left(x\right)&=
\frac{9x^2-x}{8\left(1-x\right)^3}\log x
+\frac{13x-5}{8\left(1-x\right)^2}.
\nonumber
\end{align}
The loop functions $I_i^{[7,8]}\left(x\right)$ and
$J_i^{[7,8]}\left(x\right)$ that appear at higher orders in the
MIA are related to the functions $H_i^{[7,8]}\left(x\right)$ via the
relations
\begin{align}
I_i^{[7,8]}\left(x\right)&=\frac{d}{d x}H_i^{[7,8]}\left(x\right),
&J_i^{[7,8]}\left(x\right)&=\frac{d^2}{d x^2} H_i^{[7,8]}\left(x\right).
\nonumber
\end{align}
%%
%%JF In the limit of degenerate masses we have for $I_i^{[7,8]}$,
%%
%%JF \begin{align}
%%JF I_1^{[7]}\left(1\right)&=-\frac{1}{30}
%%JF &I_2^{[7]}\left(1\right)&=-\frac{5}{36}
%%JF &I_3^{[7]}\left(1\right)&=\frac{1}{120}
%%JF \\
%%JF I_4^{[7]}\left(1\right)&=\frac{1}{36}
%%JF &I_5^{[7]}\left(1\right)&=\frac{1}{120}
%%JF &I_6^{[7]}\left(1\right)&=\frac{1}{36}
%%JF \\
%%JF I_1^{[8]}\left(1\right)&=-\frac{1}{40}
%%JF &I_2^{[8]}\left(1\right)&=-\frac{1}{12}
%%JF &I_3^{[8]}\left(1\right)&=-\frac{1}{40}
%%JF \\
%%JF I_4^{[8]}\left(1\right)&=-\frac{1}{12}
%%JF &I_5^{[8]}\left(1\right)&=\frac{7}{120}
%%JF &I_6^{[8]}\left(1\right)&=\frac{5}{48}
%%JF \end{align}
%%
%%JF whilst for $J_i^{[7,8]}$ we have
%%JF %%
%%JF \begin{align}
%%JF J_1^{[7]}\left(1\right)&=\frac{7}{180}
%%JF &J_2^{[7]}\left(1\right)&=\frac{2}{15}
%%JF &J_3^{[7]}\left(1\right)&=-\frac{1}{90}
%%JF \\
%%JF J_4^{[7]}\left(1\right)&=-\frac{1}{30}
%%JF &J_5^{[7]}\left(1\right)&=-\frac{1}{90}
%%JF &J_6^{[7]}\left(1\right)&=-\frac{1}{30}
%%JF \\
%%JF J_1^{[8]}\left(1\right)&=\frac{1}{30}
%%JF &J_2^{[8]}\left(1\right)&=\frac{1}{10}
%%JF &J_3^{[8]}\left(1\right)&=\frac{1}{30}
%%JF \\
%%JF J_4^{[8]}\left(1\right)&=\frac{1}{10}
%%JF &J_5^{[8]}\left(1\right)&=-\frac{11}{480}
%%JF &J_6^{[8]}\left(1\right)&=-\frac{7}{80}
%%JF \end{align}

%%%%%%%%%%%%%%%%%%%%%%%%%%%%%%%%%%%%%%%%%%%%%%%%%%%%%%%%%%%%%%%%%
\subsection{$\bsm$}
\label{LF:bsm}
%%%%%%%%%%%%%%%%%%%%%%%%%%%%%%%%%%%%%%%%%%%%%%%%%%%%%%%%%%%%%%%%%

The function that appears in~\eqref{bsm:EW:Z} is given by
\begin{align}
f_Z\left(x\right)=\frac{x}{4\left(x-1\right)^4}\log x
+\frac{\left(x^3-6x^2+3x+2\right)}{24\left(x-1\right)^4}.
\nonumber
\end{align}
%%

%%%%%%%%%%%%%%%%%%%%%%%%%%%%%%%%%%%%%%%%%%%%%%%%%%%%%%%%%%%%%%%%%
\subsection{$\bbb$}
\label{LF:bbm}
%%%%%%%%%%%%%%%%%%%%%%%%%%%%%%%%%%%%%%%%%%%%%%%%%%%%%%%%%%%%%%%%%

The functions that appear the gluino matching condition~\eqref{bbb:SU:glu} are
given by
\begin{align}
f_{\widetilde{g}}^{[1]}\left(x\right)=&\frac{x^2\left(x+3\right)}{\left(x-1\right)^5}
\log x
+\frac{(-17x^3+9x^2+9x-1)}{6\left(x-1\right)^5},
\nonumber
\\
f_{\widetilde{g}}^{[2]}\left(x\right)=&\frac{2x^2\left(x+1\right)}{\left(x-1\right)^5}
\log x
+\frac{(-x^3-9x^2+9x+1)x}{3\left(x-1\right)^5}.
\nonumber
\end{align}
%%

%%%%%%%%%%%%%%%%%%%%%%%%%%%%%%%%%%%%%%%%%%%%%%%%%%%%%%%%%%%%%%%%%
\subsection{Passarino--Veltman Functions}
\label{PVfunc}
%%%%%%%%%%%%%%%%%%%%%%%%%%%%%%%%%%%%%%%%%%%%%%%%%%%%%%%%%%%%%%%%%

The Passarino--Veltman functions that appear in appendix~\ref{RadCor} 
have the following form
\begin{align}
B_0\left(x,y\right)&=\eta-\log\frac{x}{\mu^2}+1+\frac{y}{x-y}\log\frac{y}{x},
\nonumber
\\
B_1\left(x,y\right)&=-\frac{1}{2}\eta+\frac{1}{2}\log\frac{x}{\mu^2}-\frac{1}{4}-\frac{x}{2\left(x-y\right)}+\frac{y^2-2xy}{2\left(x-y\right)^2}\log\frac{y}{x},
\nonumber
\\
C_0\left(x,y,z\right)&=\frac{y}{(x-y)(y-z)}\log\frac{y}{x}+\frac{z}{(x-z)(z-y)}\log\frac{z}{x},
\nonumber
\\
C_{00}\left(x,y,z\right)&=\frac{1}{4}\left(\eta+\frac{3}{2}-\log\frac{x}{\mu^2}+\frac{y^2}{(x-y)(y-z)}\log\frac{y}{x}+\frac{z^2}{(x-z)(z-y)}\log\frac{z}{x}\right),
\nonumber
\end{align}
where $\eta=\frac{2}{\epsilon}+\log 4\pi-\gamma_E$.

%%%%%%%%%%%%%%%%%%%%%%%%%%%%%%%%%%%%%%%%%%%%%%%%%%%%%%%%%%%%%%%%%
\section{Alternative forms for {\boldmath $\dlr$} and {\boldmath $\drl$}}
\label{dlrapp}
%%%%%%%%%%%%%%%%%%%%%%%%%%%%%%%%%%%%%%%%%%%%%%%%%%%%%%%%%%%%%%%%%

As discussed in section~\ref{GA}, throughout this analysis we assume
that the trilinear SUSY breaking terms are not proportional
to the appropriate Yukawa coupling. To illustrate how our results
are altered if we do make this assumption, let us consider a specific
example with relevance to models with SUSY breaking mediated
by either supergravity or gauge interactions~\cite{KV:dxy}
\begin{align}
m_{d,LR}^2=\tilde{A}_{d\,L}^{\dag}\mdbare{}^{\dag}+\mdbare{}^{\dag}\tilde{A}_{d\,R}^{\dag}.
\label{dlrapp:def}
\end{align}
The insertions $\dlr$ and $\drl$ are therefore equal to
\begin{align}
\left(\dlr\right)_{23}=&\frac{\left(\tilde{A}_{d\,L}^{\dag}\mdbare{}^{\dag}\right)_{23}
+\left(\mdbare{}^{\dag}\tilde{A}_{d\,R}^{\dag}\right)_{23}}
{\sqrt{\left(m_{d,LL}^{2}\right)_{22}\left(m^{2}_{d,RR}\right)_{33}}}
\nonumber\\
\left(\drl\right)_{23}=&\frac{\left(\tilde{A}_{d\,R}\mdbare\right)_{23}
+\left(\mdbare\tilde{A}_{d\,L}\right)_{23}}
{\sqrt{\left(m_{d,LL}^{2}\right)_{33}\left(m^{2}_{d,RR}\right)_{22}}}
\label{dlrapp:dxy}
\end{align}
It should be noted that, in addition to the off--diagonal elements
of the complex matrices $\tilde{A}_d^{L,R}$, there is a
contribution proportional to the appropriate off--diagonal
element of $\mdbare$. If we use the choice~\eqref{dlrapp:def}, one can
redefine the insertions $\dlr$ and $\drl$~\eqref{dlrapp:dxy} to be
independent of the bare mass matrix
in the following way (ignoring terms suppressed by the strange quark
mass)
\begin{align}
\left(\dlr\right)_{23}=&\frac{\mbphys\left(\tilde{A}_{d\,L}^{\dag}\right)_{23}}
{\sqrt{\left(m_{d,LL}^{2}\right)_{22}\left(m^{2}_{d,RR}\right)_{33}}},
& \left(\drl\right)_{23}=&\frac{\mbphys\left(\tilde{A}_{d\,R}\right)_{23}}
{\sqrt{\left(m_{d,LL}^{2}\right)_{22}\left(m^{2}_{d,RR}\right)_{33}}}.
\nonumber
\end{align}
This definition is largely independent of BLO corrections
and would correspond to a input value appropriate for the
iterative procedure described in section~\ref{num}.
The effects due to the bare mass matrix then appear in two
ways. The diagonal elements result in a factor of $1/\left(\BLOfact\right)$
that accompanies each factor of $\dlr$ and $\drl$. The off--diagonal
elements, on the other hand, may be included in a similar manner to
the off--diagonal elements of the $F$--terms that appear in the
squark mass matrix by altering $\epsilon_s$ with a small ($\cotb$
suppressed) correction. The only exception to this rule concerns
the corrected charged Higgs vertex, where one
must perform the substitution 
\begin{align}
\dlr\to\dlr+\frac{\left(\tilde{A}_{d\,L}^{\dag}\right)_{22}\left(\mdbare{}^{\dag}\right)_{23}+\left(\tilde{A}_{d\,R}^{\dag}\right)_{33}\left(\mdbare{}^{\dag}\right)_{23}}
{\sqrt{\left(m_{d,LL}^{2}\right)_{22}\left(m^{2}_{d,RR}\right)_{33}}}
\nonumber
\end{align}
in~\eqref{MIA:EW:GamL2}. In the limit of MFV ($\tilde{A}_{d\,L}=A_0$,
$\tilde{A}_{d\,R}=0$) we reproduce the result presented in~\cite{BCRS:bdec}.

%%%%%%%%%%%%%%%%%%%%%%%%%%%%%%%%%%%%%%%%%%%%%%%%%%%%%%%%%%%%%%%%%
\section{Supersymmetric Vertices}
\label{SUSYver}
%%%%%%%%%%%%%%%%%%%%%%%%%%%%%%%%%%%%%%%%%%%%%%%%%%%%%%%%%%%%%%%%%

Now let us present a complete list of supersymmetric vertices
required for our calculation. Throughout this section
$I,J=1,\dots,6$, $a,b=1,2$, $\alpha,\beta=1,\dots,4$ and
finally $i,j=1,2,3$.

The coupling of the gluino to down quarks and squarks is given by
\begin{align}
\mathcal{L}_{\widetilde{g}}=&\;\tilde{d}_J^{\dagger}\left(\bar{g}\right)\left[\left(G_{d\,L}\right)_{Ji} P_L+\left(G_{d\,R}\right)_{Ji} P_R\right]\left(d\right)_i,
\label{SUSYver:Glu}
\end{align}
where $\left(G_{d\,L}\right)_{Ji}$ and $\left(G_{d\,R}\right)_{Ji}$ are given by
\begin{align}
\left(G_{d\,L}\right)_{Ji}&=-\sqrt{2}g_s\left(\GdL\right)_{Ji},
&\left(G_{d\,R}\right)_{Ji}&=\sqrt{2}g_s\left(\GdR\right)_{Ji}
\label{SUSYver:GdX}
\end{align}
The couplings to up quarks and
squarks may be obtained via the simple substitution $d\to u$.

The chargino coupling to down quarks and up squarks was discussed
in section~\ref{GA:SU} whilst the coupling to up quarks
and down squarks is given by
\begin{align}
\left(C_{u\,L}\right)_{aJi}=&-g_2 U_{a1}\left(\GdL K^{\dag}\right)_{Ji}+\frac{g_2}{\sqrt{2}m_W\cosb}U_{a2}\left(\GdR\mdbare K^{\dag}\right)_{Ji}
\label{SUSYver:CuL}
\\
\left(C_{u\,R}\right)_{aJi}=&\frac{g_2}{\sqrt{2}m_W\sinb}V_{a2}\left(\GdL K^{\dag}\mubare{}^{\dagger}\right)_{Ji}
\label{SUSYver:CuR}
\end{align}

The neutralino couplings between down quarks and squarks are
\begin{align}
\mathcal{L}_{\chi^0}=&\;\tilde{d}_J^{\dagger}\left(\bar{\chi}^{0}\right)_{\alpha}\left[\left(N_{d\,L}\right)_{\alpha Ji} P_L+\left(N_{d\,R}\right)_{\alpha Ji} P_R\right]\left(d\right)_i,
\label{SUSYver:Neut}
\end{align}
where $\left(N_{d\,L}\right)_{Ji\alpha}$ and
$\left(N_{d\,R}\right)_{Ji\alpha}$ are given by
\begin{align}
\left(N_{d\,L}\right)_{\alpha Ji}=&-\frac{g_2}{\sqrt{2}}
\left(\frac{1}{3}N_{\alpha 1}^{\ast}\tan\theta_W-N_{\alpha 2}^{\ast}\right)\left(\GdL\right)_{Ji}
-\frac{g_2}{\sqrt{2}m_W\cosb}N_{\alpha 3}^{\ast}\left(\GdR\mdbare\right)_{Ji},
\label{SUSYver:NdL}
\\
\left(N_{d\,R}\right)_{\alpha Ji}=&-\frac{g_2\sqrt{2}\tan\theta_W}{3} N_{\alpha 1}\left(\GdR\right)_{Ji}-\frac{g_2}{\sqrt{2}m_W\cosb}N_{\alpha 3}\left(\GdL\mdbare{}^{\dagger}\right)_{Ji}.
\label{SUSYver:NdR}
\end{align}
The matrix $N$ diagonalises the neutralino mass matrix in the usual
manner
\begin{align}
N^{\ast} \mathcal{M}_{\chi^{0}} N^{\dag}={\rm diag} \left(m_{\chi^0_1},\dots,m_{\chi^0_4}\right).
\end{align}
The couplings to up quarks and squarks are
\begin{align}
\left(N_{u\,L}\right)_{\alpha Ji}=&-\frac{g_2}{\sqrt{2}}
\left(\frac{1}{3}N_{\alpha 1}^{\ast}\tan\theta_W+N_{\alpha 2}^{\ast}\right)\left(\GdL\right)_{Ji}-\frac{g_2}{\sqrt{2}m_W\sinb}N_{\alpha 4}^{\ast}\left(\GuR\mubare\right)_{Ji},
\label{SUSYver:NuL}
\\
\left(N_{u\,R}\right)_{\alpha Ji}=&\frac{2\sqrt{2}g_2\tan\theta_W}{3}N_{\alpha 1}\left(\GuR\right)_{Ji}-\frac{g_2}{\sqrt{2}m_W\sinb}N_{\alpha 4}\left(\GuL\mubare{}^{\dagger}\right)_{Ji}.
\label{SUSYver:NuR}
\end{align}

Let us now consider the couplings of the $W$ boson that appear
in~\eqref{RadCor:DelW}. The coupling to squarks is given by
\begin{align}
\left(W_{\widetilde{d}}\right)_{IJ}=-\frac{g_2}{\sqrt{2}}
\left(\GuL K\GdL^{\dag}\right)_{IJ}.
\label{SUSYver:Wsq}
\end{align}
The couplings to neutralinos and charginos are given by
\begin{align}
\left(W_{\chi\,L}\right)_{a\alpha}&=
g_2\left(N_{\alpha 2}^{\ast}V_{a1}-\frac{1}{\sqrt{2}}N_{\alpha 4}^{\ast}V_{a2}\right)
&\left(W_{\chi\,R}\right)_{a\alpha}&=
g_2\left(N_{\alpha 2}U_{a1}+\frac{1}{\sqrt{2}}N_{\alpha 3}U_{a2}\right)
\label{SUSYver:Wchne}
\end{align}

The couplings of the $Z$ boson to up and down squarks are given by
\begin{align}
\left(Z_{\widetilde{d}}\right)_{IJ}&=\frac{g_2}{2\cos\theta_W}\left(\GdL\GdL^{\dag}-\frac{2}{3}\sin^2\theta_W\right)_{IJ},
\label{SUSYver:Zdsq}
\\
\left(Z_{\widetilde{u}}\right)_{IJ}&=-\frac{g_2}{2\cos\theta_W}\left(\GuL\GuL^{\dag}-\frac{4}{3}\sin^2\theta_W\right)_{IJ}.
\label{SUSYver:Zusq}
\end{align}
The couplings to neutralinos and charginos on the other hand are
\begin{align}
\left(Z_{\chi^-\,L}\right)_{ab}&=
\frac{g_2}{2\cos\theta_W}\left[U_{a1} U_{b1}^{\ast}+\left(\cos^2\theta_W-\sin^2\theta_W\right)\delta_{ab}\right],
\label{SUSYver:ZchL}
\\
\left(Z_{\chi^-\,R}\right)_{ab}&=
\frac{g_2}{2\cos\theta_W}\left[V_{a1}V_{b1}^{\ast}+\left(\cos^2\theta_W-\sin^2\theta_W\right)\delta_{ab}\right],
\label{SUSYver:ZchR}
\\
\left(Z_{\chi^0\,L}\right)_{\alpha\beta}&=
\frac{g_2}{2\cos\theta_W}\left(N_{\alpha 4}N_{\beta 4}^{\ast}-N_{\alpha 3}N_{\beta 3}^{\ast}\right),
\label{SUSYver:ZneL}
\\
\left(Z_{\chi^0\,R}\right)_{\alpha\beta}&=
-\frac{g_2}{2\cos\theta_W}\left(N_{\alpha 4}^{\ast}N_{\beta 4}-N_{\alpha 3}^{\ast}N_{\beta 3}\right).
\label{SUSYver:ZneR}
\end{align}

The coupling of the charged Higgs boson to squarks is given by
\begin{align}
\left(C_{\widetilde{d}}^{S^+}\right)_{IJ}=&
-\frac{g_2 m_W}{\sqrt{2}\sin\theta_W}\left(\cosb y_{(2)}^{S^+}+\sinb y_{(1)}^{S^+}\right)\left(\GuL K\GdL^{\dag}\right)_{IJ}
\nonumber\\
&+\frac{g_2}{\sqrt{2}m_W}\left[\frac{y_{(2)}^{S^+}}{\cosb}\left(\GuL K\mdbare{}^{\dag}\mdbare\GdL^{\dag}\right)_{IJ}
+\frac{y_{(1)}^{S^+}}{\sinb}\left(\GuL\mubare{}^{\dag}\mubare K\GdL^{\dag}\right)_{IJ}\right]
\nonumber\\
&+\frac{g_2\eta_{S^+}}{\sqrt{2}\cosb\sinb}\left(\GuR\mubare K\mdbare{}^{\dag}\GdR^{\dag}\right)_{IJ}
\nonumber\\
&+y_{(2)}^{S^+}\frac{g_2}{\sqrt{2}m_W\cosb}
\left[\left(\GuL K m_{d,LR}\GdR^{\dag}\right)_{IJ}+\mu^{\ast}\cotb\left(\GuR\mubare K\GdL^{\dag}\right)_{IJ}\right]
\nonumber\\
&+y_{(1)}^{S^+}\frac{g_2}{\sqrt{2}m_W\sinb}
\left[\left(\GuR m_{u,RL} K\GdL^{\dag}\right)_{IJ}+\mu\tanb\left(\GuL K\mdbare{}^{\dag}\GdR^{\dag}\right)_{IJ}\right],
\label{SUSYver:Chsq}
\end{align}
where $\eta_{S^+}=1,0$
The coupling of the charged Higgs to the chargino and neutralinos is
given by
\begin{align}
\left(C_{\chi\,L}^{S^+}\right)_{a\alpha}=&
\frac{g_2}{\cos\theta_W}y_{(2)}^{S^+}
\left[\frac{1}{\sqrt{2}}U_{a2}\left(N_{\alpha 1}^{\ast}\sin\theta_W+N_{\alpha 2}^{\ast}\cos\theta_W\right)-U_{a1}N_{\alpha 3}^{\ast}\cos\theta_W\right],
\label{SUSYver:ChchneL}
\\
\left(C_{\chi\,R}^{S^+}\right)_{a\alpha}=&
-\frac{g_2}{\cos\theta_W}y_{(1)}^{S^+}
\left[\frac{1}{\sqrt{2}}V_{a2}\left(N_{\alpha 1}\sin\theta_W+N_{\alpha 2}\cos\theta_W\right)+V_{a1}N_{\alpha 4}\cos\theta_W\right].
\label{SUSYver:ChchneR}
\end{align}

The couplings of the neutral Higgs bosons to squarks are
\begin{align}
\left(S_{\widetilde{d}}^{S^0}\right)_{IJ}=&
\eta_{S^0}
\bigg\{
\frac{g_2}{3}\tan^2\theta_W m_W\left(\cosb x_{(1)}^{S^0}-\sinb x_{(2)}^{S^0}\right)\left(1+\frac{3-4\sin^2\theta_W}{2\sin^2\theta_W}\GdL\GdL^{\dag}\right)_{IJ}
\nonumber\\
&-\frac{g_2}{m_W\cosb}x_{(1)}^{S^0}\left[\left(\GdL\mdbare{}^{\dag}\mdbare\GdL^{\dag}\right)_{IJ}+\left(\GdR\mdbare\mdbare{}^{\dag}\GdR^{\dag}\right)_{IJ}\right]
\bigg\}
\nonumber\\
&-\frac{g_2}{2 m_W\cosb}
\bigg[
\left(\GdR m_{d,RL}\GdL^{\dag}\right)_{IJ} x_{(1)}^{S^0}+
\left(\GdL m_{d,LR}\GdR^{\dag}\right)_{IJ} x_{(1)}^{S^0\ast}
\nonumber\\
&-\mu\left(\GdL\mdbare{}^{\dag}\GdR^{\dag}\right)_{IJ} x_{(2)}^{S^0}
-\mu^{\ast}\left(\GdR\mdbare\GdL^{\dag}\right)_{IJ} x_{(2)}^{S^0\ast}
\bigg],
\label{SUSYver:Nedsq}
\end{align}
\begin{align}
\left(S_{\widetilde{u}}^{S^0}\right)_{IJ}=&
\eta_{S^0}
\bigg\{
-\frac{2g_2}{3}\tan^2\theta_W m_W\left(\cosb x_{(1)}^{S^0}-\sinb x_{(2)}^{S^0}\right)\left(1+\frac{3-8\sin^2\theta_W}{4\sin^2\theta_W}\GuL\GuL^{\dag}\right)_{IJ}
\nonumber\\
&-\frac{g_2}{m_W\sinb}x_{(2)}^{S^0}\left[\left(\GuL\mubare{}^{\dag}\mubare\GuL^{\dag}\right)_{IJ}+\left(\GuR\mubare\mubare{}^{\dag}\GuR^{\dag}\right)_{IJ}\right]
\bigg\}
\nonumber\\
&-\frac{g_2}{2 m_W\sinb}
\bigg[
\left(\GuR m_{u,RL}\GuL^{\dag}\right)_{IJ} x_{(2)}^{S^0}+
\left(\GuL m_{u,LR}\GuR^{\dag}\right)_{IJ} x_{(2)}^{S^0\ast}
\nonumber\\
&-\mu\left(\GuL\mubare{}^{\dag}\GuR^{\dag}\right)_{IJ} x_{(1)}^{S^0}
-\mu^{\ast}\left(\GuR\mubare\GuL^{\dag}\right)_{IJ} x_{(1)}^{S^0\ast}
\bigg],
\label{SUSYver:Neusq}
\end{align}
where $\eta_{S^0}=1,1,0,0$ and $x_{(2)}^{S^0}=\sin\alpha,\cos\alpha,i\cosb,i\sinb$

The couplings of the bosons to charginos and neutralinos are

\begin{align}
\left(S_{\chi^- L}\right)_{ab}=&-\frac{g_2}{\sqrt{2}}\left(
x_{(1)}^{S^0\ast} U_{a2} V_{b1}^{\ast}+x_{(2)}^{S^0\ast} U_{a1} V_{b2}^{\ast}\right),
\label{SUSYver:NechL}
\\
\left(S_{\chi^- R}\right)_{ab}=&-\frac{g_2}{\sqrt{2}}\left(
x_{(1)}^{S^0} U_{a2}^{\ast} V_{b1}+x_{(2)}^{S^0} U_{a1}^{\ast} V_{b2}\right),
\label{SUSYver:NechR}
\end{align}

\begin{align}
\left(S_{\chi^0\,L}\right)_{\alpha\beta}=&
\frac{g_2}{2\cos\theta_W}
\bigg[
\left(x_{(1)}^{S^0\ast} N_{\alpha 3}^{\ast}-x_{(2)}^{S^0\ast} N_{\alpha 4}^{\ast}\right)
\left(N_{\beta 1}^{\ast}\sin\theta_W-N_{\beta 2}^{\ast}\cos\theta_W\right)
\nonumber\\
&+\left(x_{(1)}^{S^0\ast} N_{\beta 3}^{\ast}-x_{(2)}^{S^0\ast} N_{\beta 4}^{\ast}\right)
\left(N_{\alpha 1}^{\ast}\sin\theta_W-N_{\alpha 2}^{\ast}\cos\theta_W\right)
\bigg],
\label{SUSYver:NeneL}
\\
\left(S_{\chi^0\,R}\right)_{\alpha\beta}=&
\frac{g_2}{2\cos\theta_W}
\bigg[
\left(x_{(1)}^{S^0} N_{\alpha 3}-x_{(2)}^{S^0} N_{\alpha 4}\right)
\left(N_{\beta 1}\sin\theta_W-N_{\beta 2}\cos\theta_W\right)
\nonumber\\
&+\left(x_{(1)}^{S^0} N_{\beta 3}-x_{(2)}^{S^0} N_{\beta 4}\right)
\left(N_{\alpha 1}\sin\theta_W-N_{\alpha 2}\cos\theta_W\right)
\bigg].
\label{SUSYver:NeneR}
\end{align}

%%%%%%%%%%%%%%%%%%%%%%%%%%%%%%%%%%%%%%%%%%%%%%%%%%%%%%%%%%%%%%%%%
\section{Corrected Vertices}
\label{RadCor}
%%%%%%%%%%%%%%%%%%%%%%%%%%%%%%%%%%%%%%%%%%%%%%%%%%%%%%%%%%%%%%%%%

Let us now present the full analytic expressions used in our
numerical analysis. A number of the corrections given in this
section are divergent and should be renormalized appropriately.
However, we have checked that numerically these terms are
negligible compared to the $\tanb$ enhanced corrections
discussed in~\ref{MIA} and tend to play only a minor r\^ole.
Throughout this section $i,j=1,2,3$,
$I,J=1,\dots,6$, $a,b=1,2$, $\alpha,\beta=1,\dots,4$,
$S^+=H^+,G^+$ and $S^0=H^0,h^0,A^0,G^0$. Repeated
indices that appear in the expressions below should be summed over.
We have checked are results with those that appear in~\cite{BCRS:bdec}
and find that, once one takes into account the different form
of Passarino--Veltman functions we adopt our results agree.

The self--energies that appear in the corrected
vertices presented in section~\ref{GA:EW} as well as the expression
for $\delta m_d$ are given by
\begin{align}
\left(\Sigma^{d}_{m\,L}\right)_{ij}=
&-\frac{1}{16\pi^2}
\bigg[
\mgl \left(G_{d\,R}^{\ast}\right)_{Ii} \left(G_{d\,L}\right)_{Ij} C_2\left(3\right)
B_0\left(\mgl^2,m_{\widetilde{d}_I}^2\right)
\nonumber\\
&+
m_{\chi^-_{a}} \left(C_{d\,R}^{\ast}\right)_{aIi} \left(C_{d\,L}\right)_{aIj}
B_0\left(m_{\chi^-_{a}}^2,m_{\widetilde{u}_I}^2\right)
\nonumber\\
&+
m_{\chi^0_{\alpha}} \left(N_{d\,R}^{\ast}\right)_{\alpha Ii}\left(N_{d\,L}\right)_{\alpha Ij}
B_0\left(m_{\chi^0_{\alpha}}^2,m_{\widetilde{d}_I}^2\right)
\bigg],
\label{RadCor:SdmL}
\\
\left(\Sigma^{d}_{v\,L}\right)_{ij}=
&\frac{1}{16\pi^2}
\bigg[
\left(G_{d\,L}^{\ast}\right)_{Ii}\left(G_{d\,L}\right)_{Ij}C_2\left(3\right)
B_1\left(\mgl^2,m_{\widetilde{d}_I}^2\right)
\nonumber\\
&+
\left(C_{d\,L}^{\ast}\right)_{aIi}\left(C_{d\,L}\right)_{aIj}
B_1\left(m_{\chi^-_{a}}^2,m_{\widetilde{u}_I}^2\right)
\nonumber\\
&+
\left(N_{d\,L}^{\ast}\right)_{\alpha Ii}\left(N_{d\,L}\right)_{\alpha Ij}
B_1\left(m_{\chi^0_{\alpha}}^2,m_{\widetilde{d}_I}^2\right)
\bigg].
\label{RadCor:SdvL}
\end{align}
The various supersymmetric couplings that appear in the above
expressions are given in appendix~\ref{SUSYver} whilst the
Passarino--Veltman functions are defined in appendix~\ref{PVfunc}.
$\Sigma^{d}_{v\,R}$ can be obtained by substituting $L$ with $R$
in~\eqref{RadCor:SdvL} the up quark self--energy corrections
can be obtained by substituting $u\leftrightarrow d$.

The correction $\Delta C_L^W$ that appears in~\eqref{GA:EW:KM}
is given by
\begin{align}
\left(\Delta C_L^W\right)_{ij}=&\frac{1}{16\pi^2}
\bigg\{
2 C_2\left(3\right) \left(W_{\widetilde{d}}\right)_{IJ}
\left(G_{u\,L}^{\ast}\right)_{Ii}\left(G_{d\,L}\right)_{Jj}
C_{00}\left(\mgl^2,m_{\widetilde{d}_J}^2,m_{\widetilde{u}_I}^2\right)
\nonumber\\
&+
2 \left(W_{\widetilde{d}}\right)_{IJ}
\left(N_{u\,L}^{\ast}\right)_{\alpha Ii} \left(N_{d\,L}\right)_{\alpha Jj}
C_{00}\left(m_{\chi^0_{\alpha}}^2,m_{\widetilde{d}_J}^2,m_{\widetilde{u}_I}^2\right)
\nonumber\\
&+
m_{\chi^0_{\alpha}}m_{\chi^-_{a}}
\left(W_{\chi\,R}\right)_{a\alpha}
\left(N_{u\,L}^{\ast}\right)_{\alpha Ii}\left(C_{d\,L}\right)_{aIj}
C_{0}\left(m_{\chi^0_{\alpha}}^2,m_{\chi^-_{a}}^2,
m_{\widetilde{u}_I}^2\right)
\nonumber\\
&-
m_{\chi^0_{\alpha}}m_{\chi^-_{a}}
\left(W_{\chi\,L}\right)_{a\alpha}
\left(C_{u\,L}^{\ast}\right)_{aJi}\left(N_{d\,L}\right)_{\alpha Jj}
C_{0}\left(m_{\chi^0_{\alpha}}^2,m_{\chi^-_{a}}^2,
m_{\widetilde{d}_J}^2\right)
\nonumber\\
&-
2 \left(W_{\chi\,L}\right)_{a\alpha}
\left(N_{u\,L}^{\ast}\right)_{\alpha Ii}\left(C_{d\,L}\right)_{aIj}
\left[C_{00}\left(m_{\chi^0_{\alpha}}^2,m_{\chi^-_{a}}^2,m_{\widetilde{u}_I}^2\right)-\frac{1}{4}\right]
\nonumber\\
&+
2 \left(W_{\chi\,R}\right)_{a\alpha}
\left(C_{u\,L}^{\ast}\right)_{aJi}\left(N_{d\,L}\right)_{\alpha Jj}
\left[C_{00}\left(m_{\chi^0_{\alpha}}^2,m_{\chi^-_{a}}^2,m_{\widetilde{d}_J}^2\right)-\frac{1}{4}\right]
\bigg\}.
\label{RadCor:DelW}
\end{align}
$\Delta C_R^W$ can be obtained by the simple substitution $L \leftrightarrow R$.
The correction to the left handed coupling of the
$Z$--boson~\eqref{GA:EW:ZVer} is given by
\begin{align}
\left(\Delta C_L^Z\right)_{ij}=&\frac{1}{16\pi^2}
\bigg\{
2 C_2\left(3\right) \left(Z_{\widetilde{d}}\right)_{IJ}
\left(G_{d\,L}^{\ast}\right)_{Ii}\left(G_{d\,L}\right)_{Jj}
C_{00}\left(\mgl^2,m_{\widetilde{d}_I}^2,m_{\widetilde{d}_J}^2\right)
\nonumber\\
&+
2 \left(Z_{\widetilde{d}}\right)_{IJ}
\left(N_{d\,L}^{\ast}\right)_{\alpha Ii}\left(N_{d\,L}\right)_{\alpha Jj}
C_{00}\left(m_{\chi^0_{\alpha}}^2,m_{\widetilde{d}_I}^2,m_{\widetilde{d}_J}^2\right)
\nonumber\\
&+
2 \left(Z_{\widetilde{u}}\right)_{IJ}
\left(C_{d\,L}^{\ast}\right)_{aIi}\left(C_{d\,L}\right)_{aJj}
C_{00}\left(m_{\chi^-_{a}}^2,m_{\widetilde{u}_I}^2,m_{\widetilde{u}_J}^2\right)
\nonumber\\
&-
m_{\chi^-_{a}}m_{\chi^-_{b}}
\left(Z_{\chi^-\,L}\right)_{ab}
\left(C_{d\,L}^{\ast}\right)_{bIi}\left(C_{d\,L}\right)_{aIj}
C_{0}\left(m_{\chi^-_{a}}^2,m_{\chi^-_{b}}^2,
m_{\widetilde{u}_I}^2\right)
\nonumber\\
&-
m_{\chi^0_{\alpha}}m_{\chi^0_{\beta}}
\left(Z_{\chi^0\,L}\right)_{\alpha\beta}
\left(N_{d\,L}^{\ast}\right)_{\beta Ji}\left(N_{d\,L}\right)_{\alpha Jj}
C_{0}\left(m_{\chi^0_{\alpha}}^2,m_{\chi^0_{\beta}}^2,
m_{\widetilde{d}_J}^2\right)
\nonumber\\
&+
2 \left(Z_{\chi^-\,R}\right)_{ab}
\left(C_{d\,L}^{\ast}\right)_{bIi}\left(C_{d\,L}\right)_{aIj}
\left[C_{00}\left(m_{\chi^-_{a}}^2,m_{\chi^-_{b}}^2,m_{\widetilde{u}_I}^2\right)-\frac{1}{4}\right]
\nonumber\\
&+
2 \left(Z_{\chi^0\,R}\right)_{\alpha\beta}
\left(N_{d\,L}^{\ast}\right)_{\beta Ji} \left(N_{d\,L}\right)_{\alpha Jj}
\left[C_{00}\left(m_{\chi^0_{\alpha}}^2,m_{\chi^0_{\beta}}^2,m_{\widetilde{d}_J}^2\right)-\frac{1}{4}\right]
\bigg\}.
\label{RadCor:DelZ}
\end{align}
$\Delta C_R^Z$ may be obtained in a similar manner to $\Delta C_R^W$.

Turning to the Higgs sector the vertex correction $\Delta C_L^{S^+}$
is given by
\begin{align}
\left(\Delta C_L^{S^+}\right)_{ij}=&-\frac{1}{16\pi^2}
\bigg\{
C_2\left(3\right)\mgl
\left(C_{\widetilde{d}}^{S^+}\right)_{IJ}
\left(G_{u\,R}^{\ast}\right)_{Ii}\left(G_{d\,L}\right)_{Jj}
C_{0}\left(\mgl^2,m_{\widetilde{d}_J}^2,m_{\widetilde{u}_I}^2\right)
\nonumber\\
&+
m_{\chi^0_{\alpha}}
\left(C_{\widetilde{d}}^{S^+}\right)_{IJ}
\left(N_{u\,R}^{\ast}\right)_{\alpha Ii}\left(N_{d\,L}\right)_{\alpha Jj}
C_{0}\left(m_{\chi^0_{\alpha}}^2,m_{\widetilde{d}_J}^2,m_{\widetilde{u}_I}^2\right)
\nonumber\\
&+
m_{\chi^0_{\alpha}} m_{\chi^-_{a}}
\left(C_{\chi L}^{S^+}\right)_{a\alpha}
\left(N_{u\,R}^{\ast}\right)_{\alpha Ii}\left(C_{d\,L}\right)_{aIj}
C_{0}\left(m_{\chi^0_{\alpha}}^2,m_{\chi^-_{a}}^2,
m_{\widetilde{u}_I}^2\right)
\nonumber\\
&+
m_{\chi^-_{a}} m_{\chi^0_{\alpha}}
\left(C_{\chi L}^{S^+}\right)_{a\alpha}
\left(C_{u\,R}^{\ast}\right)_{aJi}\left(N_{d\,L}\right)_{\alpha Jj}
C_{0}\left(m_{\chi^-_{a}}^2,m_{\chi^0_{\alpha}}^2,
m_{\widetilde{d}_J}^2\right)
\nonumber\\
&+
4 \left(C_{\chi R}^{S^+}\right)_{a\alpha}
\left(N_{u\,R}^{\ast}\right)_{\alpha Ii}\left(C_{d\,L}\right)_{aIj}
\left[C_{00}\left(m_{\chi^0_{\alpha}}^2,m_{\chi^-_{a}}^2,
m_{\widetilde{u}_I}^2\right)-\frac{1}{8}\right]
\nonumber\\
&+
4 \left(C_{\chi R}^{S^+}\right)_{a\alpha}
\left(C_{u\,R}^{\ast}\right)_{aJi}\left(N_{d\,L}\right)_{\alpha Jj}
\left[C_{00}\left(m_{\chi^-_{a}}^2,m_{\chi^0_{\alpha}}^2,
m_{\widetilde{d}_J}^2\right)-\frac{1}{8}\right]
\bigg\}.
\label{RadCor:DelC}
\end{align}
$\Delta C_R^{S^+}$ may be obtained via the substitution $L\leftrightarrow R$.
Finally the corrections to the neutral Higgs vertex may be written
\begin{align}
\left(\Delta C_L^{S^0}\right)_{ij}=&-\frac{1}{16\pi^2}
\bigg\{
C_2\left(3\right)\mgl
\left(S_{\widetilde{d}}^{S^0}\right)_{IJ}
\left(G_{d\,R}^{\ast}\right)_{Ii}\left(G_{d\,L}\right)_{Jj}
C_{0}\left(\mgl^2,m_{\widetilde{d}_I}^2,m_{\widetilde{d}_J}^2\right)
\nonumber\\
&+
m_{\chi^0_{\alpha}}
\left(S_{\widetilde{d}}^{S^0}\right)_{IJ}
\left(N_{d\,R}^{\ast}\right)_{\alpha Ii}\left(N_{d\,L}\right)_{\alpha Jj}
C_{0}\left(m_{\chi^0_{\alpha}}^2,m_{\widetilde{d}_I}^2,m_{\widetilde{d}_J}^2\right)
\nonumber\\
&+
m_{\chi^-_{a}}
\left(S_{\widetilde{u}}^{S^0}\right)_{IJ}
\left(C_{d\,R}^{\ast}\right)_{aIi}\left(C_{d\,L}\right)_{aJj}
C_{0}\left(m_{\chi^-_{a}}^2,m_{\widetilde{u}_I}^2,m_{\widetilde{u}_J}^2\right)
\nonumber\\
&+
m_{\chi^0_{\alpha}} m_{\chi^0_{\beta}}
\left(S_{\chi^0\,L}^{S^0}\right)_{\alpha\beta}
\left(N_{d\,R}^{\ast}\right)_{\beta Ji}\left(N_{d\,L}\right)_{\alpha Jj}
C_{0}\left(m_{\chi^0_{\alpha}}^2,m_{\chi^0_{\beta}}^2,
m_{\widetilde{d}_J}^2\right)
\nonumber\\
&+
m_{\chi^-_{a}} m_{\chi^-_{b}}
\left(S_{\chi^-\,L}^{S^0}\right)_{ab}
\left(C_{d\,R}^{\ast}\right)_{bIi}\left(C_{d\,L}\right)_{aIj}
C_{0}\left(m_{\chi^-_{a}}^2,m_{\chi^-_{b}}^2,
m_{\widetilde{u}_I}^2\right)
\nonumber\\
&+
4 \left(S_{\chi^0\,R}^{S^0}\right)_{\alpha\beta}
\left(N_{d\,R}^{\ast}\right)_{\beta Ji}\left(N_{d\,L}\right)_{\alpha Jj}
\left[C_{00}\left(m_{\chi^0_{\alpha}}^2,m_{\chi^0_{\beta}}^2,
m_{\widetilde{d}_J}^2\right)-\frac{1}{8}\right]
\nonumber\\
&+
4 \left(S_{\chi^-\,R}^{S^0}\right)_{ab}
\left(C_{d\,R}^{\ast}\right)_{bIi}\left(C_{d\,L}\right)_{aIj}
\left[C_{00}\left(m_{\chi^-_{a}}^2,m_{\chi^-_{b}}^2,
m_{\widetilde{u}_I}^2\right)-\frac{1}{8}\right]
\bigg\}.
\label{RadCor:DelS}
\end{align}
The correction to the right--handed vertex may be obtained in a similar
manner to $\Delta C_R^{S^+}$.
\end{appendix}

\end{document}